# The Information in Emotion Communication

Alison Duncan Kerr

Kevin Scharp

# Contents

# List of Figures

# Acknowledgements

For Nola, Pia, and Jasper. Immense joy.

Thank you to our families for their unwavering support.

And, to each other.

# Introduction

It hardly needs to be stressed how important emotions are to human life and to the lives of so many animals around us. Humanity has had a complicated past with its emotions, but the previous century of inquiry has seen remarkable scientific understanding of what emotions are, how they work, and what they do for us. A huge amount of that work has been on emotions as episodes that happen to individuals. So much has been learned on this front, and we applaud this research. We, however, follow a growing trend of looking at how emotions function in groups of animals as a system of communication. This too has been known for a long time, but the novel aspect of our approach is that we provide a *quantitative* theory of how information is transferred in groups of animals by virtue of their emotions and emotional capacities, like the ability to detect emotion behavior (e.g., fighting something), the ability to detect emotions on the basis of detecting behavior (e.g., fear) and the ability to express emotions in behavior. The theory presented here provides answers to questions about how much information is transferred from one point to another, or through a node in a network, or through an entire layer of a complex network. Although the theory presented here is quantitative, the mathematics involved is not complicated, and we work out many examples along the way. Most chapters end with a sequence of points for further research and a list of references.

Being able to put numbers to the information passed in emotion communication opens up entire new vistas of research directions. Evolutionary theory has long been connected to various quantitative measures of information in animal communication systems, so this theory ought to mesh well with contemporary evolutionary theory on emotions.[1] It supplants the "contagion" theory of emotion transmission, and it subsumes areas of active research like emotion elicitation, emotion expression, and interpersonal emotion regulation.

Even more important is that a quantitative theory of emotional information is badly needed to combat the vast emotion manipulation occurring online in social media. We know about many of these kinds of manipulations, and the information theory of emotion communication provides one with an array of technological and quantitative tools to detect and combat large-scale emotion manipulation on social media.

Finally, there is a wave of artificial emotion recognition systems just offshore and it will hit us in the next three to five years. The quantitative theory offered here seamlessly applies to artificial emotion recognition systems, and we devote half a chapter to discussing how to understand them in information-theoretic terms.

In sum, the book gives one a powerful new theory that can be applied to a huge swath of phenomena. In addition, the reader gets a front-row seat for the new scientific and philosophical work on emotions, elegant mathematical results, and futuristic technological advances, like designing social media networks with built in emotion communication security devices.

## 0.1 Emotions

Emotions involve many things – feelings, bodily sensations, action tendencies, appraisals, and social norms. *Feelings* are an obvious component. You know exactly what it feels like to be sad or angry or frustrated. This feeling is something in your mind that you have intimate access to nas is otherwise private in that no one else can feel it directly. Many emotions involve specific *bodily sensations* as well, like feeling hot when angry or a tight stomach when worried. Most emotional states make us likely to take certain causes of action or specific behaviors, called *action tendencies*. Fear promotes running away or fighting or hiding or showing submission. Disgust promotes avoidance or extreme care in approach. Many studies also suggest that being in certain emotional states brings with it certain tendencies of thought as well. For example, being … . *Appraisals* are specific cognitive evaluations of how some emotionally relevant aspect of one's situation (e.g., a charging lion) bear on one's goals and interests. Almost every emotion is bound up with a complex set of *social norms* that dictate when, where, and how much to show certain emotions depending on one's social class, race, nationality, gender, sexual orientation, etc.

[1] See works in Altenmueller et al (2013) for a survey specifically related to communication.

The range of theories of emotion reflect this diversity.[2] In this book, we focus only on the *communication* aspect of emotions. We take no stand on the crucial issue of *what emotions are* (69). Instead, we think that whatever emotions are, they are part of emotion signal systems in which quantities of information are transferred through networks.

Throughout the book, we use the following terms. An *affective agent* is an agent with emotional capacities. *Emotional capacities* are those sufficiently similar to things like the ability to *feel* anger, or sadness, or fear. They are assumed to include abilities to *regulate* emotions and to *detect* emotions as well.

## 0.2 Emotion Communication

The idea that emotions serve as communication systems has steadily gained traction to the point that it now sustains several thriving research programs. One that stands out is the *contagion theory of emotion*, which aims to explain the way emotions spread through populations, like fear through crowd.[3] Several quantitative models of emotion contagion have appeared, and these are mostly adapted from thermodynamics[4] or epidemiology.[5] Some criticisms of the contagion view have appeared recently as well.[6] An alternative view is that emotional capacities should be thought of as signal systems.[7] Instead of spreading pathogens, emotions can serve as signals from one entity to another.

[2] See Scarintino (2016) and the other papers in Lewis et al (2016) for a survey; see also Section 1.3 in Chapter 1 for an overview.

[3] Hatfield et al (1994), Barsade (2002), Hill et al (2010), House (2011), Tsai et al (2011), (13) Rueff-Lopes & Caetano (2012), Yin et al (2012), Fu et al (2014), Elfenbein (2014), Hatfield et al (2014), Bertozzi et al (2015), Hess & Fischer (2016), and the papers in Lehmann & Ahn (2018).

[4] Bertozzi et al (2015)

[5] Hill et al (2010) and Fu et al (2014).

Emotions, traditionally understood, are internal or first-person phenomena, and many theorists still focus on this aspect of them. However, there has emerged over the last several decades, myriad studies on the role emotions play in communication among groups.[8] The signal view treats the animal's having emotions as trnasmitters of emotion signals and animals detecting emotions in others as receivers of emotion signals. From this idea (developed in Chapter One) it is a short step to thinking of emotion messages as originating in one animal, being encoded in to that animal's behavior, that behavior being detected by another, and then decoded to arrive at the emotion message received (described in Chapter Two). And finally we can think of many transmitters and receivers all interacting in complex ways through a network of emotion communication (in Chapters Four and Five).

Although we use the plural phrase 'emotion communication systems', it is essential to keep in mind that any animal that has emotional capacities (i.e., to feel, express, and detect emotions) is part of one big emotion communication system.[9] It should be obvious that dogs and humans, for example, exchange emotion information, but imagine that tomorrow scientists discover emotions in shrimp and not long after, one could purchase shrimp along with artificial emotion detectors for them. One could certainly imagine getting startled if the shrimp emotion detector announced fear out of the blue with no obvious cause. The same goes for emotions in robots or disembodied souls, were such things to receive serious empirical support. Hence, it makes sense to think that anything that has emotions can communicate in however a rudimentary way with any other emotional being. Nevertheless, it often makes sense to restrict one's attention to a certain subclasses of such things, and then the phrase in question makes sense as well.

---

[6] Aguilar (2013), Alshamsi et al (2015), Bacaksizlar (2019, and Wrobel & Imbir (2019).
[7] Schwartz & Clore (1988), Ross & Doumachel (2004), App et al (2011), Kappas (2013), Hareli & Hess (2013), Banziger et al (2015), and van Kleef (2016).
[8] Buck (1984), Oatley & Johnson-Laird (1996), Planalp (1998), Planalp (1999), Altenmuller et al (2013), Jack & Schyns (2015), and van Kleef et al (2016).
[9] Griffiths (1997).

The phenomenon of emotion communication across individuals and species is the overarching subject, which includes all the phenomena related to how emotions occur, how they are communicated to others, and the effects of that communication. The information theory of emotion communication incorporates large areas of empirical work like that on emotion perception in animals[10] and emotion recognition by artificially intelligent algorithms across all modalities[11] It provides a framework for understanding emotion behavior[12] as the encoding of felt emotions for communication through a channel. It engages the literature on emotion elicitation and appraisal[13] by explaining it as information communicated from a situation in the world (as with direct elicitation) or information communicated from another animal (as with indirect elicitation), where this information is first detected, and then passed on (with some probability) to actually feeling the emotion. There are myriad ties to game theory[14] and statistics as well.[15]

## 0.3 Information

What is the entropy of fear? Or the channel capacity of anger? What is the transmission rate of sadness? Or the redundancy of joy? These questions might strike you as nonsensical, but in just a moment they will sound perfectly normal, and you will learn to answer some of them.

The term 'information' has been interpreted in many ways, but the mathematical theory of information—proposed by Claude Shannon in the mid-20th century and hugely influential—is based on probability.[16] Anywhere there are probability distributions, there is information in this sense. As such, the core postulates of the information theory of emotion communication are just that there are probabilities of being in a certain emotional states (individual, joint, and higher arity) and the probabilities of error in transmission of an emotion message. Transmitting an emotion message occurs when: (i) one has an emotion, and (ii) another animal with emotional capacities detects the emotion in question. With that basis, one has all the material to calculate any of the information theoretic quantities discussed below, to derive any of the hypotheses we propose, and to imagine experiments to test them.

---

[10] Schlegel et al (2012) and Shuman et al (2015).
[11] Scherer (2013), Konar & Chakraborty (2015), Burlson (2017), Cambria et al (2017), Pozzi et al (2017), Poria et al (2018), Ko (2018), Zhao et al (2018), and Shu et al (2018).
[12] Abell & Smith (2016), Elder (2017), Hommel et al (2017), and Scarintino (2017).
[13] Scherer et al (2001), Coan & Allen (2007), and Sander et al (2018).
[14] Lewis (1969)
[15] Skyrms (2010) and Boden (2018), respectively.

The information theory of emotion communication, presented here, takes the signal theory for granted. Information theory is not compatible with the contagion theory, but can explain everything the contagion theory can explain; thus, information theory subsumes the contagion view and the signal view in different ways. Some social psychologists seem to already understand emotion communication in information theoretic terms by using explicitly information-theoretic terminology like 'channel', 'encoding', and 'decoding' in describing emotion communication.[17] However, using the terms without the theory offers no explanatory power, and no proper quantitative theory has appeared. One reason is that the formal and mathematical details are non-trivial, but it is only by seeing these that one can make the predictions and explanations adumbrated below. Shannon's information theory has enjoyed wide application across many disciplines and has a long history of explanatory and predictive successes, but it has not been applied to emotions before now.[18]

## 0.4 Social Media

One obvious application for the information theory of emotions is to social networks on the internet. Social media is so dominant in societies all over the world that humanity is just beginning to understand its effects. One major result is that engaging in social media activates all sorts of emotional capacities in humans, and this should not come as a surprise. Sharing emotions is probably *the* primary activity on social media. It is, after all, what gave us the emoticon (short for 'emotion icon') and the emoji.[19]

[16] Shannon (1948).
[17] Shannon (1948), Hankerson et al (2003), Cover & Thomas (2006).
[18] Pfeifer (2006), Nemenman (2011), Adami (2012), and Lord et al (2016).

We can think of internet groups like Facebook and Twitter as complex emotional signaling systems, and we can apply the insights of information theory to them as well. Using the mathematical models introduced here, one can understand posting on social media as a kind of emotion expression, which we think of as transmitting an emotion message. Likewise, reading, watching, or listening to others on social media is a form of emotion detection, which we think of as receiving emotion messages.

Evidence of widespread abuse on social media by manipulation of emotions and emotion communication is readily available. For example, Cambridge Analytica manipulated the emotions of many in the British public to influence the "Brexit" vote to leave the European Union. And Russian intelligence manipulated the American public to influence the United States Presidential Election in 2016.[20] Another example is the manipulation of customers emotions by leaders in surveillance capitalism, like Google.[21]

To be honest with you, before researching this book, we knew that the situation was bad and that it was going to get worse, but we really had no idea how *apocalyptic* it is *right now*. The Earth is covered right now with humans whose emotions are being swayed this way and that—on purpose—to make them buy certain things or elect certain people or support certain groups. Even to destroy entire political systems like democracy. This is affecting almost everyone on Earth. *Our emotional lives are all up for sale.* Instead of doing manual labour, *your* masters make you *buy* and *vote* and *click* and *comment* to spread their messages. This is literally your life right now.

---

[19] An early emoticon is the smiley face, ':)', and its associated emoji, '☺'. *Emoticons* are sequences of letters or keyboard symbols, whereas *emojis* are images or pictures.
[20] See Kramer et al (2014), Rudder (2014), Ienca & Vayena (2018), Lee et al (2018), Bradshaw& Howard (2018), United States of America Senate Foreign Relations Committee (2018), Hindman (2018), and Mueller (2019).
[21] Wills (2017) and Zuboff (2019).

## 0.5 Emotion Security

By utilizing the information theory of emotional communication, we can make new predictions about the effects of social media, and we can also provide a new kind of security for our emotion communication systems as they function on social media. Information theory provides us with the tools to quantify the amount of emotion information coming into and going out of each particular person in a vast network. By monitoring the information theoretic properties of how people's emotions are signaled across social media, one can identify and defend against manipulation of those very people's emotions. The information theory of emotion communication provides some powerful tools to block sophisticated manipulation techniques.

## 0.6 Plan

Chapter One contains background information on emotion communication and how it shows up in related areas of scientific research.

Chapter Two provides a formal theory of emotion signal systems, which has surprising explanatory power, although it is fundamentally qualitative.

Chapters Three, Four, Five, and Six contain the quantitative theory of emotion information. Chapter Three outlines the most basic emotion communication channel. Chapter Four gives the details for how to understand emotion coding in behavior by decoding from behavior detected in an emotion communication channel. Chapter Five is about basic emotion communication networks, with multiple emotion transmitters and receivers. Chapter Six covers advanced topics in emotion communication networks like checking someone's emotional reliability, emotion deception, emotional virtues like courage, and emotion networks that have multiple layers. A major payoff is defining the *social media influence factor*, which measures how much the social media layer of emotion communication contributes to one's overall emotional state.

Chapter Seven is about emotions on social media and how the information theory of emotion communication highlights and explains several important phenomena, including artificial emotion recognition systems. One focus is how the tools of information theory might help us guard against the negative effects of social media on our emotional capacities.

Chapter Eight is on emotion security systems for social media. We show how to use information theory to design quantitative tools that allow one to monitor activity so as to detect and prevent the kinds of attacks that happened during the 2016 United Kingdom "Brexit" Vote, the 2016 United States Presidential Election, and many others as well.

# *Chapter 1*

# Emotion Communication

You have an innate communication system that you probably have never thought of as a communication system.[1] An emotion is more often considered as something we have or something that happens to us. Rarely do we appreciate the fact that we and many other animals use our emotions to communicate all sorts of things.

In this chapter, we introduce some background material about emotions, animal communication, and emotion communication in particular. Then we consider the dominant theory of how emotions spread through populations right now, called the *contagion theory*. We reject the contagion theory, and instead endorse the signal theory of how emotional capacities function in groups. At the end of the chapter, we consider a problem for the signal theory and how to overcome it.

## 1.1 Emotions in Individuals

[1] Parts of this system are surely inherited from our evolutionary ancestors, while other parts are sensitive to experience and development.

Emotions are often studied from the perspective of the individual – what does the emotion feel like? What bodily changes occur? What triggers the emotion? What action tendencies are associated with the emotion? What kind of appraisal is involved? These sorts of questions are familiar to anyone researching emotions in philosophy, psychology, sociology, linguistics, economics, history, anthropology, neuroscience, or cognitive science. Moreover, these sorts of questions can be addressed by considering isolated individuals and specific emotion episodes.

Emotions are a complex topic, studied by philosophers, psychologists, linguists, sociologists, economists, neuroscientists, and others. A rough sketch of some regularities in contemporary thinking about emotions would have to include that: emotions are felt states that have mental and physical components to how they feel, (ii) emotions are often valenced – good or bad, (iii) many emotions have evolved ways of being expressed, for example in facial patterns, (iv) many emotions have other ways of being expressed as action tendencies, for example fight/flight during fear, (v) emotions have some kind of cognitive component, for example the appraisal of being wronged during anger, and (iv) emotions have a social component, for example the role guilt plays in enforcing norms. Which of these elements is most important is a matter of dispute that won't concern us. Instead, we emphasize a new way of thinking about the nature of emotions that complements each of these elements.

The literature on emotions is vast and  omplex; no survey an do justice to it all. Still some trends are discernible. There are three central theories on the nature of emotions in psychological literature.

- *Appraisal theory*: an emotion is an appraisal of a situation for how it impacts one's goals.
- *Basic-emotion theory*: an emotion is a distinct, neurophysiological state that is hardwired into the brain as a result of natural selection.
- *Constructivism*: an emotion is an emergent mental state that arises from more primitive phenomena—e.g., neurophysiological processes or social norms.

These categories are rough because they do not exhaust all the possibilities, but they do reflect the general trends in psychology of emotions.[2]

In addition, philosophers take the central theories on the nature of emotions to be the following.

- *Cognitivism*: an emotion is a cognitive state—e.g., an evaluative judgment or propositional attitude.
- *Feeling theory*: an emotion is a bodily feeling.
- *Perceptual theory*: an emotion is a perception of some aspect of the world.

There are also theories of emotions that involve various combinations of elements of these theories.[3] Instead of presenting the details of each category, a brief overview of what they have in common will suffice to give the reader an initial orientation.

As mentioned above, psychologists and philosophers have largely reached agreement that emotions involve (or at least closely relate to) some or all of the following components: cognition-like (e.g., perception, attention, and appraisal processes), behavior (e.g., action-tendencies, reflexes, facial expression, and motivations), physiology (e.g., neurological processes in the brain and nervous system activity in the body), and experience (e.g., subjective feelings).[4] Emotion theorists disagree about whether all the components are present with each emotion, which components are fundamental, and which are derivative. For example, basic emotion theorists emphasize neurological structures and facial expressions, while appraisal theorists emphasize, well, appraisal. Presumably, if one wants to use the term 'emotion' for only one of these components, one must still admit that the other components frequently come into play when an emotion is present. We do not aim to take a stance on what exactly counts as an emotion, but we are interested in the whole array of phenomena associated with emotional episodes.[5]

---

[2] Some hold that basic-emotion theory is a species of constructivism, e.g., Avrill (2012) and Faucher (2013). Avrill and Faucher understand constructivism as dividing into three categories: biological constructivism (i.e., basic-emotion theory), psychological constructivism, and sociological constructivism. Others acknowledge that some theorists (e.g., William James) seem to blur the lines between basic-emotion theory and psychological constructivism, but claim that contemporary arguments for these theories are more distinct, Gendron & Barrett (2009) and Lindquist (2013). Still others understand constructivism and basic-emotion theories as distinct, e.g., Cunningham, Dunfield, and Stillman (2013). And there is some newer work that falls under both appraisal theory and constructionist theories, e.g., Clore and Ortony (2013). For my purposes, whether or not one theory is properly understood a competitor (basic-emotion theory vs. constructivism) or species of another (e.g., constructivism is often divided into three groups) is not important. It is relatively uncontroversial to interpret as central and distinct theories of emotions as: (1) *basic-emotion theory*: Izard (1977), Ekman (1992), Griffiths (1997), Panksepp (1998); (2) *psychological constructivism*: Lindquist (2013), Russell (2003, 2009), Cunningham and Kirkland (2012), Barrett (2013), Cunningham (2013); (3) *social constructivism*: Averill (1980); and (4) *appraisal theory*: Arnold (1960), Frijda (1986), Lazarus (1991), Scherer (1999), Ellsworth (2013).

[3] This is an obvious over simplification. Similar to psychological theories of emotions, there are blurred lines among philosophical theories of emotions (some theories do not seem to fit easily into one category: e.g., Goldie (2000) is sometimes interpreted as a perceptual theory and other times as a feeling theory). There are also blurred lines between whether a theory of emotions is categorized best in philosophy or psychology, e.g., James (1884). Understand that there can be significant differences between the theories, but some general examples are: *feeling theory*: James (1884), Lange (1922), Damásio (1994, 2000, 2003), Prinz (2004); *cognitivism*: Solomon (1980, 1993), Greenspan (1988), Neu (2000), Nussbaum (2001, 2004); *perceptual theory*: Searle (1983), de Sousa (1987), Tappolet (2000), Deonna (2006).

## 1.2 Emotions as Social

We live in a social world and we belong to the class of essentially social animals. Humans and others are members of many social groups. Think about how many of your emotional states are caused by your interaction with others (e.g., getting angry at not being invited to a party) vs how many are caused by non-social situations (e.g., fear of a spider). And think about how many other people your emotions affect. Everyone is caught up on these massive feedback loops of interacting with other people, feeling emotions because of it, acting in certain ways because of those emotions, which becomes part of the next interaction. These emotional feedback loops colour every aspect of our lives, and the habits formed by social interaction frame even our solitary moments. Think how hard it is to imagine an exchange with another human being where no emotions are expressed at all.

---

[4] See Oatley et al (2006: 27-29), Niedenthal et al (2006: 5-10), and Winkielman et al (2007: 180) for discussion.
[5] A common objection against traditional cognitivist views concerns how best to understand the rationality of emotions. The objection concerns the fact that the only method of explaining the way in which emotions are rational is through an understanding of how the respective propositional attitude is rational; but, the rationality of emotions is not merely the rationality of propositional attitudes. For discussion, see de Sousa (1987), Goldie (2000) and Elster (2004). Moreover, a common objection to feeling theories is that they cannot accommodate rationality of any kind; for discussion, see Deigh (1994). Because various definitions of emotions conflict, it might not be possible to have a theory of emotional rationality that is *completely* neutral concerning the nature of emotions. Nonetheless, my goal is to develop an account of emotional rationality that a very broad application.

Despite the common assumption that people cannot do anything about their emotions, the reality is that humans and many other animals are constantly regulating their own emotions and the emotions of others. Some of these are done consciously like counting to ten to get over a bout of anger or soothing a crying infant. Others are unconscious like mimicking the posture of the person you are talking with in a conversation. It is also clear that emotions spread through populations with amazing rapidity, and these transitions can be exhilarating or terrifying. Not just other people but the social sphere itself helps regulate our emotions through a host of norms pertaining to all aspects of emotional behaviour. One example is display rules, which prescribe when, where, and how much people can express emotions depending on their social status, economic status, gender, sexual orientation, nationality, race, etc.

Emotions are so bound up with the social that one major tradition of emotion theories in psychology holds that emotions are socially constructed. There is a huge array of theories in this tradition, but they have in common that emotional states are constituted by social norms and strategic interactions in the communities in question. This view is typically taken to contrast with the "basic emotion" view, which states that specific emotions are innate across humans or other species. Social constructivists take evidence for their view differences in emotions across cultures and the importance of context in interpreting emotion behaviour.

The theories of emotion communication we present in what follows are neutral on this issue – they are compatible with basic emotion theories and social construction theories. However, the significance of the theories we present emphasize the importance of emotions in the social sphere.

## 1.3 Emotion Communication

It is commonplace that most communication is non-verbal. Although difficult to quatify, there is clearly some truth to this bit of conventional wisdom. It is also clear that much of non-verbal communication is emotion communication. That is, much of the information communicated non-verbally is information about felt emotions or emotionally relevant information about the environment. This idea has already been introduced: emotions constitute an important *communication* system, and emotion behaviors serve as *signals* in this system. The literature on emotions has touched on this idea for many years, but it is scattered and without a central comprehensive framework to serve as a basis for research.[6]

We should be clear that emotions are signal systems in addition to being many other things at the same time. Fear, for example, benefits the person feeling it in a situation where there is danger because it helps the person deal with the danger. That need not have anything to do with the signaling aspect of fear.

Let us look at an example. You are sitting by a fire deep in the forest alone except for your trusty dog, who is sitting peacefully next to you in the firelight surrounded by vast and quiet darkness. The dog suddenly turns to look behind you into the dark, but you notice nothing. The dog begins barking at the darkness as you turn around to look intensely, still seeing nothing. The barking turns fierce and frantic with the dog's ears back and teeth showing. And then the dog abruptly turns away and bolts off in the other direction leaving you alone in the small ring of light from the campfire. How do you feel?

Just reading this passage makes most readers feel fear, so even simulating this situation mentally triggers a fear response. Moreover, this is case of emotion communication where the dog transmits information to you. One animal feels an emotion and displays emotion behaviour, and another animal detects the behaviour and identifies the emotion. In addition, it is interspecies emotion communication, which should bolster the claim that emotion communication need not be just based on a human-human model. Finally, the communication in this example is not necessarily intentional. The example still works if the dog is not trying to communicate. Emotion communication can be automatic in that neither the transmitter nor the receiver is trying to communicate; still, the communication happens anyway.

[6] See Oatley and Johnson-Laird (1996) and Planalp (1998) for example.

One distinction in the literature that matters for understanding emotion signals is that between the emotional readout hypothesis and the behavioural ecology theory. The *emotional readout hypothesis* states that emotion behaviours (e.g., facial expressions) are involuntary displays of internal processes that are connected to evolutionary processes.[7] Alternatively, the *behavioural ecology* theory has it that emotion behaviours are largely intended for achieving social motives, regardless of what the individual is feeling inside.[8] We agree with van Kleef that these two accounts are compatible so long as they are not interpreted as saying that emotion behaviours are only one way or the other.[9] Instead, some emotion behaviours really do express what is going on inside us, and others are purely for the audience. Moreover, some emotion behaviours (e.g., voice or posture) might be more amenable to deliberate control than others (e.g., facial expressions). And there is clearly feedback between displaying emotion behaviours and actually feeling the associated emotion, so it is probably not so easy to disentangle the two functions.

The next few sections cover major research traditions about the social aspect of emotions, and the next chapter develops a formal theory of emotion signals systems and signals. Readers who want to skip the background can jump to Chapter Three for the main theory.

## 1.4 Animal Communication Science

The scientific literature on animal communication is vast and stretches back over a century. It covers many species of animals (in addition to plants and bacteria), many kinds of communication methods, and it has thrived on using the framework of information theory to explain communication and the phenomena that go with it. However, there is a massive gap in this entire literature: emotions. For some reason animal communication scientists do not think of emotions as part of a communication system. The result is that the science of animal

7 Buck (1985).

8 Fridlund (1994).

9 van Kleef (2016: 21-24).

communication presents us with a ready-made framework in which to insert emotion communication systems. Let us explore a bit of that system.

One classic definition from Wilson is that animal communication occurs when "the action of or cue given by one organism is perceived by and thus alters the probability pattern of behavior in another organism in a fashion adaptive to either one both of the participants."[10] (Wilson 1975). The organism giving the action or cue is called the *sender*, while the organism perceiving the action or cue is called the *receiver*.

There are many ways of understanding senders, receivers, and messages, but regardless of how these are interpreted, communication requires some kind of psychological process in the sender and some kind of psychological process in the receiver as well. Much work on animal communication involves characterizing these psychological processes and hypothesizing about how they might have evolved. In a major new survey, "Signaler and Receiver Psychology," Mark A Bee and Cory T Miller write:

> [W]e use the phrase psychological mechanisms to refer broadly and collectively to all of the processes carried out by the neural and neuroendocrine mechanisms operating in the peripheral and central nervous systems that are responsible for transducing, coding, processing, decoding, selecting, storing, retrieving, comparing, and acting upon information in signals. The phrase is meant to encompass the entire breadth of sensory, perceptual, and cognitive processes that underlie a signaler's abilities to adaptively use the signals in its repertoire and a receiver's abilities to adaptively respond to them. This inclusive view of psychological mechanisms in animal communication spans a broad array of interrelated phenomena, such as sensory processing, perceptual object formation, categorization, social cognition, numerical cognition, learning, memory, attention,

[10] Wilson (1975: 176).

decision-making, and concepts, among others.[11]

In short, scientists studying communication systems in animals find a huge range of mechanisms responsible on each end.

One unifying theme throughout this literature on psychological mechanisms is that receivers treat messages as perceptual objects, and senders expect this as well. Bee and Miller again:

> [S]tudying the psychological mechanisms of animal communication requires that we adopt a new view of signals and signalers as perceptual "objects" .... As the basic unit of perception, objects are formed by binding stimulus features ($\mathbf{x}$, $\mathbf{y}$, and $\mathbf{z}$) into a coherent representation that can be segregated from other potential objects in the environment. Object perception arises from neural mechanisms that bind together these separate features (e.g., color, shape, size). Notably, object formation is not the same as detecting the presence of an object in the environment or discriminating between Object 1 and Object 2 based on differences in their features ($\mathbf{x}_1$, $\mathbf{y}_1$, $\mathbf{z}_1$ vs. $\mathbf{x}_2$, $\mathbf{y}_2$, $\mathbf{z}_2$). In the literature on human perception, particularly visual processing, objects are considered to be spatiotemporally bounded feature clusters.[12]

In communication systems, messages and signallers are thought of as normal perceptual objects (e.g., an Amazon package), and they are represented as collections of properties (e.g., box shaped, composed of cardboard, brown, etc.).

There are numerous fascinating topics in the animal communication literature but we mention just three more. First there is a controversy over how to explain multi-component signals, e.g., many animal communication examples involve sending a signal in several formats at once. A bird might display part of its body, make a vocal call, and perform a particular movement pattern. Any one of these would be sufficient to convey the message, but animals often use many distinct components for a given signal.

[11] Bee & Miller (2016: 4)
[12] Bee & Miller (2016: 6)

> Animal signals commonly consist of multiple components—say a sound and a display—and students of signaling have offered many perceptual and cognitive explanations for why compound signals should be more effective. Yet, the economic benefits that receivers obtain by following multiple signal components remain unclear. Superficially, it would seem that a single discriminable difference should be sufficient to discriminate between underlying states, such as high-quality versus low-quality mates. … Indeed, one lingering question about the evolution of multicomponent signals is that they often seem advantageous from a psychological perspective, but disadvantageous from an economic one that explicitly considers their benefits and costs.[13]

In the psychology literature on emotions, these are often called multi-modal, where each kind of emotion behaviour is called a *modality*. (see Section 1). We address this problem directly in Chapter Four where we show that multi-component signals are a form of *error correction*, a well-studied topic in information theory. It turns out that one can classify multi-component signals on the basis of the type of error-correction algorithm they implement.

Second is the legitimacy of information theory as a basis for the science of animal communication. Krebs and Dawkins argued in a classic paper from 1978 that the concept of a signal or information being transmitted is misleading.[14] Animal communication should instead be thought of the manipulation of one animal by another. Unless communication is understood in the these stark cost/benefit terms for individuals, it is difficult or impossible to explain the evolutionary benefit of animal communication without resorting to dubious ideas like group selection. Scientific legitimacy, they argued, demands that communication theory be founded on influence or manipulation of one animal on another rather than signals or information passed among them.

[13] Bee & Miller (2016:7)
[14] Krebs & Dawkins (1978).

Although Krebs & Dawkins generated considerable controversy, their arguments had little effect on the practice of animal communication scientists who by and large continued to use information theory to explain communication among animals. A collection by Stegmann (2013) brings together a wide range of voices on the information theory side of the debate. Its authors provide a compelling panoply of reasons to remain confident in the scientific explanatory power of information theory for animal communication.

The information vs manipulation divide in animal communication studies runs deep, and we are obviously on the "information" side. We also see one of our major theoretical competitors – the contagion theory of emotion transfer – as part of the "manipulation" side, although this identification is far from obvious by just reading the literature (we discuss the contagion theory in Section 1.8). Instead, it fits into a trend in the manipulation view. Contagion is a kind of manipulation, and it is not a kind of signal. In particular, Krebs and Dawkins argue that animal communication systems should display certain evolutionary features akin to an arms race where each side develops ways of manipulating the other and ways of blocking the other's manipulation. These manipulation techniques and "manipulation jamming" techniques (for want of a better term) are predicted to improve together. This is exactly the kind of prediction one gets from the contagion theory as well – that animals would develop emotion contagion immune systems to prevent them from "catching" the contagious emotion. In reality, that is not what we see, but we discuss this in detail below.

Finally, we have to emphasize how shocking it is that there is virtually *nothing* in this entire literature on *emotions* as communication systems. For example, in *Animal Communication Networks*, edited by prominent animal behaviourist Peter K McGregor and published by Cambridge University Press in 2005 is a wonderful collection, full of amazing work by leaders in the field. But it does not mention emotion communication at all. In fact the word 'emotion' does not even appear in the index. We say this *not* to criticize McGregor or Cambridge; for, all the other publications in this field are the same. Instead, the fact that there is a massive emotion-shaped hole in this literature is what we think is important.

## 1.5 Theory of Mind (ToM) in Psychology

Another obvious scientific literature with which to engage is the Theory of Mind (ToM) work in psychology, which is about how humans attribute unobservable mental states to one another like beliefs, desires, intentions, goals, plans, perceptions, and emotions. The term 'Theory of Mind' and the acronym 'ToM' are now used to designate both this group of phenomena and this research field within psychology, cognitive science, and neuroscience. Other terms for the same group of phenomena (and research area) are 'mentalizing' and 'mindreading' (mostly in philosophy of mind).

The 'theory of mind' terminology can be confusing because for decades the term 'theory of mind' was used to denote a particular scientific theory, namely that humans and other animals use a *theory* of how minds work in order to figure out what is in other minds. This was sometimes known as the 'theory theory', and its major alternative was the simulation theory, which held that we primarily simulate ourselves in the place of another mind in order to understand it.[15] The debate between these two traditions has largely ended in a complex result: (i) the theory theory seems to have won a rhetorical victory by having the entire area of study come to be called 'theory of mind', and (ii) there has been a synthesis of the two traditions by recognizing that *theories* are used in many ways in *simulations* and that humans and other animals use mental *simulations* as a tool for *theorizing*. The key to the synthesis is understanding *how* theorizing and simulations work together when humans and other animals attribute mental states to one another. There is still a dedicated group of simulation theorists who fight on against the theory theory, but they are now on the margin of this literature, and their attacks are mostly baseless.[16]

[15] See papers in Carruthers & Smith (1996) for a nice snapshot of this literature during the 1990s. Shanton, K & Goldman (2010) for how it developed.

Although the term 'theory of mind' gets used by Premack and Woodruff[17] for the first time in psychological literature in 1978, the idea that humans use a particular theory – akin to a scientific theory – to figure out the contents of others' minds goes back further in philosophy. Wilfrid Sellars made much of the idea in the 1950s, and Donald Davidson, David Lewis, and Daniel Dennett all emphasized it their own philosophical views in the early 1970s using terms like 'radical interpretation' and 'the intentional stance'.[18] Robert Matthews is good example of a contemporary philosopher in this same tradition.[19]

In psychology, ToM found a major topic with Wimmer and Perner's *false belief test* in 1983, which asks kids to identify how someone else would behave when they are known to be missing certain information.[20] Kids that can attribute false beliefs make accurate predictions, and kids that cannot get the prediction wrong. ToM researchers have since published hundreds of studies on false belief tests in myriad variations. Many findings suggest that there is a certain age at which children acquire the ability to attribute false beliefs, but studies differ on what that age is, and many worry that there are multiple related but distinct phenomena at work here. One suggestion is that there are two parts to ToM abilities corresponding to the familiar System 1 and System 2 of dual process theories of cognition.[21] According to this model, sometimes we attribute mental states to other people in an automatic way without being able to control it and without any obvious effort or inference (that is using System 1); other times we attribute mental states to other people in an intentional and effortful way that is slower and more under our own control (using System 2). These two abilities develop differently, which explains the conflicting results on false belief test ages. There are, of course, those who disagree, including from the current wave of opposition to dual process theories that posit distinct System 1/System 2 cognitive processes.

[16] For example, for objections that Theory of Mind is "too dualistic" or "too scientific" see papers in Leudar & Costal (2009).
[17] Premack and Woodruff (1978).
[18] See Sellars (1953), Davidson (1973), Lewis (1973), and Dennett (1978).
[19] Matthews (2012).
[20] Wimmer & Perner (1983). See also Doherty (2009), Saxe (2013), Wellman (2014), Koster-Hale et al (2017).
[21] See Apperly (2011), Meinhardt-Injac et al (2018),

A major source of evidence for ToM comes from its engagement with psychological disorders and disabilities. Simon Baron-Cohen argued that autism spectrum disorders are caused by impairments in ToM abilities.[22] That is, autistic people lack certain abilities to attribute mental states to other people. This idea that autism is caused by ToM defects is now a huge literature on its own. Schizophrenia is another disorder that ToM theorists try to explain.[23] Overall, the fact that ToM has so much to say not only about when mindreading goes right but when it goes astray as well is convincing evidence of its power as a framework for scientific inquiry in psychology and beyond.

In the last decade, ToM has undergone a quantitative revolution, and the resulting theories are called *Bayesian Theory of Mind* (BToM).[24] There are two major innovations here, both of which come from artificial intelligence and machine learning in particular. The first innovation is to use *Bayesian networks* as a mathematical model for the "theory" people use to attribute mental states to each other. Bayesian networks have many forms but the basic idea is that each possible belief is given a *credence* (or subjective probability), which explains how strongly the belief is held. Credences are often numbers between 0 and 1, where 1 is maximum strength acceptance, .5 is indifference, and 0 is maximum strength rejection. The key to Bayesian networks are the mathematical formulae that explain how the weights change in light of new evidence. When someone makes a new observation, for example, that might increase their credence in some claim like 'my son is awake'. This change reverberates throughout the entire network of beliefs according to specific principles like Bayes theorem. Bayesian networks are

---

22 Baron-Cohen (1983)

23 See Corcoran et al (1995) and Frith & Corcoran (1996).

24 See Baker, Saxe, and Tenenbaum (2011), Byom & Mutlu (2013). Devaine et al (2014), Baker et al (2017), Thornton and Tamir (2017), Gordon & Hobbs (2017), Lee (2017), Otten et al (2017), Saxe & Houlihan (2017), Wu et al (2017), Pöppel & Kopp (2018), Tan & Ong (2019), probably most familiar as machine learning algorithms, and are a major part of the explosion in excitement over artificial intelligence in the past few years.[25]

The second major innovation, which also comes from machine learning, is to treat the agent doing the mindreading as a particular kind of artificially intelligent agent, called a *partially observable markov decision process* (POMDP). These are more familiar from reinforcement algorithms in machine learning, which are responsible for such high profile successes as Google's Alpha Go beating the human Go champion Lee Sidol in a stunning victory in 2016. Reinforcement algorithms like the one at the heart of Alpha Go use complex mathematical structures to decide which actions to take in which circumstances so as to maximize their rewards.[26] Rewards (often represented by real numbers) are what the algorithm is designed to maximize over time by the actions it chooses, and many reinforcement algorithms work by estimating what the reward will be across a range of situations and trying to get to the situation with the highest estimated reward. The 'partially observable' in POMDP refers to the fact that the agent does not have direct access to all aspects of its environment. Instead, it can observe certain aspects of its environment but it has to infer information about what it cannot observe. ToM researchers treat other people's behaviour as the observable part of the environment and other people's mental states as the unobservable part. The POMDP observes other agent's behaviour and figures out what their mental states are on that basis. By modelling the mindreading person as a POMDP and the theory used by that person as a Bayesian network, researchers can run experiments to see how well these artificial agents figure out each others mental states. By varying the aspects of the POMDP and the Bayesian network, researchers can test out various theories about how mindreading works. The great irony here is that BToM scientists are now using computer *simulations* to explain how *theory* of mind works! Far from being hypocritical, this development and others (like the rise of mathematical *theories* for how agents *simulate* situations) shows how much the two opposed camps have fused into a single research programme, despite the fact that some philosophers and psychologists still think the old debate continues.

[25] See Koshki and Noble (2009) for an accessible introduction to Bayesian networks.
[26] See Barto & Sutton (2018) for an excellent introduction to reinforcement learning algorithms.

Although BToM focuses on beliefs, some work exists on other states and even emotions. For example, Desmond C. Ong, Jamil Zakic, and Noah D. Goodmanc, in their paper "Computational models of emotion inference in Theory of Mind: A review and roadmap," use the BTOM framework to categorize the kinds of reasoning that happen when people attribute emotions to each other.[27] They write:

> People also possess a rich intuitive theory of emotions that comprises conceptual knowledge about different emotional states (e.g., anger, happiness) and how they are related to their causes and effects. People ("observers") use their intuitive theory of emotion to reason about the emotional states of others ("agents") around them, and thereby decide how best to respond in social situations. Importantly, these intuitive theories comprise the observer's beliefs about how others' emotions work, which depend on the observer's past history and their subjective beliefs. Though the observer's beliefs may not necessarily reflect the reality of how emotions "actually work", these beliefs nevertheless form the basis for how the observer understands and interacts with those around them.[28]

Even if the basis for emotion attribution and assessment that people actually use in social interaction does not reflect the science of how emotions manifest in each individual mind, brain, and body, the way people attribute emotions to each other – emotions in theory of mind – determines how humans interact with each other. Even if the assumptions we make when we attribute emotions to each other are wrong, these assumptions still help govern how we treat each other and what we expect of each other.

Ong et al posit a simple intuitive theory of how emotions are cause and caused by other elements of a mental and physical agent:

---

[27] Ong et al (2018). See also Saxe & Houlihan (2017), Kosakowski & Saxe (2018), Anzellotti et al (2019),
[28] Ong et al (2018: 340).

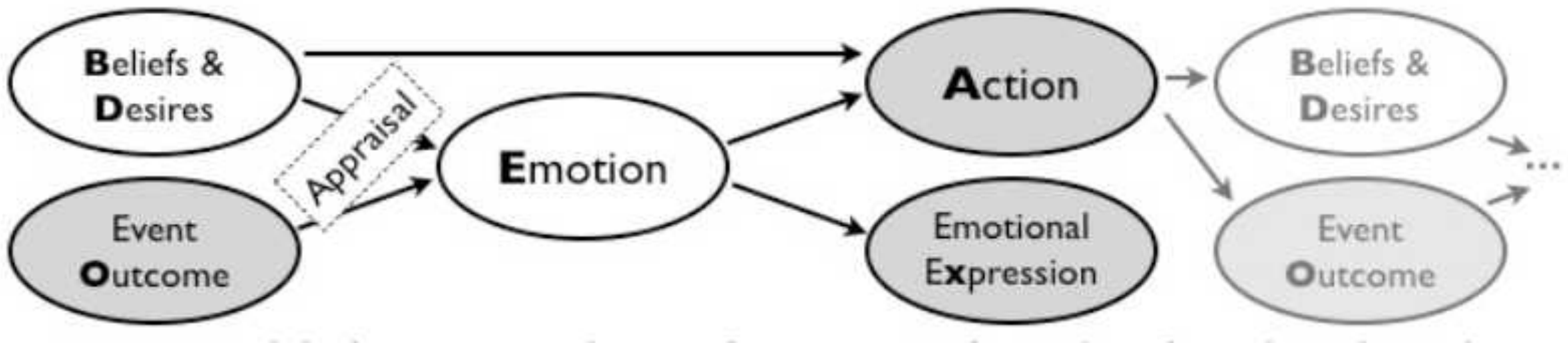

**Fig. 1: Causal interactions among emotions and other states.**

Emotions *are caused by* external events combined with the beliefs and desires of the person in question. Emotions *cause* emotional expressions (e.g., facial expressions) and actions, where actions cause further states.

Using these intuitive relationships among emotions, other mental states, and events in the environment, Ong et al classify six distinct kinds of reasoning that can occur surrounding emotions. The terms **e** (emotion), **x** (emotion expression), **a** (action), **o** (event outcome), **b** (belief), and **d** (desire), based on the above diagram.

| Categories | Description | Inferences | Examples |
|---|---|---|---|
| Emotion Recognition | Infer an agent's emotions from emotional expressions (facial expressions, body language, prosody), *or* from their actions. | $P(e\|x)$ $P(e\|a)$ | Given *smile*, Infer P(*happy*)<br><br>Given *scold-others*, Infer P(*anger*) |
| Third-person appraisals | Reason "forward" about how an event would cause an agent to feel, given also their mental states. | $P(e\|o)$ $P(e\|o,b,d)$ $P(e\|b,d)$ | Given *lose-wallet*, Infer P(*sad*) |
| Inferring causes of emotions | Reason "backwards" about the events that caused an agent's emotions. | $P(o\|e)$ $P(o\|e,b,d)$ | Given *sad*, Infer P(*lose-wallet*) |
| Emotional Cue Integration | Given multiple, potentially conflicting cues (e.g., multiple behaviors and/or causes of emotion), combine them and reason about agent's emotions | e.g., $P(e\|o,x)$ $P(e\|o,a)$ $P(e\|a,x)$ $P(e\|o,b,d,a,x)$ | Given *smile* + *lose-wallet*, Infer P(*happy*), P(*sad*), etc. |
| Reverse Appraisal | Given an event and an agent's emotions, reason backwards to mental states like beliefs and desires | $P(b,d\|e,o)$ $P(b,d\|e)$ | Given: *receive-gift* + *surprise* + *happy*, Infer *Gift-Unexpected* + *Gift-Desired* |
| Predictions (Hypothetical Reasoning) | Given an agent's emotions, predict subsequent behavior. Or, given a (hypothetical) situation, predict the agent's emotions. | $P(a\|e)$ $P(x\|e)$ | If *anger*, predict P(*scold-others*) |

**Fig 2: BToM kinds of reasoning in emotion attribution.**

The framework for understanding emotions in BToM offers researchers a plethora of new and precise hypotheses to test and novel ways of testing them with artificial agents in computer simulations.[29]

Our own information theory of emotion communication, presented in Chapters 3-6, fits nicely with the BToM framework because all the machine learning algorithms in BToM are grounded in estimating probability distributions, and information theory is also grounded in probability distributions. We illustrate how to integrate the information theory of emotion communication with Bayesian Theory of Mind (BToM) in Chapter Six.

[29] See Oakley et al (2016) for a caution about how to proceed.

## 1.6 Affective Sciences

Emotions have been a major topic in Western philosophy for millennia, but the various sciences that have emotions and other phenomena like them (e.g., moods) have for the past two decades been called 'affective science' or 'the affective sciences'.[30] These sciences are primarily psychology, cognitive science, and neuroscience, but includes sociology, economics, evolutionary biology, linguistics, computer science, and artificial intelligence.

*Psychology* is often taken to be the science of the mind and behaviour, *cognitive science* is supposed to cover everything about the mind, and *neuroscience* is the science of brains and nervous systems. These three clearly overlap, and it can be difficult to keep track of how best to classify research and to taxonomize subfields. Cognitive scientists often investigate ways of classifying or processing information that go beyond what we think groups of neurons are capable of. To this extent cognitive science is more general than neuroscience, and investigates what "in principle" is possible for cognitive systems rather than what is possible for things with 1200 $cm^3$ brains composed of neurons running on DNA. Cognitive science is closely related to cognitive psychology, but the latter includes a bigger emphasis on behaviour and interfaces with clinical psychology and personality theory more than cognitive science typically does. Cognitive scientists also favour computational models more than qualitative models, which are more common in cognitive psychology. Indeed, Theory of Mind (ToM) – outlined above – started in cognitive psychology, but it seems to us that, after it morphed into BToM with the proliferation of quantitative and computational machine learning techniques, it has shifted more into cognitive science. One sees this shift in the kinds of journals publishing these papers as well.

Often the term 'affective' is contrasted with 'cognitive' so that one might work on affective neuroscience, which would be the study of brain structures (neurons and circuits of neurons) responsible for affect like emotion, mood, feelings, pain, etc. Another important term here is 'computational', which emphasizes quantitative models where computational complexity is a key factor. Computational cognitive neuroscience, for example, studies the computational models underlying cognitive processes that might be based on circuits of neurons.[31] The boundaries between these sciences are fluid and usage of these terms can vary across traditions, so the reader should take this discussion as a guide at best.

[30] See Goldsmith et al (2002), Panksepp (2004). Vuilleumier, & Armony (2013) and Anderson (2018).

There are far too many results to survey here, but we mention that emotions have many components and much of affective science studies details of these. The amygdala is one region of the brain associated with emotions, and there is an army of scientists studying it and other areas of the brain that are active when people feel emotions. All the other things that happen in one's body when having an emotion episode are another major area of study. How appraisals fit in with other mental states and other emotion components is yet another huge tradition. All the sorts of things that can go wrong with the emotion system is another topic of colossal activity.

One area that deserves special attention is *social neuroscience*.[32] This is the study of the neuron and brain structures that are active when humans interact socially. Emotion communication is clearly is a paradigm social interaction, and we provide a quantitative theory of the information that is involved in this communication. One crucial issue in doing so is how animals represent the emotions of others. The outlines of how this works informs our own models of emotion networks in Chapter Six. In particular we make use of the valence/arousal representation of emotions to create a code for emotion states and use it to show how it makes a difference to the amount of information sent over a channel (see Section 6.10).

## 1.7 Artificial Intelligence and Computational Models of Emotions

One of the most challenging tasks in artificial intelligence is to equip artificially intelligent agents to interact with human emotions. Scientists working in this area develop algorithms that can

---

[31] See O'Reilly & Munakata (2000) for an introduction to computational cognitive neuroscience.
[32] See Verweij et al (2015).

figure out which emotional states humans are in and understand how this information can be used to predict what the humans will do so that the algorithm can plan for the appropriate response. One approach is to program artificial agents so that they have emotional states themselves. This comes in handy because many of our own emotional states call for or invite emotions from others (e.g., sadness invites sympathy). Either way, it is clear that artificial agents will need to model human emotions in some way to interact with us effectively.

Eva Hudlicka has recently summarized some of the computational requirements for artificial agents to be able to model emotions.[33] This summary is based on a large amount of technical research in artificial intelligence, machine learning, robotics, control theory, cognitive science of emotion, and neuroscience of emotion. It turns out that Hudlicka's research is even more significant because it offers the kinds of descriptions one would demand at the *computational* level for understanding emotions in humans and other animals.

Hudlicka's computational framework acknowledges a fundamental distinction between emotions and cognitive processes involving emotions:

> [F]or purposes of developing symbolic models of emotions, and for models of emotions in symbolic agent architectures, it is useful to cast the emotion modeling problem in terms of two broad categories of processes: those responsible for the generation of emotions, and those which then mediate the effects of the activated emotions on cognition, expressive behavior and action selection.[34]

We have already seen the importance of this distinction in the discussion of Bayesian Theory of Mind (BTOM), and it is at the centre of van Kleef's theory of Emotions as Social Information, discussed in the following section.

Hudlicka distinguishes two broad classes of computational requirements based on an input/output or functional way of thinking about emotions: what causes emotions and what do emotions cause? The former called *emotion generation* and the latter category she calls *emotion effects*.

[33] Hudlicka (2014). See also Petta et al (2011)
[34] Hudlicka (2014: 15).

She writes:

> The following distinct computational tasks are required to implement emotion generation via cognitive appraisal:
>
> • Define and implement the {emotion elicitor(s)}–to–{emotion(s)} mapping. Depending on the theoretical perspective adopted, this may involve additional subtasks that map the emotion elicitor(s) onto an intermediate representation (e.g., PAD dimensions; appraisal variables vectors), and the subsequent mapping of these onto the final emotion(s).
>
> • Calculate the intensity of the resulting emotion(s).
>
> • Calculate the decay of these emotions over time.
>
> • Integrate multiple emotions, if multiple emotions were generated.
>
> • Integrate the newly-generated emotion(s) with existing emotion(s) or moods.[35]

The PAD dimensions in the first requirement refers to the three dimensional emotion classification scheme based on valence (called Pleasure sometimes), Arousal, and Dominance.[36] Hudlicka continues with a list of requirements for emotion effects as well. She writes:

> The following distinct computational tasks are necessary to implement the effects of emotions across multiple modalities:
>
> • Define and implement the emotion/mood–to–effects mappings, for the modalities included in the model (e.g., cognitive, expressive, behavioral, neurophysiological). Depending on the theoretical perspective adopted, this may involve additional subtasks that implement any intermediate steps, and are defined in terms of more abstract semantic primitives provided by the theory (e.g., dimensions, appraisal variables).
>
> • Determine the magnitude of the resulting effect(s) as a function of the emotion or mood intensities.

[35] Hudlicka (2014: 16).
[36] (Mehrabian et al. 1995)

- Determine the changes in these effects as the emotion or mood intensity decays over time.
- Integrate effects of multiple emotions, moods, or some emotion and mood combinations, if multiple emotions and moods were generated, at the appropriate stage of processing.
- Integrate the effects of the newly-generated emotion with any residual, on-going effects, to ensure believable transitions among states over time.
- Account for variability in the above by both the intensity of the affective state, and by the specific personality of the modeled agent.
- Coordinate the visible manifestations of emotion effects across multiple channels and modalities within a single time frame, to ensure believable manifestations.[37]

Hudlicka completes her computational framework by listing the distinct data types (domains) that will need to be processes in order to implement the computational requirements for emotion generation and emotion effects. They are summarized in Figure 3.

[37] Hudlicka (2014: 17).

| Domain name | Description | Examples of domain elements |
|---|---|---|
| **Object (W)** | Elements of the external world (physical, social), represented by cues (agent's perceptual input) | Other agents, events, physical objects |
| **Cognitive (C)** | Internal mental constructs necessary to generate emotions, or manifest their influences on cognition | Cues, situations, goals, beliefs, expectations, norms, preferences, attitudes, plans |
| **Abstract (Ab)** | Theory-dependent; e.g., dimensions, appraisal variables, OCC evaluative criteria | Pleasure, arousal, dominance; certainty, goal relevance, goal congruence… |
| **Affective (A)** | Affective states (emotions, moods) & personality traits | Joy, sadness, fear, anger, pride, envy, jealousy; extraversion |
| **Physiology (Ph)** | Simulated physiological characteristics | Level of energy |
| **Expressive Channels (Ex)** | Channels within which agent's emotions can be manifested: facial expressions, gestures, posture, gaze & head movement, movement, speech | Facial expressions (smile, frown), speech (sad, excited) gestures (smooth, clumsy), movement (fast, slow) (represented via channel-specific primitives, e.g., FACS) |
| **Behavioral (B)** | Agent's behavioral repertoire in its physical and social environment | Walk, run, stand still, pick up object, shake hands w/ another agent |

**Fig 3: Data types for computational theories of emotion**

Notice the use of the information theoretic term 'channels' in the description of **Ex**. The reason is that computational accounts of emotions and information theoretic accounts are based on probability distributions, so they mesh well together.

In characterizing her computational framework for modelling emotions, Hudlicka suggests that we think of the study of emotions in terms of Marr's three levels of description: the computational, the algorithmic, and the implementational.[38] One can think of the neuroscience of emotion as the implementational level – these scientists are focused only on what neurons and networks of neurons can do. In the case of artificial agents, the implementational level would be completely different. The algorithmic level is the level of cognitive science of emotion. The Bayesian algorithms studied in BToM is a good example of research at this level. In the study of emotions, the computational level is rarely distinguished as its own field of study, but we agree with Hudlicka that it makes sense as a unifying perspective to think of the computational requirements on emotion systems as largely shared between natural agents and artificial agents.

[38] Hudlicka (2014: 21). Marr (1982) is the source of this distinction, but it has become part of a common framework across cognitive psychology, cognitive science, and neuroscience, in addition to being a staple of contemporary philosophy of science, philosophy of mind, and elsewhere. The "levels of description" terminology and Marr's distinction between computational, algorithmic, and implementational have an enduring influence.

One point to emphasize is how well ToM, affective science, and computational models of emotion fit together. They all utilize a similar batch of quantitative tools that come mostly from probability and statistics.

## 1.8 Emotion Contagion

As of 2020, the received view is the *contagion theory*, which holds that emotions are like pathogens and emotions spread like infections. The classic text for the contagion theory is *Emotional Contagion*, by Elaine Hatfield, John T. Cacioppo, and Richard L. Rapson, but there are hundreds of papers exploring the details of this framework.[39]

The main idea is the emotional contagion is the proper unit to utilize in analysing emotion spreading, where contagion occurs when one animal has an emotion and that causes another animal to have an emotion of the same kind. Two popular quantitative models from epidemiology, the SIS model and the SIR model have been used by emotion contagion theorists. SIR stands for Susceptible, I for Infected, and R for Recovered. Each of these quantities measures a number of animals in the population, and each one changes as a factor of time. SIS stands for Susceptible, Infected, Susceptible. The difference between the two is that in SIR models, animals cannot be reinfected, but in SIS models they can. Because we care about how well these mathematical models explain emotion transmission through groups, the SIS model is often more appropriate; a person can get sad and then feel better and then get sad again. There are many variants of these two main models, and they are the framework for quantitative investigations into emotion contagion.

[39] See Hatfield, et al (1994), Barsade (2002), Hill et al (2010), House, T. (2011), Tsai et al (2011), Coenen & Broekens, J. (2012), Rueff-Lopes & Caetano (2012), Yin et al (2012), Aguilar (2013), Elfenbein (2014), Fu et al (2014), Hatfield, et al (2014), Bertozzi et al (2015), Alshamsi et al (2015), Hess, U. & Fischer, A. (2016), Lehmann & Ahn (2018), Bacaksizlar (2019), and Wróbel & Imbir (2019).

We take the signal theory of emotions to be a competitor with the contagion theory, but that is not what van Kleef thinks. He treats them as compatible, albeit he takes the contagion theory to be mostly a metaphor rather than a genuine explanation.[40] This issue does not impact our arguments in what follows, but we emphasize that those who see emotional capacities functioning as a communication system think of emotional behaviour as messages to be interpreted by an audience, while those who see contagion think of emotional behaviour as infections.[41]

The defender of the contagion theory is likely to respond that emotions *aren't literally* contagions, emotions just behave like them in some situations. This seems sensible, but it gives everything away to the signal theory because *emotions literally are signals*. They behave like signals because they are signals. Using the contagion theory to understand complex emotion networks is like using a petri dish to simulate a microprocessor. That might work for some features, but overall, it is not up to the task.

Point 1: emotions *are not* literally contagions, but they *are* literally signal systems. When the central aspect of a theory (e.g., contagion) is an artefact, things have gone wrong.

Point 2: the best alternative to the contagion theory is the signal theory of emotions, but so far no quantitative theory of emotion signals has emerged to compete with quantitative contagion theories. That is exactly what we present in Chapters 3-6. Moreover, we can define emotion contagion in information theoretic terms (we do so in Section 6.6), so the signal theory subsumes the contagion theory, but the reverse is impossible.

---

40 See van Kleef (2016: 38-45).

41 See Bartsch & Hubner (1995) for an account of what they call emotion communication (emotion felt → emotion detected), but is really an account of emotion contagion (emotion felt → emotion felt). They survey a number of prominent theories of emotion and how well they fit with emotion contagion.

## 1.9 Emotions as Social Information

van Kleef's book assesembles the materials for what he calls the Emotions as Social Information (EASI) theory, which is comprised of the following principles (numbered formulations are quotes from van Kleef with page numbers and In Other Words (IOW) are our own gloss).

1. Emotional expressions have evolved, at least in part, because of the informational value they represent to observers, which helps coordinate social interaction. (20)

- IOW: Animals that can send information by emotions tend to perform well in evolution.

2. The social-signaling function of emotional expressions is functionally equivalent across expressive modalities in that the direction (but not necessarily the magnitude) of the interpersonal effects of emotions is similar regardless of whether emotions are expressed in the face, through the voice, by means of bodily postures, and/or with words or symbols. (27)

- IOW: When something has an emotion, the way they express it – by face or voice or body posture or emoji – does not matter for communication. The message is the same.

3. People use others' emotional expressions to infer traits and dispositions that are relevant to (social) survival and success (e.g., dominance, affiliation) and to anticipate others' behavior (e.g., collaboration vs. exploitation) as well as the trajectory of social interactions (e.g., cooperative vs. competitive). (30)

- IOW: Emotions sometimes communicate things about your society and its norms and other statuses.

4. Emotional expressions elicit reciprocal and complementary affective reactions in observers, which in turn inform observers' behavioral responses. (45)

- IOW: When one animal has an emotion, others tend to have emotions too – sometimes the same (e.g., fear → fear) and sometimes not (e.g., anger → fear, in cases where the fact that one animal displays anger – perhaps by punching a hole in the wall – makes others who see the display afraid).

5. Observers of emotional expressions may extract information from these expressions about the expresser's appraisal of and orientation vis-à-vis the situation and the people involved, which in turn shapes observers' behavioural responses to the emotional expressions. (52)

- IOW: See an animal's emotions tells you something about its mind, like what it thinks about the situation it is in.

6. Affective reactions and inferential processes have mutually influential yet conceptually distinct and empirically separable effects on observers' behavioral responses to others' emotional expressions. (55)

- IOW: having emotions and attributing emotions are distinct from thinking about emotions in oneself and others.

7. The informational value of emotions is more likely to be capitalized upon to the degree that emotions are (a) successfully encoded by the sender (as determined by the sender's emotional expressivity), (b) successfully decoded by the receiver (as determined by the receiver's emotion perception and understanding abilities), and (c) there is a match between the intensity of the sender's emotional expression and the receiver's perceptual sensitivity. (60)

- IOW: Emotion communication works best when animals act how (and how much) they feel and others can figure out how (and how much) they feel.

8. The relative influence of inferential processes (as compared with affective reactions) in shaping responses to emotional expressions becomes greater to the degree that the observer of the emotional expression is motivated and able to engage in thorough information processing; the relative influence of affective reactions increases to the degree that information processing motivation or ability is reduced. (64)

- IOW: Thinking about emotions is less important when animals think about emotions less.

9. The relative influence of negative affective reactions (as compared with inferential processes) in shaping observers' responses to others' emotional expressions increases to the degree that observers perceive the emotional expressions as inappropriate rather than appropriate, which depends on characteristics of the situation, the emotional expression, the expresser, and the observer; the relative influence of inferential processes increases to the degree that emotional expressions are perceived as appropriate. (76-77)

- IOW: Bad emotion reactions are more important when the emotion is seen as wrong and less important when the emotion is seen as right.

In sum, van Kleef uses some information theory terminology and even sees emotion expression as information encoding and sees emotion recognition as information decoding, which is a huge insight that we incorporate in Chapter 4. van Kleef also emphasizes the difference between having an emotion and inferring other things based on the having of an emotion. We have already seen this distinction play an important role in the BToM literature on computational categories of reasoning about emotions.

We take no issue with van Kleef about each of these principles, in part because of the evidence he assembles for them in the second half of the book, which includes social effects of emotions in close relationships, in groups, in conflict and negotiation, in consumer behavior, and in leadership.[42]

The only complaint we have about van Kleef's framework is that so much more is possible. What he offers is right, but it is a loose collection of claims without even a formal qualitative theory to serve as a framework for them. van Kleef also uses information theoretic terms like sender, receiver, encoding, decoding, but without the formal theory, these terms mostly just metaphors. Note that EASI theory does not have the resources to answer the question of whether emotion signals are largely indicative, largely imperative, or neither. We show in the next chapter that Lewis's theory implies that they are largely indicative and have as content their fittingness conditions.

In the next chapter, we sketch a theory of emotion signal systems, which then serves as a the basis of the quantitative theory of emotion information in the following chapters. Everything we say builds on the evidence van Kleef's assembles for his EASI theory.

## 1.10 Summary of Research on Emotion Communication

There has been a vast amount of research that is relevant to emotion communication, but many of these research programmes are insulated from one another. Animal communication science

---

[42] Comprehensive evidence covered by van Kleef falls into these categories:

- Social effects of emotions in close relationships
- Social effects of emotions in groups
- Social effects of emotions in conflict and negotiation
- Social effects of emotions on consumer behavior and
- Social effects of emotions in leadership

See van Kleef (2016: 81-193).

studies many different systems across many life forms, and a great many theories use of infomation theoretic models. There is a huge emphasis on evolution, but there is nothing on emotion communication at all.

In philosophy and psychology there is a huge amount on emotions in individuals, some on the social role of emotions, much of it about emotions as contagion, and some on emotions as signal systems. There is some work on emotions in ToM and some uses of information theory terminology in social psychology literature on emotions.

BToM is a great quantitative framework, but emotions have yet to be explored much within it. EASI is a great qualitative framework but it has no formal framework for understanding signals and it does not utilize any quantitative elements for information.

Contagion theory is the one that seems like a holdover from an earlier period that does not fit with the other major paradigms emerging from a number of spots in emotion research. BToM, affective science, animal communication science, and computational theories of emotion all rely on a quantitative mathematical framework grounded in probability theory and statistical techniques for estimating probability distributions. Contagion theory is not.

One way of thinking about a major strand in this book is that it provides an superior alternative to the contagion theory in part because the information theory of emotion communication fits seamlessly with the other major quantitative approaches to emotion phenomena, which are based on probability distributions. The basic foundation of any information theoretic framework is a theory of senders, receivers, and messages. Hence, we need to describe a signal theory of emotions to serve as this sort of framework. Chapter Two introduces the signal theory of emotions and Chapter Three begins the presentation of the information theory of emotions.

## 1.11 Extensions

There are vast literatures on emotions, but the *Stanford Encyclopedia of Philosophy* entry on emotions by Andrea Scarintino is a short survey that covers both philosophical and psychological aspects of emotion research (plato.stanford.edu/entries/emotion/). For a longer survey, start with the *Handbook of Emotions*, 4th Edition, edited by Michael Lewis, Jeannette M. Haviland-Jones, Lisa Feldman Barrett (2016). Note that the contents of the various editions of this title are substantially different from one another, so one can look at earlier editions to get a feel for the state of the conversation at earlier times (in 2008, 2000, and 1993, respectively, though only Michael Lewis and Jeannette M Haviland-Jones are editors on earlier editions). Other nice collections include *The Nature of Emotion: Fundamental Questions* 2nd Edition, edited by Andrew Fox, Regina Lapate, Alexander Shackman, Richard Davidson (2018), which is a wonderful survey with concise papers by experts on a huge range of central topics. Peter Goldie's collection *The Oxford Handbook of the Philosophy of Emotion* (2012) provides good coverage with detailed papers on a range of philosophical aspects of emotions. There are a number of psychology textbooks on emotions, but *Psychology of Emotion, 2nd edition*, by François Ric and Paula M. Niedenthal (2017) is a good place to start.

The tradition focusing on emotion communication is scattered, but the best contemporary overview is van Kleef's book, *The Interpersonal Dynamics of Emotion: Toward an Integrative Theory of Emotions as Social Information* (2016), is a spectacular resource for the cutting edge. Sally Planalp's 1999 book, *Communicating Emotion: Social, Moral, and Cultural Processes*, is an excellent survey and synthesis of research on many central issues related to emotion communication; although it is a dated now, it is still a reliable guide to the issues. A nice companion to Planalp's book is *The Handbook of Emotion and Communication: Research, Theory, Applications, and Contexts*, edited by Peter Anderson and Laura Guerrero (1998). Ross Buck's *The Communication of Emotion* from 1984 is a founding text in this tradition, but it is far enough from the contemporary scene now that it is mostly of historical interest. Buck's new book, *Emotion: A*

*Biosocial Synthesis* is well worth a look, as it engages with decades of work on emotion expressions and offers an interesting synthesis of a huge range of emotion phenomena.

Animal communication science is mostly conducted by biologists, but they rarely think of emotions as a communication system at all. Nevertheless, this is an important literature for us and anyone interested in the topic of the book. See the papers in *Animal Communication Theory: Information and Influence*, edited by Ulrich E. Stegmann (2014) for an overview of the debate and the major positions in it. *Psychological Mechanisms in Animal Communication*, edited by Mark Bee and Cory Miller (2016) is another solid collection that has more overlap with psychology and cognitive science. Older but still relevant is *Animal Communication Networks*, edited by P.K. McGregor (2005). Jack W. Bradbury and Sandra L. Vehrencamp's textbook, *Principles of Animal Communication 2nd Edition* (2011) provides a fine overview of the field.

Theory of Mind (ToM) in psychology is huge, and it has taken a sudden turn into complex technical models borrowed from machine learning algorithms. For the theory-theory vs simulation theory debate, see Peter Carruthers & Peter Smith (1996). There is no textbook on Bayesian Theory of Mind (BToM), but *Reinforcement Learning: An Introduction* by Andrew Barto and Richard S. Sutton provides a wonderful and detailed account of Markov Decision Processes and the kinds of statistical learning algorithms that are the basis for BToM. Donald Davidson's "Radical Interpretation" is a short philosophical piece on theory of mind with an information theoretic angle. For a survey of BToM, and see "Computational models of emotion inference in Theory of Mind: A review and roadmap," by Desmond C. Ong, Jamil Zakic, and Noah D. Goodman.

Affective Science has been highly influenced by *Handbook of Affective Sciences*, edited by H. Hill Goldsmith, Klaus Scherer, Richard Davidson (2002), and by Jaak Panksepp's *Affective Neuroscience: The Foundations of Human and Animal Emotions* (2004). See Patrik Vuilleumier, Jorge Armony's *The Cambridge Handbook of Human Affective Neuroscience* (2013) for a more contemporary collection. See also *The Neuroscience of Emotion: A New Synthesis* (2018) by David J. Anderson and Ralph Adolphs for a fascinating synthesis of contemporary research in a bold new direction (Drosophila have emotions?!).

Computational Theories of Emotion are mostly confined to the journals, but see *Emotion Modeling: Towards Pragmatic Computational Models of Affective Processes*, Tibor Bosse, Joost Broekens, João Dias, and Janneke van der Zwaan (Eds.) (2014), which contains the Hudlicka paper discussed in detail above. See *Engineering Computational Emotion—A Reference Model for Emotion in Artificial Systems* by M. Guadalupe Sánchez-Escribano (2018) for a comprehensive survey of all the central issues along with an innovative proposal.

For work on signal systems and information theory: see the next two chapters.

One final point to emphasize in this literature review chapter: there is too much written on emotions for any person or team of people to find it all, much less digest it. We have tried to cast our net widely, but we apologise for any obvious omissions in this chapter or as we move forward.

# *Chapter 2*

# Emotion Signal Systems

The main point of the book is to introduce the quantitative theory of information that is transmitted in emotion communication systems and apply it to social media. The quantitative theory of information depends on being able to identify senders, receivers, and signals in communication. With respect to emotion communication in particular, no one has proposed a formal but qualitative theory of emotion signal systems. So we dedicate this chapter to emotion signal systems, senders, receivers, and messages. The following chapter is the beginning of the quantitative theory of the information in emotion signal systems.

We use the term *signal theory* to cover any model that aims to explain how emotional capacities function as a system of communication by taking emotional behaviors to be social signals. The signal theory of emotions offers an interesting qualitative model for emotion communication.[43] For example, according to van Kleef, it predicts that different expressions of the same emotion in emotion behaviors (e.g., in one's face, in one's voice, and in one's body posture or movement) are functionally equivalent.[44] Different expressions of the same emotion are distinct methods of conveying the same message (in the information theory literature, this is called error-correction). Note that the contagion theory (Section 1.7) does not have this prediction.

[43] There are quantitative techniques for processing signals, but these have not been applied to emotion communication systems; see Oppenheim and Willsky (1997).

We think David Lewis's theory of signal systems is perfect for the role of a central formal framework. Lewis's *Convention*, a version of his PhD dissertation, is a landmark in 20th century philosophy. It focuses on *conventions*, which are certain kinds of regularities in populations. But along the way to developing the full theory of conventions, Lewis defines a signal system, and it is this definition that concerns us. Lewis's book is famous for its foundations of contemporary natural language semantics, and its account of conventions has proven to be a lasting achievement.[45]

The *signal theory of emotions* is that the emotional capacities of people in groups function as a signaling system for communication across the group. When one person has an emotion and expresses it, another perceives the first person's expression of this emotion and the second often comes to feel it too. The second person expresses the emotion as well and a third perceives it and comes to feel it as well. If the emotion is fear, for example, then the first person's fear might have been triggered by a nearby danger. The second and third need not have seen the danger in order to be alerted to its presence. In this way, groups can communicate information through emotion signal systems. Anyone who has felt a wave of panic sweep across a crowd knows this phenomenon well. It is a visceral feeling emanating from everyone around you that simply cannot be ignored.

The tendency for animals that have emotions to send signals with their emotions is underappreciated. There exist already vast literatures on how to analyze signal systems, and these quantitative tools can be utilized to understand the features of this signal system we call emotions.[46] Moreover, thinking of emotions as signals is superior when it comes to understanding how emotions behave on the biggest signal system in the world – the internet.

[44] See van Kleef (2016: 26-27, 201-202).
[45] Lewis (1969), which is based on his PhD dissertation.

## 2.1 Signal Systems

For Lewis, a *signal system* consists of two sets of contingency plans. A *contingency plan* is a set of hypothetical claims about what an agent is to do in certain circumstances. The *sender* and the *receiver* are the two parties in a communication system, and each one's contingency plan constitutes what she ought to do in a given situation. Lewis uses the ride of Paul Revere during the American Revolution as a central example. It involves the Sexton and Paul Revere who are coordinating to inform the American troops about a potential British attack. S1 and R1 are their respective contingency plans.

*S1 (Sexton's plan):*

If the redcoats are observed staying home, hang no lantern in the belfry.

If the redcoats are observed coming by land, hang one lantern in the belfry.

If the redcoats are observed coming by sea, hang two lanterns in the belfry.

*R1 (Revere's plan):*

If no lantern is observed hanging in the belfry, go home.

If one lantern is observed hanging in the belfry, warn that the redcoats are coming by land.

If two lanterns are observed hanging in the belfry, warn that the redcoats are coming by sea.

It should be obvious that the two contingency plans are coordinated in the sense that, together, they allow for successful communication – if the redcoats are observed coming by sea, then that is the message Revere receives and passes on. Had the contingency plans been different, that might not have been the case (e.g., if R1 included "If two lanterns are observed hanging in the belfry, go home" instead of its third clause). Lewis allows considerable latitude in formulating contingency plans, but his formalism has the benefit of providing an explicit analysis of what it is for contingency plans to be coordinated in the right way.

[46] We present and explore many of these in the following chapters.

In general, a *signal system* is defined relative to three sets: states of affairs, audience responses, and signals. Let $A$ be a set of states of affairs with $m$ members. We call A's members $a_1 \ldots a_m$. Let R be a set of audience responses with m members as well, labelled $r_1 \ldots r_m$. Let $\Sigma$ be a set of signals with n members; call its members $\sigma_1 \ldots \sigma_m$. Three functions are needed for the formal account as well. Let $F$ map states of affairs into responses such that $F(a_i)$ is the appropriate response to observing $a_i$; hence, $F$ is a function from $A$ to R. Furthermore, let $Fs$ be a function from $A$ to $\Sigma$ corresponding to the sender's contingency plan (i.e., if $a_i$ is observed, do $Fs(a_i)$), and let $Fr$ be a function from $\Sigma$ to R corresponding to the receiver's contingency plan (i.e., if $\sigma_l$ is observed, do $Fr(\sigma_i)$). Sender's plans take them from states of affairs to signals, and receiver's plans take them from signals to responses. *Fs is admissible* if and only if it is a 1-1 function, and *Fr is admissible* if and only if the range of $Fr$ is identical to the range of $F$. Overall, the signal system combines these two functions $Fs$ and $Fr$ to get a function from states of affairs to responses. If $Fr(Fs(a_i)) = F(a_i)$ for each $a_i$ in $A$, then the pair $<Fs, Fr>$ is a *signal system.* Using these definitions it is trivial to prove that all and only admissible contingency plans belong to signal systems.[47] We use this definition of signal system in all that follows.

## 2.2 Emotion Signal Systems

In *emotion* signal systems, the contingency plans link up what are known in the literature as emotion elicitation and emotion expression. *Emotion elicitation* covers the conditions that typically cause emotions, and *emotion expression* is about the sorts of behavior typically caused by emotions.[48] More specifically, the emotion sender's coordination plan can link up emotion appraisals (cognitive elements of identifying emotion-relevant states of affairs in the world) with action tendencies, while the emotion receiver's coordination plan links up action tendencies with other action tendencies. In the literature on emotions, appraisals and action tendencies have received tremendous attention, and there is a robust literature about each one.[49] This is probably the most natural way of setting up emotion signal systems on the Lewis model, but it is not the only one. For example, emotion responses might not be emotion behavior but rather emotion detection states. There are other options as well, but we focus on this basic framework in what follows because it attributes a simple signal system that is roughly the same in all individuals and explains how emotions spread through a population. See Figure 4 for a diagram.

[47] Lewis (1969: 132-133).

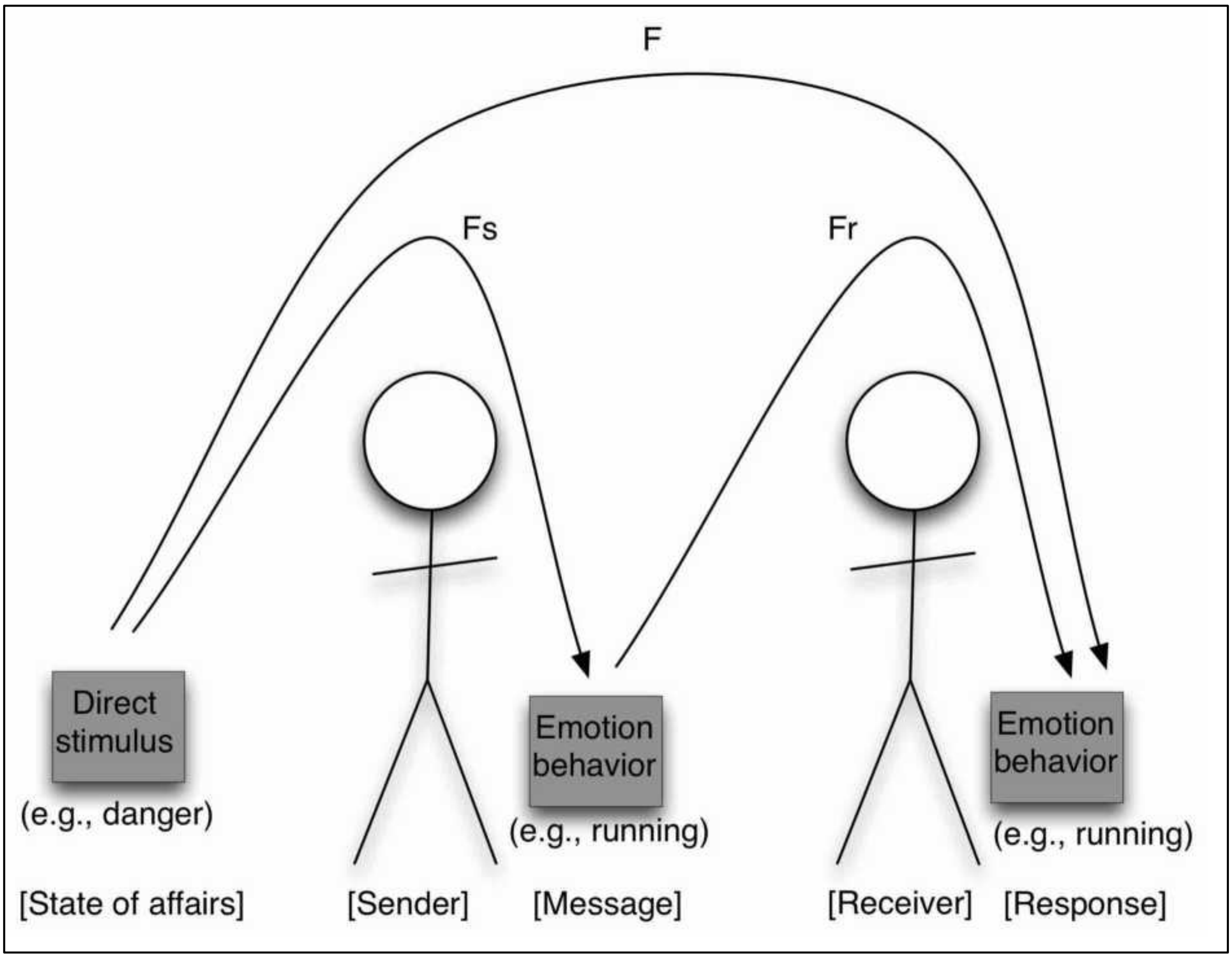

[48] See work in Coan and Allen (2007: 9-168) on the former and in Abell and Smith (2016) on the latter.
[49] For appraisal theories see Scherer et al (2001), Moors et al (2013), and Sander et al (2018) and for action tendencies see Fridja (2007), Hommel et al (2017), Scarintino (2017), and Eder (2017).

**Fig. 4: Emotion Signal System. A stimulus causes the Sender to display emotion behavior (Fs), which in turn causes the Receiver to display emotion behavior (Fr). The model treats these two processes as blackboxes and specifies the conditions under which they are coordinated in the right way to constitute a signal system in which the Receiver acquires information about the stimulus via communication.**

For example, one person, Lauren, sees a bear nearby while on a hike through a forest. Lauren thereby feels fear and begins to display fear behavior. In particular, Lauren drops her hiking pole, turns away from the direction of the bear, and starts running as fast as she can. Lauren's hiking companion, Alex, does not see the bear, but does notice Lauren's fear behavior. Immediately, Alex feels fear as well, even though he does not know what he is afraid of and does not know what Lauren was afraid of. Nevertheless, Alex begins to run too, and he runs away from the bear, who goes on with her day despite the commotion. This is the sort of situation on which theorists who emphasize the communicative function of emotions focus.

In this example case of fear from above, we could think of the contingency plans as follows:

Lauren

If danger is observed, then do fear behavior.

If fear behavior is observed, then do fear behavior.

Alex

If danger is observed, then do fear behavior.

If fear behavior is observed, then do fear behavior.

In these contingency plans, 'fear behavior' is context dependent. Some contexts call for a fight, others, a flight. And in other contexts, the proper fear behaviour might be hiding or displaying submission.

These contingency plans are rudimentary, and we consider more complex models in later sections.[50] Notice that their contingency plans are the same, which makes sense when we are dealing with emotion communication because it is natural to think that almost everyone in the population is equipped with roughly the same emotional capacities. If that is right, then we can think of the *fear signal system* as the pair <Alex, Lauren>, with the understanding that <Lauren, Lauren> is an identical signal system. We want to be clear that this is just a rudimentary model that requires substantial refinement before it could be considered to have enough power and accuracy to be fruitful. For example, it is highly doubtful that humans are set up with independent signal systems for each kind of emotion.[51] Rather, emotion appraisal systems, emotion detection systems, emotion elicitation systems, and emotion expression systems all work together in a complex constellation.

It makes sense to emphasize that a huge component of this signal theory of emotion is *emotion expression*, the formulation into observable behaviour of unobservable emotional states or aspects thereof. There is a vast and thriving literature on emotion expression that was turbocharged by Paul Ekman's work on facial expressions associated with basic emotions, but has exploded to cover many more modalities and kinds of emotions. Their survey "Expression of Emotion: New Principles for Future Inquiry," by Dacher Keltner, Daniel T. Cordaro, Jessica Tracy, and Disa Sauter, explores some of this work. One of their summaries is reproduced as Figure 5.

---

[50] For example, one can accommodate that people do not always feel fear when they detect it in others and people do not always act afraid when they feel fear.

[51] See Schlegel et al (2012) and Shuman et al (2015).

| Emotion | Face/body | Voice | Touch | Music | Dance |
|---|---|---|---|---|---|
| Amused | yes[a,b,d,i] | yes[y,z,bb] | n/a | n/a | yes[gg] |
| Anger | yes[d,w,x] | yes[y,aa,bb] | yes[dd,ee] | yes[ff] | yes[gg] |
| Awe | yes[a,c,d] | yes[y] | no | n/a | yes[gg] |
| Boredom | yes[n] | yes[aa] | n/a | n/a | n/a |
| Confused | yes[n,u] | n/a | n/a | n/a | n/a |
| Contempt | yes[v,w] | yes[y,aa] | n/a | n/a | n/a |
| Content | yes[d] | yes[z] | n/a | n/a | yes[gg] |
| Coy | yes[e,f,g] | n/a | n/a | n/a | n/a |
| Desire | yes[h,i] | no[y] | n/a | n/a | n/a |
| Disgust | yes[d,w,x] | yes[y,aa,bb] | yes[dd,ee] | n/a | yes[gg] |
| Embarrassed | yes[d,i,j,k,l] | yes[y] | no[ee] | n/a | yes[gg] |
| Fear | yes[d,w,x] | yes[y,aa,bb] | yes[dd,ee] | yes[ff] | yes[gg] |
| Gratitude | n/a | no[y] | yes[dd,ee] | n/a | n/a |
| Happiness | yes[i,w,x] | yes[aa] | yes[dd] | yes[ff] | n/a |
| Interested | yes[i,m,n] | yes[y] | n/a | n/a | n/a |
| Love | yes[d,i] | no[y] | yes[dd,ee] | yes[ff] | yes[gg] |
| Pain | yes[o,p,q,r] | yes[cc] | n/a | n/a | n/a |
| Pride | yes[a,i] | no[y] | no[ee] | n/a | n/a |
| Relief | n/a | yes[y,z,aa,bb] | n/a | n/a | n/a |
| Sadness | yes[d,w,x] | yes[y,bb] | yes[dd,ee] | yes[ff] | yes[gg] |
| Shame | yes[d,i,t] | no[y] | n/a | n/a | yes[gg] |
| Surprise | yes[w,x] | yes[y,bb,ee] | no[ee] | n/a | n/a |
| Sympathy | yes[i] | yes[y] | yes[dd,ee] | n/a | n/a |
| Triumph | n/a | yes[y] | n/a | n/a | n/a |

**Figure 5: Emotions and their expression in different modalities.**

The superscripts link to all the research on each of these entries (see their paper for references). As you can see, there is a tremendous amount of exciting work on emotion expression. We return to this topic below in the discussion of Scarintino's theory of affective pragmatics (Section 2.8).

Here is an *objection*: the Behavioural Ecology theory of emotion entails that there are no episodes of emotion expression.

Reply: We follow Scarintino on the nature of emotion expression. Behavioral ecology's argument against emotion expression sounds a lot like the ridiculous arguments given by anti-darwinians (past and present) about how eyes could never evolve. This sort of result cannot be rattled off from the armchair as Fridland does. There are no emotions so there are no emotion expressions. If that sounds plausible then the rest of the book is probably not going to agree with your sympathies, but the quantitative theory will still make good solid predictions. If emotions are real enough to support the distinction between having unobservable aspects of an emotion episode and expressing that emotion in physical behaviour, then we are in business.

Another point in reply is that behaviorial ecology can be purged of these eliminativist aspects in Fridland, and even combined with a readout view. That is exactly what Buck's new book does. He offers basic-emotion theory of certain emotions and behavioural ecology view of others. All of these emotions, for Buck, involve unobservable aspects and observable aspects, like facial expressions.[52] Thus, it makes sense to think that the criticisms of emotion expression from Fridland are idiosyncratic and not representative of the main thrust of this tradition.

## 2.3 Are Emotion Signal Systems Conventional?

One topic that comes up frequently is: according to the signal theory, are emotions conventional? It looks like the answer is clearly No. However, some theorists have claimed that emotion signal systems are conventional.[53]

Given what Lewis says about conventional signal systems vs signal systems in general, it is clear that there is no convention associated with many of the most familiar emotion signal systems. For Lewis, if a signal system is to be conventional, then the contingency plans in question must be conventions, where conventions are certain kinds of regularities in behavior. For Lewis, for a regularity to be a convention, there has to be common knowledge of the regularity, it has to be preferred by the members of the population, and there have to be viable alternatives to the regularity in question.[54]

The Paul Revere example is illustrative here. The Sextant and Paul Revere could have chosen different contingency plans to have the same effect (e.g., if two lanterns means no redcoats, and no lanterns means the redcoats are coming by sea). But most of these conditions fail in the case of emotions. For example, many emotional capacities are involuntary and hard-wired into our brains either by evolution or socialization. Moreover, there is no evidence to suggest that humans all know how our emotion communication systems work. Indeed, despite much research on emotion regulation, it seems like the apocryphal idea that "you cannot control your emotions" remains widespread. Finally, there is no reason to think that people prefer their emotion contingency plans to other options. Many people would probably choose not to feel negatively valenced emotions like sadness, loneliness, guilt, shame, and embarrassment at all. Therefore, emotional signal systems fail to satisfy many of the conditions necessary for being conventional. Perhaps other aspects of emotional communication are conventional. For example, regularities of emoji usage in social media is probably conventional. However, it is a mistake to think that all or even most emotion signal systems are conventional.

---

[52] Buck (2014).
[53] See Ross and Doumachel (2004).
[54] Lewis (1969: 58-82).

Lewis's theory of signal systems has an explicit tie to game theory because Lewis designs his account of signal systems so that each signal constitutes a proper coordination equilibrium in a *coordination problem*, which is defined by Lewis as a situation in which two or more agents need to perform actions but the outcomes of their actions depend on which actions are performed by others. A *coordination equilibrium* is a collection of action choices such that no one involved could have been better off had anyone chosen differently, and a *proper* equilibrium is where alternative choices lead to degradation for someone.[55] Although we do not pursue it here, the tight connection between signal systems and the kinds of strategic interaction studied in game theory adds considerably to the fruitfulness of the signal theory of emotion communication.

We have given a proper theory to the suggestion that emotions function as signaling systems. This theory has considerable explanatory power and significant applications. Information theory allows us to quantify certain aspects of emotional systems across groups. For example, entropy, error rates, and redundancy in emotional systems across groups can be accurately measured and utilized (see Chapters 3-6). When applied to social media on the internet, we see a new manifestation of the same emotional signaling system that evolved over millions of years. The insights gained from thinking of emotions as Lewisian signaling systems help us understand the effects social media have on us (e.g., vast changes in the speed and reach of our emotional signaling systems).

---

[55] See Lewis (1969: 5-51); see Cain (2017) for a contemporary survey of game theory and coordination problems.

## 2.4 Ekman's Seven Kinds of Information in Emotion Communication

One important issue that remains controversial is: what exactly is communicated in episodes of emotion communication? The question is, what is the message in the signal? One can imagine that, in a case of fear for example, the message is "there is danger!" This answer would be consistent with the normal conditions in which it would be fitting to feel fear: that is, in cases of danger. However, it seems equally correct to describe the message as "be scared!" or "run!" or maybe "I'm scared!" or "you should be scared".[56] Or perhaps the message is an appraisal, in the sense use by appraisal theorists.[57] Maybe the message is indeterminate between these options. It is not clear at all how to adjudicate this matter.

What does it matter? If we want to understand emotions – admitted by everyone to be an essential part of the human experience – then we have to understand how they are used to communicate. This function is too crucial and too omnipresent to not shape almost every aspect of our emotional lives. So if our emotional lives are shaped to a major degree by the emotion messages that we send and receive all day, every day, then it seems that the nature of these messages matters a great deal. If the messages are about the world, then we are swimming in a sea of representations. If instead emotion messages are commands, then we are manipulating and being manipulated. If the messages are indeterminate, then that matters too. And if there are different kinds of messages then we can track the differences and see how they correlate to other variables that scientists in the area are already tracking.

---

[56] See Jack and Schynes (2015) and van Doorn, E. A., van Kleef, G. A., and van der Pligt, J. (2015) for some of these options.
[57] Scherer et al (2001).

One important proposal for the message in emotion communication from Paul Ekman is that emotions communicate at least seven distinct kinds of information.[58] They are:

1. The antecedents, the events that brought about the expression;
2. The person's thoughts: plans, expectations, memories;
3. The internal physical state of the person showing the expression;
4. A metaphor;
5. What the expresser is likely to do next;
6. What the expresser wants the perceiver to do;
7. An emotion word.

Some of these can seem confusing out of context, but surely 1, 3, 5, and 6 are familiar. Ekman is focusing on distinctively human emotion communication, so we might dismiss 4 and 7 in general. 2 seems to confuse the emotion involved in communication with other information one might be able to infer. Recall that this is a central distinction in van Kleef, described in Section 1.10. Scarintino has a nice discussion of Ekman's views.[59]

We think that all these sorts of information are often available in emotion communication, but the question we aim to answer is: which one is central and which are auxiliary? That is, what is the message in emotion communication and what else can be inferred?

## 2.5 What is the Message?

It is a common theme in animal communication literature and emotion communication literature (remember that there is almost no overlap between these) to inquire into the nature of the message. Does these messages have conceptual content like words or sentences? If so, what kind of contents do these messages have? In particular, what kind of content is conveyed when the expression of emotion by one animal is detected by another? We end up defending the view that emotion communication messages have as their content the fittingness conditions of the emotion in question. This view has also been defended in various forms by Ekman, Scherer, and Scarintino, but our formulation and defence of the view are novel.[60]

[58] Ekman (1997).
[59] Scarintino (2019a)

We first present an argument, based on work by Lewis on signal systems, for the conclusion that the message in emotion communication is *indicative* rather than imperative and that in general the message in emotion communication is the *fittingness conditions* for the emotion in question. It is traditional to define *fitting* emotions as those that are sensitive in the right way to situations in the world. For example, fear is fitting exactly when it is about something actually dangerous; fear when one is not really in danger is unfitting.

Lewis's theory of signal systems has the power to distinguish indicative messages (e.g., "there is danger") from imperative messages (e.g., "be afraid!"). This is the familiar distinction between making declarations about the world versus issuing commands to others. In explaining the distinction, Lewis writes: "A contingency plan may or may not be *discretionary*; that is, it may or may not require an agent to deliberate about which course of action would be best for himself and his partners."[61] Lewis then gives the following examples of discretionary contingency plans for the sextant and for Paul Revere:

*Sextant´*

> If it seems best that Paul Revere should go home, then hang no lantern in the belfry
>
> If it seems best that Paul Revere should warn the countryside that the redcoats are coming by land, then hang one lantern in the belfry.

[60] Scarintion (2019a) writes, "Ekman (1997, p. 318) characterized emotional expressions as carrying "information about antecedents,"17 and (Scherer, 1988, p. 96) referred to emotional expressions as "symbols" of the "emotion-eliciting event"." And Scarintino himself endorses this kind of message as one of several emotional speech acts. We think the fittingness conditions for the emotion is the only message, but we defend that against Scarintino in Section 2.8. See also Rendal et al (2009) for a criticism of the signal view from the "manipulation" camp in animal communication science; see Krebs and Dawkins (1978) for the classic statement of the view that manipulation rather than information should be the basic explanatory unit of animal communication. See Stegmann (2014) for a nice collection of contemporary views from the "information" camp. We do not attempt any kind of general reply to the manipulation camp in this book.

[61] Lewis (1969: 144).

> If it seems best that Paul Revere should warn the countryside that the redcoats are coming by sea, then hang two lanterns in the belfry.

*Revere´*

> If no lantern is observed in the belfry, do whatever seems best on the assumption that the redcoats were observed staying home.
>
> If one lantern is observed in the belfry, do whatever seems best on the assumption that the redcoats were observed setting out by land.
>
> If two lanterns are observed in the belfry, do whatever seems best on the assumption that the redcoats were observed setting out by sea.[62]

The link between discretionary contingency plans and the imperative/indicative distinction comes in the following quote: "I suggest that if [*Fr*] is discretionary and [*Fs*] is not, then $\sigma$ is indicative; if [*Fs*] is discretionary and [*Fr*] is not, then $\sigma$ is imperative; if neither or both are discretionary, then $\sigma$ is neutral."[63] Hence, the distinction between imperative and indicative messages is explained in terms of which contingency plan is discretionary.

This is a nice idea, but right away we can see two problems with it. First, every message in a given signal system has the same status (indicative or imperative) according to Lewis's definition. But that is not right. Surely there are signal systems with both kinds of signals. Moreover, whether a contingency plan is discretionary is graded, so it makes sense to think that the distinction comes in degrees. Hence, an improved version of Lewis's account would be that: to the extent that the transition from states of affairs to messages is more discretionary than the transition from messages to responses, the more indicative the message. To the extent that the transition from states of affairs to messages is more discretionary than the transition from messages to responses, the more imperative the message. This account preserves Lewis's insight and his example, but is more general and subtle.

---

[62] Lewis (1969: 145-6).
[63] Lewis (1969: 144).

Recall that in a signal theory of emotion communication, emotion expression is the transition from state of affairs to message, and emotion elicitation covers, among other things, the transition from message to response. If we apply this desideratum to emotion messages, then the important questions are:

How much latitude do we have in emotion elicitation?

How much latitude do we have in emotion expression?

We think the answers are clear. We have to deliberate more about how to express our emotions than we have to deliberate about how to appraise our situation and detect the emotions expressed by others. Think of the case of fear again. In the example above, each person runs from the bear, but there are at least three major categories of fear expression – fight, flight, and hide – and many others as well. For example, if Lauren and Alex had deliberated a bit more, they might have remembered that one is not supposed to run from a bear, nor is one to hide or fight. Rather, one is supposed to calmly back away. Other emotions seem to have even more deliberative expression options; consider anger, sadness, joy, and disgust alone.

The process going from the *state of affairs* (through perception by the agent, through appraisal, through eliciting the emotion, through expressing that emotion) to *message* allows for less discretion than the process that leading from the *message* (through emotion recognition, through eliciting the emotion in the agent, and through expressing that emotion) to the *response*. Ultimately, we have more degrees of freedom in recognizing emotions than we do in appraising.

If that is right, then Lewis's theory implies that emotion messages are more indicative than imperative. That is, emotion messages describe states of affairs rather than prescribe behavior. Still, that leaves quite a few options including, from the example above, "there is danger", "I am scared", and "you should be scared" as glosses on the content of Lauren's fear behavior as an emotion message. Which of these indicative messages is the right one?

Further developments in the characterization of signal systems lead Lewis to defend the view that every message in a signal system has the semantic content of its truth conditions, where the truth conditions are those under which the content is true (if indicative) or satisfied (if imperative). He writes:

> It is not at all necessary to confine ourselves to conventional signalling systems in defining meaning for signals. Consider any signalling system <Fc, Fa> for a signalling problem S, whether or not <Fc, Fa> happens to be conventionally adopted in any population. If Fa is discretionary and Fc is not, the signals of <Fc, Fa> are indicative. Consider a signal m and the state of affairs s mapped onto m by Fc. We can call m a *signal in <Fc, Fa> that s* holds; and we can say that m *means in <Fc, Fa> that s* holds. … We would expect that by giving the meanings of indicative signals, we give their truth-conditions. And so we do. Let s be an indicative signal that s holds in signalling system <Fc, Fa> for signalling problem S. Then we can call s true in any instance of S in which s does hold and false in any instance of S in which s does not hold.[64]

Although it might sound odd to talk of truth in the case of emotions, Lewis is clear that he intends it to apply to any signal system.[65]

Thus, in the case of our fear example, Lewis's theory implies that the message is "there is danger". And in general, emotion messages are not about emotions themselves. Rather, they are about their fittingness conditions. For example, when communicating disgust, the message is "there is something toxic," when communicating anger, the message is "someone committed a wrong," etc.

This is a significant result in that it opens the door for techniques and ideas from semantics to be applied to emotion communication systems. If emotion messages have truth conditions that are the fittingness conditions of the associated emotion, then we have a much clearer picture of how emotion communication works.

---

[64] Lewis (1969: 147).
[65] Lewis (1969: 147-152). See de Sousa (2011) for a sympathetic account of emotional truth.

Compare the signal theory's account of emotion messages to a natural alternative, namely that *appraisals* are the messages in emotion communication.[66] By 'appraisal', we mean what those in the tradition of appraisal theory mean: an evaluation by an animal (or at least a part of an animal's cognitive system) of how a given situation impacts the animal's interests. Fundamentally, an appraisal is an evaluation of how one thing affects another.

Appraisals are not good candidates for emotion communication messages because they are essentially relative to each animal's interests. For example, Animal 1 sees a snake and appraises the snake as a significant threat to Animal 1's interests. Animal 1 then feels fear and expresses that fear in emotion behavior. Animal 2 perceives the emotion behavior of Animal 1, and then decodes it to detect the emotion felt by Animal 1. According to the appraisal view, what is detected is that Animal 1 feels fear. Animal 1's appraisal, namely that the snake is a significant threat to Animal 1's interests, is not necessarily relevant to Animal 2. Interests can and do differ from one animal to another and from one species to another. Lions and wildebeests can communicate emotionally even though they have vastly different interests.

Despite the initial implausibility, we can push on and discover further difficulties that prove decisive. Imagine Animal 2 goes on to feel fear as a result of detecting it in Animal 1. Animal 2 expresses fear in its behaviour, which is detected by a third, Animal 3. What Animal 3 detects is that Animal 2 feels fear. According to the appraisal view, the message between Animal 2 and Animal 3 is Animal 1's appraisal. It is implausible to suppose that somehow Animal 2 conveys information about Animal 1's appraisal to Animal 3, even though Animal 3 might have no idea Animal 1 exists. Because they are relative to each animal's interests, appraisals are the wrong kind of thing to be inter-animal messages.  Emotion communication systems are vastly more powerful and effective if the emotion message is independent of the messenger.

## 2.6 Signal Theory vs Contagion Theory

[66] See Section 1.2 for discussion of the appraisal theory.

Recall from Section 1.8 that the dominant theory for how emotions transfer from animal to animal is the contagion theory, which treats emotional states as analogous to pathogens and emotion transfer as infection. Several quantitative models have appeared that borrow from theories of heat transfer in thermodynamics or from models of genuine infection in epidemiology. We have encouraged the reader to think that contagion theory is based on a mere metaphor, outdated, makes poor predictions, and should be replaced by the signal theory of emotion communication.

The signal theory makes certain predictions, for example, that we ought to find emotion expression features and emotion detection features evolving simultaneously. And that is what experiments have shown after just scratching the surface [n] of this intriguing topic.

The contagion theory, on the other hand, predicts the opposite: that people should evolve defenses against emotional contagions: an emotional immune system, as it were. Indeed, we might expect to find evidence of arms races where emoters evolve better ways of defending from emotional infections while at the same time evolving better ways of infecting others. There does not seem to be evidence of any of this yet discovered.

Another problem is that the signal theory of emotion communication focuses on the process by which one animal has an emotion and another comes to detect that emotion in the first animal. The second animal, after detecting the emotion, might undergo any number of processes including the elicitation of its own emotion episode. Hence, the signal theory focuses on the *transmission process* (from (i) having an emotion to (ii) expressing it to (iii) detecting that emotion by another) is a proper part of the contagion process (from (i) having an emotion to (ii) expressing it to (iii) detecting that emotion by another animal to (iv) having that same emotion in the second animal). Thus, the contagion theory is more coarse grained than the signal theory, which focuses on more narrow phenomena. As a result, it is easy for the signal theory to make sense of emotion contagion: emotion contagion occurs when one animal communicates an emotion to a second animal and that same emotion is elicited in the second animal. However, the contagion theory cannot explain emotion communication. What exactly is the infection process by which one animal comes to be aware of an infection in another without thereby being infected? There is no such thing.

Another problem is that animals keep track not only of the emotions detected in others, but they keep track of which other animals have which emotions. Animal 1 might detect anger in Animal 2, and Animal 1 might detect joy in Animal 3, and as a result infer that Animal 3 was trying to make Animal 2 angry. That is not how infection works. We have no infection registers, which would track who infected us with which pathogen. So again, the contagion model is simply and obviously inadequate.

There are objections to the emotional contagion theory (Wrobel and Imbir) from the fact that social factors influence emotion transfer but not infections. But signals obviously do. So the signal theory solves major outstanding problems for the contagion theory.

The next problem is that contagion rarely features feedback, but emotions communicated are almost always part of evolving feedback loops. There is no good way to explain or simulate this phenomenon by using existing mathematical models of emotion contagion because once infected, the animal often either dies or has immunity. Moreover, animals do not pass different infections to communicate.

In addition, contagion theorists are impotent to explain cases of emotion transfer that are heterogeneous. For example, when one animal feels fear, expresses that fear, another animal detects that fear and goes on to feel fear itself, that is classic case of emotion contagion. Call this *homogeneous emotion transfer* because the emotion felt by the two animals is the same. In other cases, the fact that one animal has an emotion makes another animal have a different emotion (e.g., we might feel joy after scaring our son in a game). These are cases of *heterogeneous emotion transfer*. The contagion theorists do not focus on this phenomenon, but when they discuss it, they tend to say shockingly implausible things: for example, that all cases of emotion transfer are homogeneous, but it happens so quickly we do not notice it. This is radically implausible, goes against the empirical evidence we have (we are aware of no extant evidence to support it). The signal theory of emotion communication can handle these cases easily without implausible posits or assumptions. In both heterogeneous cases and homogeneous cases, emotion communication occurs; the difference is what happens after emotion detection. Our model of communication plus elicitation explains these cases flawlessly. The contagion theory does not.

Furthermore, capacities to detect emotions might be often less precise than capacities to feel emotions. For example, one might feel any of 100 intensities of fear, but be able to detect only 10. That phenomenon can only be part of the model if we focus on the emotion signals as the basic unit of emotional communication rather than emotion contagion. We see no way to explain anything like this with a contagion model. Positing infection magnitudes for how infectious an emotion episode is or how "bad" a particular infection is will not work. Slight expressions of emotion might be easier to interpret than dramatic ones, so even this modification to the contagion theory cannot explain this basic aspect of emotion communication. It might make sense to think of strong emotions as big infections and weak emotions as minor infections, but that is not point. The precision of emotion intensity and emotion detection precision would be like someone who is only able to give infections of certain sizes, and like someone who could only get infections of certain sizes. Infection transmission does not come in distinct levels of precision like this.

Recall that the defender of the contagion theory is likely to respond that emotions *aren't literally* contagions, emotions just behave like them in some situations. It is worth repeating the reply: *emotions literally are signals*. They behave like signals because they are signals. Using the contagion theory to understand complex emotion networks is like using a petri dish to simulate a microprocessor. That might work for some features, but overall, it isn't up to the task.

Throughout the book we return to this contrast between (i) the contagion theory and (ii) the signal theory of emotion communication presented here along with the quantitative theory of emotion information in the next four chapters. As we develop the quantitative theory of information in emotion communication, it will become clearer and clearer how inferior the contagion theory is for the contemporary research frameworks that have sprung up around emotions (e.g., information theoretic animal communication models, Bayesian Theory of Mind (BToM) models of emotion attribution, computational models of emotions and related processing in cognitive neuroscience, and machine learning algorithms in artificial emotion recognition systems). All of these "run" on probability distributions and ways of simulating them. Probability distributions are also the foundation for the theory of emotion information, but they have nothing to do with the contagion theory.

## 2.7 The Problem of Communication with Unfitting Emotions

The account of emotion messages one gets from the signal system theory faces a serious threat from some results in the social psychology literatures. van Kleef emphasizes that people often communicate messages other than the fittingnesss conditions of the emotion in question. For example, humans often use anger to communicate social dominance and hierarchy relationships.[67] The problem is how to accommodate this phenomenon in a theory of emotion communication. If anger sometimes communicates that one has been wronged (the standard fittingness conditions) and sometimes communicates that one is dominant (*not* the standards fittingness conditions), then how are we to make sense of this? We call this the *problem of emotion communication*.

Moreover, a phenomenon known as *display rules* insures that there is a serious problem for the signal theory as presented above.[68] These are social rules that differ from culture to culture and place constraints on the kinds and degrees of emotions that may or must be expressed in certain situations. Because of the extent and significance of display rules, one can

[67] van Kleef (2016: 103)
[68] See van Kleef (2016: 67-77); see also Matsumoto et al (2008) for a cultural survey and see Chaplin (2014), Richard and Converse (2015), Grandy and Sayer (2019), and Horner and Akiva (2019) for some recent experimental results.

imagine plausible situations in which people display some emotion behavior in some situation even though the emotion is unfitting.

A social regularity like this, according to Lewis's theory of signal systems, provides fittingness conditions for the emotion involved. Or, to put it another way, it seems like Lewis's theory forbids regular emotion displays that are emotion messages in an emotion signal system that are routinely or always unfitting. The existence and prevalence of display rules guarantee that that this looks like a major problem.

It seems clearly false to say that the fittingness conditions for anger include the extent to which one plays a dominant role within their particular population. We have failed to find anyone who says such a thing and it flies in the face of the standard account of the fittingness conditions for anger, which are that one has been wronged in some way or had one's goal thwarted.[69]

Anyone who takes the phenomenon highlighted by van Kleef seriously faces the problem of emotion communication. We explore several options for addressing this problem and recommend the idea that something like conversational implicature is the right explanation. In *conversational implicature* cases, one intentionally violates the rules of language to communicate something in a non-standard way. Likewise, in cases of unfitting emotions, animals violate the "rules" of emotions (e.g., one feels fear appropriately if and only there is danger), to communicate something (e.g., dominance) in a non-standard way.

One upshot is that emotion communication systems display something like a semantics / pragmatics distinction familiar in linguistics and philosophy of language. There is the literal content of an emotion message (the semantics), and there is what that emotion message can be used to convey (the pragmatics). This solution preserves the insight into the content of emotion messages (i.e., their fittingness conditions) while at the same time accounting for the phenomena highlighted by van Kleef and other social psychologists.

---

[69] See work in Cherry and Flanagan (2017).

A more familiar response to the problem might be to try to be more careful about the emotion episodes in question. For example, perhaps it is not anger behavior that communicates dominance, but rather something related like indignation behavior.[70] Being more careful about the fine-grained identification of the emotions involved might very well solve the problem in some cases. However, this is unlikely to be a general solution as the problem occurs across a wide range of emotions and messages.

Instead, one might try to solve the problem of emotion communication by appeal to some sort of semantic solution like ambiguity. That would be to say that there are multiple kinds of each emotion in question. In our example, it would be there is $anger_1$, which has the normal fittingness conditions and $anger_2$, which has fittingness conditions that involve dominance relations. Presumably people would tell the difference between $anger_1$ and $anger_2$ in the same way that people tell the difference between 'bank' meaning *financial institution* and 'bank' meaning *riverside*. Yet this solution seems hopelessly ad hoc. Without any sort of independent evidence for such a distinction, the ambiguity solution is not plausible.

## 2.8 Scarintino's Theory of Affective Pragmatics

Andrea Scarintino has recently proposed a framework for understanding emotion pragmatics, and it relates directly our theory of emotion signal systems, our claims about the meaning of emotion messages, and the problem of emotion communication. Scarintino makes several important points:

1. Emotions function as a system of communication, via emotion expression, emotion recognition, and emotion elicitation.
2. Within that system of communication, emotion expression can play several communicative roles and these roles are analogous to speech acts in linguistic communication.

[70] See Miceli and Castelfranchi (2017).

It should be obvious that we thoroughly agree with Scarintino on Point 1, but we are much more cautious about Point 2. Indeed, the theory Scarintino advocates includes much more than we need to solve the problem of emotion communication. Scarintino writes:

> TAP [Theory of Affective Pragmatics] has two main parts. The first is a distinction between three things we do when we express emotions that replicates, mutatis mutandis, Austin's (1962) distinction between locutionary, illocutionary, and perlocutionary acts:
>
> 1. Emotional Expression: The nonverbal behavior of expressing emotion E.
> 2. Communicative Moves: What one does in expressing emotion E.
> 3. Communicative Effects: What one does by expressing emotion E.
>
> The second part of Affective Pragmatics is an analysis of these three dimensions of emotional communication, namely, an analysis of the nature and function of emotional expressions, the nature and function of communicative moves, and the nature and function of communicative effects. (175).

We agree with Scarintino about the importance of explaining emotion communication systems in terms of their components, and distinguishing emotion expression, episodes of emotion communication, and the effects of emotion communication on the receiver. One of the most important effects we highlight is whether the receiving animal accurately recognizes the emotion being expressed by the transmitting animal. Another important effect is whether successfully detecting an emotion in the transmitting animal causes the receiving animal to have an affiliated emotion. An affiliated emotion might be the same as the one detected (e.g., when fear detection causes fear—for example, hearing your significant other scream for help) or a different emotion (e.g., when sadness detection causes anger—for example, seeing how upset your child is when no one shows up to her birthday party).

However, we disagree with Scarintino's characterization of emotional expression as essentially non-verbal. So much of the literature on emotion expression includes the way emotions are expressed in spoken and written language that it seems like a mistake to only focus on non-verbal emotion expression.[71] We understand the Scarintino is interested in how linguistic communication developed from non-linguistic communication and he sees emotion communication as an important precursor, but that is not a good reason to ignore all the ways emotions are expressed in linguistic communication. Indeed, with Scarintino's account of emotion expression it would be impossible to make sense of all the contemporary techniques for sentiment analysis in artificial intelligence, which we emphasize in Chapters Six, Seven, and Eight. Another reason to reject Scarintino's restriction to non-verbal emotion expression is that no account like his could be applied to social media to arrive at novel emotion security systems for social media, which is our main task in Chapter Eight.

Probably the most controversial and innovative aspect of Scarintino's theory of affective pragmatics is the taxonomy of communicative actions that can be performed by episodes of emotion expression. Scarintino writes:

> I argue that emotional expressions make four types of communicative moves possible, defined by their communicative points and directions of fi t (the subscript EE stands for "emotional expression"):
>
> - Expressives$_{EE}$ have the communicative point of expressing the signaler' s emotions by means of natural information transfer, and they have no direction of fit.
> - Imperatives$_{EE}$ have the communicative point of trying to get the recipient to do something by means of natural information transfer, and they have a recipient-based world-to-mind direction of fit because the recipient is responsible for changing the world so as to fit the content.
> - Declaratives$_{EE}$ have the communicative point of representing how things are in the world by means of natural information transfer, and they have a mind-to-world direction of fit because their content aims to fit what the world is like.

[71] Cowan (2018) for example.

- Commissives$_{EE}$ have the communicative point of committing the sender to a future course of action by means of natural information transfer, and they have a signaler-based world-to-mind direction of fit because the signaler is responsible for changing the world so as to fit the content. (176).

Notice that the four categories mirror four major categories of speech act by the same names (without the subscript).

The signal theory of emotion communication presented earlier in this chapter is not in tension with Scarintino's theory of affective pragmatics, except insofar as the latter is committed to non-verbal emotion expressions only. That we think is just bizarre (Scarintino admits that it does make his view different from the norm, but he thinks this is warranted by how well it explains how linguistic communication evolved from non-linguistic communication), but fortunately it is optional.

One important point in common is that there is something like a semantics/pragmatics distinction in emotion communication. However, Scarintino does not really explain how to understand the semantic content of emotion messages. That is, he does not seem to provide an explicit answer to our "what is the message?" question. Recall that our answer is that an emotion message's semantic content is its fittingness conditions (e.g., for fear, the fittingness conditions are danger, so in fear communication, the message communicated is "there is danger").

Even without an explicit account of the semantic content in emotion communication, Scarintino's theory entails that certain emotion communication systems display certain pragmatic features. With that, we agree. And his theory has the power to reply to the problem of emotion communication with unfitting emotions: *normal expressions of emotion and unfitting expressions of emotion involve different forces*. This tidy solution demonstrates the power of Scarintino's theory and its relevance (despite the mostly negative reception it received from its commentators).[72]

[72] See Bjornsdottir & Rule (2017), Fischer & Sauter (2017), Fischer (2017), Moore (2017), van Kleef (2017), and Scarintino's reply (2017b), but skip the pathetic Fridlund (2017), which is an unfair and unhinged rant that has no place in academia.

This elegant solution to the problem of unfitting emotion communication is, however, unavailable to us because we disagree with crucial aspects of Scarintio's theory: *which* pragmatic phenomena are displayed. He thinks there are at least four distinct emotion communication acts or expression forces. Force is the kind of speech act when we are talking about linguistic communication. In the case of emotions, the force of the emotion expression – according to Scarintino – helps us understand the form of the message. Assertive force, for example, delivers a message that is a declaration, which is a statement about the world, while imperative force delivers a message that is a command, which tells someone to do something. The theory of affective pragmatics says that expressions of emotion have one of the above four forces

As examples, he offers the following:

- *Expressive force*: "Bared teeth and clenched fists (in the right context) express anger; crying (in the right context) expresses sadness; upper eyelids raised and jaw dropped open (in the right context) express fear"

- *Imperative force*: "To express anger is to demand that the recipient stops what he or she is doing and takes the signaler more seriously, to express fear is to demand that the recipient helps and protects the signaler, to express happiness is to demand that the recipient celebrates a success with the signaler, to express disgust is to demand that others stay away from a poisonous substance, and so on."

- *Declarative force*: "To produce a representation through a communicative move only presupposes that I voluntarily or involuntarily provide you through my emotional expression with natural information about my being in an emotional state that represents the world as being a certain way. I should note that emotional expressions do not express beliefs despite the fact that they represent states of affairs. If I assert that the picture of a snake is dangerous, I have expressed my belief that it is. But if I respond to the picture of a snake with a fear expression, I have not expressed my belief that it is dangerous."

- *Commissive force*: "Bared teeth and and clenched fists (in the right context) convey the signaler's commitment to aggressive action; crying (in the right context) conveys the signaler's commitment to disengage with the world; upper eyelids raised and jaw dropped open (in the right context) convey the signaler's commitment to submit or escape."

Although we applaud the idea that emotion communication systems can display pragmatic phenomena, we disagree with Scarintino's claim that emotion episodes have pragmatic forces like expressives, imperatives, declaratives, and commissives. We point out two problems with Scarintino's theory of emotion forces and the we offer a more conservative explanation of the same phenomena.

One obvious problem is that the expressive force and the declarative force cannot be distinct according to the Lewis argument we gave in section 2.6. That is, we argued that the message in emotion communications is the fittingness conditions for the emotion in question; when communicating fear, the message is that there is danger, not something idiosyncratic about the emotional state person in question or anyone else. Hence, there is no purely expressive emotion behavior that is genuinely communicative. We do not want to say that every emotion expression is involved in communication. This extreme claim is not needed for our view at all. But among those emotion expressions that do serve a communicative function, there are none that are purely expressive in Scarintino's sense. If we use his example that crying expresses sadness, then we can say that when someone cries from sadness, they are crying about something. Sadness is often said to have loss or disappointment as its fittingness conditions. Hence, the message communicated in this example is "loss occurred" or "Disappointment happened". Thus, even in Scarintino's cases of expressive force, the message is exactly the same as in his cases of declarative force – the message is just the conditions of fit for the emotion in question. The point of this objection is that, at the very least, Scarintino's typology of emotion forces is incoherent and needs to be revised in light of the expressive force = declarative force argument just given.

The second objection is stronger. In the case of linguistic communication, which is the model for Scarintino's emotion forces, no utterance has more than one force. That is, if some utterance is declarative, then it is not merely expressive, it is not commissive, and it is not imperative. Likewise for the rest – each excludes the others. In other words, this classification scheme is supposed to be *exclusive*. But a little thought suggests that is not the case in practice. When Scarintino defends the commissive force, he says that, because of the connection between emotions and action tendencies, expressing an emotion frequently provides someone perceiving that emotion expression with information about what are likely courses of action. We agree that action tendencies are essential components of many important emotions, but that entails that most expressions of emotions are going to have commissive force. But most expressions of emotion are going to have expressive force – if some animal is expressing an emotion, then anyone paying attention receives the information that that animal is experiencing that emotion. Again, we already argued that all emotion messages have as their content the fittingness conditions of the emotion in question; hence, every emotion expression has declarative force.

Finally, the imperative force is going to be present in many emotion expressions as well because of the connection between emotions and desires or goals, which is emphasized by appraisal theorists, who highlight the importance of appraisals (i.e., specifying the relevance of some event for one's goals and interests). The phenomenon of "reverse appraisal" is an important topic in the neighborhood. That is people and other animals frequently infer how an animal appraised a situation based on the way that animal expressed its emotional state. And from these reverse appraisals, one can frequently infer that the person having the emotion wants one to do something in particular. For example, a person crying from sadness might want someone to comfort them, and a scared person might want to be comforted as well. How do we, the audience figure this out? Here is Scarintino's suggestion: when the sad person cries, the crying has a certain pragmatic property – the property of having indicative force – and we understand that because it has this force, we are expected to do something like comforting the person. This explanation is genuine, but costly in that it posits complex emotion forces that are expected to be displayed and identified by everyone involved in the emotion communication.

Here is another explanation: we figure out that sad people tend to appreciate being comforted because we have a Theory of Mind (ToM). Recall the summary of ToM from Section 1.7. It is because we understand the relationship between the mental state of being sad and the mental state of desiring comfort that we comfort people who are sad. How costly is this explanation? Not costly at all since we already know that we need ToM to make sense of emotion attributions at all. We hypothesize that artificial agents (modeled by Partially Observable Markov Decision Processes using Bayesian networks) could figure out that sad people tend to want to be comforted simply through normal BToM computational modeling techniques. These ToM models would explain the same phenomenon that Scarintino aims to explain with the Theory of Affective Pragmatics, but without the additional cost. We hypothesize that BToM models could predict a desire for comfort without any kind of emotion forces at all. Recall Ong et al distinguish a number of kinds of reasoning associated with attributing emotions, and one of them is the probability that the target has a certain desire given that the target has a certain emotion (i.e., $p(d|e)$, see Section 1.8). This is exactly the kind of inference or reasoning involved in these situations.

Overall, we have three criticisms of Scarintino's Theory of Affective Pragmatics: (i) expressive and declarative forces would have to be the identical, (ii) the emotion forces should be exhaustive but they are not – just about every emotion expression will have all four forces,

(iii) and there is a better explanation for all Scarintino's target phenomena, namely the signal theory of emotion communication together with ToM.

## 2.9 Emotion Implicature

Our solution to the problem of emotion communication that preserves the intuitive fittingness conditions for emotions used in those problematic communications and provides an elegant account of the message in these cases that is consistent with Lewis's theory of signal systems. We do not know whether this solution is right, and we make no effort to defend it, but seems to have a better contrast of costs and benefits than the two options canvassed in the previous sections.

We have emphasized that it makes sense to think of emotion messages as having semantic features because of their role in emotion signal systems. Emotion signals have truth conditions in Lewis's sense, and this should not be controversial. In this section, the solution we present implies that emotion messages can be *pragmatically conveyed* even if they are not *semantically expressed*. Something is semantically expressed if it is part of the truth conditions of the message. However, signal systems can be utilized to do other things. In particular, messages with certain truth conditions can be used to get across other information if the agents in question cooperate in the right way. This other information that goes beyond what is semantically expressed, is what is pragmatically conveyed.

One major phenomenon studied by linguists and philosophers of language in the category of pragmatics is conversational implicature. This occurs when a speaker intentionally violates one or more norms of conversation. Paul Grice's account has become standard; it emphasizes four rules: provide the right *Quality* of information, provide the right *Quantity* of information, provide *Relevant* information, and provide information in the appropriate *Manner*. If a speaker says something that obviously violates one of these rules and clearly intends the audience to notice the violation, then the audience tries to figure out what the speaker might be trying to express. Speakers can express all sorts of things in this way. Grice's famous example is a professor who writes a recommendation letter praising the candidate for utterly minor achievements (e.g., speaking clearly). By noticing that the professor is intentionally violating one of the rules (e.g., Relevance), the audience realizes that the professor is trying to convey that the candidate is not good without saying it explicitly.[73]

We contend that emotion signal systems can be used in a way that is very much like conversational implicature. In these cases, emotion messages that are obviously unfitting can be used to convey information without having it be a part of the fittingness conditions of the message in question. This phenomenon is rather different from cases where an emotion message is unfitting but intended to convey its usual message, as when someone feigns fear in an attempt to scare someone else. In *emotion implicature*, as we might call it, someone displays emotion behavior that is obviously unusual in some way so as to convey some other message. In the cases we care about in this Chapter, they are unusual in that they are unfitting. In the cases of anger van Greef describes and others like them, a person displays anger behavior in a situation where it is obviously unfitting so as to convey to their audience that the person in question is dominant. As long as everyone understands the fittingness conditions for anger, intentionally unfitting anger can be used to convey any number of messages, including dominance.

This account fits well with van Kleef's theory of inferential behavior in emotional communication.[74] Emotional implicature depends on inferential processes – thinking about emotional states. Hence one would expect to see it more where there is more capacity to think about emotions among the animals involved.

## 2.10 Extensions

[73] Grice (1989).
[74] van Kleef (2016: 45-55).

For advanced topics in signal systems see Lewis (1960) and Skyrms (2010). Each of these links signal systems to game theory, which is the basic formal theory of strategic interaction – where at least two things that are each rational are interacting with each other, each knowing that the other is rational and thinking ahead. See Tadelis (2016) for a contemporary introduction to game theory and see von Neumen and Morgenstern (1947).

For those interested in formal semantics see Chierchia and McConnel-Ginet (1990) for a nice introduction, and for pragmatics, see Yan Huang's *Pragmatics 2nd edition* (2015). For philosophical issues about the boundary between semantics and pragmatics, see Zoltan Gendler Szabo's collection, *Semantics Versus Pragmatics* (2005). In this chapter, we argued that emotion communication systems do not display forces (i.e., distinct speech-act-like statues for emotion expressions), but they do have other pragmatic features like implicature. It is worth listing a bunch of pragmatic phenomena as guides for empirical research on emotion pragmatics.

Assume that an agent utters a sentence, *s*, and *s* expresses the proposition P in this context.

An *entailment* of *s* is a proposition Q such that it is impossible that P is true and Q is not true.

A *semantic presupposition* of *s* is a proposition that must be true for P to be either true or false.

A *pragmatic presupposition* of the utterance of *s* is a proposition that must be in the common ground of the conversation in question for the utterance of *s* to be felicitous.

A *conventional implicature* of *s* is a proposition that is part of the conventional meaning of *s*, but not part of the truth conditions of *s*, and arises from the particular words that compose *s*.

A *conversational implicature* of the utterance of *s* is a proposition that is implied by the utterance by virtue of conversational maxims.

A phenomenon is *conventional* just in case it is generated by the conventional meaning alone.
A phenomenon is *cancelable* iff further utterances can deactivate it.
A phenomenon is *backgrounded* iff it should be part of the common ground for an utterance associated with it to be felicitous.
A phenomenon *detaches* iff it need not be present in an utterance of a truth-conditionally equivalent sentence
A phenomenon *embeds* if it is preserved when the sentence uttered is embedded under truth-functional operators.

| | *Property of* | *Conventional?* | *Cancelable?* | *Backgrounded?* | *Detaches?* | *Embeds?* |
|---|---|---|---|---|---|---|
| Entailment | Sentences | Y | N | N | N | N |
| Semantic Presupposition | Sentences | Y | N | Y | Y | Y |
| Pragmatic Presupposition | Utterances | Y/N | Y/N | Y | Y | Y |
| Conventional Implicature | Sentences | Y | N | N | Y | Y |
| Conversational Implicature | Utterances | N | Y | N | N | Y/N |

Y: Yes  N: No  Y/N: Sometimes Yes, sometimes No

**Fig 6: Kinds of Pragmatic Phenomena**[75]

Notice that in the case of emotions, there is no conventional meaning because many aren't based on conventions. Instead, one should think of conventional meaning for emotions as their fittingness conditions. The pragmatic phenomena are distinguished by five distinct properties (cancelability, etc.), and these might have similar manifestations in emotion communication

75 See Partee (2010).

networks. A distinction between kinds of implicature or between pragmatic presupposition and implicature would be especially interesting from the point of view of understanding the varieties of emotion communication and how they relate.

Finally, readers interested in the mathematics of signal processing, which are distinct from the mathematics of information theory in the next chapter, see Oppenheim and Willsky (1997). These mathematical techniques mostly involve identifying signals and transforming them to bring out certain features. Most famous are the Fourier transformations but more important today are wavelet transformations, which are more versatile.

The next chapters have examples of how to complicate the formal models contained therein, but our main way to advance over the formal but qualitative theory of signal systems presented above is to add a formal and quantitative theory of information. And that starts right now!

# *Chapter 3*

# Emotion Communication Channels

The information theory of emotion communication is presented in four chapters. The reason is that we introduce the simplest models first and then more complex ones later, in roughly three stages of complexity. In this chapter, we offer the first stage as the basic ideas of information theory, which are presented and some examples are worked out in early sections. Then we apply these ideas to emotion communication systems in particular. We focus on defining a single channel with two animals. One animal either feels fear or does not. The second animal either detects fear or detects no fear. The channel is defined by the relationship between the two. In more technical language, the channel is the conditional probability distribution that describes the fear detection probabilities, given the fear probabilities. Then, for the second stage of complexity, in Chapter Four, we introduce *coded* emotion communication channels, which are far more interesting and versatile. After that, it is coded emotion communication *networks*, which are then applied to social media in the last two chapters of the book.

## 3.1 Information Theory

Information theory is a quantitative theory of how much information is exchanged in communication. It was invented and many of its most important results were proven by Claude

Shannon in 1947 and published shortly thereafter in a sequence of technical reports and a short book, which is still a great introduction to the topic.[1]

The basis for information theory is probability. Once one specifies the relevant probability distributions for the parts of a communication process, the information theoretic features are definable in terms of it.

Information theory presupposes that one can tell what is a message, what is a source of communication, and what is the destination of communication. These are all aspects of signal systems, and there has been a qualitative theory of signal systems, as found in the previous chapter.

However, conveying more information is not necessarily a good thing – in the sphere of logic, contradictions have the most information, while tautologies have the least.[2] One wants almost all one's claims to be somewhere in between, to convey enough information without conveying too much. TMI is a real thing in information theory too.

Information theory is remarkably fruitful and hugely influential, having been applied to both honey bee dances[3] and deep space communication with satellites on the edge of our solar system[4], and so much else. Nothing like the internet as we know it would be possible without information theory. However, it has never been applied to emotion communication at all. In the rest of the book, we do just that.

## 3.2 Channels

The interesting aspects of information theory involve channels. The simplest channels involve two potentially interdependent variables, a *source* X, and a *destination* Y. X and Y each have *probability distributions* p(X) and p(Y), and they have a *joint* probability distribution as well.

---

[1] Shannon (1947) and Shannon and Weaver (1948).
[2] Barwise & Seligman (1997).
[3] Haldane and Spurway (1954).
[4] Chen and Tutlege (1976).

From this information we can derive their *conditional probabilities* (i.e., the probability that some way of being Y happens, given that some way of being Y happened, or vice versa – denoted p(Y|X) and p(X|Y), respectively). For example, X might be the capacity to feel fear, and we might think of it as having only two states, off and on. We can call these *states* $x_0$ and $x_1$, respectively. Y might be the capacity to detect fear, which likewise has two possible states, $y_0$ and $y_1$ (detect no fear and detect fear). For example, there is a cheetah who detects fear in the lion 40% and detects no fear 60%. There is a 30% probability that the lion feels fear and 70% probability that the lion feels no fear.

Let us see how the details of the cheetah and lion example might go. From the individual probabilities given above, we know that:

(i) the joint probability of lion feeling fear and cheetah detecting fear plus the joint probability of lion feeling fear and cheetah not detecting fear is 30%,

(ii) the joint probability of lion feeling no fear and cheetah detecting fear plus the joint probability of lion feeling no fear and cheetah detecting no fear is 70%,

(iii) the joint probability of lion feeling fear and cheetah detecting fear plus the joint probability of lion feeling no fear and cheetah detecting fear is 40%,

(iv) the joint probability of lion feeling fear and cheetah detecting no fear plus the joint probability of lion feeling no fear and cheetah detecting no fear is 60%,

We can place these data in Figure 7:

| | | Lion | | |
|---|---|---|---|---|
| | | Fear | No Fear | Totals |
| Cheetah | Detect Fear | ? | ? | 30 |
| | Detect No Fear | ? | ? | 70 |
| | Totals | 40 | 60 | |

**Fig 7: Lion and Cheetah incomplete probability distribution. The brackets indicate what is inside the probability distribution, whereas the totals are the sums of each row and column.**

There are multiple ways to fill in the chart, but one is in Figure 8:

| | | Lion: Fear | Lion: No Fear | Totals |
|---|---|---|---|---|
| Cheetah | Detect Fear | 10 | 20 | 30 |
| | Detect No Fear | 30 | 40 | 70 |
| | Totals | 40 | 60 | |

**Fig 8: Lion and Cheetah joint probability distribution 1**

In Figure 9 is another:

| | | Lion: Fear | Lion: No Fear | Totals |
|---|---|---|---|---|
| Cheetah | Detect Fear | 5 | 25 | 30 |
| | Detect No Fear | 35 | 35 | 70 |
| | Totals | 40 | 60 | |

**Fig 9: Lion and Cheetah joint probability distribution two**

Here is a third:

| | | Lion: Fear | Lion: No Fear | Totals |
|---|---|---|---|---|
| Cheetah | Detect Fear | 30 | 0 | 30 |
| | Detect No Fear | 10 | 60 | 70 |
| | Totals | 40 | 60 | |

**Fig 10: Lion and Cheetah joint probability distribution 3.**

Calculation of conditional probabilities are straightforward. Using the first way of filling in the joint probabilities, we get:

P(Detect Fear|Fear) = .25 (10/40)

P(Detect Fear|No Fear) = .33 (20/60)

P(Detect No Fear|Fear) = .75 (30/40)

P(Detect No Fear|No Fear) = .66 (40/60)

The second example looks like this:

P(Detect Fear|Fear) = .125 (5/40)

P(Detect Fear|No Fear) = .417 (25/60)

P(Detect No Fear|Fear) = .875 (35/40)

P(Detect No Fear|No Fear) = .583 (35/60)

And the third example is:

P(Detect Fear|Fear) = .75 (30/40

P(Detect Fear|No Fear) = 0 (0/60)

P(Detect No Fear|Fear) = .25 (10/40)

P(Detect No Fear|No Fear) = 1 (60/60)

It is the conditional probabilities that define the channel, so each of these three examples constitutes a distinct channel. We can also think about channels in terms of error rates. For the first example, the error rate is 75% for fear -- that is the chance of error in detection – and it is 33% for no fear – that is the chance of error in detection.

## 3.3 Information Measures for Channels

The *information* contained in a single random variable is given by the following formula:

$$H(X) = -[(p(x_1)\log_2 p(x_1) + \cdots + p(x_n)\log_2(x_n)]$$

if the variable X can be in n distinct states. It is more common to express this long sum with a symbol that is a bit more versatile:

$$H(X) = -\sum_{x \in X} p(x)\log_2 p(x)$$

This quantity is often called *entropy* by information theorists, and it is related to entropy as studied in other sciences.[5] It is understood as a measure of information. Different bases for the logarithm in the equation above give different units for measuring information; base 2 is *bits*. For example, a person who feels jealous 5% of the time has less information (about .29 bits) than a person who is jealous 45% of the time (about .99 bits), considering jealousy as either present or absent with no in between.

[5] See Stone (2015) for an accessible presentation.

The definition of entropy probably seems highly counter intuitive. Why use this definition with logarithms? Shannon realized that any quantitative concept of information would have something like this definition as long as it has the following features:

- *Continuity*: The amount of information associated with an outcome increases or decreases continuously as the probability of that outcome changes.
- *Symmetry*: The amount of information associated with a sequence of outcomes does not depend on the order in which those outcomes occur.
- *Maximal Value*: The amount of information associated with a set of outcomes cannot be increased if those outcomes are already equally probable.
- *Additive*: The information associated with a set of outcomes s obtained by adding the information of individual outcomes.[6]

There are many other concepts of information and many other concepts of entropy, which are ways of measuring information. We mostly stick to Shannon information and Shannon entropy, although we do consider alternative accounts of entropy from statistical mechanics in Chapter Six on complex networks and Chapter Eight on using a quantitative theory of information to detect coordinated attacks on social media.

The central features of a channel are its mutual information and its capacity. *Mutual information* measures how much knowing about one variable tells you about the other one (and vice versa—it is symmetric), which is often interpreted as the *information flow* through the channel, and has bits/second or bits/message as the unit.

$$I(X;Y) = \sum_{(x,y)\in X\times Y} p(x,y)\log\frac{p(x,y)}{p(x)p(y)}$$

The *channel capacity* is the maximum mutual information across all possible source probability distributions (i.e., probabilities of feeling fear and not feeling fear):

[6] See Shannon (1948) for a proof, and Stone (2015) for exposition.

$$C = \max_{p(x)} I(X;Y)$$

The channel capacity reaches maximum when the source probability has maximum entropy (i.e., information – remember these are synonyms in this framework), which occurs at the even distribution (the same probability for each state).

For example, let X={0,1} and Y={2,3}, where 0 is not having fear, 1 is having fear, 2 is detecting no fear and 3 is detecting fear. We might assign error rates of .1 for each of the two X states, 0 and 1. If the source probability is .8 for state 0 (i.e., X has a 80% probability of not feeling fear), then the destination probability is .74 for state 2 (i.e., Y has a 74% probability of detecting no fear which is the sum of 72% of detecting no fear when X has no fear and a 2% of detecting no fear when X has fear). The destination chance of 3 is 26% because we know there is a 100% chance Y is either in state 2 or state 3, and there is a 74% chance it is in state 2. The mutual information in this channel is .358 bits/message and the channel capacity is .531 bits/message.

To calculate the mutual information, we look at every possible state of X and of Y, there are two possible states of each variable. We need to figure out the probabilities for each of these states occurring together:

- p(0,2)= .72 (because there is a 80% chance that X is in state 0 and there is a 90% chance that when X is in state 0, Y is in state 2; we multiply these together to get 72% chance),
- p(0,3)=.08 (because there is a 80% chance that X is in state 0 and there is a 10% chance that when X is in state 0, Y is in state 3; we multiply these together to get 8% chance),
- p(1,2)=.02 (because there is a 20% chance that X is in state 1 and there is a 10% chance that when X is in state 1, Y is in state 2; we multiply these together to get 2% chance),
- p(1,3)=.18 (because there is a 20% chance that X is in state 1 and there is a 90% chance that when X is in state 1, Y is in state 3; we multiply these together to get 18% chance).

In addition to these joint probabilities, we need the individual probabilities:

- p(0)=.8 (because it was stipulated that X has an 80% chance of feeling no fear),
- p(1)=.2 (there is 20% chance of feeling fear because there is an 80% chance of feeling no fear),
- p(2)=.74 (when X=0, there is a 72% chance Y=2 (because .8 multiplied by .9 = .72)
- p(3)=.26 (because 1-.74=.26; or 100% takeaway 74% equals 26%)

The formula for mutual information in this case is:

$$I(X;Y) = \sum_{(x,y)\in\{0,1\}\times\{2,3\}} p(x,y)\log\frac{p(x,y)}{p(x)p(y)}$$

which expands to:

$$I(X;Y)$$
$$= \sum p(0,2)\log\frac{p(0,2)}{p(0)p(2)}, p(1,2)\log\frac{p(1,2)}{p(1)p(2)}, p(0,3)\log\frac{p(0,3)}{p(0)p(3)}, p(1,3)\log\frac{p(1,3)}{p(1)p(3)}$$

When we substitute in the values listed above, this becomes:

$$I(X;Y)$$
$$= \sum (.72)\log\frac{.72}{(.8)(.74)}, (.02)\log\frac{.02}{(.2)(.74)}, (.08)\log\frac{.08}{(.8)(.26)}, (.18)\log\frac{.18}{(.2)(.26)}$$

Remember, all the logorithms in our presentation of information theory are base 2, which makes our information unit the *bit*. In calculating this sum, in each of the four, we do the fraction first, then the logarithm, then multiply the result by the leading factor. If we cap our calculation at 3 significant digits past the decimal, then the first element of the sum is

$$(.72)\log\frac{.72}{(.8)(.74)} = (.72)\log(1.216) = (.72)(.282) = .203$$

The other three elements of the sum are:

$$(.02)\log\frac{.02}{(.2)(.74)} = (.02)\log(.135) = (.02)(-2.889) =. -.058$$

$$(.08)\log\frac{.08}{(.8)(.26)} = (.08)\log(.385) = (.08)(-1.377) =. -.11$$

$$(.18)\log\frac{.18}{(.2)(.26)} = (.18)\log(3.462) = (.18)(1.792) = .323$$

When we add these four results together, we get .358 bits/message, which is the mutual information between X and Y in the information channel described above.

In order to calculate the channel capacity for this channel, we need to know the mutual information in this channel for every combination of X's states. We just calculated the mutual information when p(0)=.8 and p(1)=.2, but we need to do it again for .9 and .1, and do it again for .41 and .59, and for .456 and .544, and so on for every combination. The maximum mutual information out of all of these is the channel capacity. In this case, the maximum occurs when p(0)=.5 and p(1)=.5; the mutual information in this case is .531 bits/message, so the channel capacity is .531 bits/message. With this result, we can see that the situation described above where p(0)=.8 and p(1)=.2 sends .358 bits/message, which is 67% of the capacity of this channel.

Just as entropy is a measure of information, mutual entropy is a measure of information flow. This channel sends one symbol per transmission, but the sending is just X being in some state and Y being in some state. If X feels fear and Y detects fear, then there is no error. Likewise, if X feels no fear and Y detects no fear, then there is no error. The error rate is the probability of error. There are two major kinds: false negatives (e.g., the failure to detect fear) and false positives (e.g., detection of fear when there is no fear).

Notice how error rates can sculpt the probability distribution of an emotional state in a population. For example, other things being equal, if chance of false negatives for detecting anger *increase* in a population, then that will *decrease* the average amount of time each person feels anger in that population over time. Evolutionary processes might apply to error mechanisms like this as a result of pressure in information theoretic aspects of a population's emotion communication system overall.

Using the same example, we can illustrate several other information-theoretic quantities. A *message* is just X being in either state $x_0$ or state $x_1$. So the *message information* is the same as the information in the X variable. In more complex models, the messages are more complicated so one has to calculate the information in the message independently (e.g., if the channel sent a message about the state of X over the last five moments all at once as a sequence of five 0s or 1s, or if the channel uses a code, as in the next chapter).

In order to calculate some additional information theoretic measures of this channel, recall that the conditional probabilities in this example are stipulated (by the error rates) to be:

$p(2|0)=.9$

$p(3|0)=.1$

$p(2|1)=.1$

$p(3|1)=.9$

This is the $p(Y|X)$ probability distribution; we also need the $p(X|Y)$ probability distribution, which we can calculate using Bayes Theorem[7]:

$p(0|2)=.972$

$p(0|3)=.308$

$p(1|2)=.027$

$p(1|3)=.692$

The difference between these two conditional probability distributions is crucial.

The first one – $p(Y|X)$ – is the probability that Y is in a certain state, given that X is in a certain state. For example, $p(2|0)=.9$ so there is a 90% chance that Y is in state 2 (Y detects no fear), given that X is in state 0 (X has no fear). Hence, if we know that X has no fear, then we also know that there is a 90% chance that Y has detected no fear. Using this conditional probability, we reason in the same direction as the channel: from X to Y. So this conditional probability distribution tells us probabilities for Y's states, given that we already know about X's states. In our example, it tells us about Y's emotion detections, given what we know about X's emotions.

[7] Bayes Theorem relates conditional probabilities, $p(X|Y)$ and $p(Y|X)$ to individual probabilities $p(X)$ and $p(Y)$ in the following way: $p(X|Y)p(Y)=P(Y|X)p(X)$. Because we know all these quantities except $p(X|Y)$, we can solve for it.

The second one – p(X|Y) – is the probability that X in a certain state, given that Y is in a certain state. For example, p(0|2)=.972, so there is a 97% chance that X is in state 0 (X has no fear), given that Y is in state 2 (Y detects no fear). Hence, if we know that Y has detected no fear, then we also know that there is a 97% chance that X has no fear. Using this conditional probability, we reason in the opposite direction of the channel: from Y to X. So this conditional probability distribution tells us probabilities for X's states, given what we already know about Y's states. In our example, it tells us about X's emotions, given what we know about Y's emotion detections.

These two probability distributions are called the *conditional entropy* and the *equivocation.* We can use them for information-theoretic goals. The entropy of p(Y|X) is given by the H function defined above for the entropy of a single variable. H(Y|X) is the *conditional entropy* of the channel. On the other hand, the entropy of p(X|Y) – H(X|Y) is the *equivocation.* The conditional entropy, equivocation, and mutual information are related to *joint entropy* (i.e., H(X,Y), defined in the obvious way) and the individual entropies, H(X) and H(Y), in various ways.

The *noise* in the channel is given by the error rates. In this case, there are two possible states of the source X. There is a error rate of 10% for state 0 and an error rate of 10% for state 1. These two error rates are included in the conditional entropy, H(Y|X), which is sometimes called the noise in the channel.

*Redundancy* is a measure of how much information could be compressed, so it tells us about excess information. The word is used in several different ways, but for us it is how close the entropy of a variable is to its maximum value. For example, in our variable X described above as an animal's capacity to either have fear (20% chance) or have no fear (80% chance), the entropy of X is .721 bits (calculation: -((.8)log(.8) + (.2)log(.2))= .721). The maximum possible entropy of X is 1 (i.e., the maximum occurs when there is a 50% chance of each state). The redundancy of X then is .279 (calculation: 1-(.721/1)). Redundancy of a variable is between 0 and 1; the higher the redundancy, the more extra information there is. Variables with maximum entropy have zero redundancy. Our variable X does not have much extra information. This is not very interesting for our case, but redundancy plays a huge role in coded channels (next chapter), so we will return to it shortly.

There are dozens of more complex information theoretic concepts and quantities defined in the literature that could be brought to bear on emotion communication. For example, *transfer entropy* is defined for communication systems transmitting over a period of time as the amount of information transferred (at time step t) from the past of Y to the current state of X. Transfer entropy is closely related causation and has recently attracted much attention among information theorists.[8] One nice feature is that it is not symmetric. The mutual information of X and Y – I(X;Y) – is identical to the mutual information of Y and X – I(Y;X); that is, mutual information is symmetric. Transfer entropy is not like this and so seems to capture the idea of information flowing one way but not the other. Overall, the reader should use caution: different authors use terms differently, and even basic terms like 'information' and 'entropy' have many different mathematical definitions within and across disciplines, e.g., entropy in computer science and in statistical mechanics.

## 3.4 Basic Emotion Communication Channel

The information channel, defined by a conditional probability distribution, p(Y|X) between two individual probability distributions, p(X), and p(Y), is the basic unit of information theory. As such, it is the basic unit of the information theory of emotion communication as well. In our case, both X and Y are emotion capacities in animals. X is the capacity of have an emotion, and Y is the capacity to detect that emotion. So, p(Y|X), is the probability of the Y states given the X

[8] Bossomaier et al (2016).

states. That is, the probability of each emotion detection state in Y given each emotion state in X. P(Y|X) is a probability distribution, which means it contains the individual probabilities of every possible combination of states.

With this basic unit – the basic emotion communication channel – we can see the most fundamental phenomena in emotion communication, but it is not more interesting. That comes with *coded* communication, which is the next chapter.

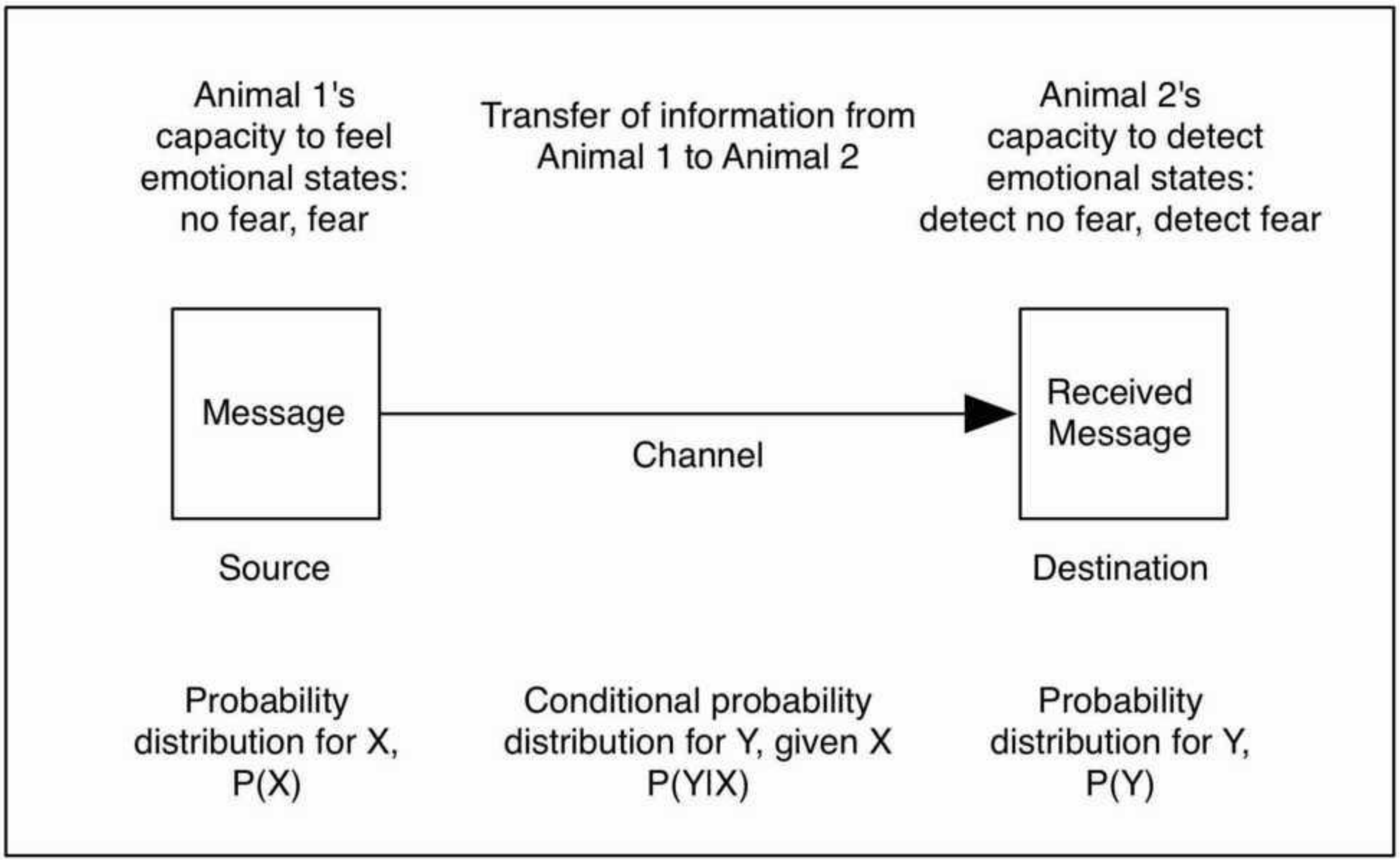

**Fig. 11. The most basic emotion communication channel, in which a simple message is sent from an animal having an emotional state to an animal detecting that emotional state. In this example, the source animal can have only two states; it can either feel fear or not feel fear. Animal one has a certain probability of being in each of these states. The destination animal can have only two states as well; it can either detect fear or detect no fear. Animal two likewise has a probability distribution for its states. The channel is defined by the conditional probability distribution, which specifies the probability that the destination animal will be in a certain state, provided that the source animal is in a certain state (for each of the possibilities).**

## 3.5 Extensions

Let the above simple emotion communication channel model be *Model 1*. The following variants on this model reduce its idealizations in various ways that have to do with emotion communication channels in general:

(a) Multiple emotion *types*. For example, three distinct emotion registers: fear, no fear, anger, no anger, sadness, no sadness. These might be felt in certain combinations but not others. The messages sent could be more complicated (e.g., 101 might signal fear, no anger, sadness).

(b) Emotion *levels* for each emotion type. For example, no fear, low fear, moderate fear, and high fear, and corresponding changes to detection at the destination. Sensitivities could differ on either end.

(c) *Memory*. Allows the destination to remember previous messages to improve communication.[9]

(d) *Feedback*. Allows the destination to communicate back to the source to improve communication.[10]

---

[9] Ahlswede, R. (2015).

[10] See Cover & Thomas (2005) and el Gamal, A. (2012).

(f) *Side channels*. These are additional channels of information flow interacting with a channel that impact the features of the channel. For example, a few years from now, your romantic partner wears an EEG hat for continuous emotion monitoring and uploading to a website; in a conversation over the phone, you cannot tell whether your partner is angry or excited, so you quickly check an app on your phone, which informs you that your partner is merely excited. You breathe a sigh of relief. These sorts of applications are already possible and modelled with side channels.[11]

(g) *continuous* variables. All the variables in Models 1 are discrete, but things are very different with continuous variables. The most notable is that one has to use an alternative account of information called *differential entropy*:[12]

$$h(X) = -\int f(x)\log f(x)dx$$

There are many other extensions, but plenty are explained in future chapters. See Shannon and Weaver (1948) for the classic treatment (almost every information theory textbook has variants of Shannon's diagrams, including the one above). See Cover & Thomas (2005) for a comprehensive contemporary treatment, Stone (2015) is a great "tutorial" introduction for general audience, and Roman (1992) and Hankerson (2003) for our favourite advanced treatments. See Bossomaier et al (2016) for an introduction to transfer entropy. See Sommaruga et al (2009) and Bowden (2018) for more general settings and the fascinating Fisher et al (2005) on information behaviour.

One topic that is sorely lacking in textbooks on information theory is dynamic theories of information theoretic features that go beyond evolutionary studies.[13] About ten years ago, researchers began to explore how to combine information theory with dynamical systems. So far, there have emerged some great results – the account of transfer entropy is one of them – but we know of no accessible general treatment. See Holiday et al (2005) for an early treatment and Bollt and Santitissadeekorn (2013) for more recent work. If we apply these frameworks to the information theoretic aspects of emotion communication systems, then we could begin to address the question: How do the information-theoretic features of emotional signal systems change over time from our evolutionary roots as hunter-gatherers, to subsistence farmers in small villages, to industrial workers in large cities, to social media on the internet, and beyond?

[11] Cover & Thomas (2005).
[12] Cover & Thomas (2005) and Hankerson, D., Harris, G., and Johnson, P. (2003).
[13] However see Kerr (forthcoming) and Cochrane (2018) for examples in philosophy.

# *Chapter 4*

# Coded Emotion Channels

A far more powerful emotion communication model incorporates encoding and decoding of messages. In information theory, *coding* is required when the source variable does not work well as a communication device. In this case, the internal aspects of emotions (i.e., their feel and what they do in the body and brain) are not a good physical basis for communication because they are not readily observable. So animals encode emotion information in observable behaviour and decode emotion information from the behavior of others. These terms, *encoding* and *decoding* are already used by some signal theorists in the literature, but we are unaware of anyone who has backed them up with a proper theory.[1] The new model, call it Model 2, contains an *encoder* between the source and the message and it contains a *decoder* between the message and the destination. As in each communication channel, the source is the emotional state of a person and the destination is the detection of that emotional state of that person by someone else.

## 4.1 Encoding and Decoding Messages

As above, let X={0,1} and Y={2,3}, where 0 is not having fear, 1 is having fear, 2 is detecting no fear and 3 is detecting fear. The channel sends one symbol per transmission, but now the encoder takes states of X as input and outputs a code in some other language. The message is sent in the *code language*. Then the decoder runs an algorithm that takes as input the units of the code and outputs a message in terms of states of Y. The code can be chosen for all sorts of reasons including *security* and *error correction*. We hypothesize that human emotion signal systems utilize complex error-correction schemes. In Model 2, the encoder is emotion expression. That is, it turns emotion states into emotion behavior. The decoder is a human sensory system, brain, and the overall algorithm implemented in them, which takes observed actions as inputs and outputs emotion detection states. In our example, there are only two emotion states, fear in X and no fear in X, and there are only two detection states, detect fear and detect no fear.

[1] Ross & Doumachel (2004), Banziger, Hosoya, and Scherer (2015), and van Kleef (2016).

There are so many complexities associated with emotion expression that the model will have to start with many idealizations. We can use a code from the language of {A, B}, where A indicates an action and B indicates no action. Our basic model might track three modalities: facial expression, voice modulation, and bodily posture. We call these words, but this is just to follow information theory terminology (e.g., particular honeybee dance patterns are also words in this sense). Likewise, we use the terms ‘codeword’ and ‘alphabet’ even though we are talking about animal behaviours that express emotions. In that case, each word from the source (i.e., 0 and 1), would be paired with a codeword. In the case of 0 being no fear, its codeword would be BBB. In the case of 1 being fear, it would get codeword AAA. The former means no fear face, no fear voice, and no fear posture. The latter means fear face, fear voice, and fear posture. In Model 2, person X always expresses fear in face, voice, and posture when afraid, and never expresses fear in any of the three modalities when not afraid. That is, the encoder is *perfect* (in information theoretic terms, it is *lossless*). Likewise, the decoder algorithm always runs properly without any performance errors, but that does not mean that it always decodes the codewords into a message identical to the source; rather, it has to deal with noise in the channel.

The channel is specified in terms of the *code alphabet*, which might be very different from either the source or destination alphabets. Let us try assigning error rates of .1 for each of the two code symbols, A and B. That is, there is a 90% chance the symbol is transmitted correctly for each of the two symbols. *Mutual information* and *channel capacity* are calculated in the same way as before (see Chapter 3). Again, the equations are:

$$I(X;Y) = \sum_{(x,y)\in X\times Y} p(x,y)\log\frac{p(x,y)}{p(x)p(y)}$$

$$C = \max_{p(x)} I(X;Y)$$

respectively. The mutual information in the example is .358 bits/message in a channel with capacity .531 bits/message.

## 4.2 Information Measures for Coded Channels

Loads of new concepts and quantities are definable in coded communication channels. For example, the *input frequencies of the coding scheme* describe the chance of each code letter occurring. If sj are source letters (for emotion states) with frequencies fj respectively, and the code assigns codeword wj to each source letter $s_j$, and codeletter $a_i$ occurs $u_{ij}$ times in word $w_j$, then the frequency $p_i$ for each codeletter $a_i$ (for i=1…m) is given by:

$$p_i = \frac{\sum_{j=1}^{m} u_{ij} f_j}{\sum_{j=1}^{m} f_j \,\text{length}(w_j)}$$

For example, in emotion communication, the code takes felt emotions to emotion behaviours through the process of emotion expression. In many animals, the codes make use of multiple modalities (e.g., face, voice, body posture, etc.). We can think of activating an emotion expression modality as a codeletter and a complete expression of an emotion in all modalities as a codeword (i.e., an ordered sequence of activation states). We can then use the above equation to calculate the frequencies of activation states.

In addition, we can calculate the *optimal* channel input frequencies and use a code that transforms source frequencies for sourceletters into channel frequencies for codeletters that are closer to optimal. If the channel is something familiar like a binary symmetric channel (i.e., same error rates in each direction), then the optimal input frequencies are .5 for each of the two codewords. It seems like one would find evidence of evolutionary pressure in this direction. We can call this the *emotion code frequency optimization hypothesis*.

Beyond the features of code letters, we can think about the code as a whole. For example, the *information rate of the code*, which describes how much information is transmitted by the code, rather than how much information is transmitted by the source, is H(X)/(average codeword length).[2] Coding for compression increases the amount of information sent in each codeword, relative to the source, while coding for error-correction decreases the amount of information sent in each codeword.

Another important measure is *distortion*. Distortion measures the distance between a variable like X and the way it is represented by a code. For example, emotion distortion would measure the difference between emotion states and their representations in behaviour. Authors frequently characterize distortion functions as measures of the cost of representing the variable in question with the code in question. Because accuracy is one obvious goal in communication, the difference between a variable and its representation is interpreted as a cost. The other major goal in communication, efficiency, is linked to accuracy in the way that distortion is linked to rates of communication. The idea is that we can decrease distortion but at the cost of decreasing the rate as well; likewise, we can increase the rate, but at the expense of increasing distortion as well. One fundamental question is how distortions and rates are related to each other, and this is known as *rate distortion theory*.

There is an entire of world of formal results on encoding and decoding, compression and error-correction, but we focus here on error correction, because it plays such a pivotal role in emotion communication systems.[3]

## 4.3 Compression and Error Correction

[2] See Hankerson et al (200?)
[3] Gill et al (2014).

Once we have a model for coded emotion communication, we can distinguish two fundamental information-theoretic phenomena: data compression and error correction. Data compression occurs when the code *decreases* the redundancy (i.e., extra information beyond the bare minimum) and error-correction occurs when the code *increases* the redundancy. These two processes pull in different directions and have different purposes.

The point of data compression is efficiency. In 2020, data compression is everywhere, but we remember using .zip programs to store and send files. There are two major kinds of data compression: *lossless*, where the original uncompressed data can be reconstructed perfectly, and *lossy*, where it cannot. Ultimately, data compression codes the same amount of information with less redundancy so that it is easier for the receiver to get the message.

The point of error-correction is accuracy. In 2020, error-correction is ubiquitous as well. It is what allows us to communicate with the voyager spacecraft even though they are outside our solar system and have miniscule transmitters. Ultimately, error-correction codes the same amount of information with more redundancy so that it is easier for the receiver to get the message right.

The two communicative values, efficiency and accuracy, which are served by compression and error correction respectively, pull in different directions, like seeking pleasure and evading pain or finding truth and avoiding falsity. Although in tension, the two fundamental processes involved – compression and error correction – are related by one of the most celebrated theorems in information theory, which states that they are independent of one another in communication channels. Also called *the source-channel coding theorem*, it states the independence of source coding for compression and channel coding for error-correction. That is, one can figure out the best compression algorithm and figure out the best error-correction algorithms, and these calculations do not infer with each other. This elegant result fails in networks where the two processes of compression and error-correction can be in open conflict.

In emotion communication, data compression and error correction pertain to the process of *emotion expression* (and of course, error correction pertains to emotion detection as well). When compression occurs, the emotion information is expressed more efficiently. When error-correction occurs, the emotion information is expressed less efficiently. For example, an animal who has an emotion might feel a certain way (e.g., scared), and feel this to a certain degree (e.g., low), and feel it about a certain object (e.g., a spider). However, that animal might only express the kind of emotion without expressing the degree or the object. This would be emotion information compression (this is lossy compression). Moreover, the animal might express this kind of emotion in several modalities (e.g., facial expression, voice, and body posture), when only one would do so that it is easier for other people to understand her emotions. This would be emotion error-correction. The compression and error correction do not work against one another in this case, since the error correction algorithm (emotion information for communication → more redundant emotion information for communication) takes as input the output of the compression algorithm (felt emotion → emotion information for communication). The source coding theorem entails that something like this is always possible with communication channels, but it fails for networks.

We continue the same example to illustrate these two processes. To see how the *error correction* mechanism works, notice that the sender sends only messages BBB or AAA. No other combinations are encoded and so no others are sent.

Imagine that the message sent is AAA and the message received is ABA. The second 'A' was turned into a 'B' by the channel noise. In other words, person Y detected fear in person X's face and posture, but detected no fear in X's voice, even though there was fear in X's voice as well.

How exactly is the receiver supposed to interpret 'ABA'? It does not fit any known message the sender would send. The answer is that the receiver's decoding algorithm is set up so that 'ABA' is decoded as mostly likely resulting from an error in transmitting 'AAA' because 'ABA' is closer to 'AAA' than it is to 'BBB'.

This notion of 'closeness' is familiar in coding theory as the *Hamming distance*.[4] As such, 'ABA' gets decoded as 'fear detected'. If 'BBA' had been received instead (i.e., detect no fear in X's face or voice, but detect fear in X's posture), then the decoder might have decoded it as 'no fear detected', but the chances of 'AAA' being received as 'BBA' are less than 1%.

Despite the risk of these sorts of errors, the code clearly increases the *reliability* of the channel. How do we measure this? If the error rates were similar (per symbol), then 0 (no fear) state has a 10% chance of being received as a 'detect fear'. That is, false positive chance is 10%. With the code, and using the Hamming distance decoder, the chance of a false positive drops to 2%. Adding more symbols to each codeword (i.e., including modalities other than just face, voice, and posture) decreases the chances of error even further.

The lesson here is that one reason animals go to the trouble of expressing their emotinos in many different ways all at once is to increase the efficiency of the emotion communication channels in which they participate. *Multi-modal emotion expression is an error-correcting code for communication.*

One of the major theorems of information theory, the Noisy Channel Coding Theorem, was proven by Claude Shannon in the original presentation of information theory (it is often called just Shannon's theorem). It states that as long as the rate of information is less than the channel capacity, one can find error-correcting codes with error rates as low as one wants (given trade-offs in rate of transmission).[5] This elegant result suggests that in emotion communication systems, the rate of information is below the channel capacity to allow for the potential for ever-better error correction. Call this the *Shannon hypothesis*.

Understanding how emotions are encoded in action as an information theoretic process opens up several other interesting lines of inquiry. One is the nature of *emotion compression*, which is how an animal converts the information in the emotion state itself into information to be transmitted via emotion behaviour.[6]

---

4 Hankerson et al (2003).

5 Shannon (1948).

One obvious suggestion is to look for Huffman compression in emotion communication systems since it is such a natural and ubiquitous coding option. *Huffman coding* algorithms assign simpler codewords to more frequent messages and more complex codewords to less frequent messages.[7] For emotions, the *Huffman hypothesis* would be that emotion behavior displays a Huffman-type pattern where simpler actions are associated with more common emotions and less simple actions are linked to less frequent emotions. Depending on how to interpret 'simple', one could interpret the Huffman hypothesis in different ways (e.g., lower energy, quicker, easier to detect, etc.). We might find Huffaman phenomena across emotion communication systems, or perhaps only for certain categories (e.g., basic emotions), or not at all.

Other emotion encoding phenomena to investigate include the extent to which emotion behavior obeys the *prefix condition* on encoding, which requires that no complete codeword appears as the prefix of another codeword. In emotion communication, this would be the requirement that no individual emotion expression forms the initial part of another more complex emotion expression. It is easy to see why this would be desirable to decrease the need to disambiguate. The question of whether natural emotion communication systems satisfy prefix conditions has never been formulated as far as we can tell; call it the *prefix hypothesis*.

Another is whether we find increases in coding efficiency by encoding several emotional states at once in individual codewords.[8] That we find this phenomenon in emotional communication systems would be the *higher-order-encoding hypothesis*. One question is whether the recent evidence supports the higher-order-encoding hypothesis.[9]

Emotion *decoding* almost certainly instantiates a variety of decoding algorithms. One familiar in the coding literature is Maximum Likelihood Decoding (MLD), which calculates the probabilities of messages sent, given that a certain message was received and picks the highest one.[10] In emotion communication, MLD would be picking the emotion that corresponds to the emotion behavior with the greatest probability given which emotion behavior was observed. Another is Nearest Codeword Decoding (NCWD) – also known as "nearest neighbor", which defines a metric (i.e., distance function) on the set of codewords and picks the destination message for the codeword nearest to the received message. In emotion communication, NCWD would be picking the emotion that corresponds to the emotion behavior most like the observed emotion behavior.

---

[6] Hankerson et al (2003).
[7] Hankerson et al (2003) and Gill et al (2014).
[8] Hankerson et al (2003).
[9] Qiao-Tasserit (2017) and Sacharin et al (2012).

Do humans use MLD or NCWD or something else? We suggest that humans do not use anything like MLD because using it requires knowing about probability distributions of emotions in the population, whereas NCWD can be used by anyone who knows how close an arbitrary action is to each kind of emotion expression.

In general, we postulate that using an emotion communication system requires as little information about probability distributions as possible, since these values can change dramatically across populations as they encounter new environments. There are a host of interesting formal results connecting MLD, NCWD, and many others waiting to be explored (e.g., syndrome decoding).

Other decoding issues include weighting modalities differently for distinct emotion-type decoding (e.g., the face matters most for joy and disgust but body posture matters more for fear and anger).[11] Another big topic is decoding algorithms that consider the reliability of the source, memory, feedback, etc. There also are many families of error-correcting codes (e.g., convolutional codes) and these are related to each other in various ways; one could investigate the variety of error-correcting codes in nature (e.g., do human emotional communication systems ever use turbo codes?).

---

[10] Hankerson (2003).

[11] as in Planalp, S. (1998).

A final topic is whether humans and other animals make what are known as hard decisions or soft decisions in decoding emotions. A *hard-decision* detection sorts input into discrete categories (e.g., detect fear vs detect no fear) using thresholds (as in the error correction example above, which uses Hamming distance), whereas *soft-decision* detection sorts inputs by using more sophisticated measures that use more of information present and are less likely to end up with ambiguities and arbitrary decoding choices. One could naturally investigate hypotheses about emotion communication systems that posit various features (e.g., that we use soft-decision encoding based on data).

One would expect from the information theory of emotion communication that emotion encoding and decoding schemes in a species would evolve together, and that is exactly what preliminary evidence indicates.[12] The same data provide initial support for the decorrelation of encoded emotion behaviors as well, which is another obvious prediction from the information model.[13]

## 4.4 Basic Coded Emotion Communication Channel

The coded basic emotion communication channel is the simplest emotion communication channel that displays realistic features like emotion expression and emotion perception. These coded channels also display central phenomena like error-correction and compression. Coded communication channels are also the subject of the most fundamental theorems of communication theory like Shannon's theorem. The basic coded emotion communication channel is displayed in Figure 12 (compare to Figure 11 in Section 3.4).

Nevertheless, coded emotion communication channels will not be sufficient for our purposes, which is to analyse the emotion information in social media. That comes with coded communication *networks*, which is the next chapter.

[12] Schyns, Petro, and Smith (2009).
[13] Schyns, Petro, and Smith (2009) and Cover & Thomas (2006).

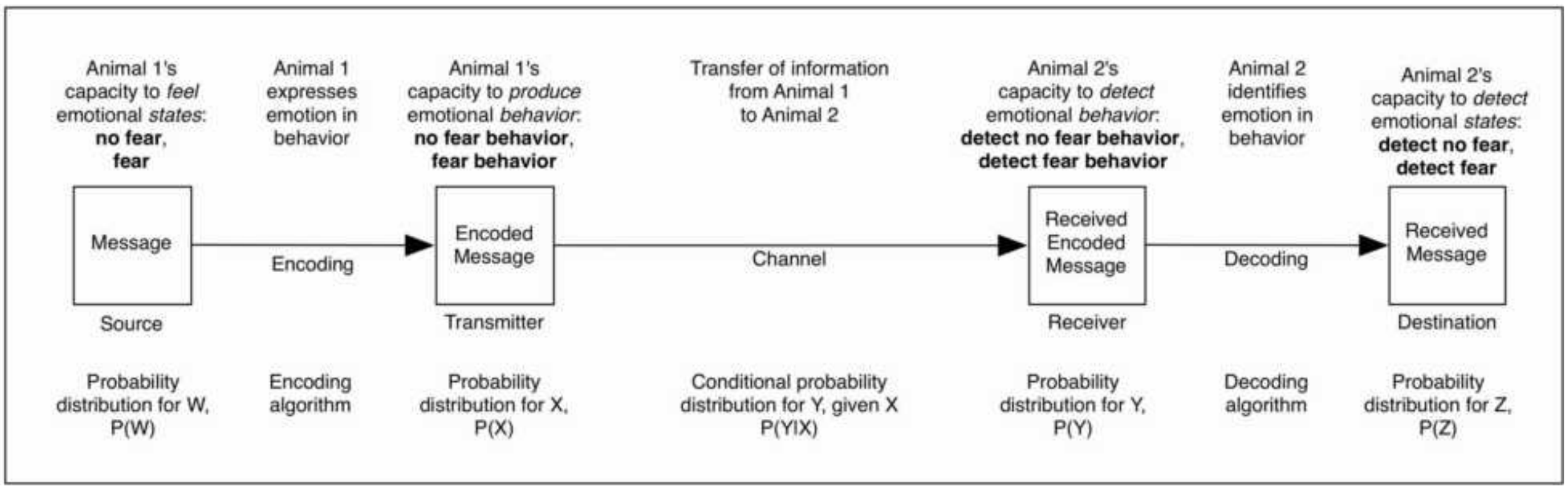

**Fig. 12. An emotion communication channel with message encoding and decoding. Each of the elements from the basic communication model are here, but the message runs through an encoder first and then is decoded before reaching the destination. In particular, the source and destination are the same. The channel is still defined by the conditional probability distribution, but now it is the probability that the receiver will be in a certain state, provided that the transmitter is in a certain state. In coded emotion communication, the transmitter is animal one's emotion behavior; in the case of fear, this would include facial expressions, tone of voice, body movements, and more. The receiver is animal two's detection of emotion behavior; in the case of fear, it would include detecting facial expressions, detecting tone of voice, detecting body movements, and more. The encoding algorithm is just the tendency of animal one to express its fear in different ways (modalities). The decoding algorithm is the tendency of animal two to detect the emotion state of animal one on the basis of detecting animal one's emotion behavior. The encoding algorithm and decoding algorithms are assumed to be perfect in this model in the sense that they introduce no new errors.**

## 4.5 Information Theory and Biology

Why care about information-theoretic features of emotion communication? Biologists have been using them for decades, in the words of Jessica Pfeifer's survey, "First, the measures abstract from the particulars of a system and, therefore, allow biologists to compare very different types of communications systems. Second, it is claimed that they provide a quantitative measure of features of communication systems that ought to be selected for. The measures thereby allow biologists to explain how the communication systems … evolved by natural selection."[14] Pfeifer goes on to present worries for doing this too hastily, and we agree. For example, emotion communication networks' effects on the evolution of a population are *not necessarily* to increase information transfer rates.

To illustrate, it is trivial in information theory to show that the closer the source probability distributions are to .5, the closer the transmission rate of certain communication channels is to channel capacity. And the closer the transmission rate is to channel capacity, the more cumbersome error correction gets. So there is good reason, especially with noisy signal systems that need lots of error correction, to have probability distributions for sources kept closer to the extremes: nearer to 0 or nearer to 1.

For example, a population develops a fear signal system, at a time when the members of that population feel fear about 30% of the time. That population might evolve toward higher thresholds for fear, so that its members feel fear closer to 10% of the time. That would allow for far greater error correction in their coding; that is, it allows for more actions to express each emotion so that emotions are easier to detect accurately. The overall effect of the lower rates of fear would be to *decrease* the information in the channel to allow for more effective information processing on either end.

Once emotion communication networks are up and running in a population, they will likely have an effect on the evolution of that population. And this effect is *not necessarily* to increase information transfer rates. To illustrate, it is trivial in information theory to show that the closer the source probability distributions are to .5, the closer the transmission rate of a communication channel is to channel capacity. And the closer the transmission rate is to channel capacity, the more cumbersome error correction gets. So there is good reason, especially with noisy signal systems that need lots of error correction, to have probability distributions for sources kept closer to the extremes: nearer to 0 or nearer to 1. For example, a population develops a fear signal system, at a time when the members of that population feel fear about 30% of the time. That population might evolve toward higher thresholds for fear, so that its members feel fear closer to 10% of the time. That would allow for far more redundancy in their coding; that is, it allows for more actions to express each emotion so that emotions are easier to detect. The overall effect of the lower rates of fear would be to *decrease* the information in the channel to allow for more effective information processing on either end. One of the beauties of the connection to information theory is this application of Shannon's famous channel coding theorem which relates channel capacity, information rate, and error correction.

[14] Pfeifer, J. (2006).

## 4.6 Extensions

Model 2 is also highly idealized; in addition to each of the *general* variants (b)-(e) of Model 1, which could be added to Model 2 as well, the following are variants specific to emotion coding that would make Model 2 more realistic:

(g) Code *complexity*. Human emotion expression alone is shockingly varied. One should add social-media-based expression modalities to the list as well; e.g., emojis[15], wearable emotion recognition systems[16], and external emotion recognition systems (e.g., at an airport or in a department store).

(h) Different *error rates* for different emotion expression modalities.

(i) Different *degrees* of expression (e.g., no frown, slight frown, full frown).

(j) Encoding the *objects* of emotions and communication (e.g., signalling what one is afraid of or what one is angry about or what one is disgusted by).

[15] Riordan (2017).
[16] Yavuz et al (2018).

(k) *Imperfect* emotion encoders and decoders. This topic goes under the heading of *lossy compression* and *decompression*, which is needed to simulate real world situations where emotion expression displays error rates (i.e., people do not always express emotions felt, and they sometimes produce emotion behavior without feeling the emotion in question), as does emotion detection from properly perceived behaviors.

(l) Emotion *encryption*. Codes can be used to hide messages from others.[17] In emotion communication systems, this would be using actions that can be detected by one person but not by another.

(m) Emotion *security* and *control*. Side channels, dynamic information flow monitoring, artificial emotion recognition systems, and artificial agents (e.g., bots) can be used to control, monitor, or secure communication channels. For emotion communication, especially over the internet, this is a vitally important topic – see Chapter 8.

For work on error correction, see Huffman & Pless (2003) and for compression, see Salomon (2007).

The standard text for information theory and evolutionary biology is Avery (2012), but see Searcy and Nowicki (2005), Yockey (2005), Oller and Grievel (2008), and Altenmuller, Schmidt, and Zimmermann (2013).

[17] Easttom (2016)

# *Chapter 5*

# Emotion Communication Networks

So far we have seen simple emotion communication channels (Chapter 3) and coded emotion communication channels (Chapter 4). We have only scratched the surface of these topics but in this chapter we expand the quantitative theory of emotion communication to networks with multiple transmitters, multiple receivers, and chains of communication.

Communication networks are not just collections of communication channels; they display unique behaviour as in the breakdown of fundamental theorems like the source coding theorem. There is at present no general theory of information in communication networks, but we do have a large number of results for simpler cases.[1] The mathematics is more complex and there are fewer general results to rely on, but this fascinating topic is essential for understanding emotion communication.

The key to *emotion networks* is that people can receive emotion information from multiple sources at the same time and they can send emotion information to multiple destinations at the same time. The latter is called *broadcasting* in network information theory and the former is called

[1] Cover & Thomas (2006) and el Gamal (2012).

*multiple access*. We can construct networks that have broadcasting from a single source to multiple destinations. Or we can construct networks that have multiple access to a single destination from multiple sources. Or we can have multiple senders broadcasting and many multiple-access destinations. It is this aspect of networks that makes them more complex than collections of simple channels. A single animal can be receiving transmissions from multiple sources and at the same time be broadcasting to multiple other destinations.

## 5.1 Mathematics of Networks

Intuitively networks have nodes and connections between nodes. In emotion communication networks, the nodes might be animals and connections might be emotion communication. We could instead think of the connections as having mathematical values like the quantity of information communicated. It seems like connections should have a direction as well, since we might want information measures like transfer entropy that are not symmetric. This fits well with our intuitive idea that emotion communication can be from X to Y but not from Y to X. In our quantitative models from Chapters 3 and 4, we enforce non-symmetric relations by stipulating that emotion communication has a variable with emotion states as a source and emotion detection states as a destination. So X can have emotion communication with Y even though Y is not having emotion communication with X. That is, X's emotion states to Y's emotion detection states is an emotion communication channel with mutual information I(X's emotion states; Y's emotion detection states). This quantity is different from I(Y's emotion states; X's emotion detection states), so our models satisfy the intuition of non-symmetric communication. Of course, I(X's emotion states; Y's emotion detection states) = I(Y's emotion detection states; X's emotion states), so if we think of X's emotion states communicating with Y's emotion detection states, then Y's emotion detection states are also communicating with X's emotion states. This counterintuitive result is one reason to prefer transfer entropy over entropy, but we do not pursue this topic.

Mathematically, a network is a pair of sets: N, a set of nodes and E a set of edges, which are connections between nodes. The edges can be symmetric or not. To allow for either possibility we can let the edges be ordered pairs of nodes so that <n0, n1> is distinct from <n1, n0>. Networks with symmetric edges can then stipulate that each collection of pairs is either in E or not in E. Networks whose edges have mathematical quantities (called weights) are represented by a set of nodes, a set of edges and a set of weights (one for each edge).

For our purposes, there are two bodies of mathematical results that are relevant to emotion communication networks. The first has nothing in particular to do with information. These mathematical quantities pertain only to the mathematics of networks in general, not to communication networks in particular. We present these results in the next section. The second set of mathematical results are about the information in networks. These results are presented in section three. The two kinds of mathematical results interact with one another because some of the structural network quantities can be defined using information as network weights, and some of the information theoretic measures presuppose certain structural quantities. As we will see in Chapter Eight, these two literatures blend together once we get to emotion network security techniques like those based on statistical mechanics and the Ising model.

## 5.2 Structural Network Measures

There is a vast literature on how to define fruitful mathematical concepts related to networks and how they can be applied in the sciences, engineering, medicine, and beyond. Here we just mention the most familiar ways of quantifying aspects of networks.[2]

- *Network size* is the cardinality of the set of nodes, and *total links* is the cardinality of the set of edges.

[2] See Bianconi (2018) for a clear and careful survey.

- The *adjacency matrix* describes the network completely by specifying the existence or nonexistence of every possible link between nodes. The adjacency matrix is often used to calculate the structural quantities of networks.
- The *degree* of a node is the number of edges in which it participates (for an undirected network). The *strength* of a node is sum of weights for edges in which it participates (for an undirected weighted network). For directed networks, we distinguish *in-degree* and *out-degree*, which are number of edges terminating at the node in question and number of edges starting from the node in question, respectively (for unweighted). In weighted directed networks, strength can be defined in the same way, but in-strength and out-strength can be defined as well. The *degree distribution* includes the frequency of each degree among all the nodes in the network (for directed networks, there are in-degree distributions and out-degree distributions). A *scale-free network* has a degree distribution in the form of a power law (i.e., for each degree k, $p(k) = ck^{-r}$ for some constant c and rate r), which are studied extensively in natural and social sciences. Scale-free networks tend to be dominated by a few nodes with very large degrees.
- The *clustering* of a node is the measure of the chance that its neighbours are connected by an edge. The clustering coefficient is between 0 and 1 for each node, and its average over all nodes is the *global clustering coefficient.* These measures correspond to some extent with the geometry of the network connections, where high clustering coefficients indicate lots of triangles and low values indicate tree structures.
- A *path* is a sequence of nodes that are connected by edges, and a *directed path* follows the direction of the edge between nodes. The *length* of a path is the number of links connecting its nodes. If we define the *distance* between any two nodes as the length of the shortest path between them, then we can also define the *diameter* and *average distance* of a network that is *connected* (i.e., any two nodes have a path). A *small-world network* is one that has a large average clustering coefficient (compared to random networks of the same size) and whose diameter is near or smaller than the natural logarithm of its network size. In small-world networks, nodes are connected by relatively short paths.

- The *correlations among degrees* of nodes classifies networks in to *assortative* (high degree nodes are highly correlated with other high degree nodes), *disassortative* (high degree nodes are highly correlated with low degree nodes), or *uncorrelated*. Each of these categories has stereotypical features. There are also potential *correlations between degrees and weights* in weighted networks. The *inverse participation ratio* of a node i is given by:

$$Y_i = \sum_{j=1}^{N} (\frac{w_{ij}}{s_i})^2$$

  If $Y_i$ is close to 1,that indicates a single node with much greater strength than the rest, while values close to $1/k_i$ indicate close to equal weights among links.

- A *community* is a subset of nodes that are more closely linked to each other than to the rest of the network. There are many definitions of communities (e.g., can communities overlap?) and many community detection algorithms in the literature. This is an active area of research in part because there is so little consensus on basic issues.

- The *centrality* of a node is a measure of its importance within the network. There are numerous ways of quantifying importance as one might imagine, but some influential ones include *closeness* of a node, which the inverse of the average distance between it and all the other nodes in a connected network. The *efficiency* of a node in a network is the sum of inverses of distances between it and all the other nodes in the network, divided by N-1. The *betweenness* of a node i is given by:

$$b_i = \sum_{r,s} \frac{n_{rs}^i}{g_{rs}}$$

  where $n_{rs}^i$ is the number of shortest paths between node r and node s that pas through node I, and $g_{rs}$ is the total number of shortest paths between r and s. The *PageRank* of a node (so-called because of its use by Google's search engine is a bit more complicated:

This is just a small taste that excludes interesting techniques like percolation and ensembles, and ignores significant phenomena like diffusion and control. Nevertheless, this sample will be enough to give the reader a flavour of this literature and introduce some important structural quantities to which we appeal later.

## 5.3 Network Information Theory

In Chapters Three and Four, we saw information channels and coded information channels and saw many examples of information theoretic measures like mutual information and equivocation. In network information theory, we consider situations where there is more than one source (multiple-access), more than one destination (broadcast), and destinations of one communication can be the sources of others (relay). Ideally, we would want to be able to measure information flow through a network, capacities of networks, and so on. It turns out that this is usually impossible because there is no general theory of entropies and capacities in networks. Many of the obvious things one would want to know are open problems. Still, there are myriad techniques for computing significant and fruitful network information measures.

Let us imagine a simple network of four nodes in the following configuration: X1 and X2 are sources, Y1 and Y2 are receivers. We can model the way X1 and X2 send information to Y1 and Y2 in various ways. One would be to have four point-to-point emotion channels. This would, however, ignore all the network effects we care about. Instead, we can think of various combinations of influences of X1 and X2 on Y1 and Y2 (e.g., the probability that Y1 will be in a certain state, given that X1 and X2 are in certain states together, or the probability that Y1 and Y2 will be in certain states together given that X1 is in a certain state). The most general is a probability distribution across joint states of Y1 and Y2, given joint states of X1 and X2. X1 and X2 are each broadcasting and Y1 and Y2 are each multiple accessing (i.e., $p(Y1,Y2|X1,X2)$).

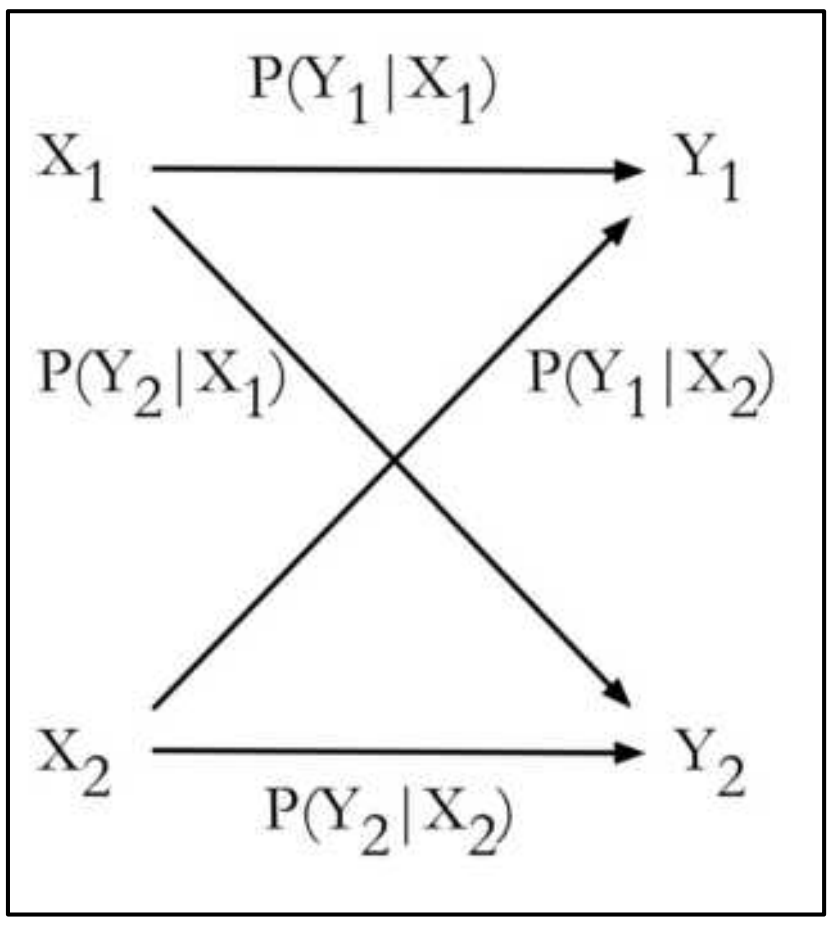

**Fig 13: Four independent channels among two sources and two destinations. This model treats the four conditional probabilities (one for each channel) as independent of one another, which is unrealistic.**

It ignores the effects that come from interactions. Instead, we need to consider a probability distribution like $P(Y_1, Y_2 | X_1, X_2)$. If each of X1 and X2 can be in state 1 (e.g., fear) or state 2 (e.g., no fear) and each of Y1 (e.g., detect fear) and Y2 (e.g., detect no fear) can be in state 3 or state 4, then we can represent it as in Figure 14. If we think of our example with the lion and cheetah from Chapter 3, we can add another source – a bear – and another destination – a Rabbit. So the Lion and the Bear ($X_1$ and $X_2$) either have fear or no fear, while the Cheetah and the Rabbit ($Y_1$ and $Y_2$) either detect fear or detect no fear.

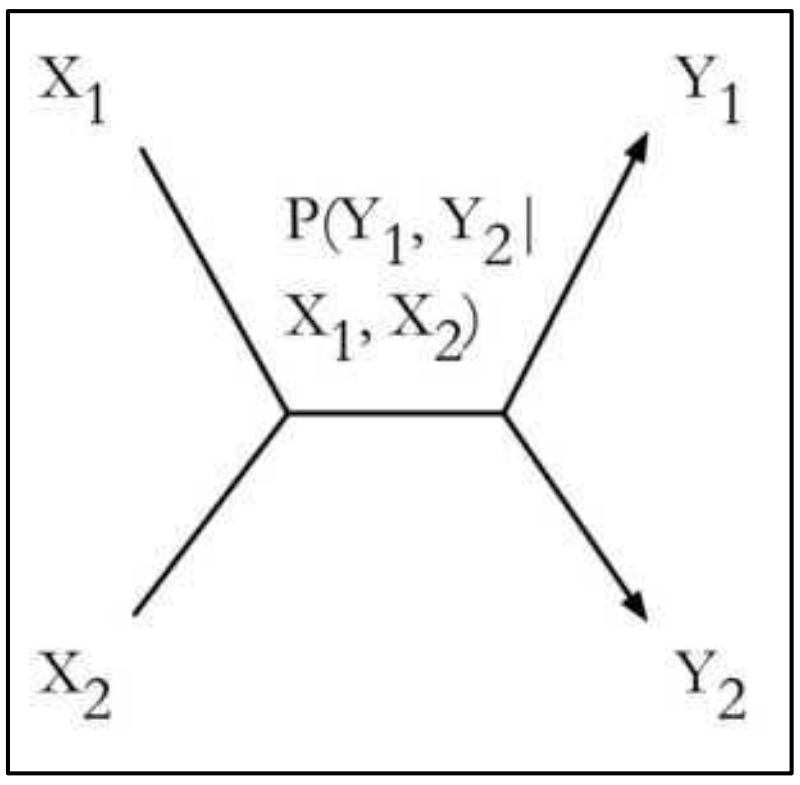

**Figure 14: A communication network with two sources and two destinations. This model uses a single complex conditional probability so that it can account for interaction (e.g., the presence of Y1 alters the way Y2 detects emotion in X1).**

It is natural to represent the data in this network as all the probabilities for each possible combination of states, which would be as a 2x2 matrix whose entries are each 2x2 matrices. All 16 joint probabilities are represented (e.g., the probability that $X_1$ has fear, $X_2$ has no fear, $Y_1$ detects fear and $Y_2$ detects no fear). In linear algebra, this structure is called a multi-dimensional array, and they are related to data structures that play an important role in machine learning and artificial intelligence research. The schema, with ? for the 16 entries, is shown in Figure 15.

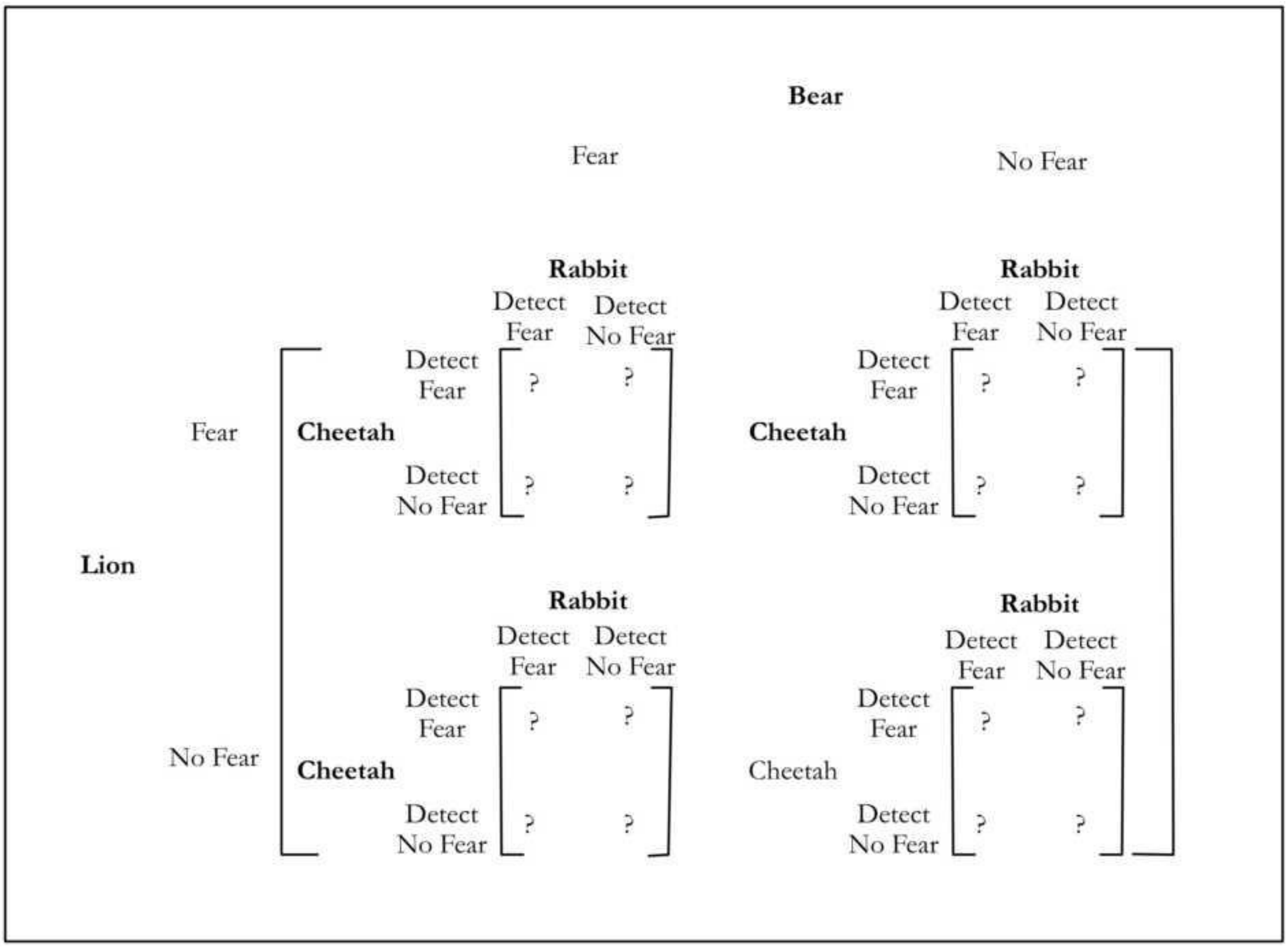

**Fig 15: Array of sixteen probability combinations for four variables.**

From this information, we can reconstruct other conditional probabilities like P(Y1|X1, X2) and P(Y2|X1, X2). For example, if we fill in the tensor in the following way as in Figure 16.

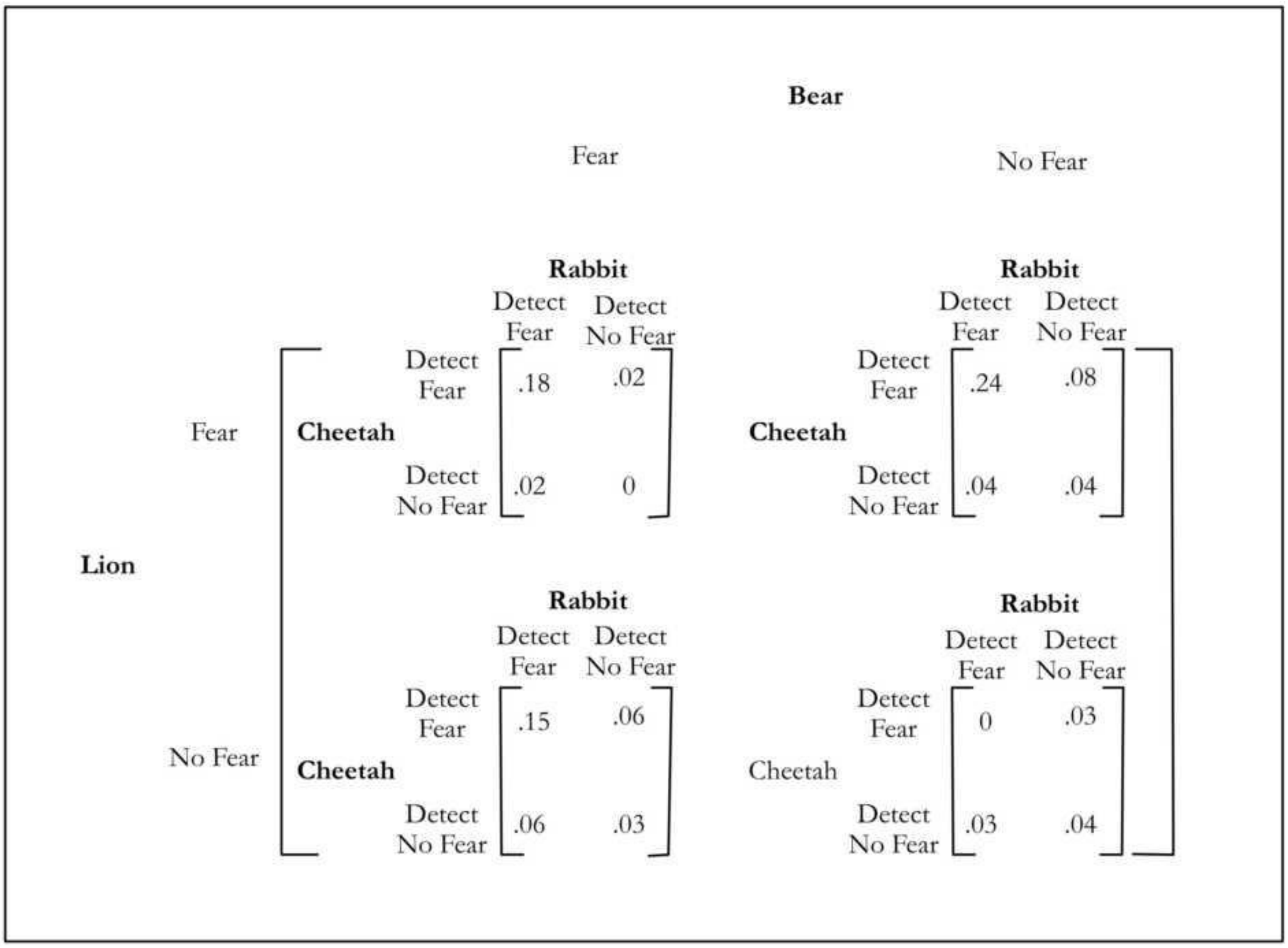

**Fig 16: Example probability distribution for lion, bear, cheetah, and rabbit.**

We can do some calculations from here. The way to read this array is that the top left entry, .18, is the probability that Y1 is in state 3 (e.g., detect no fear) and Y2 is in state 3, X1 is in state 1 (e.g., no fear) and X2 is in state 1 (e.g., no fear). So there would be an 18% chance of this combination occurring. The .04 in the bottom right is the probability that Y1 is in state 4 (e.g., detect fear) and Y2 is in state 4, X1 is in state 2 (e.g., fear) and X2 is in state 2 (e.g., fear). That is, there would be a 4% probability that this combination occurrs. Moreover, in that same matrix, we see .03 and .03 in top right and bottom left. Together these add to .06, which is the probability that only one of Y1 and Y2 detect fear and both X1 and X2 have fear. The 0 in the same matrix tells us that there is no chance that both X1 and X2 have fear but neither Y1 nor Y2 detect fear.

We can recover the chances that X1 and X2 are in various states. The matrix is as below:

| | | Lion | | |
|---|---|---|---|---|
| | | Fear | No Fear | Totals |
| Bear | Fear | .2 | .4 | .6 |
| | No Fear | .3 | .1 | .4 |
| | Totals | .5 | .5 | |

**Fig 17: Projected matrix probability distribution for lion and bear.**

There is 50% chance that X2 has fear and there is a 40% chance that X1 has fear. Conditional probabilities are also easy to calculate. For example, what is the probability that at least one of Y1 or Y2 detect fear when neither X1 nor X2 have fear? For that we look at the top left of the tensor (where X1 is 1 and X2 is 1). There is only a 20% chance of landing in this matrix at all, but when we do conditional probabilities, we assume that we are already in this matrix, and then calculate probabilities accordingly. In that matrix,{ .18.02 / .02 0}, the .02 and the other .02 are the entries we are looking for – .02 represents the chance that X1 and X2 do not feel fear and that Y1 detects fear but Y2 do not detect fear. The other .02 represents the chance that X1 and X2 do not feel fear and that Y1 detects no fear but Y2 detects fear. Together these add to .04, which is the chance that X1 and X2 do not feel fear and that exactly one of Y1 and Y2 do detect fear. To get the conditional probability we see that .04 is 20% of .2 (.04/.2). So there is a 20% chance that exactly one of Y1 and Y2 detect fear when neither X1 nor X2 have fear.

Abbas El Gamal & Young-Han Kim's network information theory textbook distinguishes between *single hop* cases where there is just one information step (i.e., sources are distinct from destinations) and *multihop* cases where there are multiple information steps (i.e., sources and destinations overlap). What distinguishes all these cases from channels discussed in Chapters 3 and 4 is the presence of multiple sources or destinations.

The two most important single-hop cases are multiple-access and broadcast. Consider a coded multiple access network with two sources, two encoding schemes, and one destination, and one decoding scheme, which is to estimate each message from each source. This structure is often called a Multiple-Access *Channel* (MAC), and many follow the convention of calling single-hop communication structures "channels." However, we follow the alternative convention (also found in the literature) of using the "channel" / "network" locutions to mark the difference between structures with one source and one destination and structures with more than one source or destination. Given the clean and elegant general theory of the former and the patchy collection of results on the latter, this is the most fundamental distinction in contemporary information theory.

Let W1 and W2 be the two sources and let X1 and X2 be the variables sending coded messages. Let Y be the variable receiving the coded messages from X1 and X1, and let Z be the destination where decoded messages arrive. The network is defined by the conditional probability distribution p(Y|X1, X2), which describes the probability of Y being in each state, given that X1 and X2 are in certain states.

We want to characterize information flow and channel capacity for this network, just as we did for channels. To do so, define the average probability of error is:

$$P_e^{(n)} = p(\left[(\widehat{M}_1, \widehat{M}_2\right) \neq (M_1, M_2)])$$

where $M_1$ is the message from W1, $M_2$ is the message from W2, and $(\widehat{M_1, \hat{M}_2})$ is the decoded two messages that arrive at Z. A rate pair R1, R2 is *achievable* iff there exists a sequence of multiple access codes, each with R1 and R2 as rates, such that the average probability of error for each code goes to zero as n goes to infinity.

In the case of communication channels, the channel capacity was a single number in bits/message, which was the maximum mutual information between the codesource and codedestination. Capacities are single numbers because there is only one rate involved. However, in multiple-access networks, there are multiple rates; hence, the capacity is a set of sequences of rates, one for each source. A set of pairs of numbers is a two-dimensional region, and we can call the *capacity region* the set of the best combinations of rates for the network in question. More precisely, the capacity region for a multiple-access network is the closure of the set of achievable rate pairs (R1, R2).

It turns out that we can characterize the capacity region for multiple-access networks in terms of the maximum values of certain mutual information measures, just as in the case of channels, in the following way:

$$R_1 \leq I(X_1; Y|X_2)$$

$$R_2 \leq I(X_2; Y\ |X_1)$$

$$R_1+R_2 \leq I(X_1, X_2; Y)$$

These mutual information measures are more complex than the channel mutual information measure defined in Chapter Three. It is worth explicitly defining the schema:

$$I(X;Y|Z) = \sum_{z \in Z} \sum_{y \in Y} \sum_{x \in X} p(x,y|z) \log \frac{p(x,y|z)}{p(x|z)p(y|z)}$$

One can prove that the capacity region for the multi-access network described above is the convex hull of the union of regions $\mathbf{R}(X_1, X_2)$ consisting of rate pairs satisfying the above inequalities over all source probability distributions for $X_1$ and $X_2$.

Broadcast networks have multiple destinations. Consider a coded broadcast network with one source, one encoding scheme, two destinations, and two decoding schemes, each of which is to estimate the message from the source. Let W be the source and let X be the variable sending coded messages. Let Y1 and Y2 be the variables receiving the coded messages from X, and let Z1 and Z2 be the destinations where each decoded message arrives. The network is defined by the conditional probability distribution p(Y1, Y2|X), which describes the probability of Y1 and Y2 being in each state, given that X is in a certain state. We stipulate that the message sent by X is supposed to be received by Y1 and Y2 without any private messages.

Again, we want to characterize information flow and channel capacity for this network, and define the average probability of error is:

$$P_e^{(n)} = p\left(\widehat{M_1} \neq M\right) \text{ or } p(\widehat{M_2} \neq M)$$

where M is the message sent from W, $\widehat{M}_1$ is the decoded message received by Z1 and  and $\widehat{M}_2$ is the decoded message received at Z2. A rate pair R1, R2 is *achievable* iff there exists a sequence of multiple access codes, each with R1 and R2 as rates, such that the average probability of error for each code goes to zero as n goes to infinity.

There is no known way to characterize the capacity region of a broadcast network in as nice a way as in the case of multiple-access networks. Still there are myriad techniques for estimating portions of the capacity region, and many specific cases of broadcast networks can be solved exactly.

Now that we have multiple-access (many sources, one destination) and broadcast networks (one source, many destinations), which are the most familiar single-hop networks, we can get a bit more complicated with multiple-hop networks. These are networks in which a single node can be a destination and a source. The most basic of these is the *relay*. A relay is essentially two channels connected together in sequence. Let U be the relay source sending the original messages, V be the coded source sending coded messages, W be the relay receiving coded messages, X be the relay sending recoded messages, Y be the coded destination receiving recoded messages, and Z be the relay destination, receiving decoded messages. An uncoded message comes from U (which just means U is in a certain state), passes through an encoder to arrive at V (which again, is just V being in some state), which sends it along the communication channel and its noise to arrive at the relay, W. The relay can just send along the same coded message, but it could alter the message instead. The relay, X, sends along the relayed message to the coded destination Y, where the recoded message is decoded and the estimated original message is delivered to the destination at Z. The relay does not decode the message. It merely passes along the coded message or some processed version of the coded message. Notice that the physical relay involves two information theoretic nodes: W and X. W is a receiver and X is a transmitter. These two components are treated as distinct, even though the relay encompasses each one. It is this feature that makes the relay a multi-hop case.

Average probability of error for relays is defined in the obvious way as the probability the decoded message is not identical to the original source message. Achievable rates are defined in the same way, and the capacity (not capacity region, because there is only one rate describing the system) is the supremum of all achievable rates. There is no known general way of charactering the capacity, but there are many ways of estimating aspects of it. For example, using the *cutset method*, which looks at every way of cutting all the edges leading from sources to destinations, provides an upper bound on the capacity:

$$C \leq \max_{p(v,x)} \min[I(V,X;Y), I(V;W,Y|X)]$$

These mutual information measures are complicated because they involve joint distributions and conditional distributions as their factors, which we also saw above.

We have already seen examples of relays both in the characterization of signal systems in Chapter Two and in the description of direct elicitation channels as sequential combination of a pre-appraisal channel with a post-appraisal channel. In the first case, each animal serves as a relay in virtue of its contingency plan, which was assumed to be the same throughout the population. In the second case, the relay was the appraisal mechanism, which receives information from the environment (e.g., that there is a spider) and transmits an appraisal of that information (e.g., that there is danger). In each of these cases, we were ignoring coded communication, so the issue of relays did not come up. Note also that our argument that the message in emotion communication is the fittingness conditions of the emotion involved ignores the possibility of relays that might process coded information without decoding it. It is unclear whether genuine relays exist in real emotion communication systems. Perhaps one might be the propensity for some animals to mimic the posture of the people near them. This posture mimicking might be performed automatically without interpreting the posture being mimicked. If so, then posture mimicking might have a relay structure. To the extent that posture is an expression of emotion, posture mimicking relays might need to be elements in quantitative emotion communication models.

When we describe more complex networks, we use the mathematical tools from the previous two sections. A *multi-cast network* is a weighted directed a-cyclic graph G={N, E, C}, where N is the set of nodes, E is the set of edges, and C is the set of weights for each edge, just as in Section 5.1 and 5.2. Here, each node is a source / destination pair, each edge is a communication channel directed from source to destination, and each weight is the information through the channel in question. Each channel is assumed to be noiseless in the simple case, but this idealization can be dropped later. In a multi-cast network, there is a unique node that is only a source, and there are multiple nodes that are only destinations. A *unicast* network is a multi-cast network with just a single destination node.

We want to understand the behaviour of the network as messages are sent by the source node and ultimately received by the destination nodes after passing through the other nodes in the network. The nodes that are not sources or destinations are assumed to be relays with their own relay encoders; they do not decode and then encode the messages. Notice that this multicast network has no broadcasting at all, and there is no interference in the multiple-access elements. There is also just one overall source in the network and it sends only one message.

If we define the average probability of error across the destination nodes in the usual way, then there is a single rate for the network as a whole. As such, network has a capacity, which is the supremum of the achievable rates. For each destination node j, define a *cut* as a partition $(S, S^C)$ of N such that the source node is in S and j is in $S^C$. The *capacity of the cut* is given by:

$$C(S) = \sum_{\substack{(k,l)\in E \\ k\in S, l\in S^C}} C_{kl}$$

The capacity of the cut is the sum of capacities of edges that cross the cut from S to $S^C$. We can then define the capacity of a multi-cast network as:

$$C = \min_{j\in D} \min_{\substack{S\subset N \\ 1\in S, j\in S^C}} C(S)$$

This result is known as the network coding theorem, and it can be extended in various ways to networks with broadcasting and cycles. The method of cuts can be used to get bounds on capacity regions in more complex setups like having noise, multiple sources, multiple messages, interference, and cooperation.

New concepts in information networks include *interference*, where the actions of multiple emotion sources impact the probabilities of detection for each source.[3] For example, detecting anger in one animal affects one's ability to detect joy in another. There is also *cooperation*, where sources work together to improve communication. For example, two people coordinate their expressions of frustration so that it is more likely that some other person will detect the frustration in each of them than it would have been had they not coordinated their behaviour. That such phenomena are present in nature we can call the *emotion communication interference hypothesis* and the *emotion communication cooperation hypothesis*, respectively.

## 5.4 Information Theoretic Results: Channels vs Networks

[3] el Gamal (2012).

Breakdowns of theorems like the source coding theorem mean that emotion information compression and emotion information error correction, which are independent in a single channel, interact in networks. That is, when dealing with channels, compression (how much of the emotion information in the animal is communicated by that animal's behaviour) and error-correction (how much extra emotion behaviour beyond the bare minimum for the information involved) are separate. The forces pushing these quantities in various ways act independently. One's choices about compression do not affect one's choices for error correction, and vice versa. Not so with networks. Decisions about how to correct errors in emotion communication networks can affect decisions about how to compress emotion information, and vice versa. There might not be ways of optimizing each independently and so tradeoffs and compromises would be the rule. One might expect to see network effects on error correction codes (i.e., emotion behaviors and behavior detections) and on emotion compression (i.e., information lost from emotion state to emotion behavior) in emotion communication systems.

Another major breakdown is already apparent – Shannon's theorem for channels states that as long as the rate is below the channel capacity, any level of accuracy can be achieved through error correction, where the capacity is defined as the maximum mutual information. We have seen that there is often no known way to characterize network capacity regions in terms of mutual information of variables involved.

## 5.5 Animal vs Register Interpretation of Nodes

When developing the theory of emotion communication networks, we face a fundamental problem. There are various ways to interpret a basic network that depend on disassociating *animals* and *detectors*. One choice is to think of X1 and X2 and Y1 and Y2 as individual people or animals. But then the detection states in Y1 and Y2 are not detecting individual emotions in X1 and X2 individually. Rather, they are each detecting something like an overall emotional state in the environment (which includes the combination of X1 and X2). Perhaps some animals rely on such detection systems. The other choice is to think of Y1 and Y2 as separate *emotion registers* in a single person or animal, where Y1 is set to measure X1 and Y2 is set to measure X2. This is a more familiar way of interpreting the information theory formalism from Models 1 and 2 of thinking about the variables as individual *detectors*. However, this interpretation is no longer an emotion network in the sense of multiple *people* both broadcasting and multiple accessing. The choice between these two ways of understanding emotion communication networks formally is fundamental. Everything that follows depends on it. We choose the second option for the simple reason that we think animals and people in particular have individual emotion registers that track the emotions of specific animals in the environment. One of our emotional capacities is to be able to keep track of the emotions of multiple independent animals at once. We do not think that humans and animals are limited to an overall emotion assessment of the entire environment. Option 1: You do not know how anyone feels but you know that the overall emotion conveyed to you by everyone. Option 2: You can know how each other animal feels, but you have no direct measure of the overall emotional state in your environment.

The choice of option two for interpreting emotion communication networks means that our model will be more realistic but at the cost of being more complex. Ultimately, a rather complex constellation of networks will be necessary to make sense of how many people, each with many emotion registers, keep track of the emotions of many people, while at the same time broadcasting to many people about their own emotional states. That sounds complicated, but you do it every time you are among a group of a few or more people.

One thing the reader must keep in mind is the difference between a *network of animals* (several animals broadcasting emotions and several animals multiple-accessing emotions from them) vs an *information theoretic network* (a conditional probability distribution with more than one individual probability distribution involved on each side). Because we have gone with option 2 when interpreting information theoretic networks, there will be situations where we have an information theoretic network but not a network of animals. Networks of animals always require information theoretic networks, but even with a single animal multiple-accessing from two broadcasting animals (not a network of animals), we need an information theoretic network with multiple sources and multiple destinations.

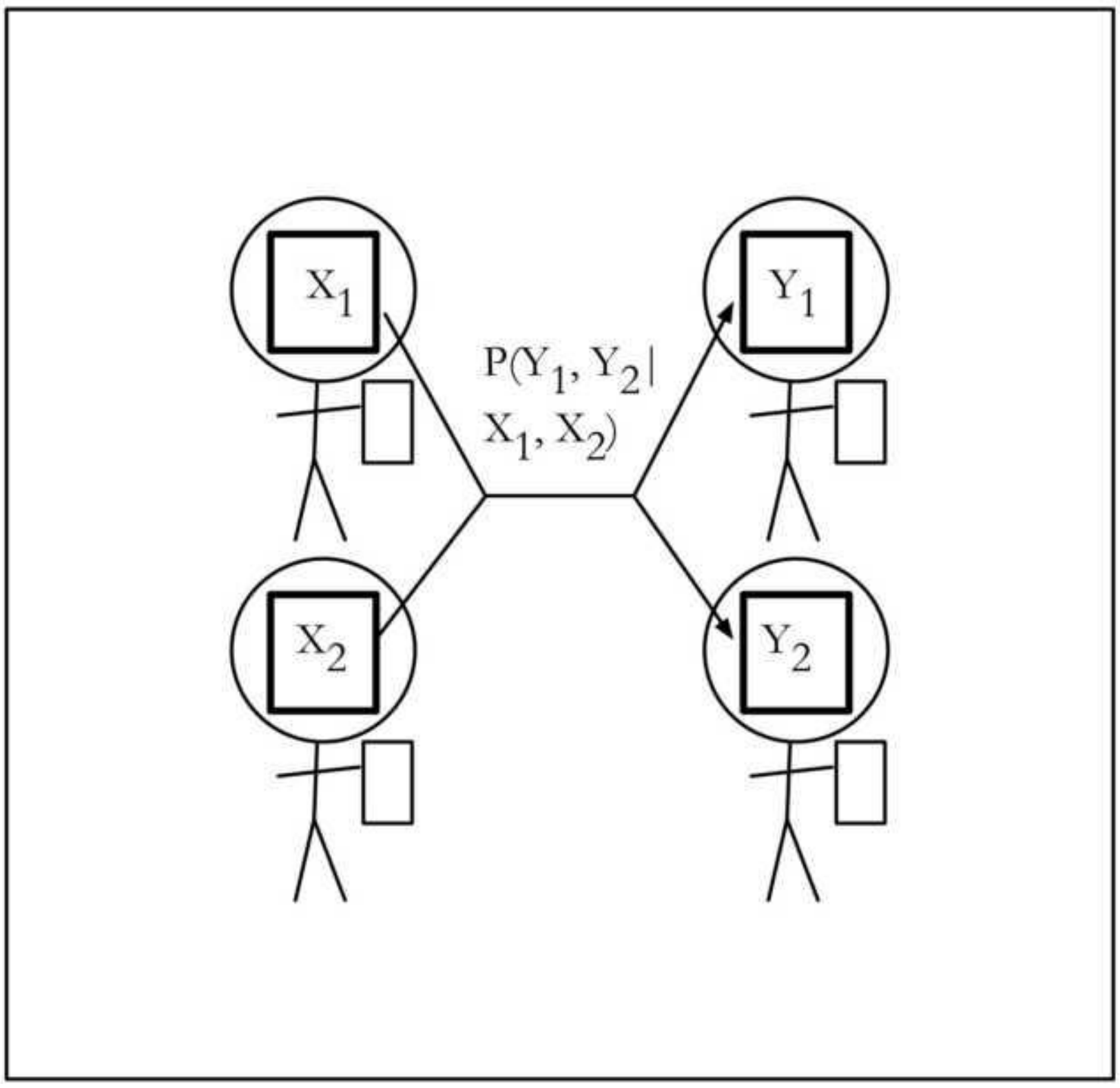

**Fig 18: Option 1 for interpreting a network – "Animal" interpretation. Communication network with destination nodes as individual animals. X1 and X2 are capacities to feel an emotion (e.g., capacity to feel fear or no fear) and Y1 and Y2 are capacities to detect an emotion (e.g., capacity to detect fear or no fear). Option one treats destinations (Y1 and Y2) as "overall" emotion detectors, not targeted to individual animals. Y1 is the overall emotion signal in that animal's environment and Y2 is the overall emotion signal in that animal's environment.**

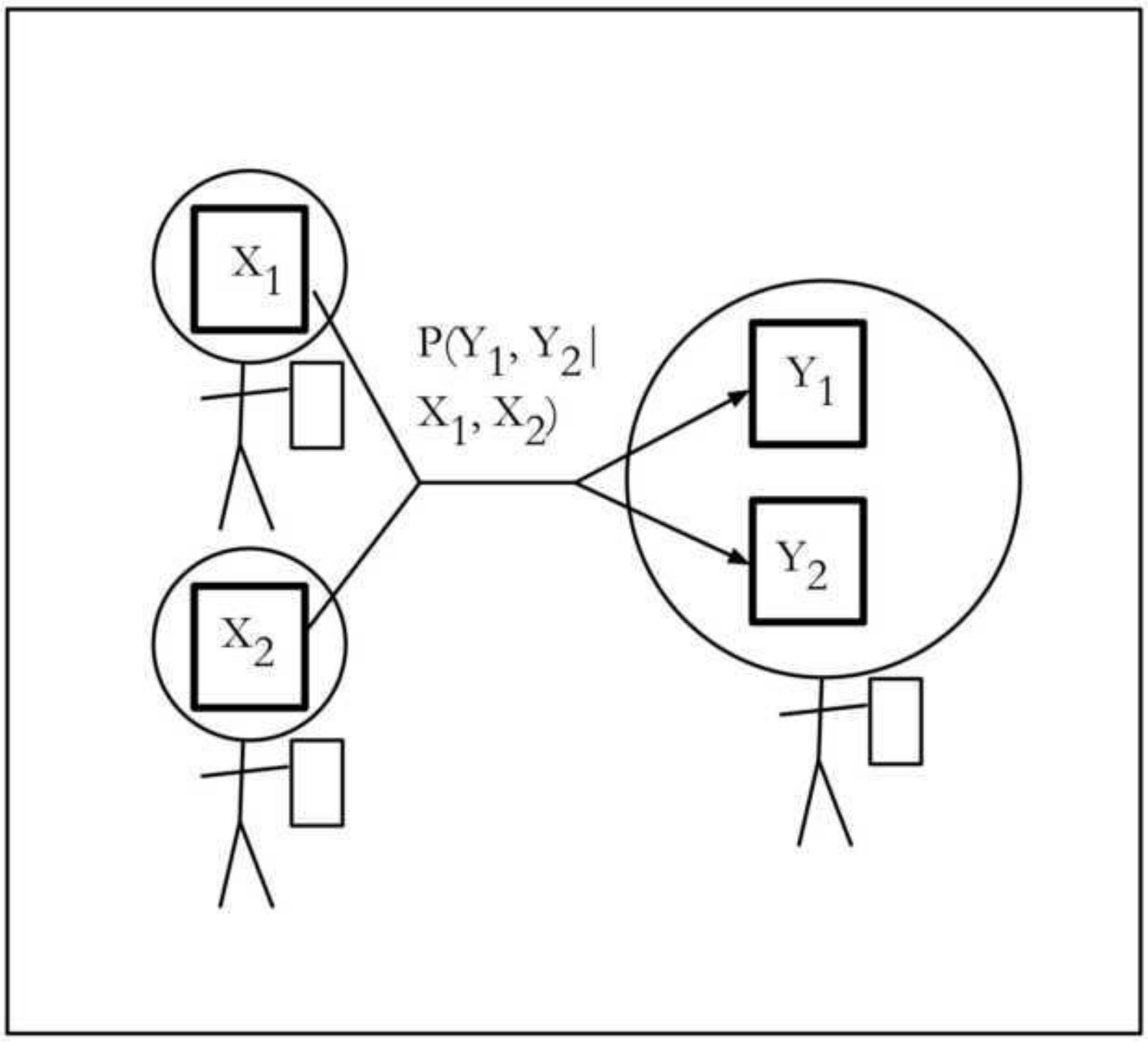

**Fig 19: Option 2 for interpreting a network – "Register" interpretation. Communication network with destination nodes as registers in a single animal. This is an information network, but only one animal is receiving information. Option two treats destinations (Y1 and Y2) as individual emotion detectors that are targeted to individual animals. Y1 detects the individual emotion signal from animal X1 and Y2 detects the individual emotion signal from animal X2.**

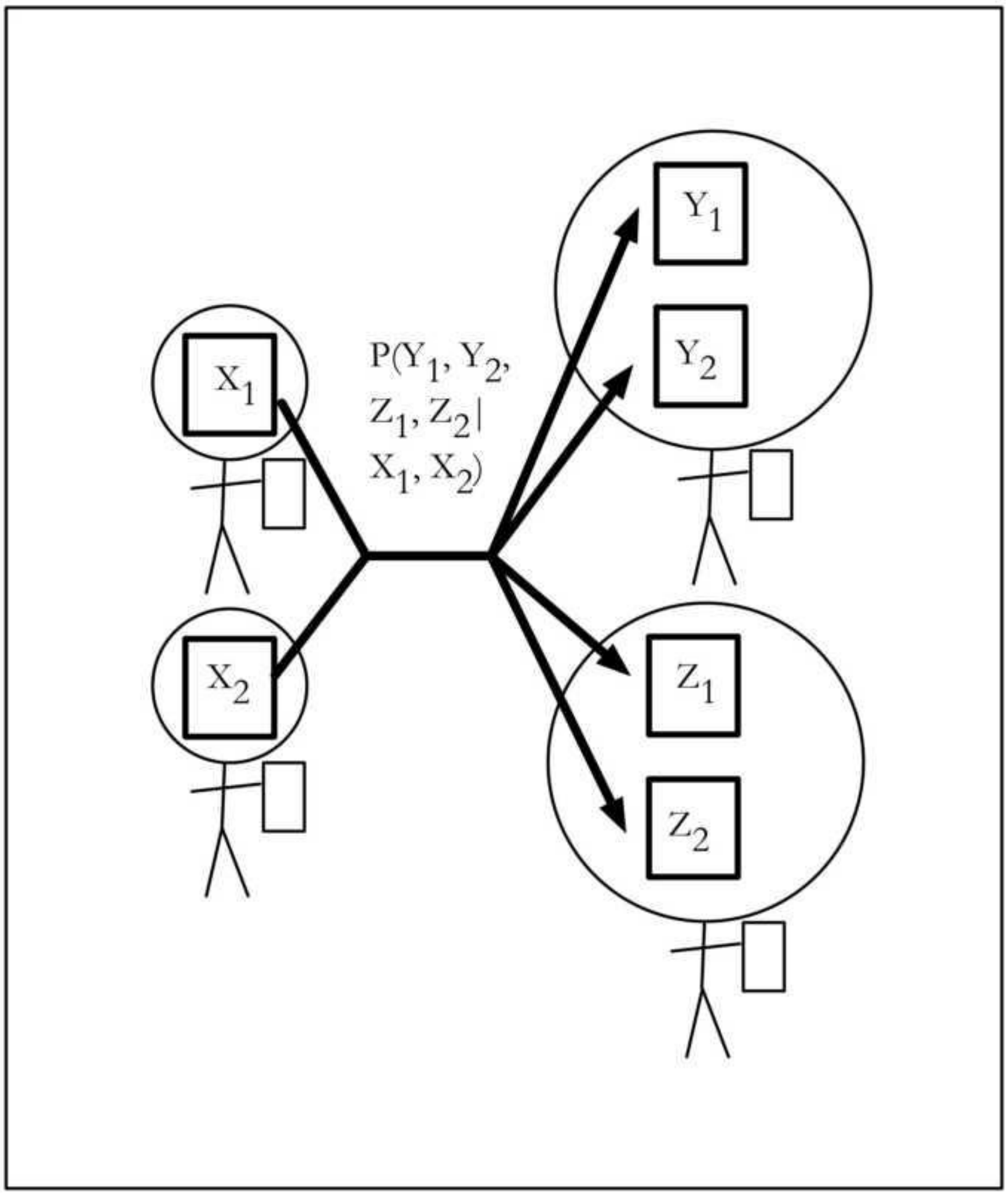

**Figure 20: Network with multiple receiving animals. Multiple receiving animals according to option two interpretation, where each animal has multiple detecting registers. X1 and X2 are capacities to feel an emotion (e.g., capacity to feel fear or no fear) and Y1 and Y2 are capacities to detect individual's emotions. Option two treats destinations (Y1, Y2, Z1, Z2) as individual emotion detectors that are targeted to individual animals. Y1 is the capacity to detect fear or no fear in X1. Y2 is the capacity to detect fear or no fear in X2. . Z1 is the capacity to detect fear or no fear in X1; Z2 is the capacity to detect fear or no fear in X2. Y1 and Y2 are inside a single animal, while Z1 and Z2 are inside a different animal. Y1 and Y2 allow animal Y to keep track of X1 and X2 separately. Likewise, Z1 and Z2 allow animal Z to keep track of X1 and X2 separately.**

Now that we have decided how to empirically interpret our mathematical formalism for networks, we can look at the basic set up and use some examples to illustrate how to calculate a few information-theoretic quantities of networks.

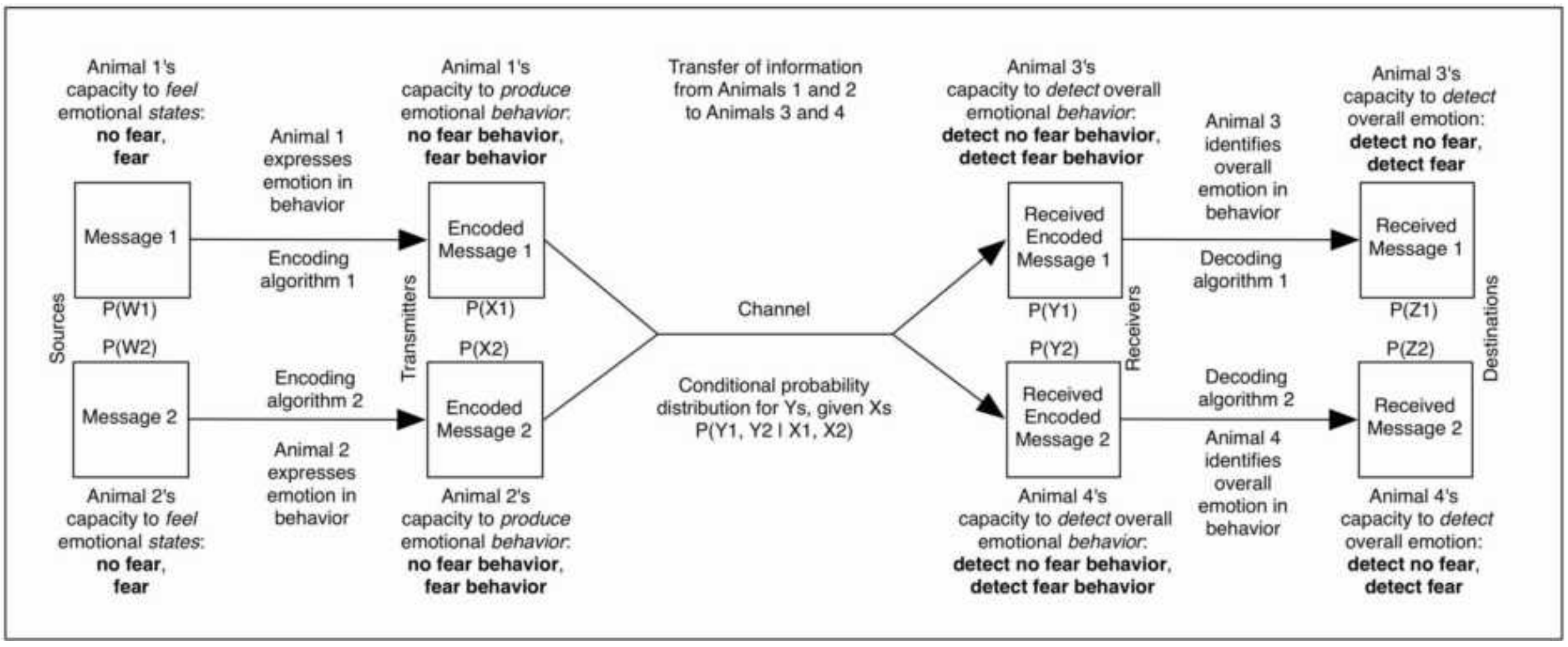

**Fig. 21. Coded emotion network – "Animal" interpretation. A coded emotion communication network with two sources and two destinations. Each source encodes a message independently, and each destination decodes a message independently. The channel is defined as a conditional probability distribution, which specifies the probability that receiver one and receiver two will be in certain states, provided that transmitter one and transmitter two are in certain states. The emotion communication network displays broadcasting (sending a message to multiple sources) and multiple-access (receiving a message from multiple sources). Emotion communication networks can be interpreted in two major ways. In this figure, each destination is a distinct animal. Animals three and four each detect emotion behavior, but they do not distinguish between animal one and animal two – the just detect general emotion behavior in their environment and each one detects emotional states in their environment. Whereas in Figure 21, the destination animal detected fear in the source animal, in this model, each destination animal detects fear or no fear generally in its environment.**

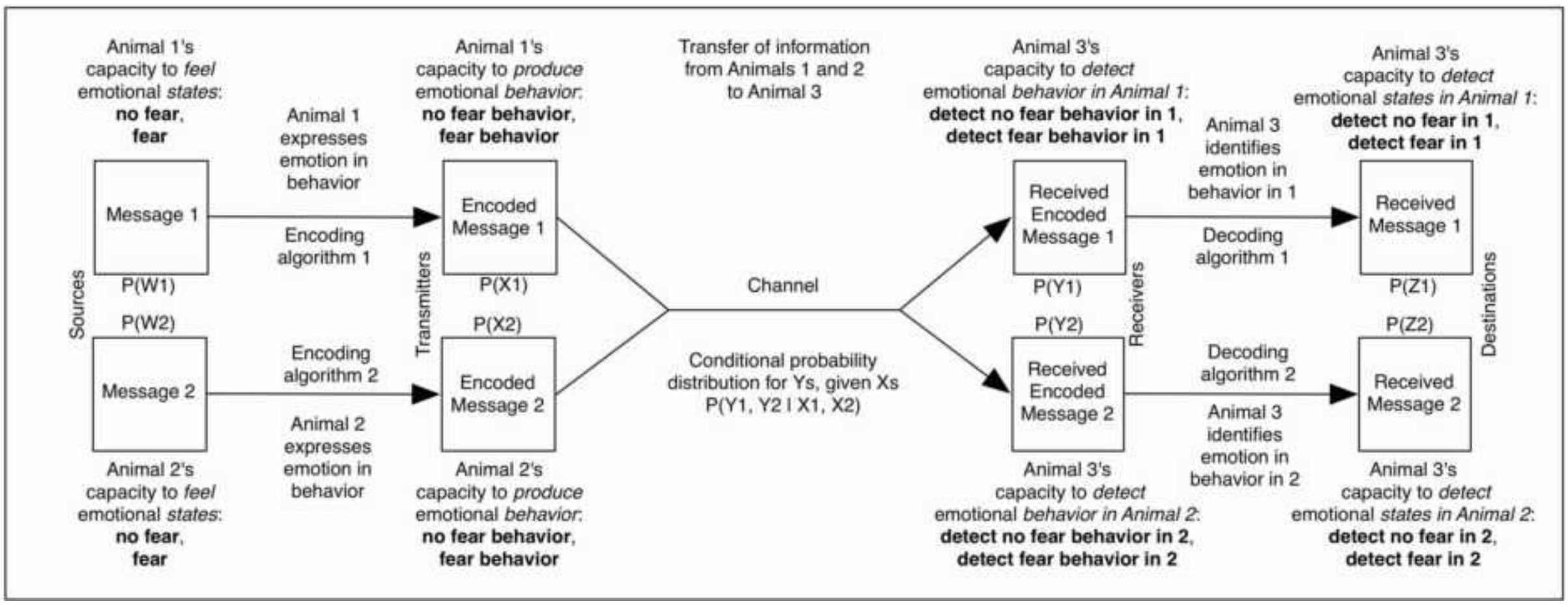

**Fig. 22: Coded emotion network – "Register" interepretation. A coded emotion communication network with two sources and two destinations, just as in Figure 21, but here the destinations are interpreted differently. All the destinations are within a single animal (animal three). Each destination tracks the emotional state of a distinct source animal. In this sense, the receivers and the destinations are registers that allow animal**

**three to receive independent emotion messages and keep track of which animals they are from. Whereas in Figure 21, the destination animals each detected fear or no fear in their general environment, in this model, animal three detects fear or no fear in animal one and detects fear or no fear in animal two. This interpretation is more fruitful than the one in Figure 21, and we use only it below.**

Because we have chosen the "register" interpretation which allows each animal to keep track of emotion information from multiple sources separately, the structural features of emotion communication networks can be characterized in more detail. In the rest of this section, we focus on a network of animals, each with a number of emotion detection registers, but we keep track of separate steps in the network. Each step represents one level of emotion communication. As such, we can think of the network as a *message network* – the path taken by an emotion message, where we keep track of how many communication steps have occurred since the original source.

In a message network, two fundamental parameters must be specified. The first is the number of emotion detection *registers* in each animal. For humans, Dunbar's number of a 148 is probably a good estimate. The other parameter is how many other animals have the opportunity to engage in emotion communication. In face-to-face communication, this number is fairly small – how many animals can observe another animal well enough to detect its emotion behaviour. In social media communication, this number might be the entire network. We can call this the *accessibility* parameter. Once we specify the register and accessibility of an emotion communication network, we can see that if accessibility is greater than register, then there will be more than one step of communication that displays *edge spreading*, where the number of edges coming in to a node is less than the number of edges coming out of a node. Only once we reach *register saturation*, do the nodes in the network stop showing edge spreading. For example, if accessibility is 2 and register is 3, then each animal has three emotion detection registers but each animal's expression of emotion can be detected by only two other animals. In this case, we can plot the number of animals and number of information theoretic nodes as a function of communication steps, assuming animals communicate only with other animals in the previous step or in the next step:

| | | | | | | | | |
|---|---|---|---|---|---|---|---|---|
| Step | 1 | 2 | 3 | 4 | 5 | 6 | 7 | … |
| Animals | 1 | 2 | 4 | 8 | 16 | 32 | 64 | |
| Nodes | 1 | 2 | 8 | 24 | 48 | 96 | 192 | |
| Inactive | 3 | 4 | 4 | 0 | 0 | 0 | 0 | |
| In/animal | 0 | 1 | 2 | 3 | 3 | 3 | 3 | |
| Out/animal | 2 | 4 | 6 | 6 | 6 | 6 | 6 | |
| In/Out | 0 | .25 | .3 | .5 | .5 | .5 | .5 | |

The Animals value goes up according to accessibility, and Nodes measures how many information theoretic nodes there are at that step, by which we mean the number of active registers. Inactive is the number of inactive registers at each step. In/animal is the number of communication channels going into each animal. It starts at zero for the first step with only one animal serving as a source, but not a destination, and it goes up by one each step as the animals in each step detect emotions from more other animals until it reaches register, which is the maximum number of other animals one can track individually. In this case, register=3. Out/animal is the average number of communication channels leaving each animal, which is the number of other animals detecting emotions in that animal. Out/animal starts at 2, which is the accessibility parameter, and goes up until register saturation. It stops increasing after that because there are only so many animal registers in the next step that can be used to detect emotions in animals at the current step. This is a key insight – that the number of emotion registers in each animal sets a limit to not only the incoming connections but to the outgoing connections as well. In/Out is the average number of in-edges /out-edges for each animal. For us, it measures the ratio of incoming emotion communication (detected in others) to out going emotion communication (detected by others). For emotion communication networks understood in step-wise fashion like this, the In/Out value is always less than one – that is, each animal is being detected by more other animals than it is detecting animals' emotions. Prior to register saturation, the In/Out ratio is much lower, reflecting greater edge spreading. It increases until it hits a value of .5, which depends both on the register parameter and the accessibility parameter.

The importance of register saturation in message networks (which occurs at step n=register+1) for understanding emotion network structure and dynamics is crucial. Once we hit that register saturation, the Out/animal value can differ across animals within a step because there are too many animals at each step for all the animals at the next step to communicate with all of them. So, to understand network structure after register saturation, we need to specify how the animals select which animals to communicate with from among those animals that are accessible.[4] This *selection algorithm* might differ from animal to animal and from context to context. The choice of selection algorithm will certainly affect the distribution of Out/animal values across the population of animals at a step. Some selection algorithms will result in a few animals with high Out/animal and many animals with low Out-animal, whereas some will result in more uniform distributions of Out/animal values among animals at a step.

Register saturation, and the resulting plateau for In/Out value also means that emotion message networks are unlikely to be scale-free because they have natural upper bounds on In/animal with the number of registers, and on Out/animal as well. The selection algorithms used in a population can shift the distribution of out-degrees, but not in-degrees. In face-to-face communication networks, the connected sub-networks are small and the accessibility value is small, and emotion message networks involved probably do not usually extend past the register saturation point, so their In/Out values would tend to be low and selection algorithms would have little effect. On social media, networks are huge, and accessibility values are huge, so their emotion message networks would quickly zoom past register saturation, In/Out values would tend to plateau, and selection algorithms would dominate.

[4] This is another instance of the frame problem – how to focus only on the relevant bits from among a vast possible space of information.

The importance of selection algorithms, especially on social media, should highlight a little noticed phenomenon: emotion detection rules. Psychologists already study *emotion display rules* in detail; these are societal rules governing emotion displays, also known as emotion expression. There is a variety of them, and they can be specific to age, gender, social status, community, and many other features. Emotion detection rules would be the opposite: societal rules governing emotion detection, also known as emotion perception. These rules would specify which other people's emotions get noticed. In information theoretic terms, emotion display (i.e., emotion expression) is encoding, and emotion detection (i.e., emotion perception) is decoding. So these two would be emotion encoding rules and emotion decoding rules. If animals have emotion detection registers, then there have to be all sorts of rules for which emotion registers get used for which purpose. As you walk down the street in a city, you can communicate emotionally with hundreds of people in an hour. The same registers must be reused over and over for this. Are there rules for how that works? Are there registers that are reserved only for close family or friends? Do different emotion detection registers use different decoding algorithms? What are the connections between display rules and detection rules? It would be surprising if these two kinds of rules were not correlated in various ways.

## 5.6 Qualitative Simple Emotion Communication Network

We are going to describe a situation where there is emotion communication through a network. Throughout the rest of the chapter, we develop the tools to introduce a mathematical model for this very situation (it comes in Section 4.13, and is illustrated in Figure 23).

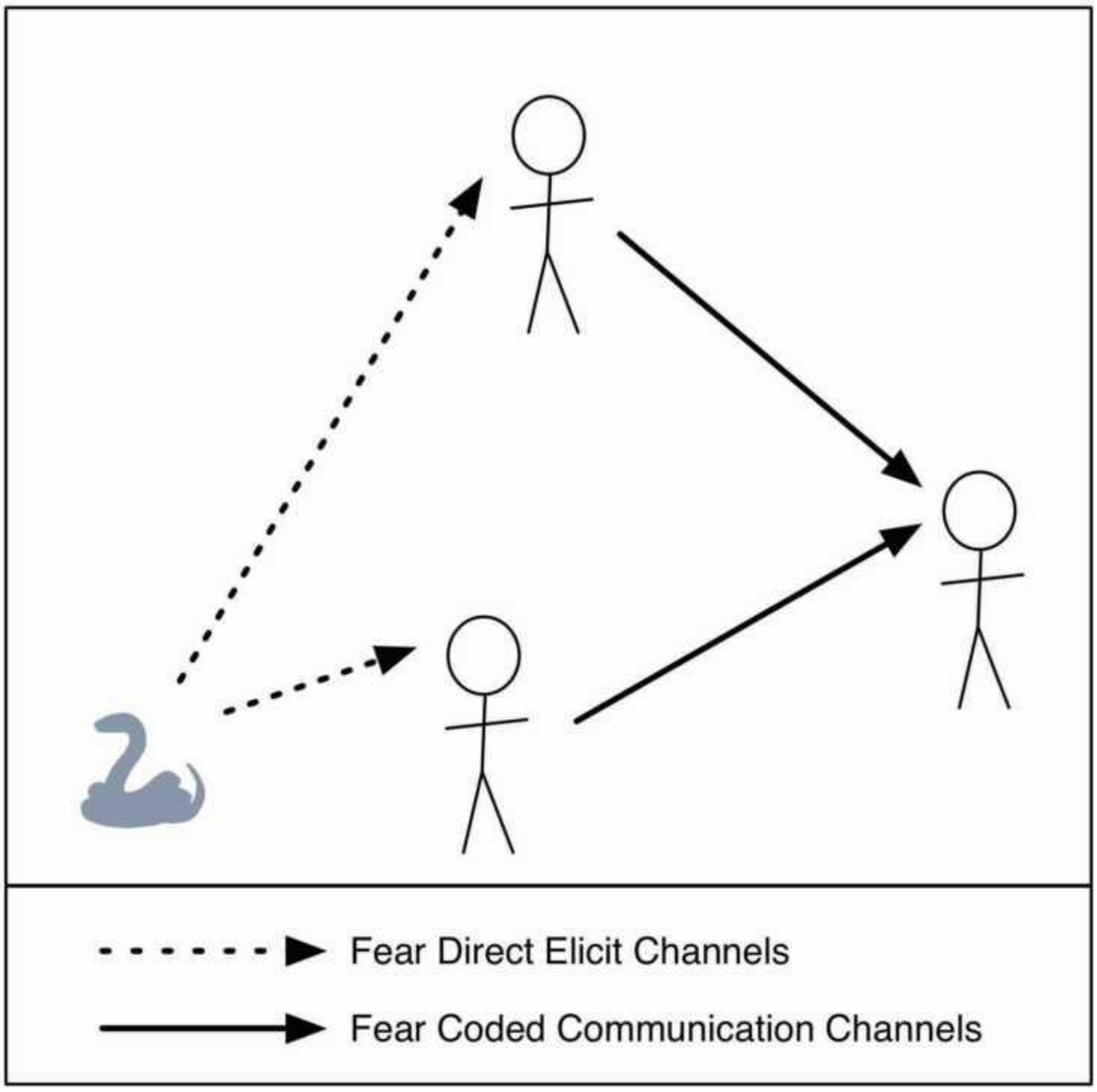

**Fig. 23. A basic emotion communication network in which a dangerous situation conveys information to two animals, which in turn convey information to a third animal. The only worldly situation is the presence of danger (snake), and there are two direct emotion elicitation channels going to two distinct animals. There are also two coded emotion communication channels. If the network is working properly, then the third person detects fear in each of the other two people by virtue of detecting their fear behavior and thereby acquires information about the dangerous worldly situation indirectly via emotion communication. The presentation is highly idealized since each node is a blackbox. Figures 24 through 30 present increasingly complex models of what is depicted here as a basic node in an emotion communication network.**

## 5.7 Emotion Elicitation Channels

Our job right now is to describe all the parts of our qualitative simple emotion communication network (illustrated in Figure 24) with enough detail and in the right way so that we can assemble a mathematical of this emotion communication network. We first work on the spider causing fear in Animal 1 and in Animal 2. These processes occur when something in an animal's environment causes them to have a certain emotion because the animal observes or comes to know about it. This process is known in the lit3ratures as emotion elicitation. We recognize two fundamental kinds of emotion elicitiation. *Indirect* emotion elicitation is when another animal's emotion behaviour causes one to have an emotion. This is an essential elicitation in emotion communication. *Direct* emotion elicitation is everything else—when something causes an emotion and it is not the emotion expression of some other animal.

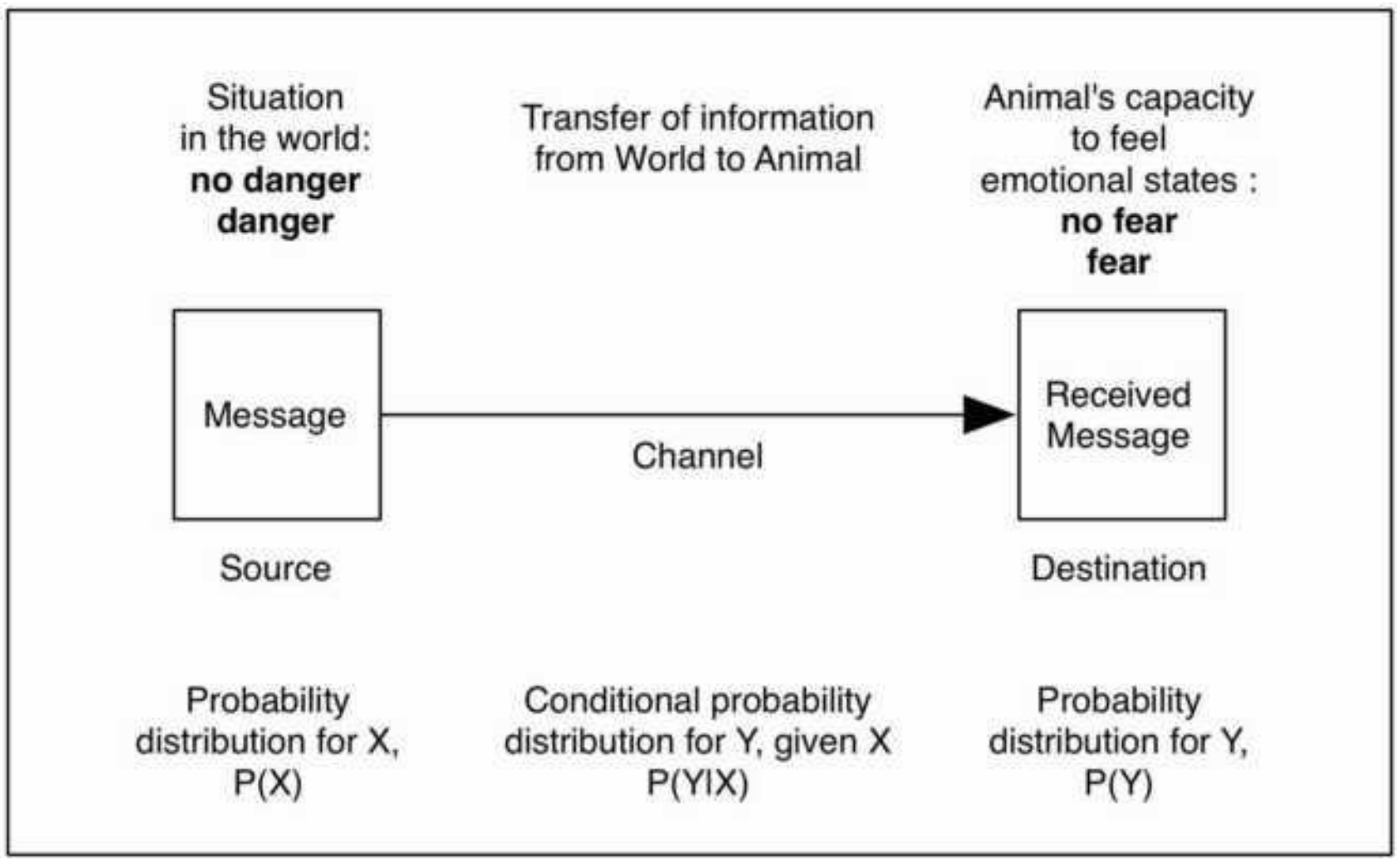

**Fig. 24. A direct elicit emotion channel.**

This is not a communication channel, but rather an *elicitation* channel. Its source is a situation in the world (here the presence or absence of danger), and its destination is an animal's capacity to feel emotional states (here fear or no fear). As usual, the channel is defined as a conditional probability distribution. Although this is not a communication channel, elicitation channels play a vital role in emotion communication networks, and we need to understand how they work to understand that role. In *direct* elicitation channels, information flows from the world to an animal's capacity to feel emotions. In indirect elicitation channels (below in Figure 25), information flows from one animal's detection of emotions to that same animal's capacity to feel emotions. Each type plays a different role in emotion communication networks (Figure 30) and multi-layer networks (Figure 39).

Now we complicate the direct elicitation channel so as to distinguish the role of emotion appraisal, which has attracted so much attention in the literature.

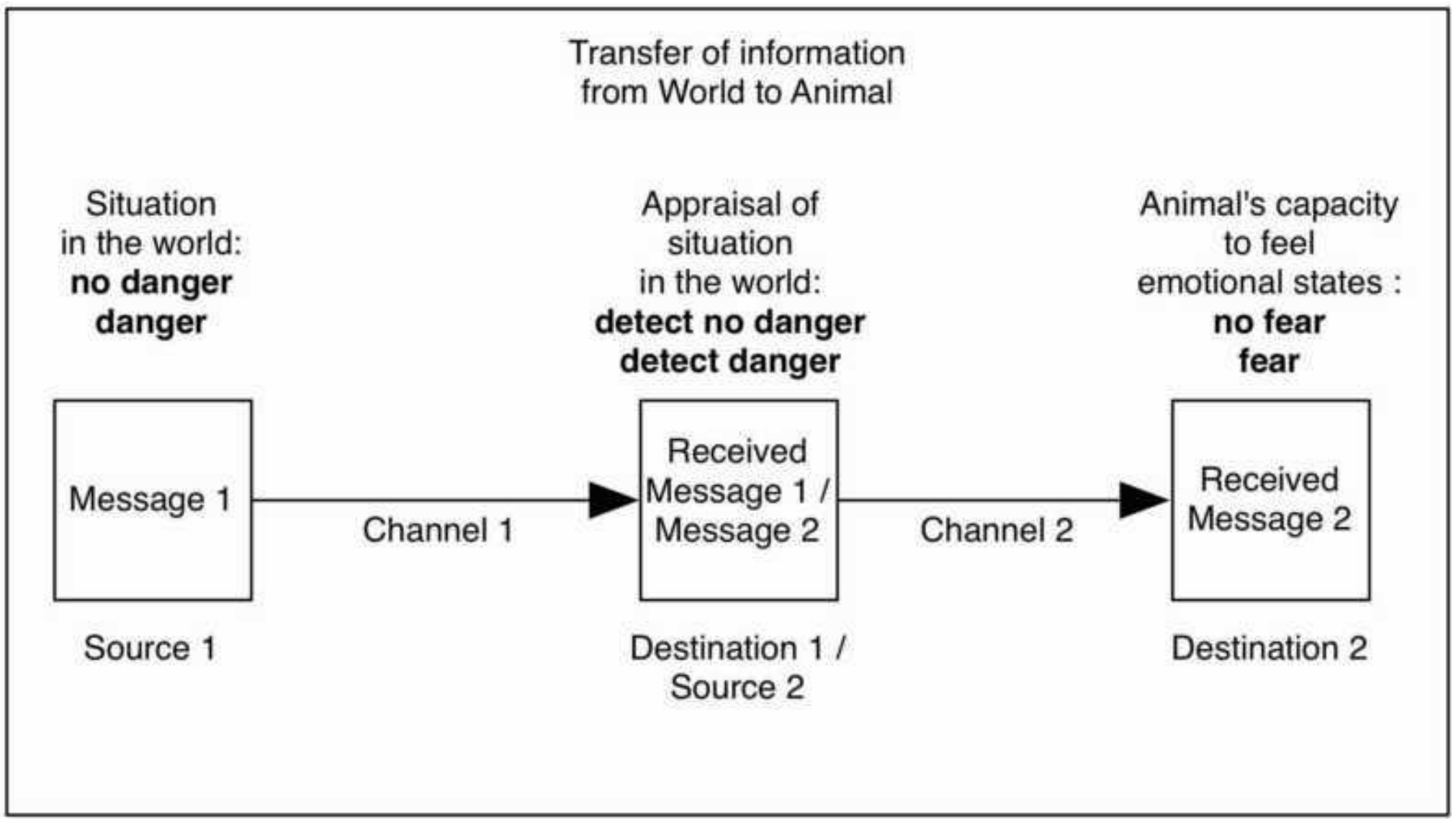

**Fig. 25. Multi-channel direct elicitation network A more complex version of the same direct elicitation process from Figure 24. In this case, there are two channels chained together (called a relay). Source 1 is the same as the source in Figure 24 and Destination 2 is the same as the destination in Figure 24. The middle node is the animal's appraisal of the situation. An appraisal is an evaluation of the bearing that the situation in the world has on the animal's interests. In this case, it also has only two states – detecting danger and detecting no danger. In the relay, each channel is defined by a distinct conditional probability distribution, and one can combine them to define a distribution over the whole, which would be equivalent to the channel in Figure 24. Two distinguish**

**the parts, Channel 1 is an *appraisal* channel and Channel 2 is an *appraisal elicitation* channel.**

## 5.8 Emotion Pre-Appraisal Channel

The appraisal theory of emotions typically emphasizes the appraisal process as essential (and maybe as the only essential element) of emotion phenomena more generally. We are neutral on this matter, but the fact that appraisals play such a big role in the literature on emotions, especially in humans, it is good reason to focus on the information theory of appraisals. In our model, we call the appraisal channel the information channel whose source is something in the world and whose destination is an emotion appraisal by some animal. Appraisals can be influenced by a variety of other variables (e.g., beliefs, desires), but for our purposes, we do not need that level of detail.

## 5.9 Emotion Post-Appraisal Channel

The emotion appraisal elicitation process begins when some animal has an emotion appraisal and ends when that same animal has an emotion. Often we think of the appraisal as causing the emotion. For example, when a spider causes fear in an animal, we have two channels – the appraisal channel and the appraisal elicitation channel. The *appraisal channel* begins with the spider and ends with the animal's appraisal (e.g., danger). The *appraisal elicitation channel* begins with the animal's appraisal (e.g., danger) and ends with that same animal's emotion (e.g., fear). Pre-appraisal and post-appraisal channels. Together they make up the direct elicitation channel (source: worldly object or event; destination: capacity to feel emotion in an animal).

## 5.10 Emotion Indirect Elicitation Channel

Indirect emotion elicitation is the process involved when emotion communication causes the receiving animal to have a certain emotion. The first half of this process is emotion communication, with which we are already familiar. It starts with one animal having an emotion and ends with another animal detecting that emotion. The second half of the process is *indirect elicitation*. It begins with emotion detection and ends with having a certain emotion.

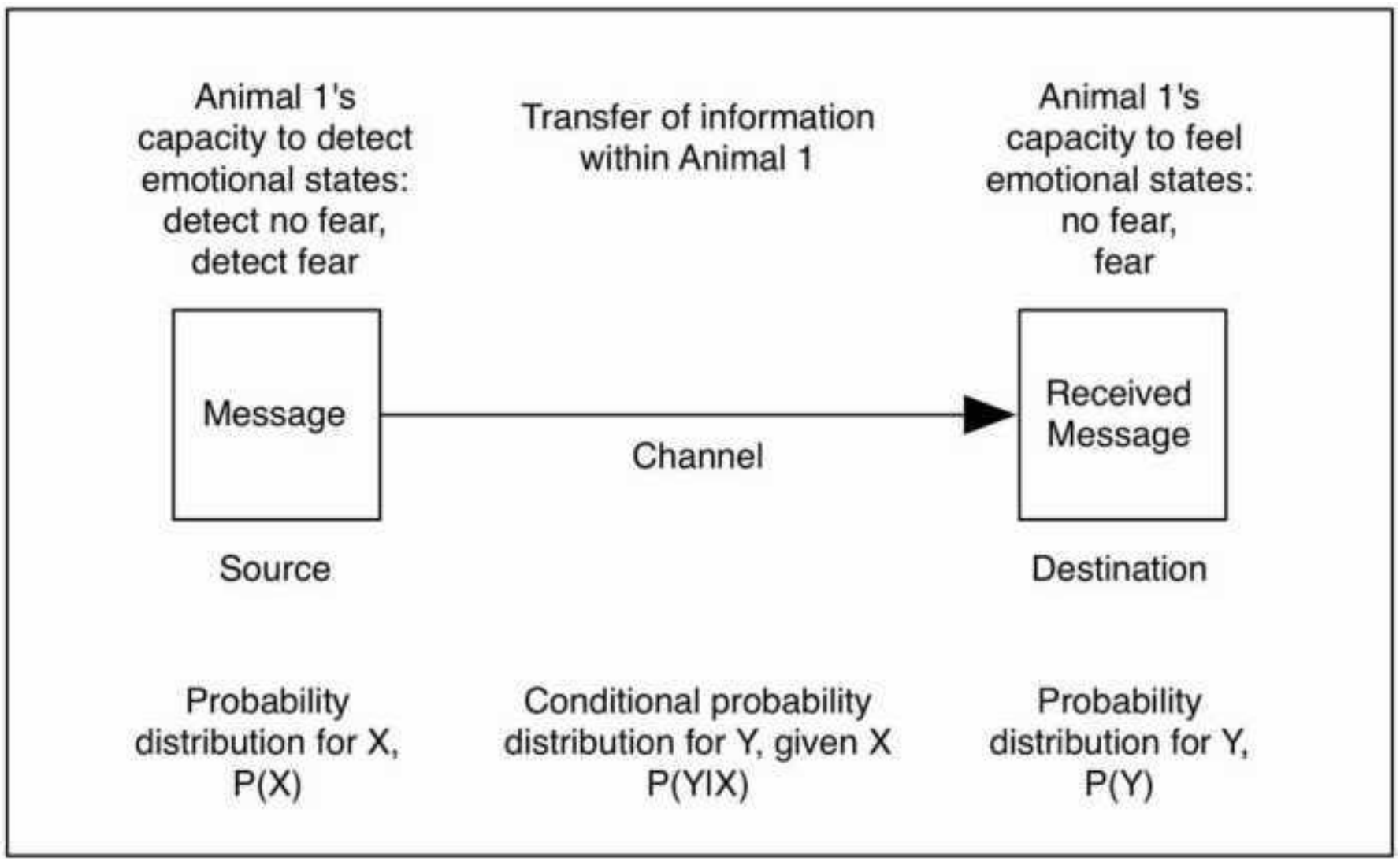

**Fig. 26. An *indirect* emotion elicitation channel. This is not a communication channel, but rather an *elicitation* channel. Its source is animal 1's capacity to detect an emotion (here the presence or absence of fear), and its destination is that same animal's capacity to feel emotional states (here fear or no fear). As usual, the channel is defined as a conditional probability distribution. Indirect elicitation channels play a vital role in emotion communication networks, in which they link emotion communication channels to one another and to direct emotion elicitation channels.**

## 5.11 Complex Nodes

So far we have developed some channels and simple networks, which will serve as parts of our models of complex networks. One additional task is developing models of the nodes (animals) in an emotion communication system. Only by developing models of what happens between the inputs (emotion behavior detection, appraisals) and the outputs (emotion behavior). That is the task of this section.

We know that we need a way of bringing together the emotions detected in others so that they might cause the animal to feel a certain emotion. Moreover, information about emotions detected in others has to combine with information about the animal's environment to have an overall effect on the animal's emotional state. We already know that the animal's emotional state is then part of the outgoing communication network whose next step is encoding that emotional state in behavior.

Below in Figures 27, 28, and 29, we present increasingly complex node diagrams. The first one has just the incoming communication network with two registers, the outgoing emotion network, and the indirect elicitation channel that connects them.

The next one has a more complex incoming communication network with three registers, and it has three direct appraisal channels. Another novel aspect is that the appraisal registers and the emotion detection registers are inputs to a single elicitation network. The other elicitation channels and networks we have seen have been either direct (emotions caused by the world) or indirect (emotions caused by others). This one incorporates each kind. The point of having an elicitation network like this is to model situations where the emotions being detected in others and the appraisals have an interdependent effect on the animal's emotional state. For example, detecting fear in other people might increase the chances that an appraisal of danger will be taken seriously, or vice versa.

The third diagram is more complex only in that it has a single integrated input rather than distinct appraisal channels and communication network. The point of having an input network like this is to model situations where emotion behavior being detected in others and the situations in the world have an interdependent effect on the animal's appraisal registers and emotion detection registers. For example, detecting fear behavior in other people might increase the chances that an observed situation will be appraised as dangerous, or vice versa.

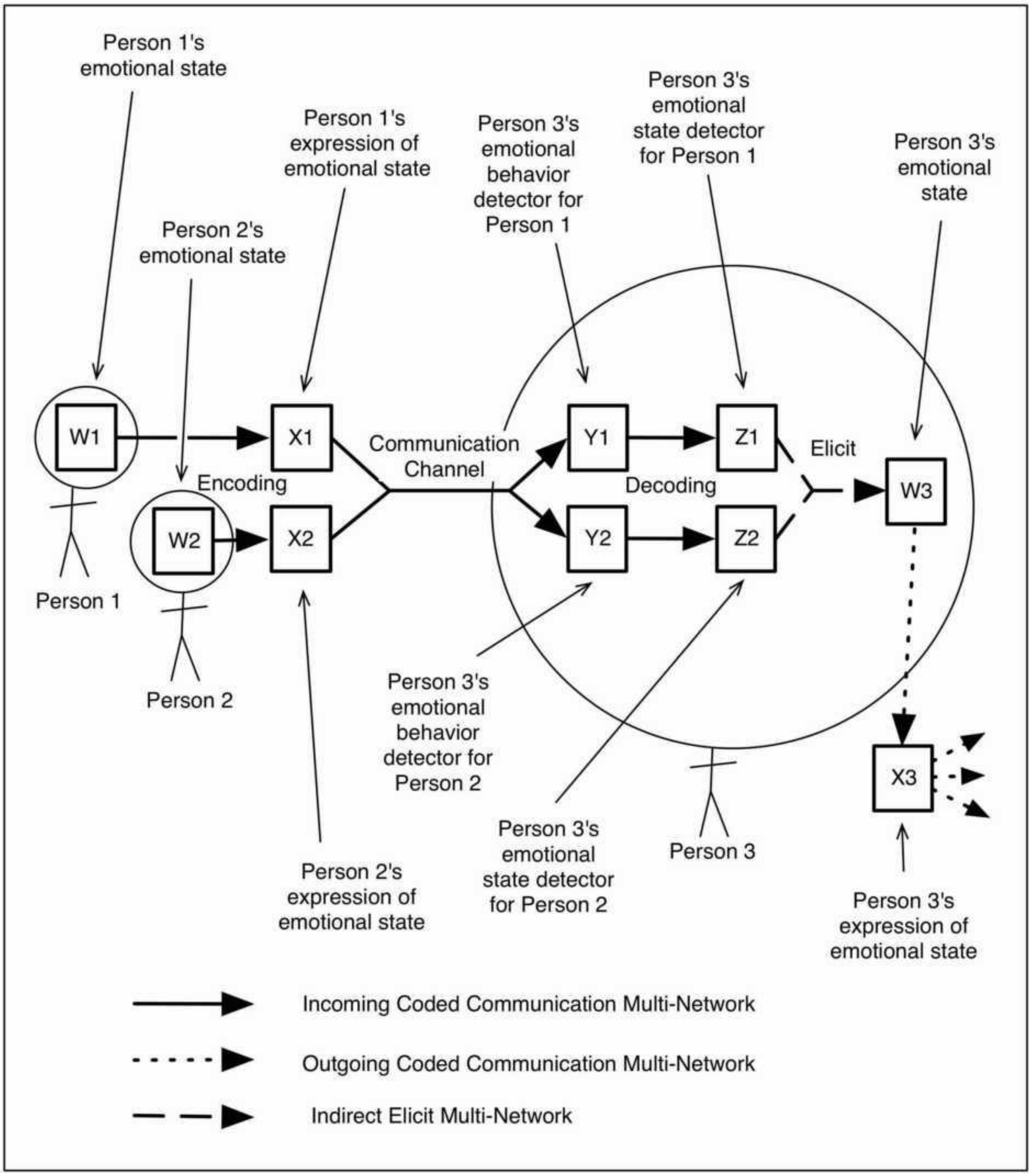

**Fig. 27. Complex node with registers and indirect elicit. A single node from Figure 23 in detail with just the communication channels and indirect elicitation channel. Person 3 is the node. The incoming communication network is just like that in Figure 22. It has**

**Person 1's emotional state (W1) and Person 2's emotional state (W2) as sources, those people's emotion behavior (X1 and X2) as transmitters, the two emotion behavior detectors within Person 3 as receivers, and the two emotion detectors within Person 3 as destinations. As usual, the emotion states are encoded in behavior, and detected behavior is decoded into detected emotion. The outgoing communication network has the same structure, but only W3 and X3 are depicted here, which would be a single source and its transmitter for other people to detect. A single indirect elicitation channel is located within Person 3, and it commects up the incoming emotion communication network with the outgoing emotion communication network. In this model, Person 3 is a *two-register communication node*.**

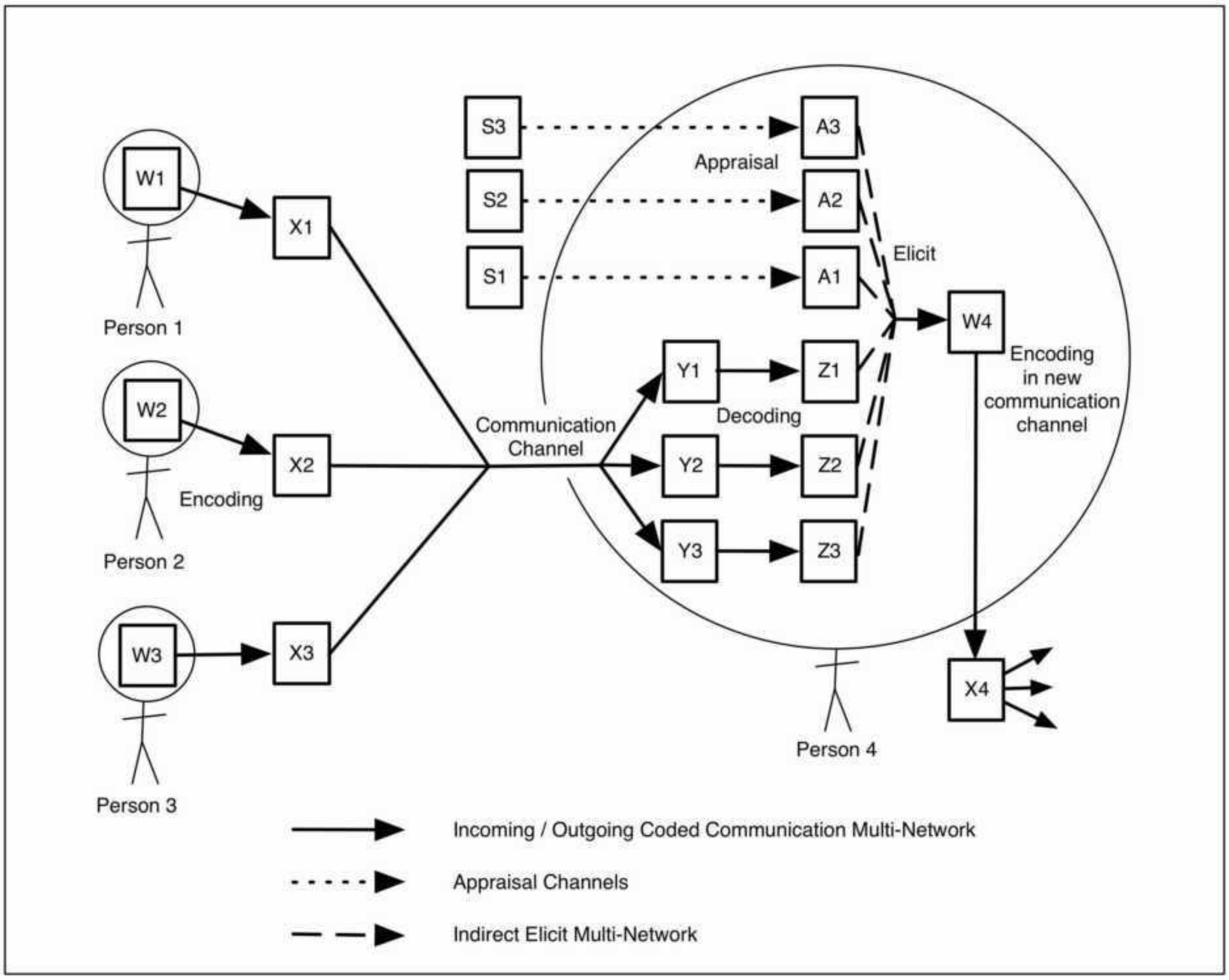

**Fig. 28. Complex node with appraisal, registers, and indirect elicit. A single node from Figure 23 with more complex detail than that depicted in Figure 27. Here, there is an incoming communication network with three sources (emotion states), three transmitters (emotion behaviors), three receivers (individual emotion behavior detectors), and three destinations (individual emotion detectors). There is also an outgoing emotion communication network, but only W4 and X4 are depicted here (just as in Figure 27). In addition, there are three appraisal channels, which means that Person 3 can keep track of three distinct situations in the world (e.g., an barking dog, a disgusting mess, and a loving partner all in the same room). As in Figure 27, there is a single indirect elicitation channel located within Person 3, but here it brings together the appraisal registers from the three appraisal channels and the three emotion behavior detectors from the emotion communication network. Technically, it is a**

**multiple access network, and it allows us to model the fact that one's appraisals, together with the emotions detected in others, combine to produce the emotion one feels. In Person 3, the indirect elicitation channel connects up the incoming emotion communication network and the appraisal channels with the outgoing emotion communication network. Notice that this depiction would have been impossible if we had been satisfied with the crude direct elicitation channel in Figure 24 rather than the more nuanced depiction in Figure 25. In this model, Person 3 is a *three-register appraisal / three-register communication node*.**

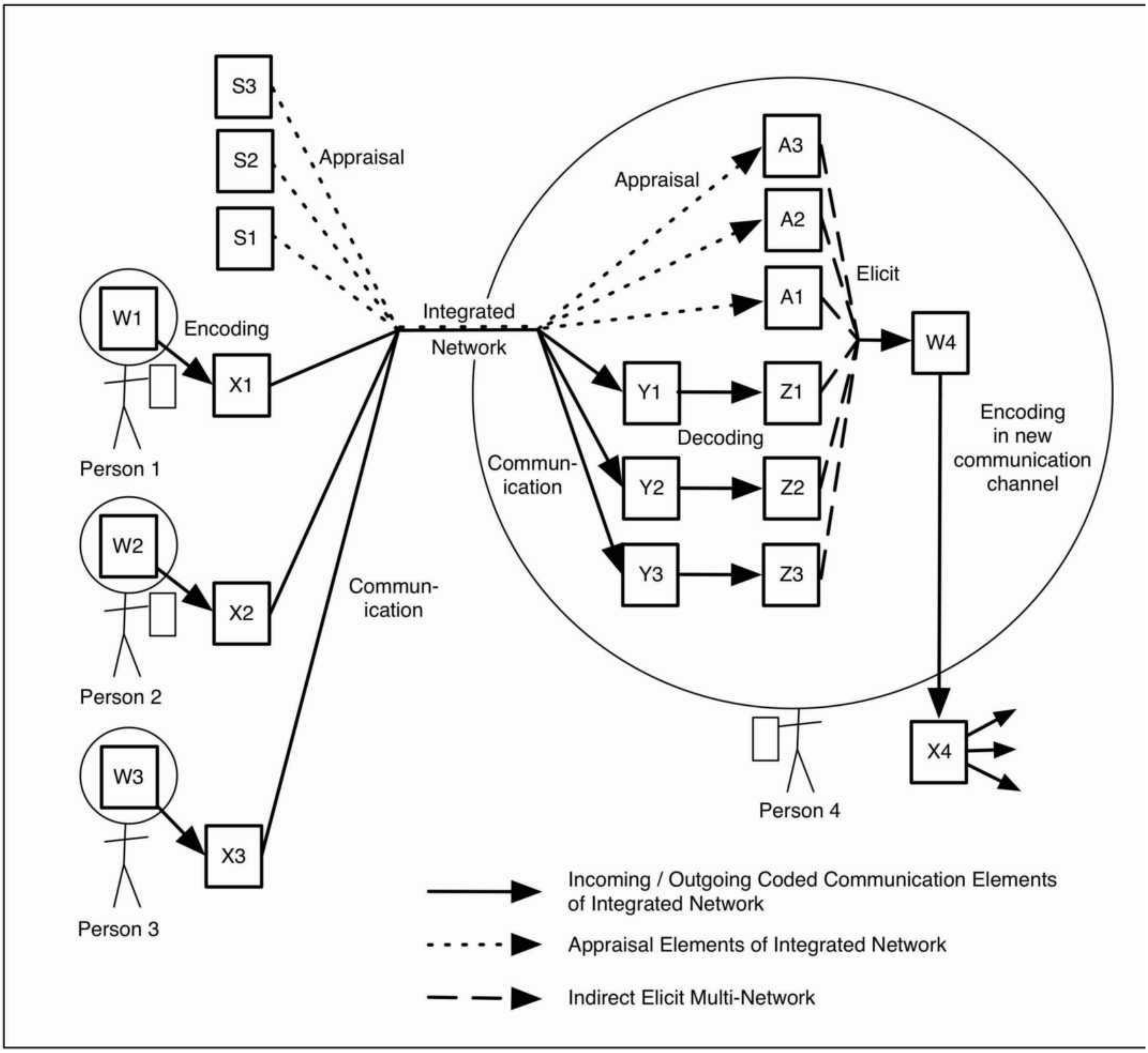

**Fig. 29. Complex node with integrated appraisal and communication. A single node from Figure 8 with more complex detail than that depicted in Figures 27 and 28. Here, the number of registers in the incoming communication network is the same as in Figure 28, as is the number of appraisal registers. The difference is that the emotion behavior from Persons 1-3 together with the situations in the world form the inputs to an integrated network that has as outputs three appraisal registers and three emotion behavior detection registers. This allows us to model the fact that situations in the world and the emotion behavior of others together impact how we appraise those situations**

**and how we detect that behavior. Just as in Figure 28, in this model, Person 3 is a *three-register appraisal / three-register communication node*.**

## 5.12 Quantitative Emotion Communication Network

Now that we have developed complex nodes and the emotion elicitation channels, we can combine them with our coded emotion communication networks from Chapter 3 to get a quantitative model of the simple communication network depicted in a qualitative way in section 4.7. It is shown in figure 30.

The nodes have only two emotion communication registers and a single appraisal channel. For simplicity, they are shown as distinct (like in Figure 28) rather than integrated (like in Figure 29).

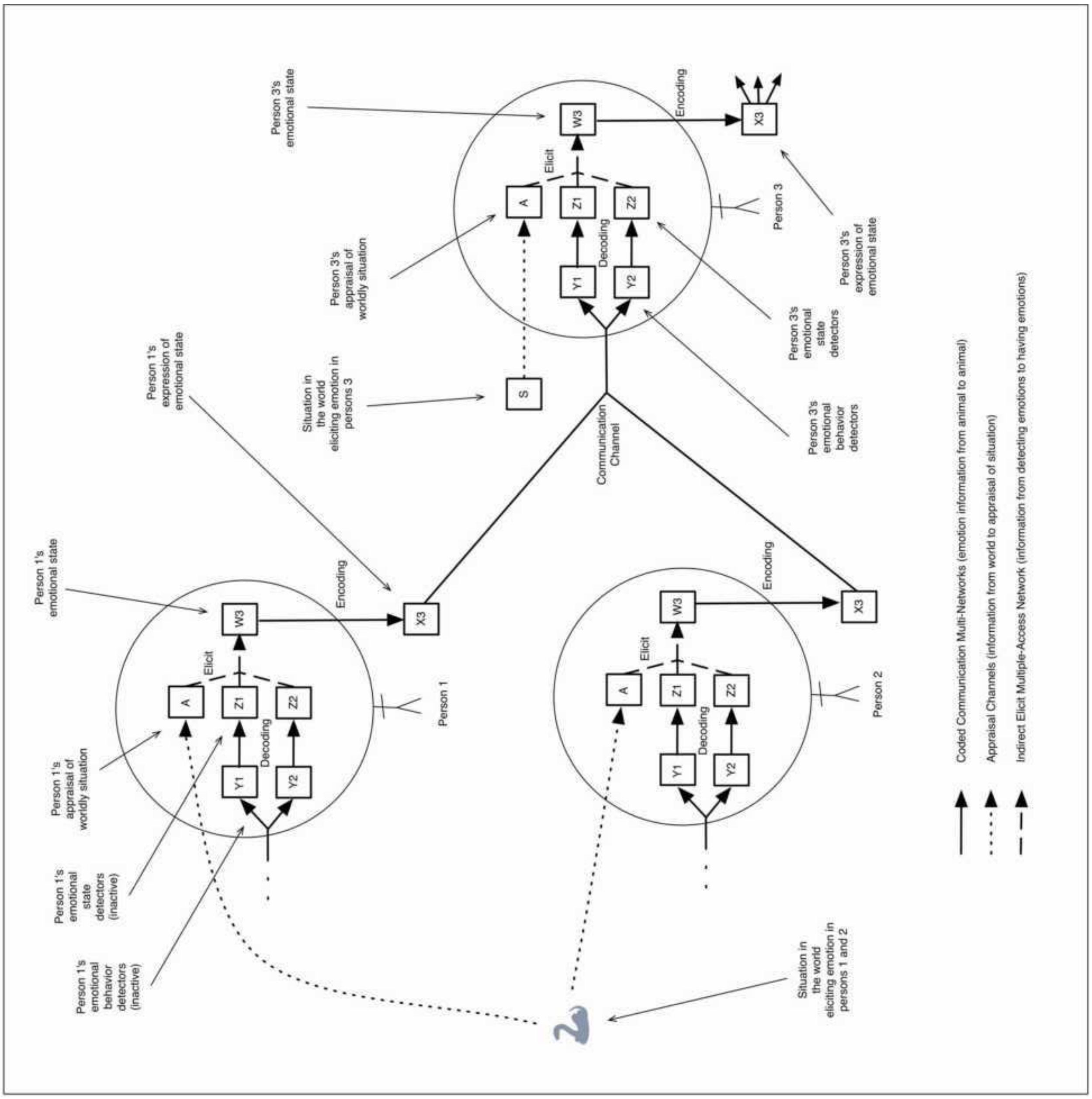

**Fig. 30. Quantitative simple emotion communication network. The same emotion communication network depicted in Figure 23, but now with complex nodes. Here each node is a one-register appraisal/two register communication node. The indirect elicit channel brings together appraisal and emotion behavior detection, as in Figure 28, but the appraisal channels are distinct from the communication channels for simplicity of the depiction, so the nodes are not as realistic as the one depicted in Figure 29. Here we can see exactly how information flows through two appraisal channels into Person 1 and**

**Person 2 respectively and into each indirect elicit channel to arrive at an emotion state in each of these two people. The appraisal channels are defined by the conditional probability that Person 1 or Person 2 will detect danger, provided that there is or isn't danger in the situation. The indirect elicit channels are defined by the conditional probability that Person 1 or Person 2 will feel fear, provided that they did or didn't detect danger. At this point, the information enters a coded communication network with Person 1's emotion state and Person 2's emotion state as sources, and the two registers for detecting emotions within Person 3 as destinations. Person 3's two emotion behavior detectors and two emotion detectors permit tracking of emotion behaviors and states in Persons 1 and 2 independently. Person 3 also has an appraisal register (just as in Persons 1 and 2). Finally the information enters Person 3's indirect elicitation channel and ends with Person 3's capacity to feel emotions. In cases of successful emotion communication, information about the danger in the situation is transferred to Person 1 and to Person 2, who then transfer it to Person 3. Although Person 3 does not directly detect the danger in the situation, information about it is communicated to Person 3 via the emotions felt and expressed by Persons 1 and 2.**

Obviously, we have not done any actual calculations on this network, but with the mathematical details from Section 5.4 and the surrounding theory in the references, anyone interested could make up all sorts of models for this very network and calculate a vast array of information theoretic quantities.

## 5.13 Extensions

Network information theory is still so young that there is no quantitative theory that works for even basic measures in general. There are plenty of good models for a wide range of special cases, and it would make sense to explore all sorts of these with respect to emotion communication. The network chapter in Cover and Thomas is a good introduction and ? is a good textbook.

We saw channel coding in Chapter 4 but there is also network coding theory, which looks at codes that work by spreading parts properly through a population rather than encoding/decoding as in a channel. Might humans or other animals use network-coded emotion communication? Even if we do not take advantage of this now, could we learn to? Might AI agents with artificial emotions be able to use it?

Special emphasis on the phenomena that show up only in networks would make sense. These include cooperation and interference.

Phenomena studied by social psychologists would be great to investigate in a quantitative framework as well. These include emotion detection, emotion expression, and emotion elicitation. Each of these topics has an extensive literature and the information framework places each of them in a more general quantitative whole. No doubt, each one of these phenomena displays cooperation and interference. Using the information framework to calculate and predict how emotions spread through a communication network should be a priority, given the importance of emotion contagion models in the literature. Another major topic is emotion regulation, which is surely a some constellation of feedback networks, and interpersonal emotion regulation in particular, which occurs when one person or group of people attempts to influence the emotional states of another.

# *Chapter 6*

# Advanced Emotion Communication Networks

This Chapter displays some of the power of the quantitative theory of information in emotion networks developed in the previous chapter by outlining some applications. Emotion regulation is a hot topic right now, and interpersonal emotion regulation is at the forefront of work in this area. Another major topic in animal communication is deception (Section 6.5). It is important for the information theory to be able to explain everything the contagion theory can explain, so we develop this direction a bit in Section 6.8. Sections 10 through 13 are aimed at theories of emotion multi-level networks, which have several distinct levels of communication. The aim of doing this is to distinguish how the individual levels influence individual animals and to understand how the levels interact with one another. Social media is the dominant force in 21st century culture worldwide. We show how to use the information in emotion multi-layer networks to define a *social media influence factor*, which quantifies how much social media is affecting your emotional life (Section 6.13).

## 6.1 Emotion Expression, Perception, Elicitation, and Appraisal

These four categories are each recognized as major independent topics of research in cognitive psychology, cognitive science, and neuroscience find a quantitative home in the information theory of emotion communication. Each of these topics looks quite different in the information theoretic framework. Roughly, they are interpreted as:

- *Emotion Expression* is *coding* emotion information from emotion states to associated emotion behaviour. There are three major topics here: compression, error-correction, and encryption. Compression is how the information in the emotion state (unobservable) is transformed into the information in the associated emotion behaviour (observable). Error-correction is how much extra information is sent via emotion behaviour. Encryption is how the message sent via emotion behaviour is coded to avoid eavesdropping.
- *Emotion Perception* (also called *emotion recognition* and *emotion detection*) *is decoding* emotion information from emotion behaviour to detection of the associated emotion state. There are numerous topics here, but most involve categories of decoding algorithms and how emotion detection is related to other mental states like emotional states themselves.
- *Emotion Elicitation* (sometimes called *emotion induction*) breaks into two categories, indirect and direct, and is part of a larger category, emotion change. Direct elicitation is when an animal's emotions are caused by the world, not part of emotion communication. Indirect elicitation is when an animal's emotions are caused by detecting emotions in other animals through emotion communication. The important process to emphasize is changes in emotional state; elicitation – causing an emotional state to occur – is only one side of the process of emotion change, the other being delicitation – causing an emotional state to stop.

- *Emotion Appraisal* might very well break into two categories as well. Direct appraisal is the focus of appraisal theories of emotion, but indirect appraisal would be the evaluation of an emotion message with respect to one's beliefs, desires, goals, and plans. Thought of this way, indirect emotion appraisal is part of the process of indirect emotion elicitation.

Just as important as these information theoretic interpretations of these major research areas is the information theoretic view on how they are related to one another as well. Both of these insights are illustrated in Figure 31.

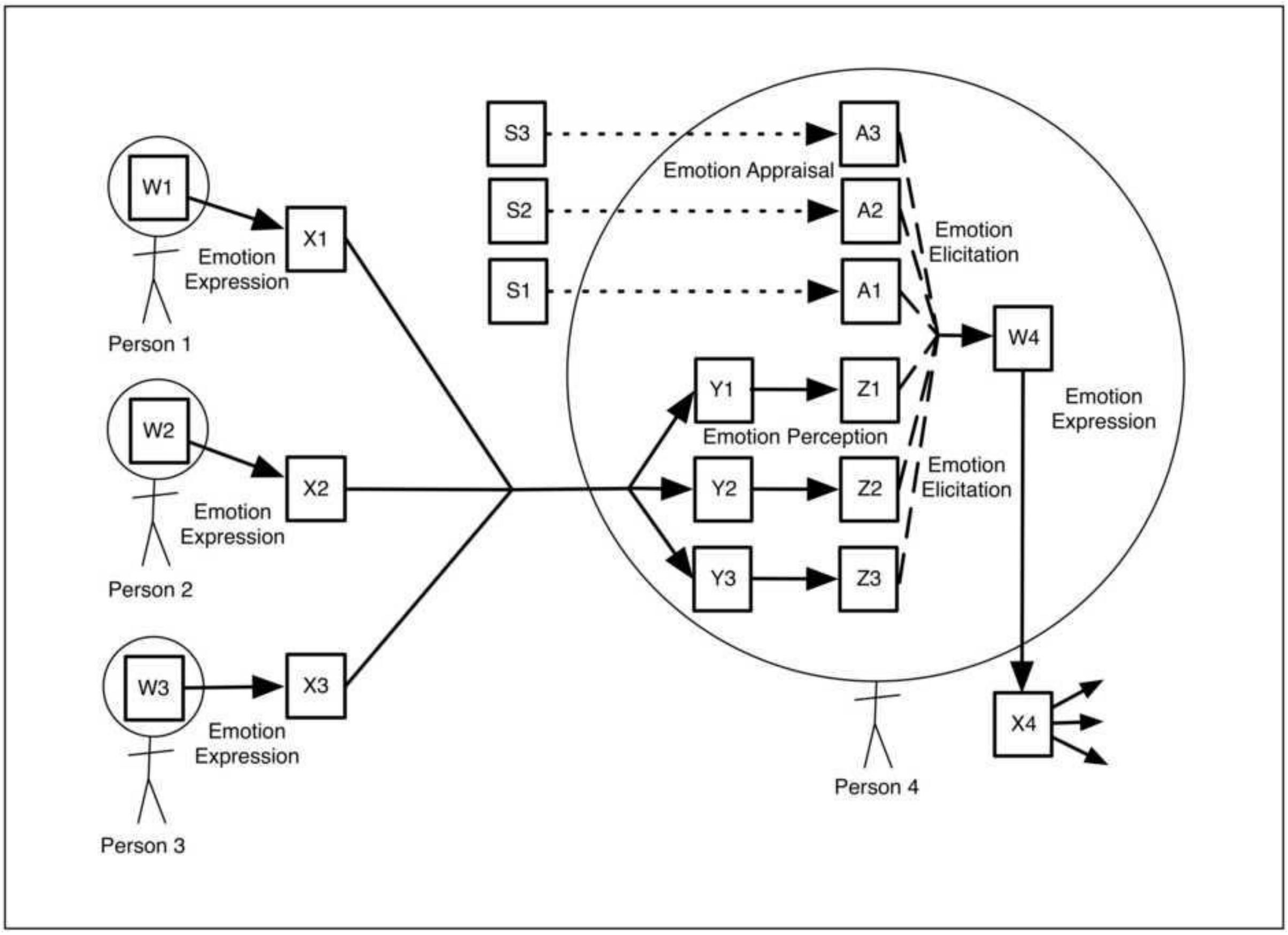

**Fig 31: Emotion research topics plotted on emotion communication node**

It should be obvious that in emotion communication networks, there are huge loops of emotion expression, emotion perception, emotion appraisal, emotion elicitation, emotion expression, and so on (track it from left to right in Figure 31). Each of these processes is seriously impacted by its place in this emotion communication cycle.

## 6.2 Emotion Regulation

People often try to either produce certain emotions (e.g., watching a critically acclaimed comedy) or avoid certain emotions (e.g., refusing to look at a horrible traffic accident). We frequently criticize one another's emotions; when someone's emotion gets it wrong, we usually make note of it. For instance, a mother tells her son that he should not fear going down to the basement after dark because it is not actually dangerous. A daughter tells her father that his anger is inappropriate because she *did* actually return his phone call. A guy tells his friend to stop getting so angry any time someone ogles his sister. Most people perform these types of assessments and evaluations of emotions on a daily basis.

Functional individuals are already controlling their emotions in many ways both implicit and explicit. These results should remove a major obstacle to thinking of how we interact with our emotional episodes as subject to rational assessment—namely, the idea that emotions just happen to us. The emotion regulation literature supports the claim that we are regulating our emotions almost continuously.

Regulating emotions well is one way to display excellence in one's affective capacities. To regulate them well certainly involves the ability to up-regulate emotions (i.e., increase the intensity, duration, or frequency of an emotion) taken to be positive in some way and down-regulate emotions (i.e., decrease the intensity, duration, or frequency of an emotion) taken to be negative in some way. Emotion regulation can be done either automatically or voluntarily—either method is possible in the display of excellence in one's affective capacities.

It is clear that we have plenty of indirect control over our emotions. For example, many studies show that simply altering one's facial expressions has a significant effect on one's emotional states.[1] Imagine that I feel angry when I actually see Billy wrong me in some relatively uncontroversial way (meaning, that there is no real question about whether or not he wronged

[1] Strack, Martin, and Stepper (1988). To further this comparison: we do think that, at times, we can even have *direct* control over our emotions. And, there is growing evidence in psychology to support this position, see Barrett, Ochsner, Gross (2006), and Cunningham, Dunfield, and Stillman (2006).

me—he clearly and obviously wronged me). Now, imagine that I do not want to be angry with Billy anymore. There are all sorts of methods for regulating my anger. I can take a deep breath. I can relax my shoulders. I can think of other things. I can even reappraise the extent to which I was wronged in the situation. These activities can have an obvious and relatively automatic effect on my anger.

Emotion regulation may concern whether one has an emotion or not, when one has an emotion, how strong or weak an emotion is, how long an emotion lasts, and how one expresses an emotion. Emotion regulation can occur automatically and unconsciously or controlled and consciously. In this section, I present a few different views on emotion regulation.

James J. Gross and Ross A. Thompson present five kinds of emotion regulation: situation selection, situation modification, attention deployment, cognitive change, and response modulation. Situation selection is where one figures out which situations one should and should not be in such that one seeks out some situations and avoids others, e.g., "avoiding an offensive coworker, renting a funny movie after a bad day, or seeking out a friend with whom we can have a good cry." This is a forward-looking method of emotion regulation in that the agent takes action before the emotion is elicited. Situation modification is where one changes the features of one's situation in an attempt to control which emotion it elicits, e.g., "When conservative in-laws visit, situation modification may take the form of hiding politically incendiary art work." This method of emotion regulation involves acting on one's external environment. Attention deployment is where one changes one's attention in a particular situation. There are four distinct methods of changing one's attention. One can use distraction where one changes one's focus to a different aspect of a situation. One can use concentration to emphasize a particular aspect of the situation. One can ruminate on an event—one attends repetitively to one's feelings and their consequence. And, one can withdraw attention, e.g., when one covers one's ears and repeats, "I can't hear you!" or simply puts on headphones.

There are also several different kinds of cognitive change, where an agent changes how she appraises the situation she is in such that it adjusts the emotional significance. One can regulate one's emotions through cognitive change by doing things like down-grading (e.g., telling one-self that it could be worse) or by re-appraising the emotional impact of the situation (e.g., rather than thinking of the weather as partly-cloudy, thinking of it as partly sunny). And, lastly, one regulates one's emotions through response modulation by doing things like decreasing or changing the expression of one's emotion (e.g., biting one's tongue). Each of these five methods of emotion regulation can, at times, be carried out either involuntarily and unconsciously or voluntarily and consciously. While Gross and Thompson's account is extensive, it is not necessarily exhaustive—it seems that there may be types of emotion regulation that do not fit cleanly within the categories that they list.

## 6.3 Quantitative Interpersonal Emotion Regulation (QIER)

People often try to either produce certain emotions or prevent certain emotions in others. In the psychology literature, this is called *interpersonal emotion regulation*. It is attracting increasing attention in psychology.[2] This is the process by which animals influence one another's emotional states.

Karen Niven recently summarized four essential features of interpersonal emotion regulation to serve as a guide for researchers in this exciting field. They are:

1. It is a form of emotion regulation. "[I]t is about changing or maintaining a state in line with some kind of reference goal. Support for this perspective is provided by research into the brain regions that are recruited during the process of IER. For example, a recent fMRI study reported that performing IER activated brain areas including the inferior frontal gyrus and pre-supplementary motor area, which have been previously implicated in other forms of regulation, such as dieting and thought suppression."

[2] Grecucci (2015).

2. It has an affective target state. "[T"he state that is being regulated is a feeling state. This distinguishes IER from other processes whereby the state being regulated is cognitive (e.g., impression management) or behavioral (e.g., peer pressure)."
3. It is a deliberate process. "IER can be distinguished from a multitude of processes that, on the face of it, appear to be quite similar. For example, we frequently leave those we interact with feeling the same way we do without any idea that we are doing so, as a result of mimicry and facial feedback (i.e., emotional contagion), or our personality (i.e., affective presence). IER is different from these processes because, like other deliberate processes, it is intentional, controlled, resource-intensive, and engaged with conscious awareness."
4. It has a social target. "[T]he target state is social in that it belongs to someone other than the regulator."[3]

Niven summarizes these basic principles and how they relate to similar phenomena in a nifty diagram, reproduced as Figure 32.

[3] Niven (2017: 89, 89, 90, 90, respectively).

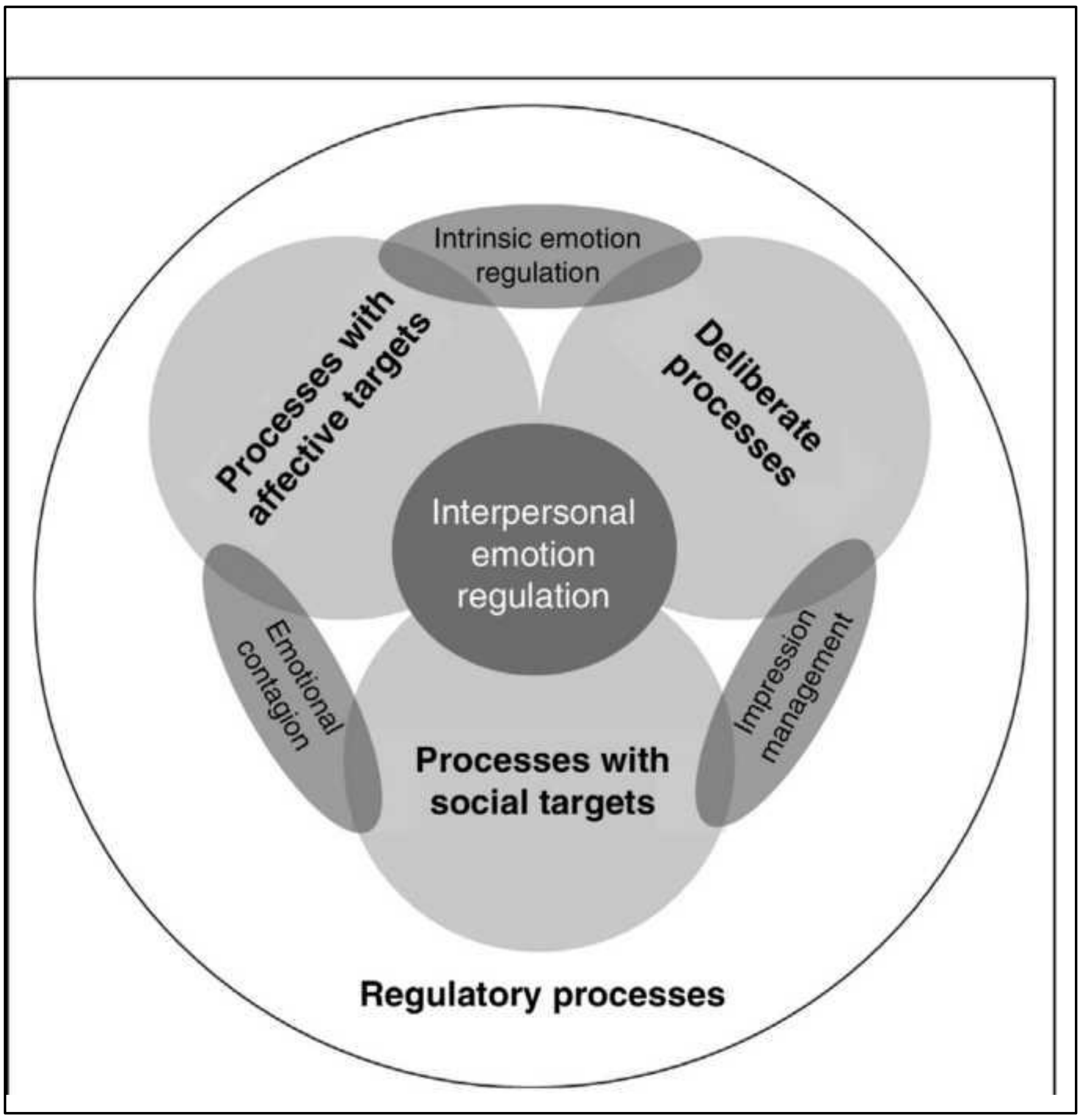

**Fig 32: Interpersonal emotion regulation processes**

It is important to note that interpersonal emotion regulation does not require emotion communication; for example, a father might change the music so his crying daughter can hear the ocean waves in an attempt to sooth her. That is interpersonal emotion regulation, but it involves no emotion communication from father to daughter (assuming the daughter is not old enough to understand that her father changed the music in reaction to her crying). However, emotion communication is often at the centre of interpersonal emotion regulation that involves many episodes of communication back and forth subtly responding to one another.

Regulatory processes are often fruitfully modelled quantitatively by control theory and machine learning.[4] Control theory focuses on how to control physical systems. In one of the central textbooks in the field, Dorf and Bishop summarize it well:

> Engineers create products that help people. Our quality of life is sustained and enhanced through engineering. To accomplish this, engineers strive to understand, model, and control the materials and forces of nature for the benefit of humankind. A key area of engineering that reaches across many technical areas is the multidisciplinary field of control system engineering. Control engineers are concerned with understanding and controlling segments of their environment, often called systems, which are interconnections of elements and devices for a desired purpose.[5]

'Control' usually means *the ability to steer the system to any desired state*, and this definition is implicit in the quote above as well. Control theory utilizes a wide range of quantitative models including machine learning algorithms (i.e., algorithms that improve themselves as they encounter data) for many aspects of regulatory processes, but we know of no one who is using it to explain interpersonal emotion regulation. The reason is obvious – until now there has been no quantitative theory of emotion communication. Emotion communication does not even show up on Niven's diagram. But with the information theory of emotion communication, a quantitative theory of interpersonal emotion regulation is now within reach.

A quantitative revolution could overtake the field of interpersonal emotion regulation in the same way that BToM has overtaken ToM. The Partially Observable Markov Decision Processes (POMDP) that are at the centre of how BToM model's rational agents could easily be used for a quantitative theory of interpersonal emotion regulation. In BToM, the unobservable states are the mental states of other agents, and in quantitative interpersonal emotion regulation models, these would be emotional states of others. The major difference would be that the

[4] See Dorf & Bishop (2017) for a popular textbook on control theory. We discuss it in more detail in Chapter 8.
[5] Dorf and Bishop (2017: 2)

POMDPs in BToM only try to figure out the mental states of others, so their only actions are mental state attributions. In Quantitative Interpersonal Emotion Regulation (QIER), the POMDPs would also be able to take actions in order to change the emotional states of others. Often, changing the emotions of others involves adopting certain emotional states oneself, or taking certain courses of action (e.g., comforting the sad). Just as in BToM, the agents in QIER would be rewarded by how well they reach their goals. In BToM, the goal is to attribute the right mental states to the targets. In QIER, the goal is to change the target's emotional states in certain ways.

## 6.4 Emotion Communication in Theory of Theory of Mind (ToToM)

Recall from the overview of Theory of Mind (ToM) research in Chapter One that these researchers focus on understanding how we attribute mental states to one another on the basis of behaviour and shared environment. In Chapter Two, we appealed to analyses of emotions in Bayesian Theory of Mind (BToM) research to explain certain aspects of emotion behaviour without appealing to emotion forces, as Scarintino does. Here, we bring together the information theory of emotion communication with Theory of Mind.

Our main point is that ToM research needs to account for emotion communication because it is through observing such communication that humans (and perhaps other animals) learn much about social norms, cues, and statues. That is, when humans learn, in part, by watching other humans communicate through emotions. This capacity requires more than just ToM, which targets a single person. Rather, it requires the ability to attribute *the attribution of* emotions, not just the ability to attribute emotions themselves. A three year-old, Maria, needs to be able to tell that Robert thinks that Melinda is angry, and Maria needs to be able to coordinate this attribution of an emotion attribution with other attributions like an attribution of anger to Melinda. The point is that ToM will not do. By integrating emotion communication with ToM, we arrive at *Theory of Theory of Mind*: our ability to attribute the ability to attribute mental states to others to others. The ability to attribute emotion communication to other pairs and networks of animals is one central part of the Theory of Theory of Mind (ToToM) ability. How would it work?

The Bayesian turn in ToM (called BToM) is promising in our opinion, and it makes sense to use this framework for information in emotion communication sense both frameworks are based on the mathematics of probability distributions. Recall that BToM uses Bayesian networks to model the theory that a person uses to attribute mental states to others, and it uses Partially Observable Markov Decision Processes to model how that person interacts with its environment, including the behaviour of other people (observable) and the mental states of other people (unobservable). BToM focuses on the kinds of algorithms that would allow a POMDP agent using a Bayes net to figure out the mental states of another person.

Bayesian Theory of Theory of Mind (BToToM) would focus on an agent (call it Agent 1), modelled as a POMDP using a Bayes net. Agent 1 is to attribute emotion communication to others, call them Agent 2 and Agent 3. Focus to start on Agent 3 having an emotion and Agent 2 detecting that emotion. Agent 1 can just attribute an emotion to Agent 3, but understanding what Agent 1 attributes to Agent 2 is the key to BToToM. Agent 1 attributes the exact same ability to Agent 2 that we researchers attribute to Agent 1! In other words, Agent 1 treats Agent 2 like a POMDP using a Bayes Net. How does Agent 1 do this? In order for *us* to understand how Agent 1 does this, we need to simply move up a level. That is, we can use a POMDP to model Agent 1, but instead of the unobservable states being mental states of some other agent, the unobservable states of the POMDP themselves describe another POMDP and Bayes net. That is, what is observable, is the behaviour of Agent 2 and Agent 3 in their environment. What is unobservable is the POMDP and the Bayes net that Agent 1 thinks Agent 2 uses to understand Agent 3. Call the POMDP used to model Agent 1 a *Second Order* POMDP. An augmented Bayes net can still work for Agent 1, but it needs to govern not only the attribution of mental states to Animal 2, but also the attribution of a Bayes net to Animal 2.

Imagine that you observe your partner watching as your partner's boss gets angry with one of your partner's colleagues at a party. How does this work in ToM? You are attributing something (call it X) to your wife who is attributing something (call it Y) to your partner's colleague who is attributing anger to your partner's boss. We have already seen Y – it is the capacity for ToM that is being attributed to your partner's colleague. Your partner is treating their colleague as if the colleague has

It might be surprising but there is considerable empirical evidence that demands a theory like BToToM to explain it. For example, Repacholi and Meltzoff summarize several experiments that suggest we learn a huge amount from watching others' emotion communication. That is, we are emotional eavesdroppers from a young age, and this might be an essential aspect of learning how to be a part of human societies. They write:

> Emotional eavesdropping is important in everyday life. One can learn a lot about the social and physical world by observing how people emotionally respond, even when the emotions are not directed at oneself (hence the term "eavesdropping"). Indeed, emotional eavesdropping will often enable infants to avoid the negative outcomes that might otherwise arise if they simply explored new objects and tried new actions without taking into account the emotional reactions of others in the environment.[6]

If these conclusions hold up, then they demand that we have a theory not only of how we attribute mental states to one another, but a theory of how we attribute the capacity to attribute mental states to one another. Attributing mental capacities (e.g., the capacity to attribute mental capacities at all) demands a different kind of theory than the basic BToM. It demands something like BToToM.

## 6.5 Eavesdropping and Emotion Encryption

[6] Repacholi & Meltzoff (2018: 66).

A perennial topic in communication studies is *eavesdropping* – decoding messages that were meant for someone else. And another is *encryption* – coding messages so that they cannot be decoded by anyone except the intended destination. Eavesdropping examples also play a big role in the development of philosophical and formal semantics by charting the role of communicative intentions and communicative context.[7]

In information theoretic terms, eavesdropping is modelled in simple terms by what is called a *wiretap channel* (but it is a network in our sense). It is a broadcast network with one source and two destinations (one destination is intended, the other is the eavesdropper). If they are X, Y, Z, respectively, then the network is determined in the usual way by the probability distribution p(Y, Z|X).

In this wiretap channel, we can define the *secrecy capacity*, which is the maximum amount of information that can be communicated to Y without being intercepted by the eavesdropper, Z. To define it, we need to first define the *leakage rate* for the code as:

$$R_L^{(n)} = \frac{I(M; Z^n)}{n}$$

The average probability of error for this secrecy code is defined in the usual way for messages sent and estimated messages from the legitimate destination. A pair of rates R, $R_L$ is *achievable* iff as n increases to infinity, the leakage rate approaches RL and the average probability of error approaches zero. The capacity region, which is called the *rate-leakage region* for wiretap networks is the closure of the set of achievable rate pairs. The *secrecy capacity* of the wiretap network is then given by:

$$C_s = \max_{p(u,x)} (I(U; Y) - I(U; Z))$$

where U is a set with cardinality less than or equal to that of X.

The definition of secrecy capacity allows us to characterize how much information can be passed securely even in the presence of an eavesdropper. All sorts of real-world features affect the secrecy capacity, including the detection capabilities of the intended destination and the eavesdropper. There is a whole world of results on information theory and encryption, but the interested reader will have to investigate these. The link to emotion communication should be obvious. It seems clear that emotion encryption is a genuine phenomenon; think of a teenager who communicates with friends on social media by using obscure emojis that their parents will not be able to interpret. The parents are the eavesdroppers in this case, and it would not be difficult to calculate the security capacity of this network.

[7] See Egan (2009).

Another application is in culturally specific emotion expressions – when you are foreigner in another culture, chances are good that the people around you are communicating with their emotional capacities in ways you do not notice. As you get used to the culture, you pick up on these emotion communication idiosyncrasies better. You are now equivalent to an eavesdropper on a wiretap channel with encrypted emotion communication. As you get better and better at picking up on these emotion communication episodes, the people in the community might begin to see you as one of them; if so, then you gradually loose your status as eavesdropper as you become an intended destination in their encrypted emotion communication systems.

One issue with when the wiretap network is a good model is how much the intentions of the sender matter. Above we described one destination in the network as "intended," but is this necessary? We have emphasized that intentions are not all that relevant to many aspects of emotion communication. The scared dog does not have to intend to communicate with you in order for it to communicate with you. So we could extend the wiretap model to include a broader range of cases, perhaps where the source and destination are each evolutionarily equipped to communicate. Another direction would be to focus on communication that involves normal or appropriate functioning. Then the intended destination would be anyone whose normal functioning involves communicating with something like the source. That would fit the example of cultural variation above as well.

## 6.6 Deception

In emotion communication systems, we can think of emotion deception as one of two kinds – either having an emotion and intentionally not expressing it or not having an emotion and intentionally behaving as if one has it. Neither of these can be defined in purely information theoretic terms because of the term 'intentional', which indicates purposive behavior. Nevertheless, we can see that emotion deception is ubiquitous (we are writing this in the UK where the British "stiff upper lip" is a familiar trope). Any time you have been upset and tried to hide it or pretended to be happy at a social function, you have engaged in emotion deception. One major problem with this account is that it predicts that animals without the advanced cognitive capacities to engage in intentional behavior cannot display deception, but it seems like deception is a widespread phenomenon throughout the animal kingdom. This problem has led many animal communication theorists to seek an alternative account of deception that avoids this problem.

Brian Skyrms, in his engaging and wide-ranging book, *Signals*, offers an elegant and influential account of deception. He writes:

> If receipt of a signal moves probabilities of states it contains information about the state. If it moves the probability of a state in the wrong direction—either by diminishing the probability of the state in which it is sent, or raising the probability of a state other than the one in which it is sent—then it is misleading information, or misinformation. If misinformation is sent systematically and benefits the sender at the expense of the receiver, we will not shrink from following the biological literature in calling it deception.[8]

[8] Skyrms (2010: )

For Skyrms, deception occurs when: (i) one animal conveys misinformation to another, (ii) this occurs systematically, and (iii) it benefits the sender at the expense of the receiver. Each of these aspects can be manifested in an animal that has no ability to form complex intentions.

In information theoretic terms, misinformation is easy to characterize because it depends only on probability distributions in sender and receiver. Conveying misinformation *systematically* is a matter of it being a *pattern* of misinformation rather than a single instance. Finally, benefitting the sender at the expense of the receiver could be defined in all sorts of ways, but it can also be explained quantitatively in terms of utility, which is familiar from decision theory and game theory.

There is a rich literature about each of these aspects of deception and how best to define deception in signaling systems. We are not going to survey them here, but we wanted to note the importance of this issue and say a bit about how it impacts emotion communication systems.

There seem to be two major kinds of emotion deception – behaving as if one has a certain emotion when one does not (acting angry when one is not), and not behaving accordingly when one does have a certain emotion (being angry without expressing it). Because emotion expression is gradable – one can express an emotion to a greater or lesser degree – these two kinds of emotion deception come in degrees as well. An additional complication is that when one assesses the extent to which a person expresses a certain emotion or feels a certain emotion, one has to take into consideration the relevant display rules, which govern which animals may or must express which emotions on which occasions (e.g., do not be happy at a funeral, or women in western societies cannot display anger in the workplace). As such, we have a choice to make: imagine a person who feels a certain emotion to a large degree but does not express it because the relevant display rules forbid it. Is this emotion deception? Or should we reserve this term for cases where the relevant display rules permit the emotion to be expressed? Likewise, we might want to take display rules into consideration for the opposite: imagine a person who acts sad at a funeral even though they are not sad, and contrast that with a person who acts sad at a party even though they are not sad. The former is following relevant display requirements, while the latter is not. These are at least different kinds of emotion deception, but we can imagine theorists wanting to not call the former deception at all.

## 6.7 Reliability

Reliability assessments are absolutely crucial to navigating signal systems. We each keep track of each other's reliability with respect to emotional capacities. For example, you known which of your family members are overly dramatic so that you might not trust what they say about an incident involving another family member. Recognizing that your dog always over reacts to men with beards would be another example.[9]

In signal systems, there are two basic kinds of reliability: being a reliable reporter of situations in the world vs. being a reliable communicator. The former is being good at introducing messages into the signal system and the latter is being good at passing on messages that are already in the system. Call the former *direct reliability* and the latter *indirect reliability*.

In emotion communication systems, direct reliability is responding with the right emotion and right emotion behavior in response to what is happening in the world around you. For example, being scared and backing away from a poisonous snake, and not being scared or running away from a garden hose. In the node diagrams, direct reliability pertains to appraisal, elicitation, and expression (encoding).

In this section, we focus on the kinds of networks needed for one node to assess the direct reliability of another node. That is, emotion communication between animals where one animal decides whether another animal is directly reliable – whether that animal is a good *reporter*. Call the animal being assessed Animal 1 and the animal doing the assessing Animal 2. What is needed for Animal 2 to decide whether Animal 1 is reliable? At a minimum, Animal 2 needs to have access to both the relevant situation in the world and access to Animal 2.

[9] In philosophy, the topic of reliability is connected to justification and knowledge; see Nagel (2016) for an overview. There is also work on emotions being justified or warranted, and even on what it takes for a communicated emotion to be warranted. This topic, emotion testimony, is relatively unexplored, but see Kerr (forthcoming).

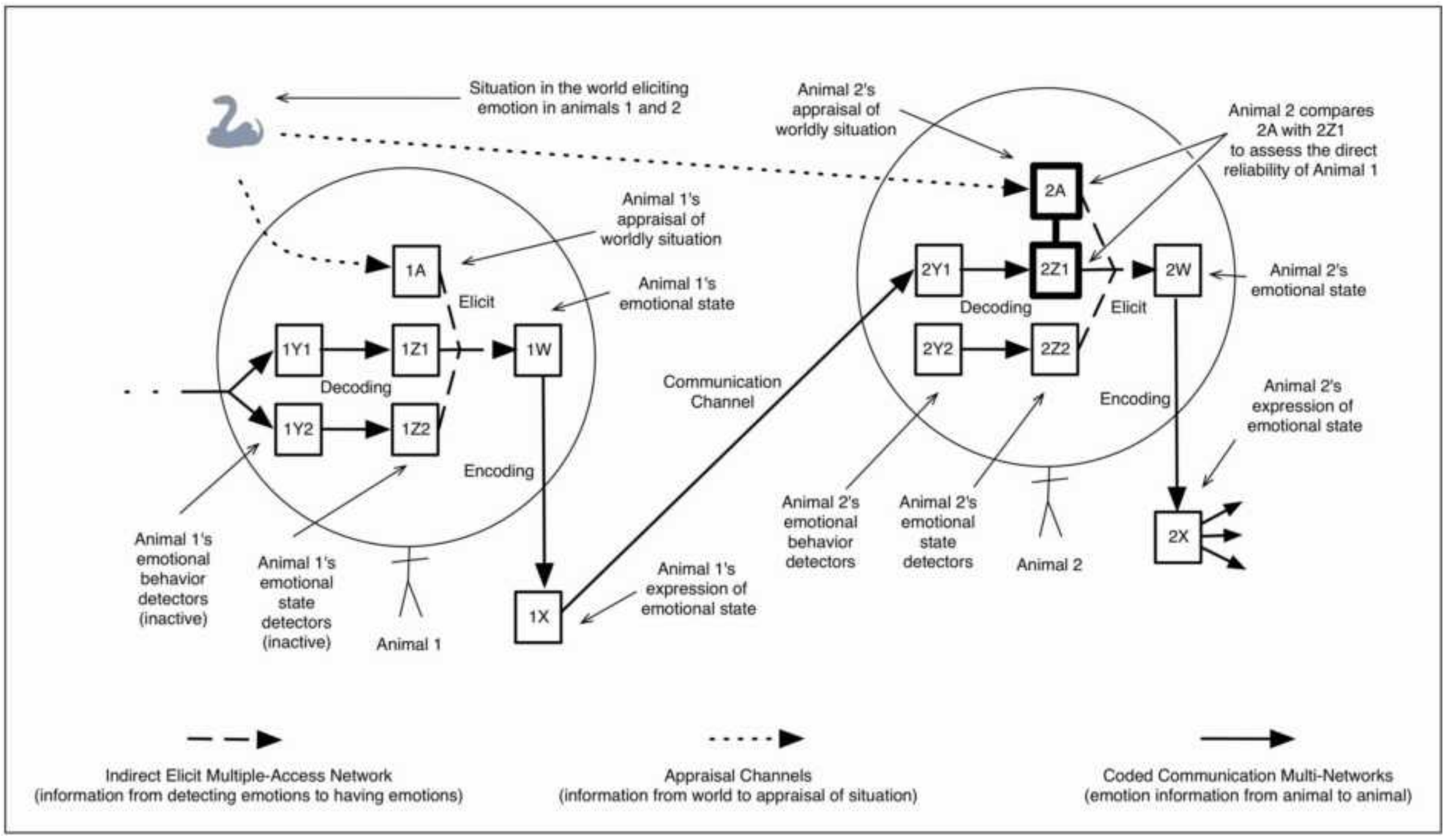

**Fig 33: Direct reliability network**

Direct emotional reliability is being good at reacting to environmental stimuli appropriately. Indirect emotional reliability is being good at passing along emotion messages. In information theoretic terms, this is being good at detecting emotion behavior, being good at decoding that emotion behavior to detect emotions, being good at assessing that information and allowing it to influence one's own emotional states, and being good at expressing one's emotional states in outgoing communication networks. In the node diagrams, indirect reliability pertains to detection, decoding, elicitation, and expression.

For example, Animal 1 sees a spider, feels fear, and expresses that fear in behaviour. Animal 2 observes Animal 1's behaviour but not the spider, and interprets Animal 1's behaviour as an expression of joy. Animal 2 then feels joy as a result and expresses joy. Animal 3 observes the entire process – spider, Animal 1, Animal 2 – and correctly decodes the emotion behaviour of both Animal 1 and Animal 2. Animal 3 then issues an indirect reliability assessment about Animal 2 – they are not indirectly reliable.

Because reliability assessments depend on comparing two inputs, and each input might be more or less removed from direct emotion stimuli, we can see at once that there is complex hierarchy of reliability assessments. Define the following terms: the *ordinal* of an emotion episode is its place in order from direct emotion stimulus where an directly elicited emotion is *ordinal 0*, an emotion elicited from ordinal 0 emotion expression is *ordinal 1*, and so on. In the direct reliability situation from the previous section, an animal compares its own appraisal of a direct stimulus (the spider) with the emotion expressed by the other animal. The other animal's emotion is ordinal 0. The animal doing the comparison might not be having an emotion at all, but we can assume that the appraisal is normally part of an emotion episode, so normally direct reliability assessments will be comparison of two ordinal 0 emotion episodes. The example describe in this section of indirect reliability assessment involves comparison of its own assessment of the situation with an ordinal 1 emotion (the scared animal 1 has an ordinal 0 emotion and the joyous animal 2 has an ordinal 1 emotion). We can imagine extending this particular hierarchy indefinitely: comparison of ordinal 0 with ordinal 2, comparison of ordinal 0 with ordinal 3, etc.

In addition to this hierarchy of reliability assessments, we can imagine others as well. If a situation unfolds as above with spider, fear, and misinterpreted joy, but instead of observing the whole process, Animal 3 only sees the fear display of Animal 1 and Animal 2's resulting joy behaviour. Animal 3 could still evaluate Animal 2's joy as unreliable, but this would be comparing an ordinal 1 emotion (the fear in Animal 1) with an ordinal 2 emotion (the joy in Animal 2). Again, one could imagine a hierarchy of these reliability assessments as well: a comparison of ordinal 1 with ordinal 2, comparison of ordinal 1 with ordinal 3, comparison of ordinal 1 with ordinal 4, etc.

Extending these considerations we can see an infinite matrix of reliability assessment kinds based on the ordinals of the emotions compared.

*Assessed* Animal's Emotion Ordinal (*super*script)

| | | 0 | 1 | 2 | 3 | 4 | … |
|---|---|---|---|---|---|---|---|
| *Assessing* | 0 | Direct | $I^{1}_{0}$ | $I^{2}_{0}$ | $I^{3}_{0}$ | $I^{4}_{0}$ | … |
| Animal's | 1 | $I^{0}_{1}$ | $I^{1}_{1}$ | $I^{2}_{1}$ | $I^{3}_{1}$ | $I^{4}_{1}$ | … |
| Emotion | 2 | $I^{0}_{2}$ | $I^{1}_{2}$ | $I^{2}_{2}$ | $I^{3}_{2}$ | $I^{4}_{2}$ | … |
| Ordinal | 3 | $I^{0}_{3}$ | $I^{1}_{3}$ | $I^{2}_{3}$ | $I^{3}_{3}$ | $I^{4}_{3}$ | … |
| (*sub*script) | 4 | $I^{0}_{4}$ | $I^{1}_{4}$ | $I^{2}_{4}$ | $I^{3}_{4}$ | $I^{4}_{4}$ | … |
| | … | … | … | … | … | … | … |

**Fig 34: Dual hierarchy of reliability assessments**

An example of an $I^{0}_{4}$ reliability assessment would be an a spider scares Animal 1 who communicates the fear to Animal 2, who gets scared and communicates the fear to Animal 3, who gets scared and communicates the fear to Animal 4, who gets scared and communicates fear to Animal 5. Animal 5 observes a different animal – Animal 6 – who feels an inappropriate emotion – maybe lust – as a result. Animal 5 does not observe the spider (else its emotion state would not be ordinal 4), but it does observe Animal 6's lust behaviour. Animal 5 could determine on this basis that Animal 6 is unreliable in this case. With all the complex loops and networks of emotion communication over social media, it would not surprise us to see a decent number of these reliability assessments showing up in the wild.

## 6.8 Contagion Channels and Networks

We return once again to the contagion theory of emotion transmission. In Chapter 1, we outlined the theory, and in Chapter 2 we contrasted it with the signal theory of emotion communication. We take the contagion theory to be obviously inferior to the signal theory, so if a choice has to be made between them, then the signal theory should win. Here we aim to show that such a choice is not necessary. The information theory of emotion communication developed in Chapters 3-5 can explain anything the contagion theory can explain.

Subsuming emotion contagion can be accomplished by defining emotion contagion channels where the source is one animal with emotional capacities either having or not having an emotion and the destination is another animal with the same emotional capacities either having or not having that same emotion. The conditional probability defining the contagion channel is the probability distribution over the second animal's states, given the first animal's states. For example, it includes the probability that animal two feels the emotion, given that animal one feels the emotion.

Now we have a new quantitative theory of emotion contagion – we can measure information rates and noise levels to predict how fast an emotion travels through a crowd of animals through emotion contagion, and these calculations would be independent of the two major ways of modelling emotion contagion right now – thermodynamic models which treat emotions like heat and infection models which treat emotions like pathogens. We have not run any tests on information models of emotion contagion, but we expect that they have the flexibility to deal with some of the major objections reported in Chapter 2.

Our information theory of emotion contagion can also deal with some unknown but devastating problems, namely cooperation, competition, and interference in emotion contagion networks. Emotion contagion networks are contagion situations where some animals infect multiple others (broadcasting) and where some animals are infected by multiple other animals (multiple-access). Our information theoretic model for emotion contagion can explain these phenomena if they are observed, but the traditional contagion model cannot.

Once we have the tools of information theory, we can see that the contagion channel is not a good way to model emotion communication. The main problem is that there are two separate processes, communicating emotional information from one person to another, and whether that second person comes to feel the same emotion in question as well. This should be intuitive – in order to receive a signal, one need not be in the same state as the source. It is far better to take emotion contagion to be a complex process composed of emotion communication and emotion elicitation. Indeed, we can decompose the emotion contagion channel into an emotion communication channel and an indirect emotion elicitation channel. Why is this decomposition better? Two reasons: (i) the main reason is that emotion contagion is clearly a compound process, and theories that treat it as such are more precise than those that do not, and (ii) taking emotion contagion to be primitive destroys the connection between emotion expression and information coding. We can try to make a coded emotion contagion channel, but it does not work. Assume that the source has an emotional state and encodes it in emotion behaviour. So far so good. The destination is another animal having an emotional state. How exactly are we to set up a decoding scheme that takes emotion behaviour as input and felt emotion as output? Only part of this process is decoding (i.e., the part at the end of the emotion communication channel and before the indirect emotion elicitation channel). Emotion behaviour just is not decoded into *emotions felt*; rather, it is decoded into *emotions detected*. Once we see that the contagion channel factors into two separate channels – communication and indirect elicitation – and that only the communication channel is coded, we can see that *coded* emotion contagion channels cannot be basic.

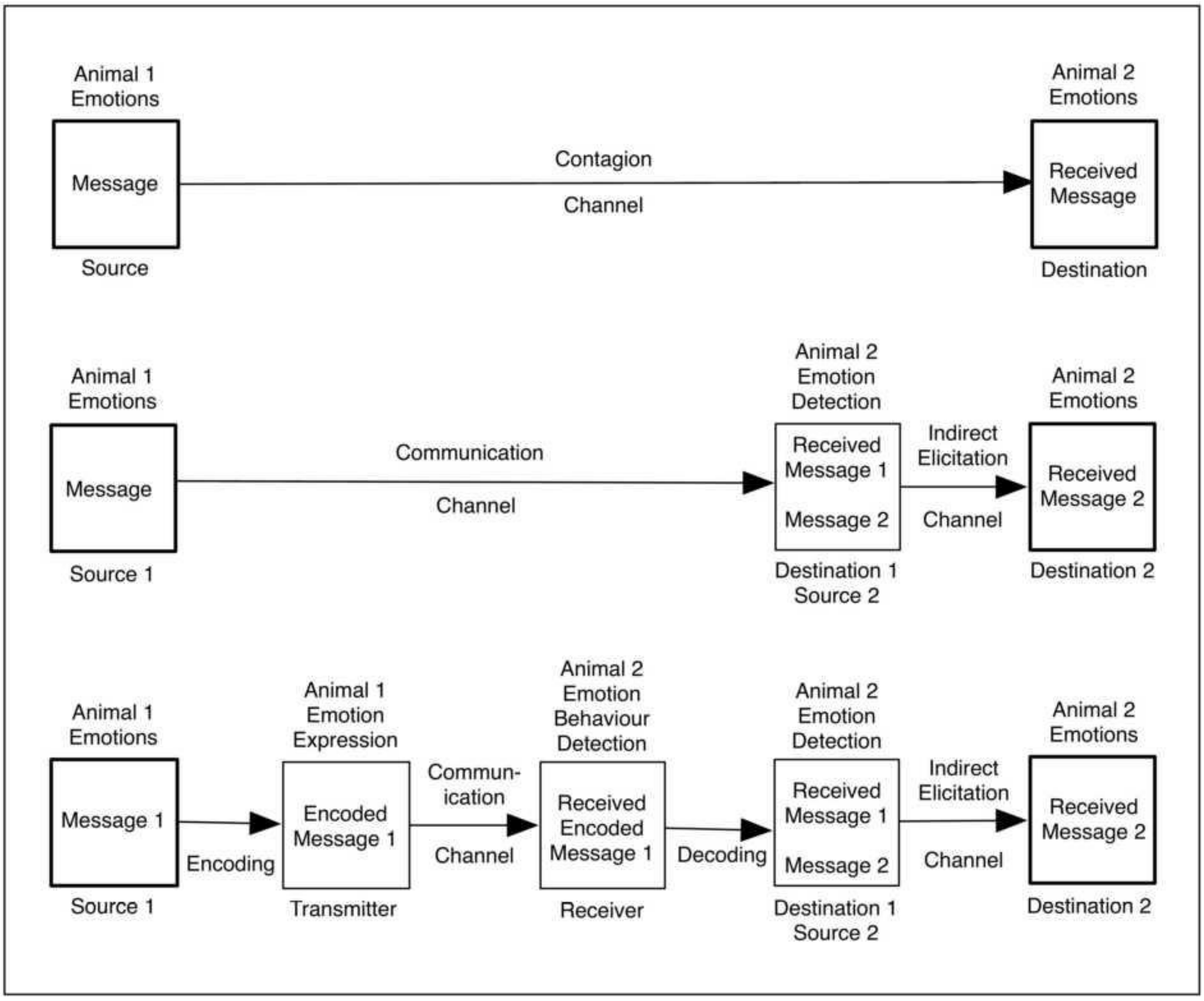

**Fig 35: Contagion channels of varying complexity**

Overall, we have good reason to think that information-based contagion theory is the best option and it treats contagion as a two-step process: communication and elicitation. In information-theoretic terms, only communication is coded; elicitation is not coded. Thus, emotion contagion is a process that begins with a coding step but does not end with a decoding step. As such, it cannot be the basic unit in an information-theoretic framework.

## 6.9 Courage and Other Virtues

Aristotle emphasized courage as one of the principle virtues in his *Nicomachean Ethics* from the 3rd century BCE.[10] His view was that courage is the middle between the extremes of rashness one the one hand, and cowardice on the other. Other ancients thought that courage was strength of soul. One perhaps surprising application of the information theory of emotion communication is to traditional virtues like bravery or courage. Traditionally, courage has been thought of as doing what one has to do when one feels fear (Google tells us "the ability to do something that frightens one."). Courage is clearly a kind of emotion regulation, despite not being thought of that way in most contexts. In more explicit terms, we can say that:

> Animal is *courageous* if and only if Animal feels fear and Animal blocks some of the fear action tendencies that might prevent Animal from achieving its goals while feeling fear.

[10] Aristotle (350BCE/1965).

For example, a soldier feeling fear on a battlefield and having the urge to run away but stifling that urge so as not to be punished. That is an account of fear that appeals to emotion regulation terminology, like blocking action tendencies. However, we do not think that this captures everything important about courage. Saying that courage involves blocking aspects of fear expression is right, but it leaves out the social aspect of courage, which requires an account of emotion communication systems.

In our information theoretic framework, blocking fear action tendencies, which figures prominently in the above definition of courage, is inhibiting fear emotion expression. That is, it means decreasing the accuracy of fear encoding in outgoing fear communication. Each animal that has emotions has the capacity to be part of an emotion communication network, in part, because it expresses its emotions. In information theoretic terms, expressing one's emotions is encoding an emotion message in a coded emotion communication network. Of course, expressing emotions in behavior does not require an audience, but it is always available for an audience.

In all our models so far, we have assumed that the encoding is perfect, but this is an idealization that can be relaxed. In the information theory literature, perfect encoding of information in a coded channel is analyzed as lossless data compression (where the original can be reconstructed perfectly), whereas imperfect encoding (where information is lost) is analyzed as lossy data compression. This way, encoders can be assumed to be perfect. In other words, the message goes from the source, through encoding and is transmitted, but how one divides up this process depends on one's interests. In a single channel, compression and error correction can be distinguished (via a famous result of Shannon's). In these cases, the original message is compressed, then encoded. In this process, all the noise is assumed to be in the compression, not in the encoding. It might be that future investigations

## 6.10 Using Emotion Taxonomies to Code Messages

Another obvious application for the information theory of emotion communication is to show the communicative power of emotion taxonomies. An emotion taxonomy uses underlying psychological features or processes to explain how emotions are related to one another.[11] For example, some theorists suggest that emotions can be modelled as combinations of valence and arousal.[12]

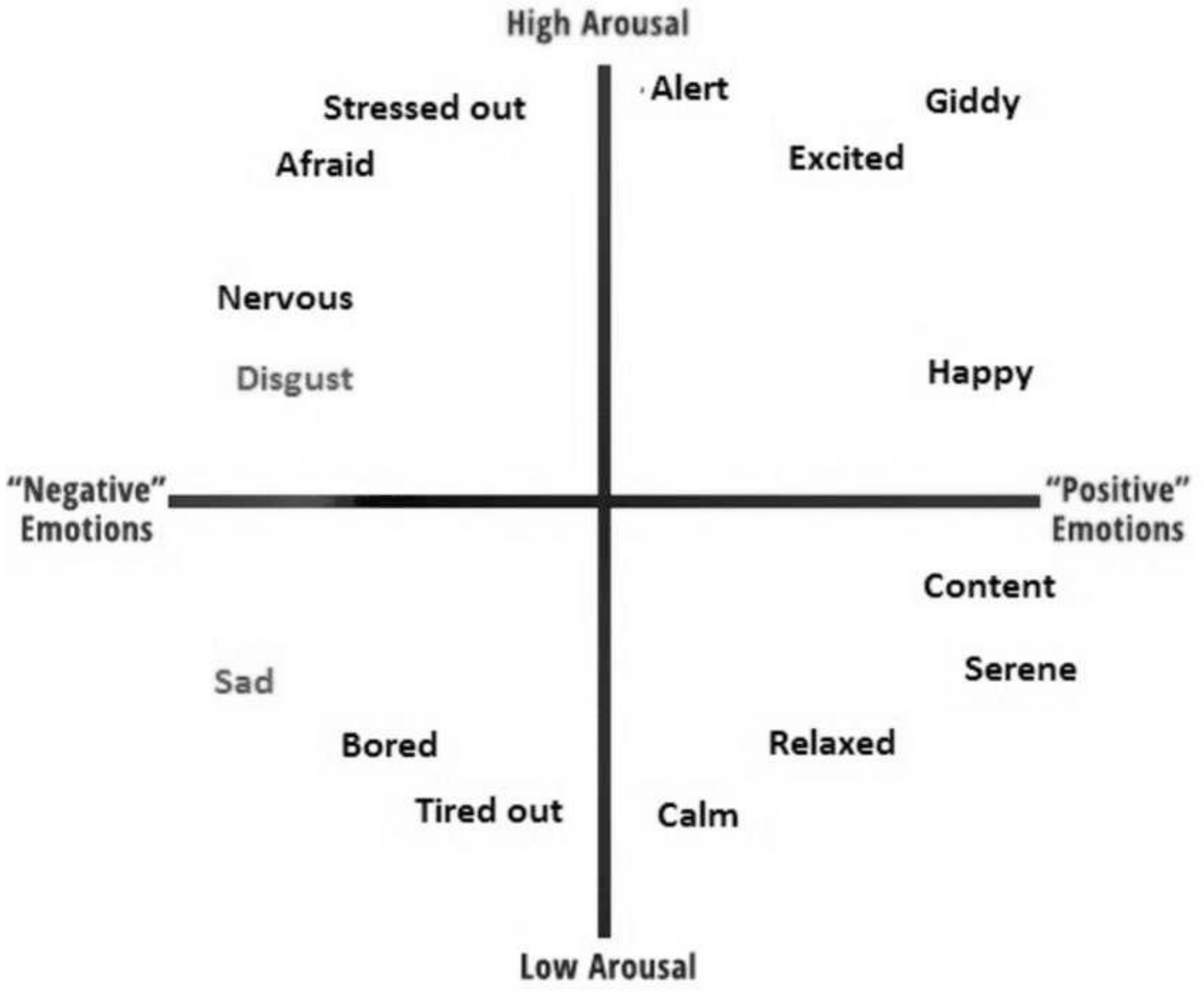

[11] For example, Bänziger et al (2009), Schlegel et al (2014), Lykousas et al (2019).
[12] James & Barrett (1999).

**Fig 36: Valence/arousal emotion taxonomy**

If it is right that emotions are identified in relation to one another by valence and arousal, then emotion communication systems could take this into consideration in the mechanisms of expression and detection (i.e., encoding and decoding algorithms). Artificial emotion recognition systems already do this[13], and some are about as good as an average human right now.[14]

Here is a toy example: let E be the coded emotion communication channel where W is the capacity to have any one of the fifteen emotions listed in Figure 36, X is the code for each of these emotion states in behavior, Y is detecting the emotion codes in behaviour, and Z is the capacity to detect the fifteen emotion states. Let AV be the coded information channel where W is the capacity to have a degree of two magnitudes: arousal and valence. We can assume that each comes in five degrees 0, 1, 2, 3, 4, 5, where 3 is neutral, 0 is min and 5 is max. X is the code for each of these magnitudes, Y is detecting the magnitudes expression in behaviour, and Z is detecting the magnitudes. We can assume that the AV scheme is as follows:

[13] For example, Li et al (2018)
[14] Compare to data in Burleson (2017).

| | | |
|---|---|---|
| Stressed: 5, -1 | Disgusted: 0, 4 | Content: 3, 4 |
| Bored: 1, 1 | Tired: 0, 2 | Happy: 3, 5 |
| Sad: 0, 1 | Calm: 0, 4 | Excited: 4, 3 |
| Afraid: 1, 4 | Relaxed: 1, 4 | Giddy: 4, 4 |
| Nervous: 1, 3 | Serene: 2, 4 | Alert: 5, 3 |

In model E, we can calculate how much information is transmitted and compare it to how much information is transmitted in model AV.

W=<a,v> where a, v =0,1,2,3,4 or 5. and X= <A, V> where A, V=0,1,2,3,4,5.

Every unit of time in system E, X sends a single codeletter: A-O, whereas in system AV, X sends a codeword consisting of two codeletters A-F.

We can assume that when Y receives the emotion message in System E, the received message might differ from the sent message by some error rate. Assume that the sent letter is received 90% of the time and the other 10% of the time, a random codeword is received. In System AV, we can assume that the sent word is the same as the received word 90% of the time, but the other 10% it is an arbitrary codeword that is received.

To have an actual example, we can let X have an equal chance of being in any particular emotional state over time (1/15), but it is always in exactly one emotional state. In system AV, this assumption means X is always in an arousal/valence combination that represents one of the above emotions. More realistic would be that the errors in AV are not random, but rather a 3 on either scale is more likely to be mistinterpreted as a 2 or a 4 than as a 1 or a 5. If we make those twice as likely, then we can see major differences.

In addition, the AV scheme has a built in error-correction mechanism. If Y receives a nonsense message like 0,0 or 5,5, which do not correspond to emotions, then Y knows an error has occurred and can try to fix it by looking at genuine codewords that are close to the nonsense message received. Contrast that with the E scheme, in which Y has no way of detecting errors at all, much less fixing them. The obvious question is: do humans or other animals take advantage of emotion taxonomies in emotion communication? Showing that this occurs would support psychological constructivist accounts of emotion, which hold that emotion states are complex combinations of underlying psychological states like arousal and valence.

## 6.11 Complex Networks

There are far more complex network configurations than we can explore, but several stand out as important in their own right and because they come up in the rest of this chapter. The first kind of network on which to focus is *cyclical networks*, where information travels in loops. One common method for dealing with them is to generate a time-expanded acyclic network from the cyclic network data. In the time-expanded acyclic network, each node of the original is depicted and repeated n times. Each of the connections in the original graph is represented as a connection from one level of the time-expanded graph to the next lower level nodes. Because each node is repeated at each level of the graph, there are no loops in the time-expanded network. Thus, one can use acyclic methods to investigate the cyclic graph, simply by focusing on its time-expanded acycylic graph. Note that this is exactly the same technique used to analyze recurrent neural networks in machine learning.

A particular cyclic network has attracted attention, in part, because it is a simple case and in part because it shows up so often in real communication networks: *the two way channel.* It consists of two nodes sending information back and forth between them, just like in a conversation between two people. However, from an information theoretic point of view, this is a network with four variables, W and X are sources and Y and Z are destinations, and the channel consists of the conditional probability distribution p(Y,Z|W,X). Each node consists of a source, a coded source, a destination, and a coded destination. Of these only the coded source, W or X and the coded destination, Y or Z is listed as variables. X sends message $M_1$ to Y and W sends message $M_2$ to Z. Each node tries to communicate with the other and provide feedback to the other, all on the same channel, so these attempts can interfere with one another, just like in real life. Think how many times in conversations two people start trying to talk at the same time and then they have to figure out who is going to talk and who is going to wait. The capacity region is the closure of the achievable rate pairs, where these are defined in the usual way in terms of the average probability of error:

$$P_e^{(n)} = p[(\widehat{M}_1, \widehat{M}_2) \neq (M_1, M_2)]$$

It is not known in general how to characterize the capacity region for two-way channels, but there are some bounds and some limiting cases have solutions.

In emotion communication networks with animals as complex nodes, the two-way channel would be much more complex. A simplified version of it appears in Figure 37.

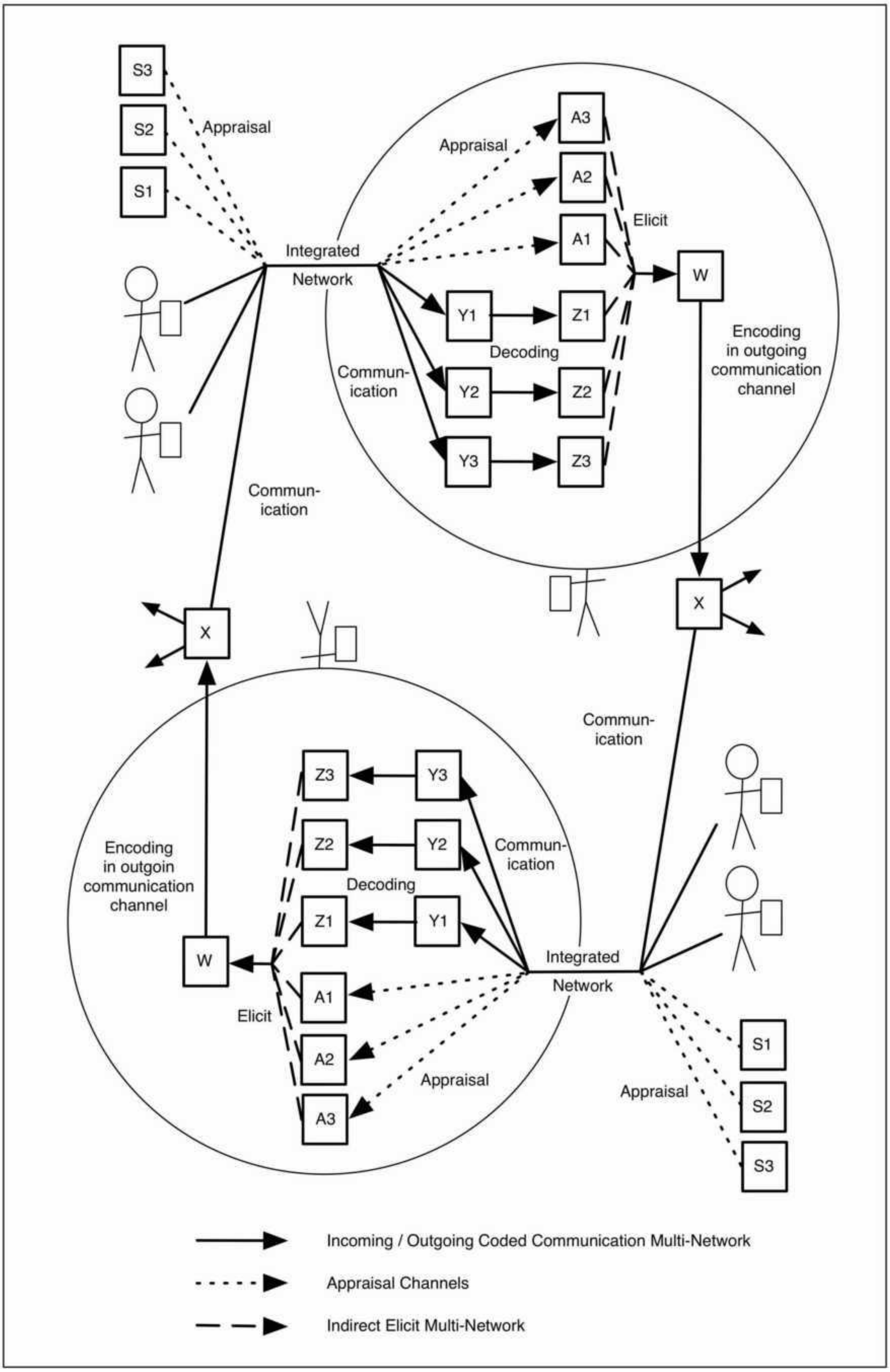

**Fig. 37. A two-person cyclic network with complex nodes and integrated inputs. Each person is depicted as a three-register communication node (as in Figure 29). There are two distinct communication networks that are parts of the overarching network. Each node has three appraisal channels, but these are integrated as in Figure 29. The roles of Person 1 and Person 2 exactly analogous. The overall network is *cyclic* in the sense that it has loops in the flow of information.**

A final class of networks deserves to be mentioned as well, and these are all the complex networks that do not have simple names or types. Because there is no general theory of information, rates, distortions, and capacities, for networks, we do not understand these networks in their full generality. However, we can still estimate various aspects of these networks by breaking them down into parts that we do understand and then trying to estimate how the parts interact. This can be done in a piecemeal way or it can be done systematically. The most obvious way is to think of a network as all its individual edges – that is as all the individual communication channels between nodes. One of course misses out on broadcast and multiple-access phenomena. These can be added in several ways, but one popular way is to use hypergraphs, which are just like graphs but can have edges connecting any number of nodes to any number of other nodes, rather than just one to one. In a hypergraph, one-to-one edges are still communication channels, but one-to-many edges are broadcast networks and many-to-one edges are multiple-access networks. The technique is imperfect because one cannot replace all the parts of the network in this way and still have a coherent mathematical structure. Some networks would end up with edges pointing to edges, which does not make sense for hypergraphs. Overall, we are building network models in the best way we know how – by using ingenuity and all the tools available to us.

## 6.12 Multi-Layer Networks

One of the most important directions to emphasize is examining *networks with different kinds of edges and different kinds of nodes*. One can then investigate how the different populations of nodes or edges interact with one another and how they influence the dynamics of the entire network. The kinds of nodes or edges can be thought of as existing in distinct layers. We can characterize these layers and the interactions between them mathematically, and thereby capture an amazing amount of structure and complexity in the network. These are called *multi-layer networks*. The study of these kinds of networks has exploded in network science, with applications all over the sciences, engineering, and medicine.

We need to use multi-layer networks to model emotion communication systems for two major reasons. First, emotion communication networks are embedded in other kinds of information transfer systems like the channel that has emotion detection as a source and emotion elicitation as the destination (i.e., moving from detecting an emotion in someone else to feeling it oneself). Second, emotion networks themselves display phenomena at multiple different levels as well.

We saw bits of network science in Chapter Five where some of the structural features of networks were listed. With multi-layer networks, there are ways of expanding them. The information theory of multi-networks is in early days, but there are plenty of interesting results, most of which use techniques from statistical mechanics. Below, we use a combination of structural and information-theoretic mathematical tools to model the relative influence of face-to-face contact versus voice/video/letter contact versus social media contact on our overall emotional lives.[1] An interesting measure that results is the Social Media Influence Factor (SMIF). The next section introduces a complex emotion communication system with multiple people communicating over different media types with several kinds of emotions. We then build a mathematical model of this system as a multi-layer network.

[1] De Domenico et al (2013), Boccalettia et al (2014), and Kivelä et al (2014). See Bianconi (2018) for the best overall summary we have found.

Recall that a *network* is defined as a triple of sets (N, E), where N is a set of nodes or vertices, E is a set of edges, which are pairs of nodes. If the edges have directions, then E is a set of ordered pairs of nodes, and the network is called *directed*. If the edges have weights (real numbers associate with how strong the edge is), then the network has an additional set, which contains a weight for each of the edges in E. This sort of mathematical object is also called a *graph*, and it is the topic of graph theory.

A *multi-layer network* is a triple of sets (Y, L, I), where Y is a set of non-negative integers up to the number of layers in the multi-layer network, L is an ordered sequence of networks, one for each layer listed in Y, and I is a set of bipartite networks that characterize the interactions between networks. Each network in L consists of a set of nodes and a set of edges. Each network in I has two sets of elements, one from one layer and one from another, with the edges between them. There is a bipartite network in I for every pair of networks in L. If L1 and L2 are in L, then there is a bipartite network in I for which one set contains the nodes of L1 and the other contains the nodes of L2. Directed multi-layer networks are those for which the edges in each layer are directed.

There are so many kinds of multi-layer networks that it is difficult to formulate general results for them, but we care about a particular kind of multi-layer network called a *multiplex* network. A multiplex network is a multi-layer network where: (i) there is a 1-1 mapping for all the nodes in each layer (called *replica* nodes) and all the interlinks are between replica nodes only. Muliplex networks model situations in which the same entities (the nodes in each layer) are related to one another in different ways (distinguished by the layer). Although the layers in a multiplex network contain distinct nodes, they are understood to be replicas in the sense that a node in layer one is some entity in the network, the replica nodes for it in other layers are that very same entity. The layers represent only different relationships between entities, not different entities. We focus on multiplex networks in what follows.

The first goal is to generalize the structural network measures from Chapter Five to multiplex networks. Very briefly, they generalize in the following ways:

- *Network size* is the cardinality of the set of nodes, and *total links* is the sum of the cardinalities of the sets of edges for each layer. The aggregated network for a multiplex network is a regular network that has an edge for edge in each layer of the multiplex network.
- The set of *adjacency matrices* describes the multiplex network completely by specifying the existence or nonexistence of every possible link between nodes in every layer (one matrix per layer). The adjacency matrices are used to calculate the structural quantities of multiplex networks. They are the algebraic engine of multiplex network science.
- The *degree* and the *strength* of a node are the same as in networks, but they are relative to a level. Aggregated degree and strength are for the aggregated network (i.e., it collects all the edges for all the layers into a single network. The *degree distribution* is relative to a layer; there is also the average for all the layers and for the aggregate. The definition of *scale-free network* works for levels within a multiplex network.
- The *clustering* of a node is relative to a node, but now it can be more complicated by including interlevel links. As above, there is also average per level and for the aggregate.
- A *path* can be restricted to within a level or allowed to traverse between levels. The *length* of a path and the *distance* between two nodes are the same. The *diameter* of a level and of the network are distinct measures. The definition of *small-world network* works for levels within a multiplex network.
- The *correlations among degrees* of nodes are the same, but there can be correlations among the degree distributions within levels as well.

- A *community* is relative to a node, but these are more complex because one can define multi-layer communities as well.
- The *centrality* of a node is relative to a layer, but many of the centrality measures can altered to take advantage of multiple layers or all the layers.

There are all sorts of new measures that can be defined on multiplex networks, but we do not need to explore those.

The information theory of multi-layer networks is in its infancy, as one might imagine, so there are few central results and no general framework for investigating these features. One tool that is used increasingly in the study of the information-theoretic aspects of networks and multi-layer networks is the concept of entropy from statistical mechanics. The standard definition of *entropy in statistical mechanics* is:

$$S = -k \sum_i p_i \ln p_i$$

where k is the Boltzman constant (i.e., exactly $1.380649\times10^{-23}$ $m^2$ kg $s^{-2}$ $K^{-1}$ as of 2019), ln is the natural logarithm function, and $p_i$ is the probability that the system in question is in a particular state, for every possible state i. *Entropy in information theory* (from Chapter Three) is:

$$H = -\sum_i p_i \log_2 p_i$$

where $p_i$ is the probability that the variable in question is in a particular state, for every possible state i.

Remember that $\log_2$ is the base-2 logarithm function, which gives us the unit of *bits* for entropy measurements. Changing the base just changes the units; changing it to ln, which is base e (i.e., 2.71828…) and gives entropy in *nats* instead of bits. This is akin to switching between feet and metres by using a conversion factor (e.g., 1 meter = 39.3701 inches). So this difference between the two definitions of entropy is not a significant one. The only other difference is the Bolzmann constant, which does make the two quantities defined, S and H, distinct physical quantities with distinct units of measurement. The Shannon entropy is measured in bits or nats or other information unit, wherease thermodynamic entropy is measured in energy per unit of temperature (e.g., joules/kelvin), which is very different. There is still considerable controversy over how to understand this relationship. One idea is that the two notions are related by the Boltzmann constant in just the same way that space and time are related by the speed of light, but exploring this interesting possibility would take us too far afield from emotion communication. We would not have thought, at the beginning of our inquiry, that how best to model emotions might turn on how to understand the Boltzmann constant, but here we are.

Either way one thinks of the relationship between S and H, experts in information theory are increasingly taking definitions of entropy from statistical mechanics to be relevant to understanding the information theoretic features of networks and multi-layer networks. Our example applications will not require any of these measures, so we will not pause to consider them, but anyone seriously interested in the information in emotion communication networks and multiplex networks ought to be familiar with this tradition.

## 6.13 Qualitative Emotion Communication System

In the remainder of this chapter we follow the same strategy as in Chapter 5: we first present a qualitative model and then develop the mathematics to describe it. Here the qualitative model is an emotion network with several distinct layers. Intuitively, each layer corresponds to a distinct medium through which emotion communication passes. We distinguish face-to-face communication and communication over social media. We also distinguish a third layer by which emotions are elicited directly from the environment (one can think of this as the ordinal 0 layer). Our qualitative model has nine people, each of whom can have joy, fear, disgust, or nothing, four emotion stimuli, spider, pizza, and poop, which are assumed to be dangerous, joyous, and disgusting, respectively. There is also one artificial emotion recognition system, and five of the people have access to social media (depicted by holding a cell phone). Ultimately, we want to be able to calculate how much influence social media has on each person's emotions.

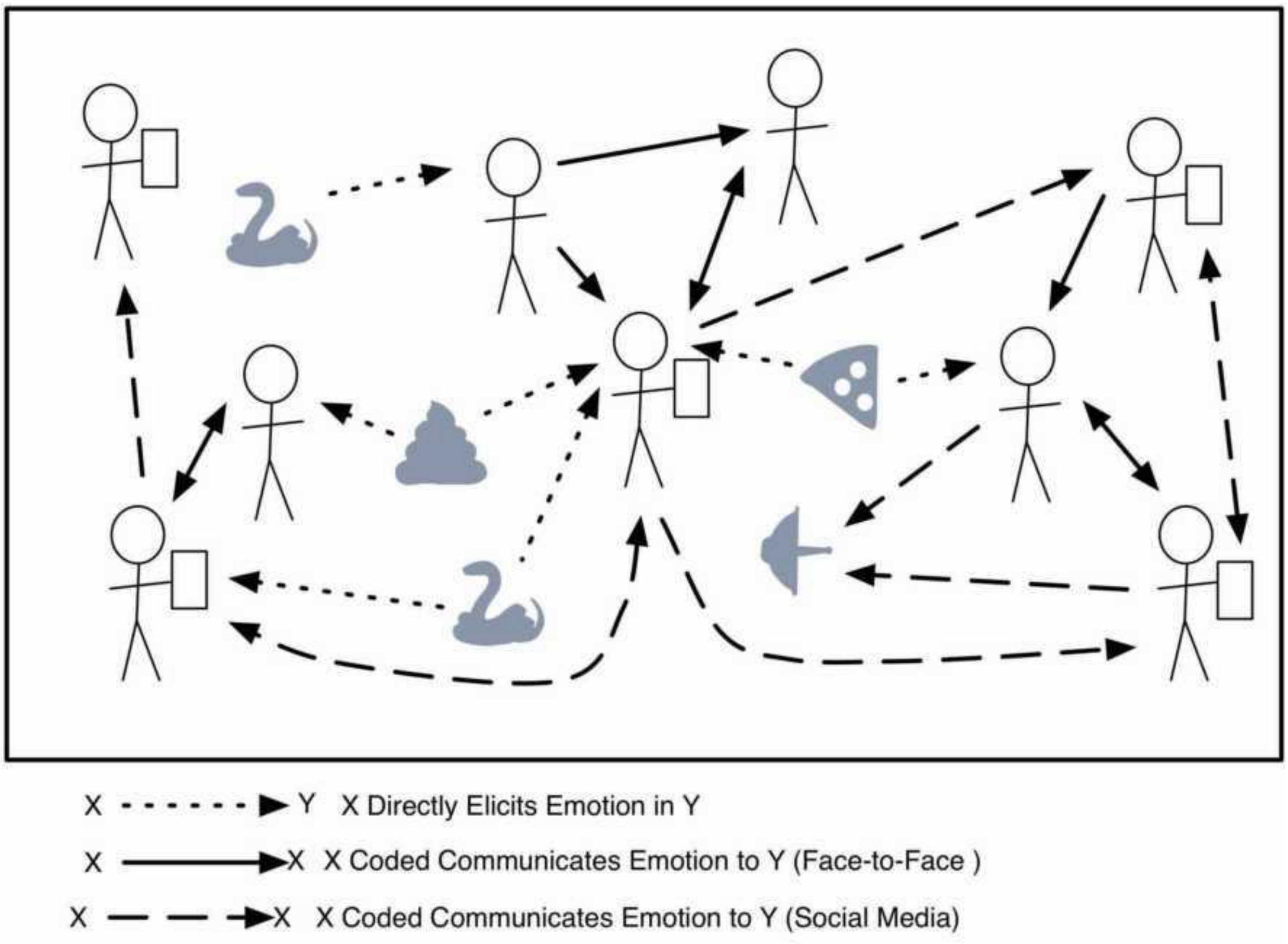

**Fig. 38. A complex emotion communication system. It is depicted as a network with three different kinds of edges, which can be thought of as a network with layers. Each layer is a distinct complex network, but the layers interact with one another. There are three kinds of worldly situations (dangerous, disgusting, and an joyful), and each of the people involved is assumed to be able to feel and express at least three kinds of emotion (fear, disgust, and joy). There are appraisal channels, in which information flows from worldly situations to people. There are two kinds of emotion communication links depicted. Some are face-to-face communication networks, where participants can perceive one another's faces, voices, and bodies. Others are social media communication networks, where participants can perceive only text, emojis or other digital outputs from the other participants. An important part of social media emotion networks are artificial emotion recognition systems, and one of these is depicted in the multi-layer network as**

**well. This figure is intended to be analogous to Figure 23, where the nodes are depicted as blackboxes with no internal structure. However, we now have the tools to model complex multi-layer networks like this with complex nodes like those depicted in Figures 27, 28, and 29. For example, the cyclic three-person network depicted in Figure 38 can be understood as a part of the multi-layer network depicted here.**

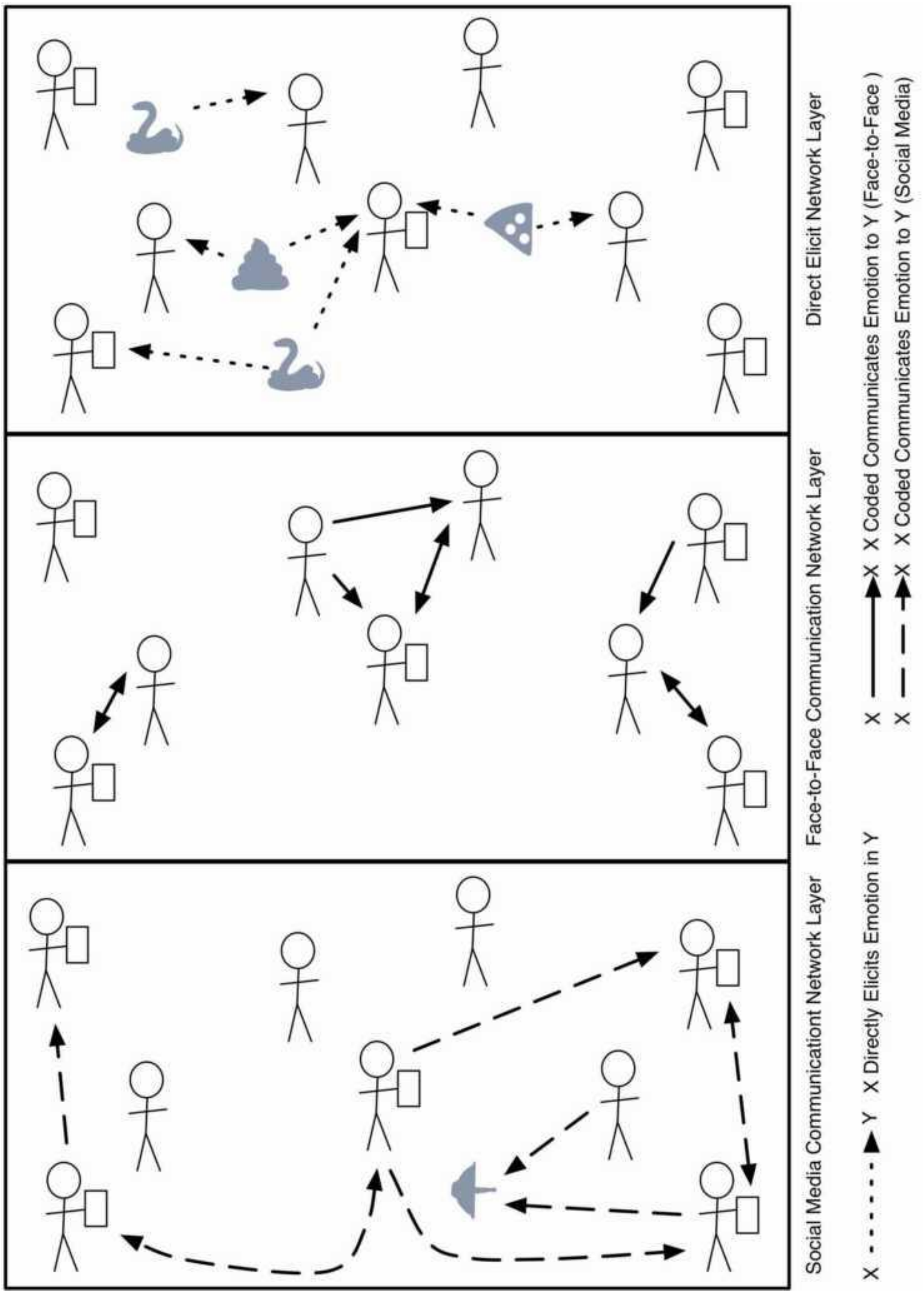

**Fig. 39. The three layers of the network depicted in Figure 38 are shown separately so as to distinguish their relative influences on the emotional phenomena of each person in**

**the network. Panel A shows the appraisals and appraisal elicitations that stem from perceiving the respective stimuli (dangerous, disgusting, and joyful). Panel B shows the face-to-face emotion communication links, and Panel C shows the social media emotion communication links. The point of distinguishing between these two kinds of emotion communication by putting them in different layers is to be able to measure the relative influence of each kind of emotion communication. Planalp (1998) argues that emotion communication is a far greater influence on our emotional states than direct emotion elicitation, but her analysis occurred before social media. We hypothesize that, for some people today, social media dominates the other two layers together in terms of influence on their emotional states.**

## 6.14 Quantitative Emotion Communication Multi-Layer Network

The simple qualitative multi-layer network for emotion communication that was depicted in Section 6.12 has three layers: direct elicitation, face-to-face emotion communication, and social media emotion communication. More complex models might have more layers (one for written communication, one for telephone, one for Facebook, one for Twitter, etc), but we treat all communication that is not face-to-face as social media. We also treat artificial emotion recognition systems as social media, even if they are sending data to a government rather than to a social media corporation.

Each of the three layers can be understood as an emotion communication network using methods of the previous chapter. That much is not new, but it is helpful to go through some of the structures found within each layer because these are a bit more complex than those illustrated in the previous chapter (but no different in kind). All of novelty in multi-layer networks is in the relationships between the layers, to which we turn in the next section.

The *direct elicitation layer* is the least complex because it involves broadcasting and multiple-access but not both together. We already have the mathematics for information networks from the previous chapter and we can use that to model the elicitation layer quantitatively. The direct elicitation layer has four fundamental sources (two spiders, one pizza, and one poop). Although there are nine people on each layer, only five of them are involved in direct elicitation. Only one of these people is engaged in multiple-access, but three out of the four sources is broadcasting to multiple destinations. The network on this layer can be decomposed into two independent sub-networks, but no further.

The *face-to-face communication layer* is more complex than we have seen because of the kind of network involved. Eight of the nine people are engaged in face-to-face communication. The network on this layer can be decomposed into three independent subnetworks, but no further. The subnetwork on the left side has two people communicating through their emotions back and forth. This sort of information network with loops has not been encountered in the book before now. We include a diagram to illustrate the information theoretic properties of this subnetwork. In the middle subnetwork, there are three people intercommunicating, which is depicted in Figure 40 below.

.

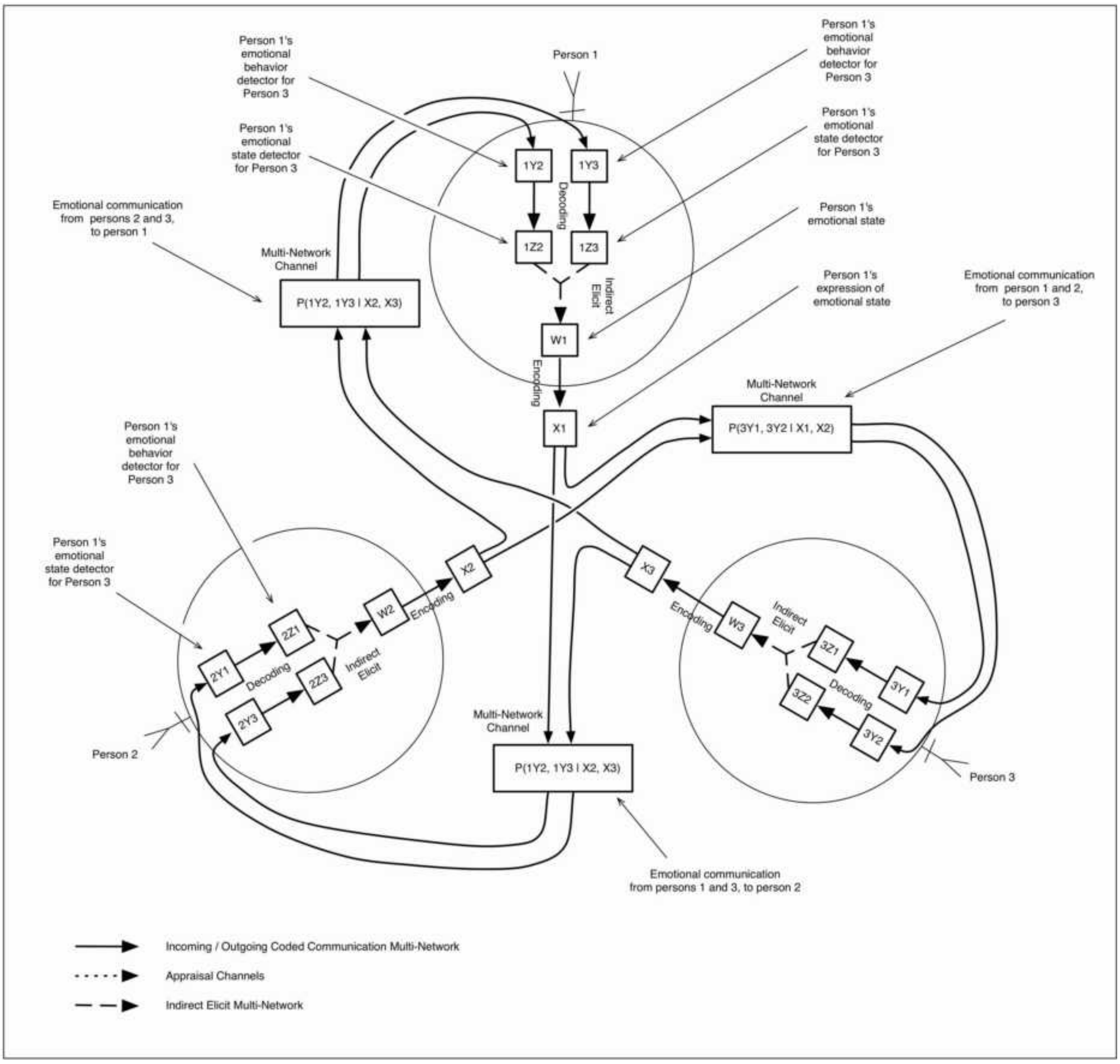

**Fig. 40. A three-person cyclic network with complex nodes. Each person is depicted as a two-register communication node (as in Figure 27). There are three distinct communication networks that are parts of the overarching network. Each of these is depected as a distinct probability distribution. For simplicity, none of the nodes have appraisal channels, but a more realistic model would include these and integrate them with the communication channels, as in Figure 29. All the aspects of the diagram that involve Person 1 are labeled as such. The roles of Person 2 and Person 3 exactly analogous. Note the importance of an indirect elicit channel within each person. These three elicit channels connect the three communication networks to one another. The overall network is *cyclic* in the sense that it has loops in the flow of information. For**

**example, Person 1's emotional state (W1) is expressed in Person 1's emotional behavior (X1), which is detected by Person 2 (2Y1) and decoded so that Person 2 can detect Person 1's emotional state (2Z1). This in turn influences Person 2's emotional state (W2), which is expressed in Person 2's emotional behavior (X2), and this is detected by Person 1 (1Y2). Person 1 decodes Person 2's emotion behavior to detect Person 2's emotional state (1Z2), which then influences Person 1's emotional state (W1), which completes the loop.**

The *social media layer* is also complex. Only five of the nine people have access to social media (depicted by holding a cell phone), but six of the nine people are involved in the social media layer – that extra person has no cell phone but is being monitored by an artificial emotion recognition system, depicted as a satellite dish in the diagram. Having ones's emotions identified by an artificial system, from an information-theoretic point of view, is communicating one's emotions to a destination, but in the cases the destination is artificial rather than an animal. Engaging artificial emotion recognition systems is one way – often involuntary – of participating in social media, even if one has no conventional internet access. The network on this layer cannot be decomposed into independent subnetworks – every node is involved with every other node indirectly. There is nothing new about the structure of this network beyond the elements introduced in Figures 37 and 40 to handle loops in information structure.

The three layers are connected by the people, which are the same in each of the three layers. Notice how the layers allow for communication that would be impossible from within a layer. Sending a signal from the right most person to the leftmost person would be impossible in the face-to-face layer alone, but once social media is in play, this kind of communication is commonplace on the internet (e.g., you get outraged reading a story of injustice committed on a person you do not know who lives in some faraway place, and you have access to this emotion information simply because they are a friend of a friend on Facebook). Information about the danger of a spider might directly elicit an emotion which is then communicated face-to-face, which is then communicated over social media, which is then communicated face-to-face. This sort of message would involve each of the three layers, and each of the communication steps depends on connections between the layers (e.g., the person who saw the spider is identical to the person involved in face-to-face communication with someone else).

Let us offer a mathematical model of this emotion communication system. Let M=(Y, L, I). Y={1, 2, 3} since there are three layers: the elicitation layer, the face-to-face communication layer, and the social media communication layer. L=<L1, L2, L3>. There are nine people, four elicitors, and one artificial receiver, for a total of 14 nodes total. We are using a multiplex network as our mathematical model, so each node exists in each layer of the network (or, to be more precise, each node in the first layer has a replica node in every other layer, and every node in every other layer is a replica node. For each layer $L_i$ = ($N_i$, $E_i$), the set of nodes $N_i$={$P1_i$, $P2_i$, $P3_i$, $P4_i$, $P5_i$, $P6_i$, $P7_i$, $P8_i$, $P9_i$, $E1_i$, $E2_i$, $E3_i$, $E4_i$, $R_i$}. For example, $P1_1$ is Person 1 in layer 1, $P1_2$ is Person 1 in layer 2, and $P1_3$ is Person 1 in layer 3. Each person is a vector of three nodes, one from each layer (e.g., Person 7 is < $P7_1$, $P7_2$, $P7_3$>.

The edges for each layer are distinct and can be read off the diagram. If Stimulus 1 is the leftmost spider, Stimulus 2 is the poop, Stimulus 3 is the rightmost spider, and Stimulus 4 is the pizza, then we can characterize the elicitation layer edges as $E_1$={<S1,P2>, <S2, S6>, <S2, P7>, <S3, P5>, <S3, P7>, <S4, P7>, <S4, P8>}. E2={<P5, P6>, <P6, P5>, <P2, P3>, <P2, P7>, <P3, P7>, <P7, P3>, <P4, P8>, <P8, P9>, <P9, P8>} and E3={<P5, P1>, <P7, P5>, <P5, P7>, <P7, P4>, <P4, P9>, <P9, P4>, <P7, P9>, <P8, R>, <P9, R>}. Notice that we have made use of the fact that this is a multiplex network by assuming 'P9' in layer 3 is $P9_3$ without having to use that cumbersome notation.

The multiplex network also needs to have its interconnections, I, specified. In the general multilayer network case, I would be an ordered sequence of three networks, one for each pair of layers. Because this is a multiplex network, there are the same nodes (or replica nodes) on each level and there are no other interlayer connections beyond those connecting the replica nodes together. So we do not really need to go into detail specifying I in this case.

So far we have described a directed multi-layer network M with three layers, fourteen nodes shared amongst the layers, and edges for each layer. However, this emotion communication system is weighted in the sense that each of these emotion communication links has certain information theoretic properties, like the rate or the capacity. So M needs a weighting parameter W, which is a sequence of sets $W_i$ of weights for each of the edges in $E_i$ for each layer i.

Let each person have four possible emotion states, joy, disgust, fear, or nothing, and each person has the ability to express each emotion in three modalities, face, voice, and body. We can label each of these twelve behaviours in the obvious way: <joy, disgust, fear, nothing> X <face, voice, body>. So *joy-face* and *nothing-body* are two examples. Assume that each person can be in only one emotion state and express only one emotion state at a time. People always express their emotions in all three modalities and they never any exhibit emotion behaviour without the accompanying emotion. We can therefore think of the uncoded emotion message as a single symbol and the coded emotion message as a triple of symbols, one for each modality. So there is a four symbol code-alphabet for this code (J, D, F, N) and the order of the symbols in the message determines the modality, so a message of FFF would indicate fear-face, fear-voice, and fear-body. Each coded emotion message sent is a triple of the same symbol.

Let each person have two emotion detection registers, each with the usual two-part structure: an emotion behaviour detector and an emotion detector, connected by a decoding algorithm. The emotion behaviour detector has a state for every kind of emotion behaviour it can detect; assume it has one for each of the twelve behaviour states. The emotion detector has a state for each kind of emotion plus an additional error state for when the decoder malfunctions. Each person's emotion behaviour detector receives a triple of symbols. They might not all be the same symbol because of noise in the channel. The decoder then delivers a single emotion detection symbol to the emotion detector or an error symbol if the decoder could not figure out the coded message. Let each person have some selection algorithm for deciding whose emotions to register from among the other people it might communicate with.

Let each person have two appraisal registers for recording emotion appraisals. An appraisal is not a kind of emotion communication, but instead it brings in information from the world (e.g., there is a spider) and produces an appraisal of it based on the animal's interests (e.g., there is danger). Each person has an internal information -theoretic structure that connects their appraisal registers and their emotion detection registers with their emotion state. We have been calling this an *elicitation* network, and that name is appropriate.

Each of the stimulus nodes has no emotions, but could be thought to have two possible states – active and inactive. A spider in active state is dangerous but in inactive state is not. After all, spiders behind glass are safe to be near and pizza in the hands of one's enemy does not bring joy. We can assume that the pre-appraisal channels have a certain amount of noise, but the appraisal registers have four possible states: detect danger, detect disgustingness, detect joyfulness, and detect nothing. We can interpret the stimuli's activation state differently for each kind of stimulus: the spider can be either dangerous or not, the poop can be either disgusting or not, and the pizza can be either joyful or not.

The artificial emotion detection system functions just like an emotion detection register in a person – it has a two-part structure: emotion behaviour detection, which is decoded to arrive at emotion detection. Messages arrive at the behaviour detector as triples and decoded messages are a single emotion. We could assume that it has limited capacities, but we might as well give it the capacity of a human register, so it the behaviour detector has twelve states and detector has four.

This completes the information-theoretic description of the multi-level network. Notice that the mathematical description of the network given above treats people as individual nodes.

This idealization can be helpful, as it allows us to use network structural measures on whole people. However, just as in the case of the two-way network (above), it will be essential to look a the multiplex network of people at a different resolution where people are networks, not nodes. We will not bother to provide the details of this more detailed network. I would be nice to have names for them, so micro-network for the detailed one and macro-network for the one with people as nodes. Instead, consider some structural features of the macro-network:

| Id | Label | InDegree | OutDegree | Degree | Eccentricity | Closness-centrality | Betweeness-centrality | Eigencentrality | Clustering | Authority | Hub | Modularity | PageRank |
|---|---|---|---|---|---|---|---|---|---|---|---|---|---|
| 0 | P1 | 1 | 0 | 1 | 0 | 0 | 0 | 0.232989 | 0 | 0.151025 | 0 | 1 | 0.052042 |
| 1 | P2 | 1 | 2 | 3 | 3 | 0.45 | 9 | 0.004774 | 0.333333 | 0 | 0.35933 | 0 | 0.037324 |
| 2 | P3 | 2 | 1 | 3 | 3 | 0.421053 | 0 | 0.277454 | 0.5 | 0.245451 | 0.272351 | 0 | 0.068283 |
| 3 | P4 | 2 | 2 | 4 | 2 | 0.75 | 4 | 0.736082 | 0.5 | 0.188473 | 0.17003 | 2 | 0.090975 |
| 4 | P5 | 3 | 3 | 6 | 3 | 0.533333 | 20 | 0.403871 | 0.083333 | 0.286611 | 0.426184 | 1 | 0.11255 |
| 5 | P6 | 2 | 1 | 3 | 4 | 0.363636 | 1 | 0.237763 | 0 | 0.283084 | 0.101565 | 1 | 0.060615 |
| 6 | P7 | 6 | 4 | 10 | 2 | 0.666667 | 47.5 | 0.430029 | 0.071429 | 0.76856 | 0.33332 | 0 | 0.151635 |
| 7 | P8 | 3 | 2 | 5 | 2 | 0.75 | 2 | 0.834552 | 0.25 | 0.259737 | 0.1177 | 2 | 0.105964 |
| 8 | P9 | 3 | 3 | 6 | 1 | 1 | 13.5 | 1 | 0.25 | 0.220078 | 0.198542 | 2 | 0.136125 |
| 9 | S1 | 0 | 1 | 1 | 4 | 0.333333 | 0 | 0 | 0 | 0 | 0 | 0 | 0.020177 |
| 10 | S2 | 0 | 2 | 2 | 3 | 0.473684 | 0 | 0 | 0 | 0 | 0.372666 | 1 | 0.020177 |
| 11 | S3 | 0 | 2 | 2 | 3 | 0.5 | 0 | 0 | 1 | 0 | 0.373916 | 1 | 0.020177 |
| 12 | S4 | 0 | 2 | 2 | 3 | 0.5 | 0 | 0 | 0 | 0 | 0.364392 | 0 | 0.020177 |
| 13 | R | 2 | 0 | 2 | 0 | 0 | 0 | 0.851797 | 1 | 0.112065 | 0 | 2 | 0.103779 |

**Fig 41: Structural features of emotion communication multi-layer network**

These are all calculated from the macro details listed above (the aggregate network) using the free program Gephi after a couple of minutes inputting the data. Notice that Person 9 gets highest closeness centrality, but Person 7 gets highest Betweenness centrality. PageRank is fascinating because it lists the four people as the most important nodes: Person 5, Person 7, Person 8, and Person 9. But the next most important node – more important than any of the other five people – is the artificial emotion recognition system! It ranks very highly because it is connected to especially highly ranked nodes.

In addition to the measures for nodes, we can list some measures for the entire (agglomerated) network:

- Average degree: 1.79
- Network diameter: 4
- Density: .14
- Modualarity: .36

- Average clustering: .29
- Average path length: 2.12

These measures indicate that the aggregation network for this multi-layer network is *not* a small world network and is *not* a scale-invariant network.

So far, we have only utilized network measures that are independent of the quantitative theory of information in emotion communication. In order to generate information-theoretic measures, we need far more details about the information theoretic nodes. In particular, we need probability distributions – *lots of probability distributions*. A distribution across the four emotional states for every person's emotional states would be a start, but each stimulus needs its own distribution across its two states as well. Every edge in the multiplex network needs its own conditional probability distribution to define the channel. But here we reach a problem. The network displays considerable multiple access and broadcasting. If we think of each incoming emotion communication link for a person as an independent channel, then we cannot account for their interactions like interference. Likewise, broadcasting emotions to multiple people cannot be accounted for in a network with independent outgoing channels from each person. Instead, we need to use a more complex mathematical framework, which is common in network information theory: the hypergraph. In a graph, an edge can have only one source node and only one destination node, but in a hypergraph, edges can connect up any number of sources and destinations. It is common to use one-many edges to model broadcasting and many-one edges to model multiple access with interference. One could imagine filling in each of the broadcast edges and the multiple-access edges in the hypergraph so that it matches exactly the structure of our example emotion communication system. It turns out that this innocent vision is a mere pipedream. There are several major problems, but we highlight two.

Some people have too many connections for the number of registers they have (e.g., Person 7), so there needs to be a way of specifying how three incoming potential connections are filtered into two emotion detection registers. There is no way using independent channels for us to model how people like this select whose emotions to track, nor can we do it with independent one-many edges since there are three possible sources and two possible destinations. One way to do this is to have an edge with three sources and two destinations. It would need a more complicated probability distribution than we have seen so far: p(Destination 1, Destination 2 | Source 1, Source 2, Source 3). So in order to model selection algorithms we need many-many edges.

The second problem is much worse. It can turn out to be impossible to *both* replace the collection of edges going out of a single node that has Out-degree >1 with a single one-many edge *and* replace the collection of edges going into a single node that has In-degree >1 with a single many-one edge. That there are network for which it is impossible to include all broadcast edges (one-many) and multiple-access edges (many-one) at the same time in a network. The problem is that the resulting hypergraphs do not accurately reflect the underlying information theoretic connections or they have artifictual cycles. That is, it makes edges point at edges rather than just at nodes. Take for example a simple network with three nodes, 1, 2, and 3, and three directed edges: <1, 2>, <1, 3>, and <2, 3>. Node 1 is broadcasting, node 3 is multiple-access, and node 2 has a single input and single output. Let us replace the graph G consisting of this set of nodes and edges, with a hypergraph H such that each broadcast node is the input to a one-many hyperedge and each multiple-access node is the input to a many-one hyperedge. H has hyperedges <1 →2,3> and <1, 2→3>. The problem with this setup is that communication from 1 to 3 is counted twice, once in each hyperedge. We could try to fix this by using a single big hyperedge <1, 2 → 2, 3>, but now the problem is that this hypergraph has a cyclic edge (i.e., an edge with overlap between its sources and destinations), but our underlying network has no cycles. Moreover, when we posit a conditional probability distribution for this hyperedge, we get p(2, 3 | 1, 2), which is nonsensical unless we distinguish various times and allow earlier states of 2 to affect later states of 2, just as when we generate acyclic graphs for cyclic ones. Again, this is a gross mischaracterization of our underlying system, which is distorted by treating node 2 in this way. So we are forced to choose between two different hypergraph representations : (i) <1 →2,3>, <2, 3>, and (ii) <1, 2→3>, <1, 2>. The problem, of course, is that neither of these take in to account all the interference patterns, for example, in (i) the fact that node 3 is taking in information from 1 and 2 and so these might interfere with one another is impossible to model. The other option, (ii), misses that node 1 is sending information sent to 3 to 2, so these might interfere with one another.

What are we to do? One solution is to estimate which aspect is less important and pick the model that ignores it. Or one could try to use both models instead. Other, more complex options would involve looking for specific network coding schemes that might be effective and well-understood for the specific network in question. Even more involved solutions would pick tools from statistical mechanics (e.g., ensembles) rather than some kind of mutual information among variables.

However, one chooses to deal with the problem of understanding information flow through networks, assigning probability distributions to all the relevant variables and conditional probability distributions to all the edges would not necessarily tell us the weights of the edges in this multiplex network. The problem is that as information flows among the people, it changes their emotion states and their emotion state probability distributions. These in turn change the distributions for emotion expression, which changes the information flow though the channels. This too changes the distributions of emotion states, and we are back to familiar territory. There are clearly feedback loops that show up when we consider how this network develops over time. So any weights we might assign to the edges of this network would almost certainly not be in equilibrium with the probability distributions involved. Thus, any example assignment of weights would not be illuminating unless we could track it over time to see how it develops.

## 6.15 The Social Media Influence Factor

There are a number of information theoretic measures for how the layers relate to one another, but we are looking for a relatively simple measure of how each layer affects each node. Since we only have three layers in our model, this will be somewhat crude, but it illustrates the point: the information theory of emotion communication is powerful enough to quantify the influence social media has on a person's emotional life. Mathematically, this would be a measure of how much emotion information from social media influences which emotional state the person is in, compared to face-to-face emotion communication and direct emotion elicitation from stimuli. It turns out that there is no simple formula for this measure, and the topic has not been explored much. Probably what will happen is that this area will develop a number of distinct families of measures for influence, just as in the case of other measures like communities, centrality, and importance.

One easy way to get something like a layer-influence measure is to look at all the in-degrees for a node in each level and take the fraction from the level in question. For example Person 7 has aggregate In-Degree of 6, but it breaks down by layer as: Layer 1 (stimulus) is 3, Layer 2 (face-to-face) is 2, and Layer 3 (social media) is 1. So Person 7 has 1/6 of her aggregate In-Degee accounted for by social media. This is probably far too easy since the InDegree does not measure influence on emotion *state*, but rather influence on emotion *detection*. We want to know how much of your emotional life is influenced by social media. That is, how much of your probability distribution for your emotion states is influenced by your engagement with social media?

There is another nearby measure that would be interesting as well, which answers: how much of your probability distribution for your emotion states is influenced by the presence of social media? The first measure is clearly zero if you have no social media engagement at all. However, the second measure might have a positive value for you even though you have never been on social media. For example, your partner looks on Facebook, sees a post about immigrants, and gets angry. Your partner's anger might make you angry or it might make you sad or it might make you feel a million other things. The point is that if social media had not existed, then you would not feel any of those things because your partner would not have been angry right then about immigrants. So social media is affecting your emotional states even if you do not engage with it. Call the first measure the Direct Social Media Influence Factor and the second, the Indirect Social Media Influence Factor. The former measures how much *your social media* affects your emotional life, while the latter measures how much *social media* affects your emotional life. "What would your emotional life be like if *you* didn't use social media?" vs "What would your emotional life be like if *no one* used social media?" These are clearly distinct, but equally interesting.

We illustrate three routes to defining these measures that seem promising. The first option is to use a method called *percolation*, which looks at how networks behave when parts of them are disabled or deleted. Just as traumatic brain injuries have been a godsend to neuroscience and cognitive science, we can learn a lot about a network by breaking parts of it and seeing what happens. We can use percolation by investigating an alternative multiplex network where the node in question is isolated in the social media level. That is, all the node's edges in the social media level are deleted (yes, even the out-edges – it can happen that I am sending you emotion information over social media and you are sending me feedback face to face, but this social media influence on me channel will not be affected by deleting only my in-edges). Then we see how the alternative multi-network evolves over time and estimate an alternative probability distribution for the node in question (e.g., even mutual information, which is not officially a metric on probability space, but is often treated as a nice way to measure differences between probability distributions). We can then compare its real probability distribution over time to its alternative over time, and use any number of statistical methods to measure how different they are and this would effectively be a measure of how much being on social media is influencing the node's emotional life. Even though we haven't nailed down all the details for how to compare the real and alternative distributions, we can call this the *node-* *percolation influence factor for a node*, and when that level is the social media level, it becomes a social media influence factor. Indirect influence is measured by deleting not only that node's edges in the level, but the entire level. Again, we let the alternative multiplex network develop over time and measure the probability distribution for the node in question. We can call this the *level-percolation influence factor for a node.*

This first option is mostly cheating because it does not offer an algorithm, so these are not really measures (e.g., what if the probability distributions never settle down at all?). Instead, we can define influence in an information theoretic way. Let X1-Xn be the variables representing each animal's emotion state in the multiplex network. Assume that level 1 is the level in question whose influence is to be measured. Intuitively, we want to measure the information flow from all the other animal's emotion states to the animal in question's emotion states, but only in layer 1. Of course, we are not measuring the influence of the other animal's emotion states in layer 1 on our animal's emotion states in layer 1; instead, we want a measure of influence on our animal's total emotional life, which we are taking to be a probability distribution (perhaps varying over time, or averaged over time). To do this, define the restricted probability distribution for each node in the multiplex network, which is its probability of communicating an emotion message *in that layer*. For layer j and each node i, it is $p_j(X_i)=p(X_i=x$, where x is communicated via layer j or x is the state of not having a j-relevant state). That is, each $p_j$ might need an additional state to indicate that nothing is being communicated if $X_j$ does not already have such a state.

Now we want to know the influence of each one of these restricted probability distributions on our target animal's emotion state probability distribution. One way to do that measurement theoretically is via the mutual information of two distributions X and Y, which is written as "I(X;Y)" and is a measure we have seen a large number of times already. Using mutual information would be equivalent to saying "this measure tells you how much extra information is in the target node beyond the information we already have about all the other nodes' behaviour in layer j" where j is the layer whose influence is being measured. The equation would be:

$$S(i,j) = I(X_{1j}, \dots, X_{i-1j}, X_{i+1j}, \dots X_{nj}; X_i)$$

Where S(i,j) stands for the *subservience of node i to layer j*, and $X_{ab}$ is the probability distribution for $X_a$ restricted to layer b. Because subservience is an information-theoretic measure, we cannot calculate it for the example multiplex emotion communication network above.

We can normalize this measure so that the total influence is 1. If the social media influence factor (SMI factor) is greater than .33, then it is responsible for more than 1/3 of the person's emotional life. If the SMI factor is great than .5, then it is responsible for the majority of the person's emotional life. A person with no social media accounts who never accesses the internet might have an SMI factor close to zero, whereas someone who spends almost all their time indoors on a computer with many social media profiles and very little face-to-face interaction with other people might have an SMI factor above .9. It is important to recognize that even someone who has no social media accounts at all but still uses Google has a significant social media influence factor simply because of the surveillance capitalism framework in place. Unless you are taking extreme measures, your behaviour is being tracked across the internet – every site, every click, every video, everything you see and do is being recorded and sent to several companies including Google and Facebook, even if you have no accounts with these companies. This information is used to decide which advertisements you see online, and so have considerable influence over your emotional state. In 2019, adults in Western democracies spent almost six hours a day online, on average. Because of surveillance capitalism, much of that time will count toward your SMI factor, even if you never go on Facebook, Twitter, or any other obvious social media website.

## 6.16 Extensions

This chapter had the last part of our information theory of emotion communication. But there is so much more to be done with it. As just a taste, consider an emotion communication system with 100 animals and each animal has the emotional capacities to feel dozens or even hundreds of emotions. We could represent each animal's emotional state as a continuously changing point (or even several points) in an emotion space.

Compression and encoding are each continuous as is emotion expression. Emotion detection too is a continuous function with thresholds for positive and negative identifications. Emotion behaviour detection and emotion detection involve continuous probability distributions as well. Each emotion has a continuous intensity too, as do emotion expressions, behaviour detections, and emotion detections.

Each of these 100 animals has a dozen communication registers and a dozen appraisal registers with an integrated input network. The emotion communication among the population can be divided into several layers (e.g., face-to-face, voice telephone, video conference, social media) based on which emotion behavior modalities (i.e., kinds of emotion behaviour, like running away from something or crying or writing a snide remark as a Facebook status). Each layer displays complex networks with memory, feedback, side-channels, multiple coding and decoding algorithms, and network coding.

If all the relevant probability distributions are available, then one could calculate all sorts of complex information theoretic measures, even ones that depend on treating the whole system as discrete. If one is trying to figure out how and what to measure in real animal populations so as to construct a model like this, then it makes sense to work backwards from which information theoretic values one wants (e.g., transfer entropy), to which aspects of the real population need to be measured in order to calculate it.

If you have understood the basic content of chapters 2-6, then you now have the expertise to do the research and learn what you need to put together a model like this. Much simpler models than this would still be dramatically more powerful than the explicit models we have seen in this book. Most of these could be put together from the models and extensions presented in chapters 3-6.

Other obvious directions would be to construct dynamic information theoretic models that explain how emotion communication systems change over time. Once we see how emotion networks function at a moment, we naturally wonder: how do the information-theoretic features of emotional signal systems change from our evolutionary roots as hunter-gatherers, to subsistence farmers in small villages, to industrial workers in large cities, to social media on the internet, and beyond?

Linking the mathematical theory of emotion communication presented here with game theory and evolutionary biology (and evolutionary game theory!) would be a big success. The link to game theory is through the theory of signal systems presented at the end of Chapter One: information theory presupposes that one can already identify sources, messages, and destinations, but this sort of thing requires a formal theory of signal systems; we borrow Lewis's formal theory of signal systems for this purpose, and it is explicitly based on game theoretic principles. In particular, each signal constitutes a *proper coordination equilibrium* (i.e., a collection of action choices such that someone would have been worse off had anyone chosen differently) in a *coordination problem*, (a situation in which two or more agents need to perform actions but the outcomes of their actions depend on which actions are performed by others).[2] Skyrms develops this connection between signal systems, evolution, and game theory in myriad fascinating ways, but he does not mention emotions. More low hanging fruit.

The final application of the information theory of emotion communication is to social media in the final two chapters. In the next chapter we detail some trends in social media and some examples of emotion manipulation on a massive scale across social media. In the final chapter, we use the information theory of emotion communication to devise security techniques for emotion communication systems that can detect and prevent the major kinds of abuses we see right now.

[2] See Lewis (1969: 5-51); see Cain (2017) for a contemporary survey of game theory and coordination problems.

Those researchers who want a direct line into the relevant information theory mathematics should look at the first eight or ten chapters of Cover and Thomas, then El Abbas and Kim on network information theory, then Bianconi on multi-layer networks.

## *Chapter 7*

# Emotions on Social Media

It is should be uncontroversial at this point to say that the internet is the most directly significant invention in human history, in the sense that it changed our lives more and in a shorter time than any other. Being a part of the generation with analogue childhoods and digital adolescences, we feel the paradigm shift first-hand. One of the most significant aspects of the internet is social media.

All kinds of information are exchanged on social media but the information via emotion communication is one of the most significant classes. Now we put the tools developed in chapters 3-6 to our final application: understanding online, how it is being manipulated on a vast scale, and what we can do about it.

Before we start, it makes sense to be clear about the phrase ‘emotion manipulation’. By this we mean any process by which one or more agents causally affect one or more distinct agent’s emotions or emotion capacities. The effects in question might be positive or negative on the targets. The emotion manipulation might be intentional or unintentional. The difference between unintentional emotion manipulation and mere coincidence is sometimes difficult to discern, but it surely involves finding patterns of behaviour and identifying relevant causal mechanisms.

## 7.1 Social Media

What is social media? There is widespread disagreement from laypeople to experts and the phrase seems to change its meaning about as fast as social media itself morphs from one incarnation to the next. In the introduction to her prominent collection of work on social media from 2012, Diane Rasmussen Neal writes:

> Social media – and its sister term, 'Web 2.0' – are difficult to define, because there is little agreement about what they mean. My view, in the simplest terms possible, is that these phrases refer to the many easy-to-use services that anyone can use to interact with other people online. For example, when you watch and/or comment on a YouTube video, 'Like' a friend's Facebook update and read your colleague's blog, you are using social media. (xxiii 2012).

Social media here is defined as essentially interacting directly with other people on the internet by reading, listening, watching or writing, recording, videoing.

We essentially agree that this is the core of social media, but we argue that it is important to take a wide view of social media. Obvious exemplars at the beginning of 2020 in Western democracies are Facebook, TikTok, Twitter, Reddit, and Instagram. But we include much more than websites or apps where users upload personal information. If you are using any normal web browser (e.g., Chrome, Firefox, Explorer) then any number of corporations including Google and Facebook have programs running on your computer tracking everywhere you go so that advertisements in your web browser are customized just for you. This is social media too, even if it does not feel like it to the user. For example, Google uses cookies to track everything you do online and compares what you do to what other people have done that are likely to buy certain kinds of products when shown certain kinds of advertisements. Simply by searching, typing, and clicking, you are participating in social media with hundreds of millions of other people without really realizing it. Every query in a search engine, every click on a link, every video clip – you are sharing it with everyone else by way of the products you buy and the advertisements they see.

Another almost invisible aspect of social media is artificial emotion recognition systems. Many of these in 2020 still use video, but an explosion of new systems using all sorts of modalities will be a reality in the next few years – auditory, tactile, and olfactory information along with the ability to scan brains efficiently and scan bodies remotely will mean people in 2030 will be constantly tracked in most public places if the trend continues. Imagine that as you walk down a public street with shops, one has a simple device that listens to your footsteps, identifies your likely age, sex, gender, and emotional state and switches the advertisements on the video screens facing the street to ones that appeal to someone of your age, your sex, your gender, and in your emotional state.

Two doors down, a state of the art video system scans your face to identify age, sex, gender, sexual orientation, class, emotional state, mood, personality features, hunger and thirst levels, and myriad other information about you. It also uses a spectrometer to detect a large number of chemicals – called Volatile Organic Compounds (VOCs) – in the air given off by your body including alcohol, marijuana, arousal, pheromones, and hundreds of other chemicals to pin down your emotional and personality states more precisely. Based on your VOCs, the machine notices that you probably have throat cancer, that you are hung-over from last night's alcohol consumption, and hundreds of other features of your life. It offers you a refreshing fruit and energy drink advertisement.

Meanwhile, above you, government radio frequency sensors bounce electromagnetic waves off your exposed skin to identify all sorts of information about you from emotional state to sexual arousal and this can be used to track your movements through the city without knowing who you are, even after your haircut and when you changed clothes following your workout. All of these interactions with artificial recognition systems are forms of participation with social media, in our wide sense.

To be sure, social media utilizes familiar emotion communication techniques like text, audio, images, and video, but it has brought with it new kinds of emotion communication techniques as well, like emoticons and emojis. Sending emoticons and emojis are a new kind of emotion expression that have come with social media. Even the familiar modalities like text are swept up in novel emotion action tendencies. For example, typing something morose into your Facebook status might become a standard way of expressing sadness, even though you do not say anything to others about it face-to-face.

Two trends deserve emphasis. First is that people spend more time on the internet over time and on social media in particular. By this we mean people spend more and more time on stereotypical social media like Facebook and Twitter. As more people join and develop more online friends, they spend more time keeping track of their lives and posting as well. Social media sites also get better over time at keeping users attention longer and attracting new users. Facebook's newsfeed algorithm and Youtube's recommendation algorithm are famous examples, but these are just the tip of the iceberg.

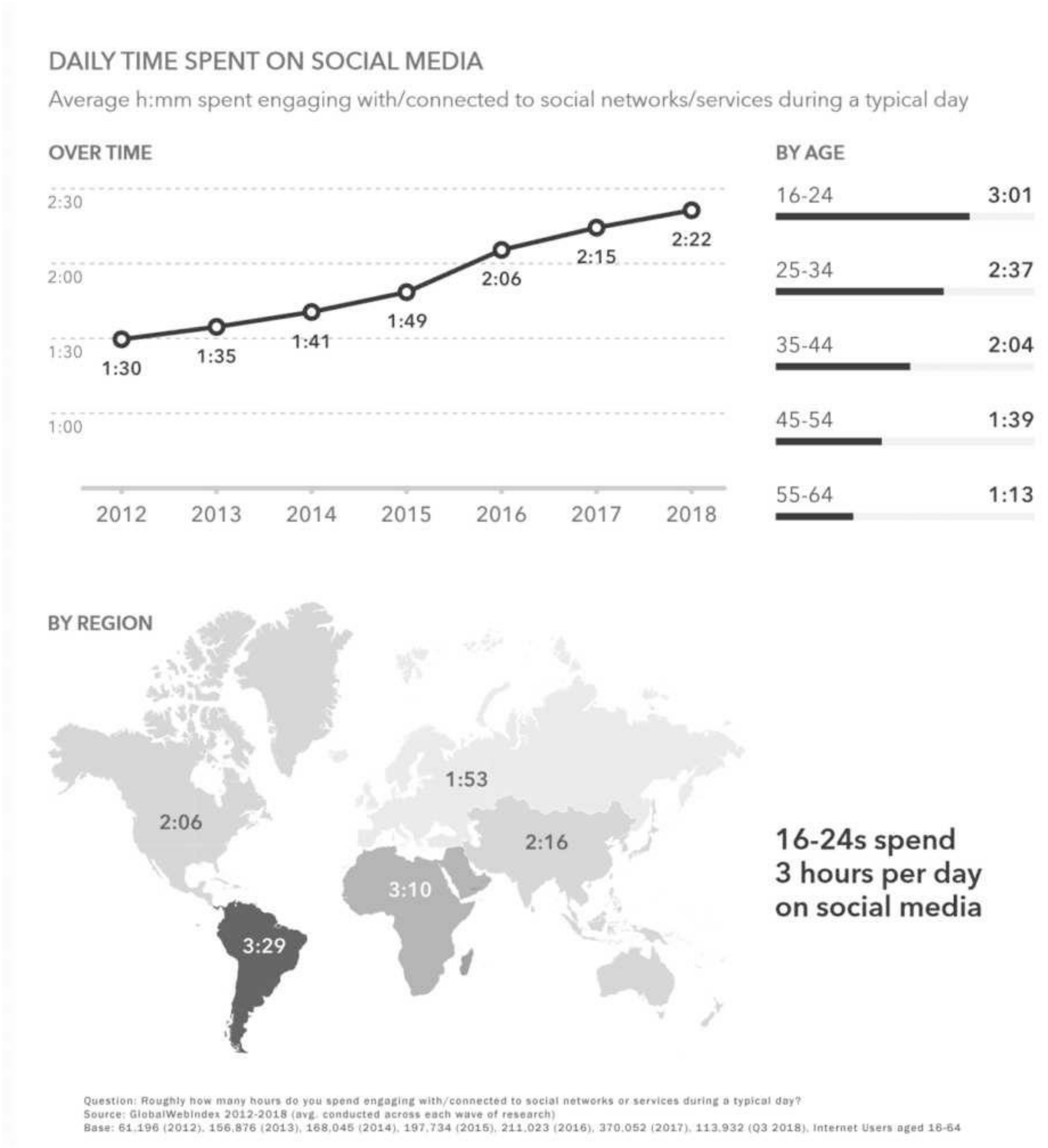

**Fig 42: Daily time spent on social media worldwide**[1]

The second trend is that which technologies count as social media has changed as well. As Google and other companies using advertising cookies to track users everywhere online, it transforms the entire internet into social media. Perhaps the only escape from the social media takeover is the dark web, which uses Tor to enforce anonymity. The rise of surveillance capitalism has transformed search engines, for example, from something that was outside social media to something that is very much a part of social media. It is not direct – other Google search users cannot see your queries – but it is indirect in that your queries together with the products you buy affect the advertisements seen by other users Google tracking cookies. This social media takeover of the internet is one of the key features of surveillance capitalism, described in more detail below.

[1] https://www.digitalinformationworld.com/2019/01/how-much-time-do-people-spend-social-media-infographic.html

## 7.2 The Information Theory of Social Media

There has already developed a loose set of scientific standards for studying social media.[2] The heart of most of these approaches is network science, which we saw in Chapter 5 and 6. Few, if any, are based on information theory. Many emphasize the importance of emotions and emotion communication but none do so in a quantitative way.

We can think about the emotion communication channels, networks, and multi-layer networks explored in previous chapters as a framework for understanding certain functional aspects of social media. For example, posting text, an image, audio, or video on social media is a form of emotion expression. It is just about impossible to post something without that action fitting into some stereotypical emotion expression pattern. Not only explicit posts, but engaging at all with the internet is now a form of emotion expression. Every click and every query express your emotional state to Google's machine learning algorithms that keep each and every one of us under constant surveillance. If your laptop is watching you through the camera and listening to you through its microphone, then you are participating in social media even when you are not on the internet at all (yes, this is common, even Mark Zuckerberg covers his laptop camera and microphone).

One major effect of social media, therefore, is to increase the scope of emotion expression. Many more things count as emotion expression now than they did in previous years for two reasons: there are more things one can do now because of technology (e.g., sending an emoji), and even the mundane things we used to do (e.g., using a search engine) now count as expressing emotions on social media. It is likely that humans have never used skin reflectance properties in emotion communication, but these and other physical changes accompanying emotional state will turn into explicit expressions of emotions as our artificial emotion recognition systems and the machine learning algorithms that power them get better and better.

[2] See Ortega, J. L. (2016), Dey, N., Babo, R. Ashour, A, Bhatnagar, V, Bouhlel (eds) (2018), Dey, N, Borah, S. Babo, R. Ashour, A. (2019), and McMahon, C. (2019).

Another major effect of social media is that we engage emotionally with so many more people. Even people who live in large cities engage in emotional communication face-to-face with less than a few hundred people on average throughout the day. And most people over the past several millennia have lived in much smaller communities where one might engage emotionally with only a hundred people over a lifetime. That is the sort of environment in which our emotional resources were refined by evolution and the same is true of our hominid ancestors. Today, even a solitary person alone in a basement can engage emotionally with thousands of people. By "engage" we mean such a person can detect the emotions of thousands of different people in a single day spent online, even without posting anything. Of course, Facebook's algorithm is keeping track of what you look at and using this information to figure out your emotional state so as to better sell you its advertisers' products.

In sum, social media vastly expands our emotion communication networks in several ways. It expands our emotion expression behaviours in two ways – it gives us new kinds of emotion expression behaviour (e.g., posting) and it changes certain old behaviours into emotion expressions for emotion communication (e.g., searching). It also expands our emotion detection by providing an inexhaustible line-up of personal information about other people who we cannot interact with face-to-face.

We can be more precise by considering the function of coding in emotion communication. Recall that emotion expressions *encode* our unobservable emotional states in our observable behaviour. Likewise, emotion perception *decodes* emotion behaviour observed in others to detect their unobservable emotional states. The changes brought by social media just described are encoding and decoding changes. Having more kinds of emotion expression behaviour requires more complex emotion coding schemes. It also requires more complex emotion decoding schemes. In addition, increasing the ways of expressing emotions contributes in complex ways to the kinds of error-correction techniques at play in emotion communication systems.

One final thing to note about social media is that in the multi-layer network described in the previous chapter, the social media layer connects independent subnetworks of face-to-face emotion communication. Even people who are not online are swept up in the wave of emotion information emanating from social media. As long as one person in your face-to-face group is on social media, they are communicating emotions from social media to you, even if you are not online. The surprising behaviour of multi-layer networks compared to simple networks is still being explored, but we can be sure that adding this layer to our emotion communication systems is having a profound effect on emotion communication worldwide.

## 7.3 Emotional Effects of Social Media

There have been a wide array of studies on the effects of social media, but this area of research is hampered because the research is spread over many individual disciplines in which terms are used in different ways.[3]

From an information theoretic point of view, we want to focus on two aspects of social media with respect to emotions: expressing our emotions (source to message) and detecting emotions in others (message to destination). Emotion communication networks are characterized by broadcast (sending messages to multiple destinations) and multiple-access (receiving messages from multiple sources). That is, sending emotion information to many others and receiving emotion information from many others.

[3] See Kušen, et al (2017) and des Mesnards (2019) for example.

One way we have already emphasized in the previous section to think about social networks is that they vastly increase *broadcast power* and *reception power*. Consider the reception side. There is evidence that reception of social media and the inevitable comparisons that go with it increases the chances of mental illness.[4]

The other side of excessive emotional reception is excessive emotional broadcasting. This topic is much less explored, but one can see instantly that it is significant. On social networks like Facebook, Twitter, and Instagram, when one is posting regularly every day, one is emotionally broadcasting to thousands or even millions of people. From the first-person perspective, it is as if there are hundreds of people in your home with you watching you all the time: when you eat, when you sleep. Even alone on the sofa, your mind's emotional capacities are working as if you were in the middle of a massive crowd with all eyes on you.[5]

There are numerous problems that might be caused by this phenomenon, but we focus on the following train of thought.

1. Regular use of social networks provokes excessive conformity to emotion display rules.[6] The display rules on social networks are idiosyncratic (e.g., many can identify with the stereotypical cringe induced by reading social network posts from one's parents), and ever changing.[7] Navigating this environment requires significant emotional labour investment.

2. Excessive conformity to emotional display rules causes a kind of fatigue associated with emotional labour.[8] One sees this in employees who are forced to adopt certain emotion behaviour as part of the job (e.g., "service with a smile").

3. Excess emotional labour contributes to mental and physical health effects.[9]

---

[4] O'Keefe & Clarke-Pearson (2011).
[5] Horner, C. G., and Akiva, T. (2019).
[6] Horner, C. G., and Akiva, T. (2019). Grandey, A. A., & Sayre, G. M. (2019).
[7] Matsumoto et al (2008), Safdar et al (2009), and Richard, E. & Converse, P. (2016).
[8] Bono & Vey (2005) and Grandey, A. A., & Sayre, G. M. (2019).

Put these three pieces together – (i) excessive conformity to display rules cause emotional fatigue, (ii) emotional fatigue causes mental health problems, and (iii) regular use of social networks provokes excessive conformity to display rules. We hypothesize that regular social network use ought to have similar effects to those we see in people who perform extensive emotional labour.

For emotion communication, the increase in broadcast power and the increase in multiple-access power brought by social media probably have serious negative effects, regardless of the content of emotion messages being communicated. If we add to those effects the result of being saturated by hateful, angry, sad, despondent, and lonely emotion messages, we can expect even more serious effects. These might take generations to be felt fully because social media is so young. Consider the following crude chart of emotion network size and the length of time to get familiar with its effects:

| Basic societal unit | Duration | Emotion connections / Emotion network size |
|---|---|---|
| Extended family | 200,000 yrs | 10s / 10s |
| Village | 10,000 yrs | 100s / 1000s |
| City | 2000 yrs | 100s / 10,000s |
| Social Media | 20 yrs | 1000s / 1,000,000,000s |

**Fig 43: Emotion communication stages in the development of humanity**

These are obviously very rough estimates based on what is available,[10] but even if each one is way off, the trend is still unaffected because of how severe it is—the difference between a city network of 80,000 and a city network of 500,000 is negligible to a network of 3,000,000,000 (.002% vs .01%).

The changes in both emotion connections and the network size with the switch to social media are massive even in comparison to the agricultural revolution and the urban revolution (0-

[9] Jeung et al (2018).
[10] See Christian et al (2013) or Harari (2014) for example.

1 orders of magnitude/2 orders of magnitude for the former vs 1 order of magnitude / 10,000 orders of magnitude for the latter). Those earlier revolutions took at least dozens or hundreds of generations, whereas the social media revolution has occurred within a single lifetime (ours!).

## 7.4 You are in an Abusive Relationship

Imagine this situation: you are in the market for a car. You are also in a romantic relationship, but this car is for you. You are buying the car with your own money. Your partner, who you love and trust, knows all kinds of intimate and potentially embarrassing details about your life. Imagine that your partner goes behind your back and tells the owner of the automobile store a bunch of your secrets in exchange for money. The salesperson at the shop then uses the information your partner provided to manipulate you into spending more on a car than you otherwise would have.

Is that abusive? We think that most people would be so upset upon finding out about the partner's betrayal that they would end the relationship. We sure would. Nevertheless, this is exactly what Google and Facebook are doing to you and everyone else, even if you do not have a page on their social networks. They track what you do online and in your apps, build a profile for you based on this information, and they use this profile to customize which advertisements you see as you surf the web on your browser.

It is hard to find clear definitions of emotional abuse, but here is Sandra K. Burge in a piece published in an academic medical journal:

> Of all forms of abuse, emotional abuse is an especially gray area. How does one identify it? Pence and Paymar[11] describe the key element in abusive relationships as a pattern of one partner's power over the other. To maintain control, the more powerful partner must suppress the other's attempts to act or think independently or to detach from the partner. In abusive relationships, several strategies maintain

[11] Pence and Paymar (1993).

> power and control: verbal insults or humiliation, intimidation, threats, economic control, isolation, male privilege (in male-against-female abuse), minimization, and using children. These strategies are almost always found in physically abusive relationships; however, even when no hitting occurs, these behaviors can control the other partner. For example, the more powerful partner, A, can suppress B's contrary opinions with insults, criticism, or public humiliation. A can counter B's criticisms with threats or intimidation. To prevent B from making independent financial decisions, A can control family finances and discourage employment by B. If B's friends disapprove of A, A can undermine those relationships and socially isolate B. B's attempts at self-improvement through further education or advanced employment can be met with ridicule or accusations of bad parenting, family neglect, or sexual infidelity. Emotional abuse gets less attention from health professionals, perhaps because the damage is less obvious or the aggression is more subtle. The article by Wagner and Mongan7 is the first to demonstrate that emotional abuse is associated with poorer health status and functioning and thus deserves intervention by health care personnel.[12]

It seems like emotional manipulation is at the heart of emotional abuse, all that is needed is to stipulate that the emotional manipulation is harmful or constitutes mistreatment for it to be emotional *abuse*.

Are your partner's actions in the above scenario emotionally abusive? We think the answer is clearly Yes. Not only would you feel bad for spending more money than you wanted to spend (and this could have all sorts of ripple effects, like you now cannot afford the vacation you planned next summer, which would be upsetting, and so on), you would also feel bad for not sticking with your plan, you would feel bad for letting the salesperson talk you into spending too much, you would feel bad for not having better negotiating skills, etc. These results are probably the most damaging – feeling like you are just terrible at negotiating will surely have all sorts of effects, possibly for years to come. This is a form of gaslighting where the manipulation can influence not just the person's emotions (making them feel worthless), but the person's emotional capacities as well (making them unable to feel confidence in negotiations). Almost all of these negative emotional effects are based on your appraisal of the situation as: you failed in a negotiation with an arbitrary stranger who had no advantage in the interaction. Of course, this appraisal is false. The salesperson had a massive amount of information about you and could utilize this information to manipulate you in to feeling good about purchasing an overly expensive car and feeling bad about not purchasing it. Therefore, this case is clearly one with explicit and deliberate emotional manipulation that constitutes mistreatment or results in harm.

[12] Burge (1998. See Wagner & Mongan(1998).

Does the same verdict hold for Google and Facebook? Yes. They are mistreating people by the ways in which they collect personal data and use it for monetary gain. In addition, they use their massive profits to lobby governments to change legal systems so that people have fewer rights over their data. One might think that because users agree to this treatment, it is not emotional abuse. Of course, that does not matter. What matters is that it is emotional manipulation that is mistreatment or harmful. What Google and other surveillance capitalist corporations are doing is clearly mistreatment and harmful and much of it works via emotion manipulation. Just to be clear – it is not the advertisements themselves that constitute emotional abuse. Rather, it is the fact that advertisements are specifically targeted at you using information about you that was collected without your explicit knowledge. When you download WhatsApp or sign up for a Gmail account, you sign away your rights to your information, and virtually no one reads the agreement. Many people when told of what they have agreed to, would not make the same decision if fully informed. So in the imagined scenario, it would be like your partner asking to use your phone and then searching through all your conversations.

## 7.5 Artificial Emotion Recognition

You walk into a department store on Monday after having terrific sex with your boyfriend that morning. You are in a great mood. The cameras at the entrance of the department store analyse your facial expression and your posture to identify your emotional state. Once the store's systems know that you are experiencing joy, they direct the screens you are approaching to switch to an advertisement that has a high probability for people who are experiencing joy. As such, you have a higher probability of noticing the ad, of thinking about that product, and of purchasing that product. The next week, after a fight with your brother on a gaming group chat, you come back to the same store, but this time the emotion recognition system detects sadness and anger in your face, posture, and movements. It casually switches the advertisements as you walk past to ones targeted specifically to people who are sad or angry. Again, you are more likely to notice, consider, and buy.

In reality, the emotion recognition systems can identify hundreds of subtle emotions. A standard library, the Voghte data set, has 706 emotional categories. So these artificial emotion recognition systems are as discriminating as many people. Accuracy rates vary, but Marechal et al (2019) offers the current estimates.

Coming soon there will be more advanced new emotion expression modalities like always-on emotion detection (based on wearable EEG technology) that people can use to monitor their own emotions and share this information with whomever else they choose.[13] Also noteworthy are artificial emotion recognition systems that rely on video or even RF waves[14] that pick up your body's vital signs like heart rate from a distance. We count all these technologies as part of social media.

One might wonder why people choose to buy and wear emotion recognition items, but many are advertised as promoting health by allowing one to monitor one's emotions. This seems odd, since you can already monitor your emotions, but surely some people are better at this than others. Another draw is the futuristic aspect of the technology. The first person to have a portable EEG implanted in her skull will surely get her 15 minutes of fame. And there is the lure of being able to sit next to someone in silence, where each of you have personal emotion recognition systems and some way of receiving information from a network in one's head. The two would be able to send and receive information about each others emotional states continuously, and that seems like it has endless possibilities.

[13] Scherer K. R., Meuleman B. (2013), Konar, A. & Chakraborty, A. (2015), Harley, J. M., (2016), Burleson, W. (2017), Pozzi et al (2017), Cambria et al (2017), Zhao et al (2018), Shu et al (2018), Poria et al (2018), Ko (2018), Li et al (2018).
[14] Ko (2018).

We anticipate a range of radical new technologies for artificial emotion recognition, especially those using remote sensing techniques (remote sensing is the science of measuring from afar, especially with respect to planets. These systems would be far away from the people whose emotions are being measured so that the people in question would not know about it. Moreover, many of these techniques have the almost magical power of seeing through walls. For example, some systems might measure the galvanic skin response and electrodermal activity (electrical properties of the skin like resistance, potential, impedance, and admittance). Others might measure the electromagnetic field of a person (the study of electromagnetic fields around living objects is called bioelectromagnetics). There are a huge number of other kinds of techniques that might be adapted for remote artificial emotion recognition.

As an example, consider EQ-Radio, which uses RF signals that reflect off a human, carrying information back that can be analysed. In the case of EQ-Radio, they distinguish two sources of information, heart rate and breathing rate. The nuances and features of these signals can tell one about the person's emotional state by using the latest machine learning algorithms to select features and classify. See Fig 44 for an illustration from their website.

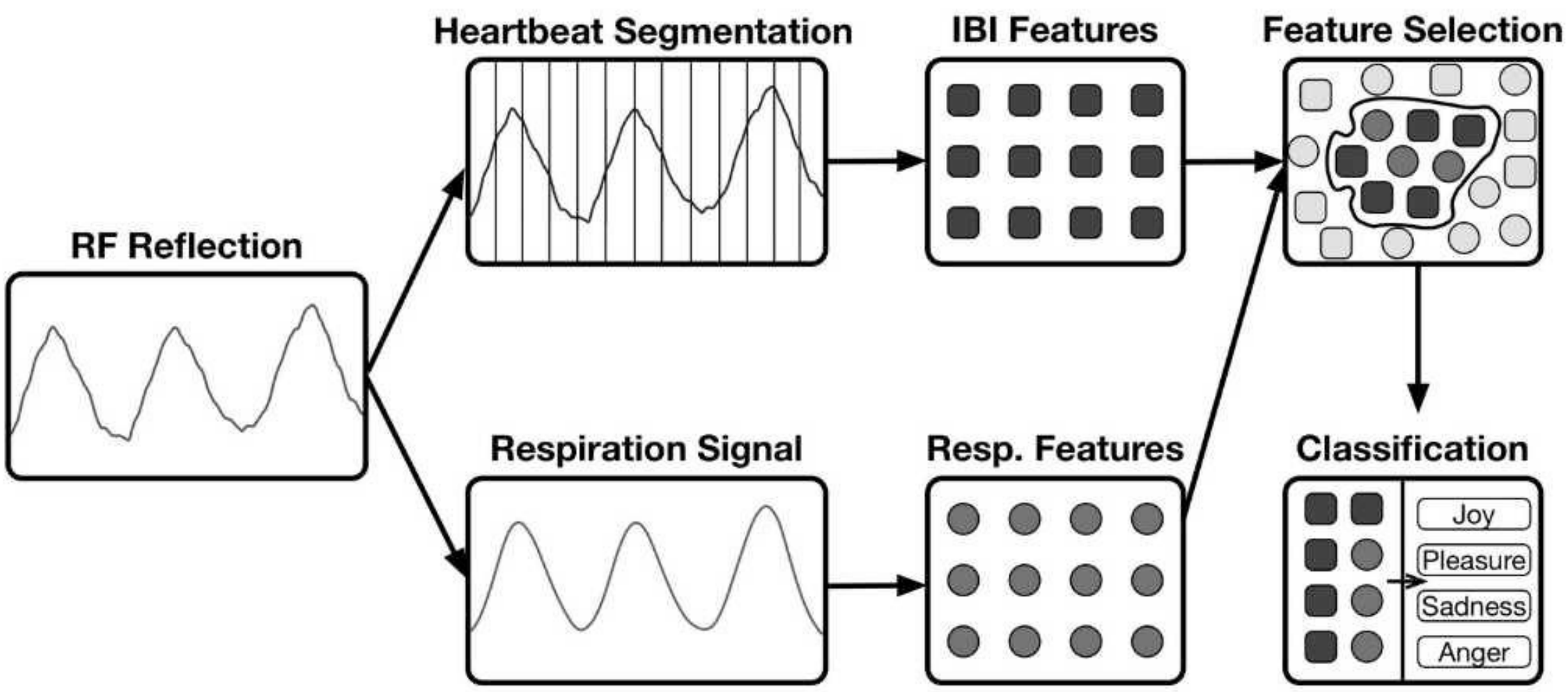

**Fig 44: EQ-Radio system of remote emotion recognition transmits a radio frequency (RF) signal and receives the signal back after it bounces off a person. The reflected signal carries information about the person's body. Algorithms extract heartrate signal and respiration signal from the reflected RF signal. These are broken down and together fed into a machine learning algorithm that identifies the emotions in the target person. The RF signals can be sent, received, and processed by a small machine from far away without the person noticing. It even works through walls; anywhere a wifi signal can travel, this system can work.**[15]

## 7.6 Information Theory of Artificial Emotion Recognition

Many artificial emotion recognition systems operate as *side channels* in emotion communication networks. Side channels are well-studied and there is considerable literature that could be applied directly to artificial emotion recognition, but we will not pause to do so.

The deeper point is that these new technological ways of detecting people's emotions are effectively new emotion expression modalities as well, from the perspective of information theory. That is, if an artificial emotion recognition system can identify your emotions by monitoring your brain waves, then brain waves now count as a kind of emotion expression. The person having the emotion expresses it in many traditional ways – face, voice, touch, posture, movement – and some non-traditional ways like brain waves. As these kinds of emotion recognition systems become more widespread, the phenomena they use for emotion communication (e.g., brain waves) become part of our emotion communication systems as established ways of expressing oneself, with all that comes with it like deception, eavesdropping, cooperation, interference, competition, and so on.

[15] http://eqradio.csail.mit.edu/ See also Raja & Sigg (2017).

Access to big data allows for new emotion decoding techniques that would otherwise be infeasible – e.g., because they are based on precise estimates of probability distributions in a population. We can expect the rise of big data applications that focus on emotion recognition to use these and other indirect techniques. From an information theoretic point of view, this change is an increase in the diversity of decoding techniques, but that is not all because these new techniques use new sources of data, they are opening up new channels and networks of communication. That is, from an information theoretic perspective, these new data are new kinds of emotion expressions that are then being communicated (inadvertently of course) to the artificial emotion recognition systems. In other words, these new data sources are entirely new modes of emotion expression akin to growing an additional face.

## 7.7 Emotion Recognition Jamming

As soon as enough people get upset about having their emotions monitored every time they are in public or at least in certain public places like department stores, *emotion recognition jamming* technologies will have a market.

Jammers are well known in communications theory.[16] In information theoretic terms, a jammer increases the noise in a channel. That is, it increases the probability that errors in message reception occur. In emotion communication networks, jammers would make it more difficult for artificial emotion recognition systems to identify your emotions. It might be that additives to makeup or lotion could interfere with how light is reflected from the face. Perhaps something like an invisible malicious QR code would be possible to wear on one's forehead. This field is just starting so it is unclear right now which directions will be most profitable.

[16] McEliece R.J. (1983) for a classic take and Poisel (2011) for a contemporary textbook.

## 7.8 Advertising, Propaganda, and Coordinated Inauthentic Behaviour

One natural response from defenders of surveillance capitalism is that it is just effective advertising. The same goes for the behaviour of Cambridge Analytica or the Russian consortium – why isn't this just savvy political marketing? The response should be obvious – emotion manipulation on social media is different because it is undetectable by its victims. Advertising can be identified by almost all agents. In order to make this point more compelling, consider the difference between advertising and propaganda.

*Advertising*, even mass advertising that uses psychological theory to optimize the way it manipulates people, predates the internet by decades. Almost any advertising will involve emotion manipulation, and the more advertisers know about human psychology, the more effective their advertising becomes. By and large Western culture has grown accustomed to this form of advertising and sees it as a by-product of a healthy capitalism. There are forms of advertising that "cross the line" and these include subliminal advertising, false advertising, and many others. The problem in each of these cases seems to be that the advertising was disguised to the point of being undetectable. Presumably, the common sense thinking here is that if you can identify advertising then you can factor that part of your experience accordingly – everyone knows not to trust everything the salesperson says to you, and only InCels think that friendly restaurant servers genuinely like them. Mass advertising online has been very similar to mass advertising offline until the rise of surveillance capitalism, which fundamentally changed the way advertising is targeted.

We have admitted that it involves emotion manipulation but the public attitude seems to be accepting of this, provided that the advertising can be identified by those to whom it is targeted. Propaganda, on the other hand, seems to be something that the public is opposed to. One of the most influential treatments of propaganda is in *Manufacturing Consent*, by Herman and Chomsky, which was made into a popular film as well. They focus on the media and argue that mainstream media is a form of propaganda. Their "propaganda" model is instructive for our purposes because of how it labels propaganda, not as much for how it describes the media. They write:

> A propaganda model focuses on this inequality of wealth and power and its multilevel effects on mass-media interests and choices. It traces the routes by which money and power are able to filter out the news fit to print, marginalize dissent, and allow the government and dominant private interests to get their messages across to the public. The essential ingredients of our propaganda model, or set of news "filters," fall under the following headings: (1) the size, concentrated ownership, owner wealth, and profit orientation of the dominant mass-media firms; (2) advertising as the primary income source of the mass media; (3) the reliance of the media on information provided by government, business, and "experts" funded and approved by these primary sources and agents of power; (4) "flak" as a means of disciplining the media; and (5) "anticommunism" as a national religion and control mechanism. These elements interact with and reinforce one another. The raw material of news must pass through successive filters, leaving only the cleansed residue fit to print. They fix the premises of discourse and interpretation, and the definition of what is newsworthy in the first place, and they explain the basis and operations of what amount to propaganda campaigns. (1988: 2)

For them, 'flak' is all the negative consequences aimed at anyone who disagrees with or undermines the goals of those in charge of the media. Herman and Chomsky focus on television, radio, and print news in the time before the internet, so their discussion can feel a bit antiquated today. Nevertheless, we can see that for them, propaganda is not pushing any particular agenda. Rather, it is the structure and function of the media that makes what they do propaganda. The fact that stories are filtered for content based on the interests of those in charge of the media companies makes what they peddle propaganda. Or, rather, it is the fact that these news agencies present themselves as objective but in reality are seriously biased that makes what they produce propaganda.

Herman and Chomsky write a new introduction for the book in 2000, and they are still satisfied with the propaganda model:

> In short, the changes in politics and communication over the past dozen years have tended on balance to enhance the applicability of the propaganda model. The increase in corporate power and global reach, the mergers and further centralization of the media, and the decline of public broadcasting, have made bottom-line considerations more influential both in the United States and abroad. The competition for advertising has become more intense and the boundaries between editorial and advertising departments have weakened further. Newsrooms have been more thoroughly incorporated into transnational corporate empires, with budget cuts and a further diminution of management enthusiasm for investigative journalism that would challenge the structures of power. (2000: xvii)

They see the problem only getting worse, but this was published before the advent of surveillance capitalism, which seems to have exacerbated the problem in two obvious ways. First, it affects not just news, but *everything* we see online including advertisements, internet search results, and ultimately which data we can see and which we cannot. Second, all internet content ends up being filtered not just for a mass public but for *you in particular*. Only what has the best chance of really impacting *you* gets through; this is based on all the information you provide about yourself just by being online at all.

In a recent collection on themes from *Manufacturing Consent*, Alan MacLeod proposes to update their propaganda model. He writes:

> Whilst digitalization has substantially altered the news media on the one hand, the industry's institutional arrangements have remained in place on the other. Access to the

> news media is still regulated by markets and thus requires significant purchasing power and capital investments. Advertising sponsorship still constitutes the main revenue source of the media. Professional journalistic ideology and underlying market pressure continue to incentivize the sourcing of powerful societal actors. Ideology has shifted from Cold War to "free market" and other schemas. Flak is widely used by powerful actors who put pressure on the news media and public intellectuals to abide by the dominant agenda. Hence, the PM's news filters should still be applied as analytical categories to understand news selection and production processes. Additionally, I have suggested to refine and expand the PM to account for context-specifics. Different political and media systems warrant changes and additions to the individual news filters. There is clearly a need to conduct more research in order to incorporate agency-structure dynamics: in some circumstances, agency may allow for more open reporting when journalists bypass news filters. In other cases, journalistic participation in elite networks or collusion with intelligence services may reinforce the PM's filters. And finally, I have suggested to add sexism and racism as news filters to the PM. A class-based model aimed at critically assessing news media performance should certainly account for gender and race. (Macleod 2019: 18).

Hence, MacLeod finds Herman and Chomsky's model holds up pretty well even after thirty years of colossal technological and cultural changes.

A more recent detailed treatment of propaganda comes from Jason Stanley, who offers a different take on what is problematic about propaganda. Just like Herman and Chomsky, Stanley thinks that the essence of propaganda is deception. Stanley sees deception in advertising as well, which differs a bit from what we said above. He offers the following definition:

> *Advertising*: A contribution to public discourse that is presented as an embodiment of

> certain ideals, but in the service of a goal that is irrelevant to those very ideals.[17]

It seems to use that advertising is presented as trying to sell goods and services. Yes, it does invoke ideals like the good life or health which have nothing to do with the product, but as long as it is detectable as advertising, the public can interpret it accordingly.

Stanley's distinction between supporting propaganda and undermining propaganda is important, since propaganda could be in support of some positive goals.

> Supporting Propaganda: A contribution to public discourse that is presented as an embodiment of certain ideals, yet is of a kind that tends to increase the realization of those very ideals by either emotional or other nonrational means.
>
> Undermining Propaganda: A contribution to public discourse that is presented as an embodiment of certain ideals, yet is of a kind that tends to erode those very ideals.[18]

The problem with supporting propaganda is that it uses non-rational means rather than rational ones, but the problem with undermining propaganda is worse, and it is this category on which Stanley focuses. He offers a number of illustrative examples, of which the following is an excerpt:

> According to James Hoggan, in his book Climate Cover- Up, the American Petroleum Institute created a team to assemble a "Global Climate Change Communication Action Plan." According to Hoggan, "The document plainly states that its purpose is to convince the public, through the media, that climate science is awash in uncertainty." Stephen Milloy was a founding member of that team. Hoggan reports that Milloy now appears on Fox News as a "junk science expert." Milloy has "spent his entire career in public relations and lobbying, taking money from companies that include Exxon, Philip Morris, The Edison Electric Institute, the International Food Additives Council, and Monsanto in return for his work declaring environmental concerns to be 'junk science.' " Milloy's assertions are presented as

[17] Stanley (2015: 56).
[18] Stanley (2015: 24).

> embodying the ideals of scientific objectivity. However, anyone not convinced by the ideology of the corporate- funded anti– climate science movement would recognize that they clearly conflict with the ideals of scientific objectivity.[19]

Again, the main problem is the structure of undermining propaganda – it is not really self-defeating because it never really endorses its publicized goals – but there is a conflict between its stated goals and the results of its methods, which go against those goals. It does not matter what the goals are; undermining propaganda is about structure, not content.

We see the same pattern in a distinctively contemporary phenomenon, what Facebook calls *Coordinated Inauthentic Behaviour.* According to Nathaniel Gleicher, Head of Cybersecurity Policy at Facebook, coordinated inauthentic behaviour is, "when groups of pages or people work together to mislead others about who they are or what they are doing."[20] He adds that this phenomenon has occurred for centuries (probably as long as there have been humans, in our view). Facebook's policy on coordinated inauthentic behaviour states that

> But most of the content shared by coordinated manipulation campaigns isn't provably false, and would in fact be acceptable political discourse if shared by authentic audiences. The real issue is that the actors behind these campaigns are using deceptive behaviors to conceal the identity of the organization behind a campaign, make the organization or its activity appear more popular or trustworthy than it is, or evade our enforcement efforts. That's why, when we take down information operations, we are taking action based on the behavior we see on our platform — not based on who the actors behind it are or what they say.[21]

The key idea that the problem in these cases is independent of the *content* is crucial. The problem is not that they are sharing false information or any kind of information at all. The problem is *the way* the information is being shared: inauthentically.

---

[19] Stanley (2015: 33).
[20] https://about.fb.com/news/2018/12/inside-feed-coordinated-inauthentic-behavior/. See also: https://about.fb.com/news/2019/10/removing-more-coordinated-inauthentic-behavior-from-iran-and-russia/
[21] ibid

One recent and frightening example is described by Graphika and the Atlantic Council's Digital Forensics Research Labas, in the following way.

> On December 20, 2019, Facebook took action against a network of over 900 pages, groups, and accounts on its own platform and on Instagram that were associated with "The Beauty of Life" (TheBL), reportedly an offshoot of the Epoch Media Group (EMG). These assets were removed for engaging in large-scale coordinated inauthentic behavior (CIB). The takedown also encompassed a set of Vietnamese assets that were linked to TheBL and to EMG. […] TheBL Facebook page claimed that "Truth in Content is our purpose," but Facebook said that the operators behind this network "made widespread use of fake accounts — many of which had been automatically removed by our systems — to manage Pages and Groups, automate posting at very high frequencies and direct traffic to off-platform sites. Some of these accounts used profile photos generated by artificial intelligence and masqueraded as Americans to join Groups and post the BL content. […] [T]he assets in TheBL portfolio spent over 9 million US dollars on advertising and amassed about 55 million followers.[22]

The Beauty of Life (TheBL) network is exactly the kind of operation used in the Russian attack. Many of these Coordinated Inauthentic Behaviour attacks start by putting out innocuous content for months before changing to content reflecting their real purpose. In TheBL case, many of the account had profile pictures generated by artificial intelligence. These use deep fake machine learning algorithms to generate photo-realistic faces that cannot be traced to anyone in particular or anywhere else on the internet. Expect this trend to intensify.

Attacks like the Cambridge Analytica attack on UK elections or the Russian attack on US elections are clearly coordinated inauthentic behaviour. However, it is not clear that they count as undermining propaganda in Stanley's sense, because the ideals they purported to serve did not conflict with the ideals served by their behaviour. Cambridge Analytica used personality profile pages and then just gained access to direct user data. None of this was undermined by the fact that "Leave" won. Russia used all sorts of targeted advertisements to make specific people in specific districts enraged, and many of these ads involved race, gender, or immigration. It is not obvious that Trump being elected hurt these ideals.

[22] Nimmo et al (2019).

One big question is: do we want social media to block propaganda or coordinated inauthentic behaviour or something else? We think that coordinated inauthentic behaviour mostly covers undermining propaganda as well (of course, the latter is just a proper subset of the former since we think some cases are CIB but not undermining propaganda). The trick is to block coordinated inauthentic behaviour while allowing advertising. But there is a bigger problem, which has already been mentioned, namely surveillance capitalism, to which we turn next.

## 7.9 Surveillance Capitalism

*Surveillance capitalism* is a form of economic behaviour that uses mass surveillance to predict human behaviour in the interest of more effective advertising.[23] Humans are quickly gaining the capability to identify and classify the majority of human actions, spoken words, written words, and audio / video recordings. Soon these capabilities will expand from actions and speech to emotional states, cognitive states, and other personality aspects.[24] Many of these capabilities are powered by machine learning algorithms, which comb through these vast quantities of data and change themselves in light of processing it.[25] That is exactly what a machine learning algorithm is – it learns by changing its own parameters as it encounters more data.

---

[23] Wills (2017) and Zuboff (2019).

[24] See Ko (2018) for a survey on emotion recognition and Kosinski and Wang (2018) on sexual orientation detection.

[25] *Artificial intelligence* is the discipline devoted to creating an artificial intelligent agent, something that can do all the things a human can do, but hopefully much better. *Machine learning* is a branch of artificial intelligence that studies algorithms that display remarkable intelligent behavior without themselves being agents. See Russell & Norvig (2009) on artificial intelligence and Alpaydin (2014) on machine learning.

The leading corporations employing the surveillance capitalist model, like Google and Facebook, collect this information about everyone so that it can be processed and sold to advertisers. Advertisers, in turn, can sell their goods or services more effectively and make greater profits by understanding every conceivable minute detail of their customers' lives. That is the essence of surveillance capitalism. Before surveillance capitalism, online advertising targeted particular websites just like offline advertisers targeted particular magazines, television slots, or roadside billboards. After surveillance capitalism, online advertising targeted particular people or very specific groups of people. According to Zuboff, surveillance capitalists are not just trying to describe our behaviour, they are trying to change it. Their goal is massive manipulation of beliefs, desires, plans, emotions, and anything else that might affect which products we buy. It is the fact that legal systems in Western democracies have allowed surveillance capitalist corporations to diminish individual rights over personal information that permits surveillance capitalism. And it is the existence of surveillance capitalism that permits other kinds of massive emotional manipulation in the interest of political powers.

## 7.10 Cambridge Analytica

*Cambridge Analytica* is a company that has been at the heart of two major political events recently: the 2016 UK Brexit vote to leave the EU and the 2016 US Presidential Election. As a result of scandals surrounding these cases, Cambridge Analytica has come to be the posterchild for every kind of dystopian malfeasance and moral corruption imaginable. It looks now like Cambridge Analytica played a contributing role in Donald Trump's election. It also played a major role in the Brexit vote in the UK, but this has received very little press coverage, especially in the UK.

Cambridge Analytica's overarching strategy is a clear emotion manipulation program using social media to achieve a political outcome. It has officially ceased operations, having declared bankruptcy in 2018. However, it is very closely connected to a dizzying array of companies and individuals that are pursuing the same kind of work all over the world. Even though Cambridge Analytica is no longer functioning, its data and its methods have exploded into a new generation that aggressively manipulate people's emotions on social media for political ends.

The history of Cambridge Analytica goes back at least to 1990, when advertising executive Nigel Oaks founded the Behavioural Dynamics Institute (BDI), which emphasized academic work on the psychology of persuasion, social influence, and digital communication.[26] In 1993, he started Strategic Communications Laboratories (SCL) to apply and profit from advertising techniques based on the psychological and communications research done at BDI. SCL has numerous subsidiaries and affiliated companies, many with 'SCL' in the name (e.g., 'SCL Elections', 'SCL Digital', 'SCL Insight', 'SCL Analytics', 'SCL Defence'). These SCL companies spent a considerable period teaching military-style psychological operations to intelligence officials and military operatives from many different countries.

A number of these SCL companies were brought under a single umbrella called SCL Group in 2015, which was owned by SCL Elections. See Fig 45 for a diagram. At the same time, a new SCL Group business was incorporated – Cambridge Analytica – which was aimed at the American election market and led by Alexander Nix and Stephen Bannon (who went on to co-direct Trump's 2016 presidential campaign), and financed by right-wing billionaire Robert Mercer. By this point, the SCL Group and Cambridge Analytica were clearly using techniques from military psychological operations including propaganda and disinformation campaigns. This is a real case of military technology developed for warfare being sold to the highest bidder for use on the general population. This technology was then used in the 2016 UK Brexit campaign to promote leaving the EU and in the 2016 US presidential campaign to promote Trump.

[26] See Magee (2020) for a detailed history.

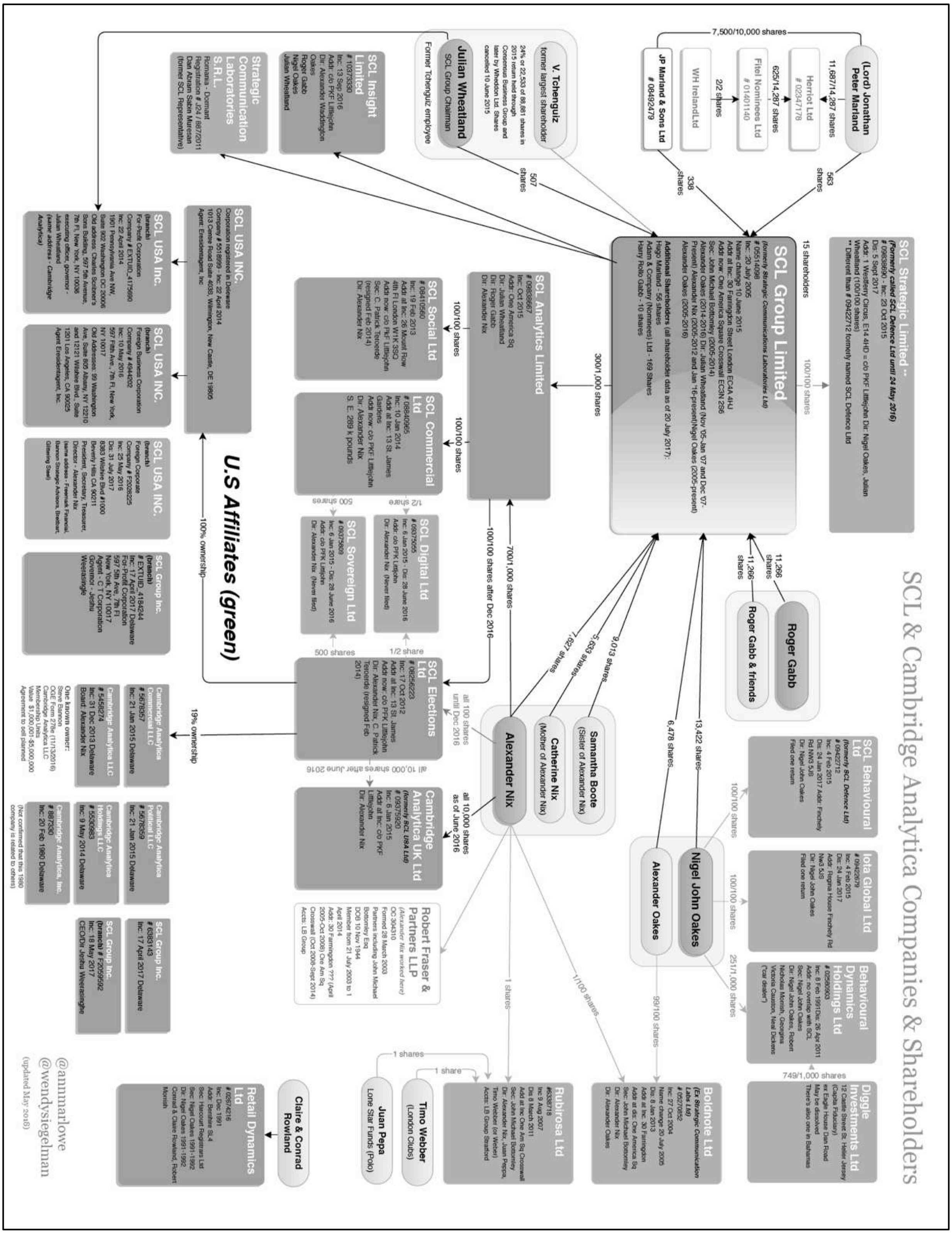

**Fig 45: SCL and Cambridge Analytica Companies and Shareholders**[27]

After the 2016 US presidential campaign, information came to light about Cambridge Analytica's deceptive and illegal practices of stealing user data from Facebook and employing it to manipulate voters. Eventually, in 2018, Cambridge Analytica and its parent company, SCL Group, filed for bankruptcy. However, a new company, Emerdata, that was established in 2017 by Mercer and others connected to Cambridge Analytica and SCL, took control of finances and holdings of Cambridge Analytica, SCL Group, and other related companies, which includes any datasets and proprietary algorithms. See Fig 46 for details on how it is related to other entities mentioned here.

[27] https://medium.com/@wsiegelman/scl-companies-shareholders-e65a4f394158

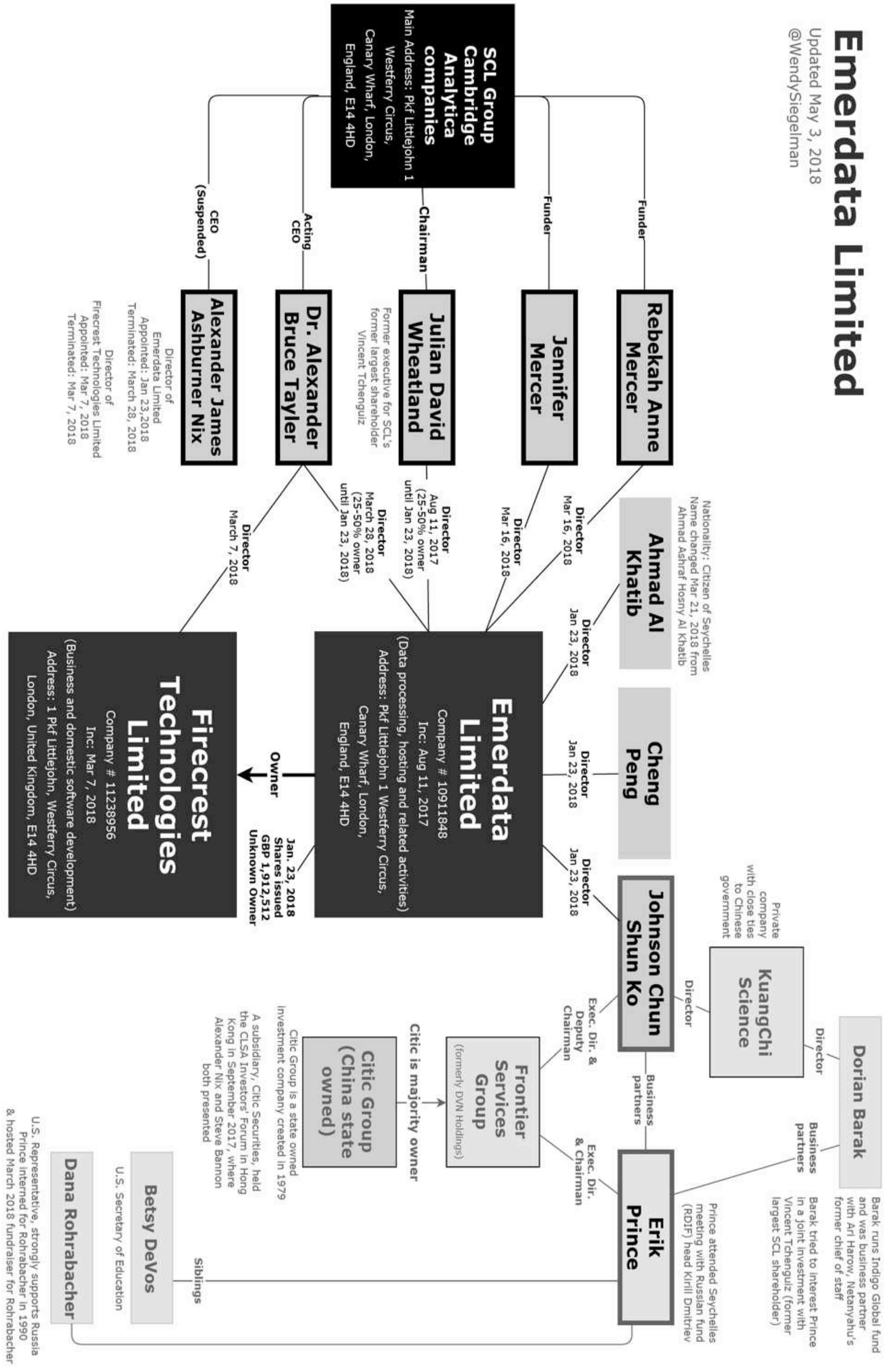

Emerdata Limited
Updated May 3, 2018
@WendySiegelman
SCL Group Cambridge Analytica companies
Main Address: Pkf Littlejohn 1 Westferry Circus, Canary Wharf, London, England, E14 4HD
Funder
Funder
Chairman
Acting CEO
CEO (Suspended)
Rebekah Anne Mercer
Jennifer Mercer
Julian David Wheatland
Former executive for SCL's former largest shareholder Vincent Tchenguiz
Dr. Alexander Bruce Tayler
Alexander James Ashburner Nix
Director of Emerdata Limited
Appointed: Jan 23,2018
Terminated: March 28, 2018
Director of Firecrest Technologies Limited
Appointed: Mar 7, 2018
Terminated: Mar 7, 2018
Director Mar 16, 2018
Director Mar 16, 2018
Director Aug 11, 2017 (25-50% owner until Jan 23, 2018)
Director March 28, 2018 (25-50% owner until Jan 23, 2018)
Director March 7, 2018
Ahmad Al Khatib
Nationality: Citizen of Seychelles
Name changed Mar 21, 2018 from Ahmad Ashraf Hosny Al Khatib
Director Jan 23, 2018
Cheng Peng
Director Jan 23, 2018
Johnson Chun Shun Ko
Director Jan 23, 2018
Emerdata Limited
Company # 10911848
Inc: Aug 11, 2017
(Data processing, hosting and related activities)
Address: Pkf Littlejohn 1 Westferry Circus, Canary Wharf, London, England, E14 4HD
Owner
Jan. 23, 2018 Shares issued GBP 1,912,512 Unknown Owner
Firecrest Technologies Limited
Company # 11238956
Inc: Mar 7, 2018
(Business and domestic software development)
Address: 1 Pkf Littlejohn, Westferry Circus, London, United Kingdom, E14 4HD
KuangChi Science
Private company with close ties to Chinese government
Director
Director
Dorian Barak
Barak runs Indigo Global fund and was business partner with Ari Harow, Netanyahu's former chief of staff
Barak tried to interest Prince in a joint investment with Vincent Tchenguiz (former largest SCL shareholder)
Business partners
Business partners
Exec. Dir. & Deputy Chairman
Exec. Dir. & Chairman
Frontier Services Group
(formerly DVN Holdings)
Citic is majority owner
Citic Group (China state owned)
Citic Group is a state owned investment company created in 1979
A subsidiary, Citic Securities, held the CLSA Investors' Forum in Hong Kong in September 2017, where Alexander Nix and Steve Bannon both presented
Erik Prince
Prince attended Seychelles meeting with Russian fund (RDIF) head Kirill Dmitriev
Siblings
Betsy DeVos
U.S. Secretary of Education
Dana Rohrabacher
U.S. Representative, strongly supports Russia
Prince interned for Rohrabacher in 1990 & hosted March 2018 fundraiser for Rohrabacher

**Fig 46: Emerdata Limited and Cambridge Analytica**[28]

In addition, a new generation of similar companies have started up, often by people who worked for Cambridge Analytica. These include Auspex International, Data Propria, Parscale Digital, Cloud Commerce, IDEIA, AggregateIQ, Tovo Labs, and many others. Consequently, the public should expect that there is an army of corporations selling military grade psychological operations and disinformation techniques to sway elections, and these are being sold to anyone who can pay the price in every corner of the world.

Even though we know this is happening, we certainly do not understand the full extent of it; probably many of the big players in this market are unknown to anyone but insiders. Brittany Kaiser, a former Cambridge Analytica employee turned whistleblower, has provided a trove of documents on the activities of these companies in elections around the world. She commented, "The documents reveal a much clearer idea of what actually happened in the 2016 US presidential election, which has a huge bearing on what will happen in 2020. It's the same people involved who we know are building on these same techniques. … There's evidence of really quite disturbing experiments on American voters, manipulating them with fear-based messaging, targeting the most vulnerable, that seems to be continuing. This is an entire global industry that's out of control but what this does is lay out what was happening with this one company." Some these documents are already available, and more are slated to drop throughout 2020. See https://twitter.com/hindsightfiles for the document dump, which has files on eight countries as of January 2020.

How did they do it? Almost all the work done by Cambridge Analytica and related companies is based on a 2013 paper by Michal Kosinski, David Stillwell, and Thore Graepel called "Private traits and attributes are predictable from digital records of human behaviour."[29] The authors showed that one can use machine learning algorithms to predict a person's personality traits from seemingly innocuous data on social media. Kosinski and coauthors showed that they could use some very basic algorithms (dimensionality reduction and logistic regression) to identify key personality characteristics from Facebook "likes" with considerable accuracy. These theorists (and almost everyone applying their work) use the standard Big Five theory of personality, also known by the acronym OCEAN, which characterizes people's personalities as a combination of values along five dimensions: Openness, Conscientiousness, Exroversion, Agreeableness, and Need for safety. Indeed the Cambridge Analytica copycat, IDEIA, offers a nice graphic to understand the OCEAN model, depicted in Fig n. This graphic is part of their "pitch pack" to get clients to purchase their services.

[28] https://medium.com/@wsiegelman/chart-emerdata-limited-the-new-cambridge-analytica-scl-group-63283f47670d
[29] Kosinski et al (2013).

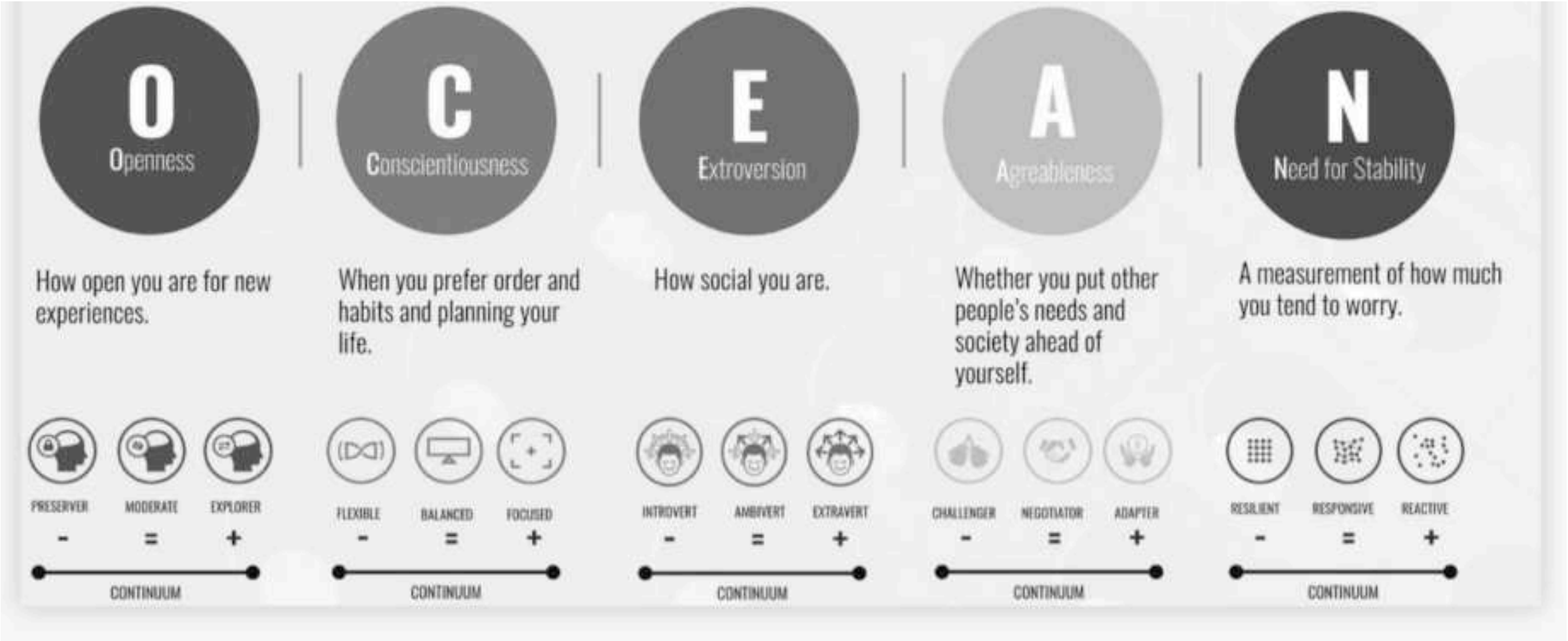

**Fig 47: Diagram of the OCEAN model of personality from the Cambridge Analytica successor, IDEIA' pitch to potential election meddlers.**[30]

Their paper showed that machine learning algorithms could be a powerful tool in identifying crucial but hidden personality traits of social media users. These are depicted in a chart borrowed from their paper, in Figure 48. As you can see, these are shockingly accurate for being based solely on Facebook "likes". These key personality traits could be used to manipulate people in to buying certain products or voting for a certain candidate or supporting a certain group. And that is exactly why the SLC Group approached these researchers to work for them. Kosinski and his collaborators, who worked in the Psychometrics Centre at Cambridge

[30] https://qz.com/1666776/data-firm-ideia-uses-cambridge-analytica-methods-to-target-voters/

University, refused the offer, but another psychologist, also at Cambridge, Aleksandr Kogan was not so ethical. Hence new company was born and named *Cambridge Analytica*, for which Bannon has taken credit.

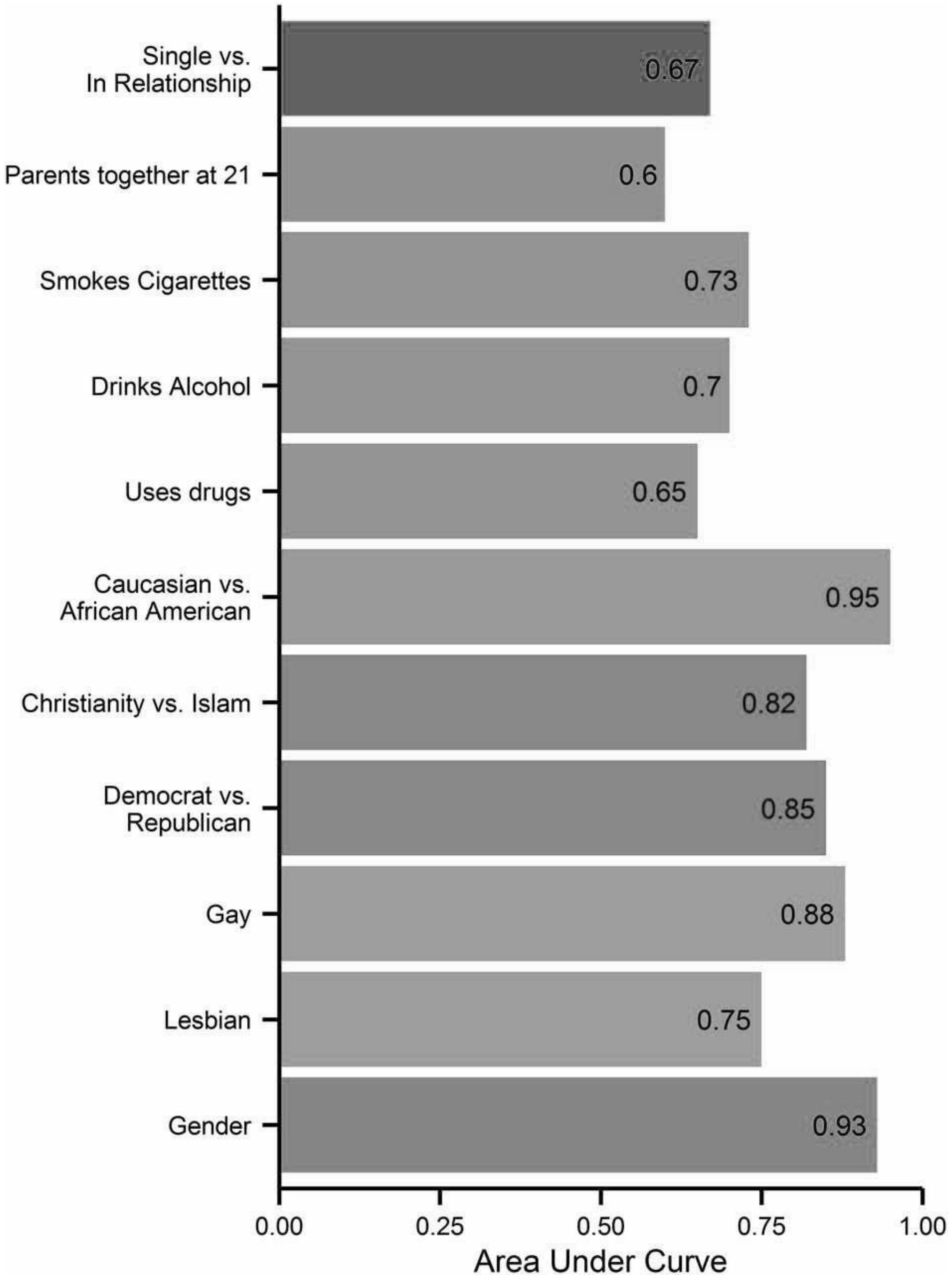

**Fig 48: Accuracy of personality trait predictions from Facebook likes by Kosinski et al (2013).**[31] **1 (far right) means the algorithm was perfect at guessing that trait. The value**

[31] Kosinski et al (2013).

**listed for each trait is how good the algorithm was a guessing that trait (e.g., .67 is 67% right).**

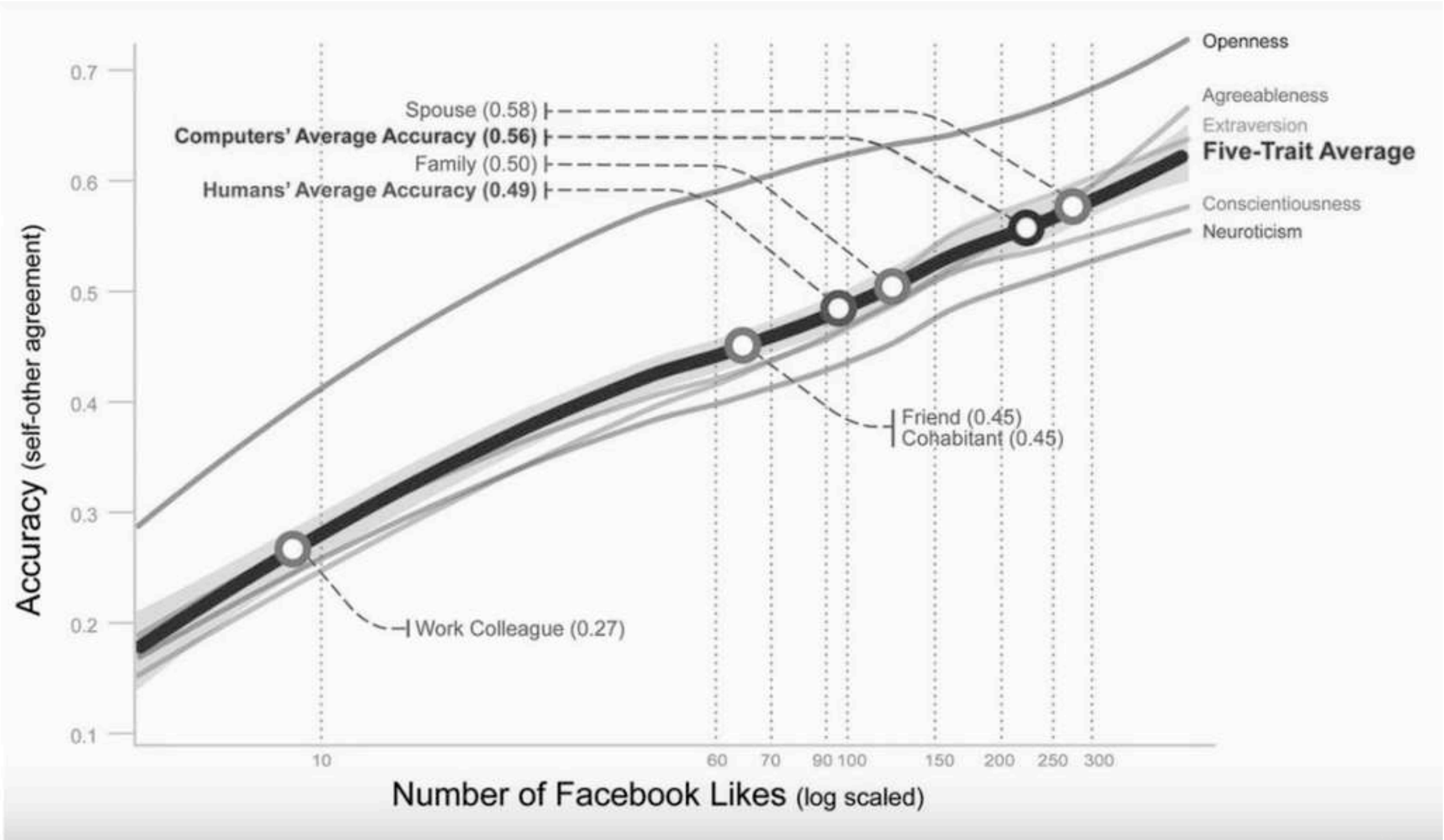

**Fig 49: Accuracy of OCEAN personality component predictions from Facebook likes based on LASSO algorithm in later paper.**[32]

Kogan created an app called "This is Your Digital Life," which offered Facebook users personality profile based on their profile information. The app also scraped the data from all profiles of all friends of the person using the app ('scraping' is collecting data from a website using algorithms—it might be authorized or unauthorized). But following this procedure, Cambridge Analytica was able to collect a massive amount of data on millions of voters. The company claimed to have an average of 5000 data points on every single voter in the United States. Because of the network properties (e.g., how connected it is), they only needed 200,000 people to use the app in order to get data on a reported 87,000,000 people, but the company claimed to have enough data to construct detailed profiles on all 300,000,000 potential voters in the United States. That is a staggering example of how susceptible networks are to being observed or controlled by way of relatively few important nodes (only .00067% of the nodes in this case). Facebook has supposedly fixed the problem that allowed the app to do this, but it went against Facebook's terms of use anyway. There are allegations made that employees of Facebook worked directly with Cambridge Analytica in some capacity, but these have not been substantiated.[33]

[32] Youyou et al (2015).

They then used the personality profiles to identify which people to target (*persuadables* in the parlance of our times), how to target them, and what to persuade them to do. Different groups of persuadables get pitched different ads, depending on the personality traits involved. These categories might be very precise with hundreds or thousands of distinct ads. And different personality types might be more effectively motivated to do different things. For example one trait might effectively get a certain percentage of people to plan to vote for a certain person (if triggered by the right ad), while another trait might prevent a person from voting at all if triggered effectively. It is important to note that these targeted advertisements were not just on Facebook. The targeting campaign builds a profile of the person, their personality, and their presence across all media, so that *they can be targeted across the internet*, day and night, no matter whether they are on their phone waiting for a bus, watching their flatscreen at home or on their laptop at work. The profile is trans-media and the advertising is too. Moreover, certain kinds of advertising might be specifically designed for a certain person on a certain platform (e.g., comments on Reddit posts vs snapchat ads). An important lesson is that one unhealthy social network like Facebook can infect all the others and in fact the internet itself.

Any suggestion that we try to stop our information from leaking like this is clearly not going to work. The box is open. There is already enough data available to these companies to have personality profiles on 99% of the humans currently alive on Earth. That cannot be undone. Armies of people are currently competing to use all sorts of private information about you as a person to manipulate you into doing a wide array of behaviours. We have to find ways to prevent them from exploiting your personality profile that already exists in the databases of corporations and governments all over the world right now. And we will see several in the next chapter.

[33] Amer & Noujaim (2019).

In the Brexit Vote, which preceded the US election by about five months, Cambridge Analytica was a major influence. Profiles were created of UK voters using the same methods. These people were then shown all sorts of political ads favouring the "Leave" side; some of these were factual, many were not. Most of these advertisements were aimed not at providing evidence that "Leave" is the better choice, but rather appealing to the emotions of the people targeted. One favourite tactic was to inflame anger toward immigrants.[34] Whereas the Trump campaign reportedly paid almost five million dollars for their work on the 2016 campaign, Cambridge Analytica was reportedly not paid for their work on the Leave campaign. Instead, Robert Osborn has reported that:

> Robert Mercer, co-owner of right-wing news organisation Breitbart, allegedly directed his data-analytics firm Cambridge Analytica to provide expert advice to the Leave campaign. Mr Mercer, whose firm was paid £4.8m by the Trump campaign to persuade swing voters, offered his firm's help to Ukip leader Nigel Farage for free, Leave.eu communications director Andy Wigmore told The Observer.[35]

This point is corroborated by whistleblower Brittany Kaiser's account as well.[36] Unfortunately, the fact that the Brexit vote was manipulated in this way and that millions of pounds worth of military grade psychological operations tactics were donated for free to the Leave campaign is still not commonly known in the UK and not discussed in UK media.

---

34 See B

35 Osborn (2017).

36 "Chargeable work was completed for Ukip and Leave.EU, and I have strong reasons to believe that those data sets and analysed data processed by Cambridge Analytica … were later used by the Leave.EU campaign without Cambridge Analytica's further assistance," Kaiser wrote in a letter to Damian Collins. (Hern 2019)

There have been a number of critics of the idea that the tactics used by firms like Cambridge Analytica are ineffective and that the hype about them is way overblown. All of these criticisms we have seen offer armchair explanations like "it's too hard to change people's minds" or "the meddlers didn't really know what they were doing".[37] We have two replies. First, regardless of whether Cambridge Analytica and related firms are accomplishing their goals, their tactics are surely manipulating people's emotions. Again and again, the experts and Cambridge Analytica's own employees describe what Cambridge Analytica actually did as tantamount to emotional abuse on a vast scale. Propaganda expert, Emma Briant, in written testimony to the House of Commons about election meddling on social media states:

> My findings reveal that Leave.EU deployed its cynical and calculating strategies using borrowed methods of Cambridge Analytica (CA), to win *at all costs* despite violence unfolding before their eyes. Leave.EU sought to create an impression of 'democracy' and a campaign channeling public will, while creating deliberately 'provocative' communications to subvert it and win by channelling hateful propaganda.[38]

She goes on:

> Trump channelled resentment and fear on immigrant scapegoats as 'Drug dealers, criminals, rapists' and leveraged a Muslim 'artificial enemy' (Interview: Oakes/Briant, 24th November 2017 - this interview excerpt has been published in parliamentary evidence) in a manner that SCL CEO Nigel Oakes compared coldly to Hitler's propaganda against Jews. Interestingly, Leave.EU's Communications Director Andy Wigmore also mentioned the Nazis, and how Goebbels' propaganda strategy has value in a 'pure marketing sense'.[39]

Jonathan Albright, Director of the Digital Forensics Initiative at the Tow Center for Digital Journalism, Columbia University, writes: "This is a propaganda machine. It's targeting people individually to recruit them to an idea. It's a level of social engineering that I've never seen before. They're capturing people and then keeping them on an emotional leash and never letting them go."

[37] Resnick (2018). See Kalla and Broockman (2017) for a survey cited by Resnick.
[38] Briant essay 2. See https://www.parliament.uk/business/committees/committees-a-z/commons-select/digital-culture-media-and-sport-committee/news/fake-news-briant-evidence-17-19/
[39] ibid

The second point is that we have good evidence that personality-based microtargeting *does* work. Political strategist David Goldstein did an experiment that was reported on the podcast Planet Money (#915: How to Meddle in an Election). He did exactly what Cambridge Analytica does with personality-based microtargeting, but he used six US Senate voting districts in Alabama during the 2017 election. The candidates were republican Roy Moore and Democrat Doug Jones. Goldstein, himself a longtime democratic strategist, worked to elect Jones. For 10 days before the election, he targeted persuadable Republications with automatically collected stories from around the internet about how Republicans were likely to write in other candidates, in an effort to make them stay home; he also targeted persuadable Democrats with automatically collected uplifting stories about how signification voting is. Three districts were control; three got the microtargeting advertisements. Jones won, and when the voting data were analysed by the data scientists on Goldstein's team, the results were that their efforts produced a positive difference of 4% in Democratic turnout in his test districts compared to the control districts. Moreover, he reported negative differences of 2.5% for moderate Republications and 4.4% for conservative Republicans. These are huge numbers that would make the difference in most elections if done statewide. And this by a small team using less than $100,000 for only 10 days. Alex Goldmark, the host of the podcast ran the numbers himself and found the positive difference for Democrats to be more like 2% when accounting for additional factors. But that is still enormous. He quips: "one of the campaigners I talked to also told me that they've heard of other tests and other studies from other campaign groups that are not ever going to be made public, and in a lot of ways, this isn't so far out of line with them." And there are other quantitative reasons to think that personality-based microtargeting works, but for those we need to look at the next case.

## 7.11 Russian Active Measures and the Current Cyberwar

*Russia's role in the United States 2016 Presidential Election* is another clear case of using emotion manipulation techniques for political ends. The United States Senate Select Committee on Intelligence issued a Report on Russian Active Measures Campaigns and Interference in the 2016 U.S. Election. The public version (i.e., redacted) of Volume 2 of that report, called "Russia's Use of Social Media," begins with a paragraph that looks like this:

> In 2016, Russian operatives associated with the St. Petersburg-based Internet Research Agency (IRA) used social media to conduct an information warfare campaign designed to spread disinformation and societal division in the United States.[40] [redacted] Masquerading as Americans, these operatives used targeted advertisements, intentionally falsified news articles, self-generated content, and social media platform tools to interact with and attempt to deceive tens of millions of social media users in the United States. This campaign sought to polarize Americans on the basis of societal, ideological, and racial differences, provoked real world events, and was part of a foreign government's covert support of Russia's favored candidate in the U.S. presidential election. [redacted]

[40] footnote 1 in the text reads: For purposes of this Volume, "information warfare" refers to Russia's strategy for the use and management of information to pursue a competitive advantage. See Congressional Research Service, Defense Primer: Information Operations, December 18, 2018.]

The report goes on to detail how the IRA manipulated Americans on social media to benefit Donald. Although we know that the Russians' favoured candidate was Trump, and their efforts were to support his campaign, the report does not say so here, although it does elsewhere.

We have many public records of United States intelligence agencies stating that Russia was trying to help Trump. The Intelligence Community Assessment (ICA) from 2017 on Russian influence in the 2016 election is especially helpful. It states: "We also assess Putin and the Russian Government aspired to help President-elect Trump's election chances when possible by discrediting Secretary Clinton and publicly contrasting her unfavorably to him. All three agencies agree with this judgment. CIA and FBI have high confidence in this judgment; NSA has moderate confidence," (ii). The US Intelligence agencies agree that the Russian attack was, in part, to get Trump elected.

A report from the Rand Corporation summarizes what happened in the graphic depicted in Fig 50, which is based on the same information source (US Senate Report, Intelligence Community Assessment, Mueller Report).

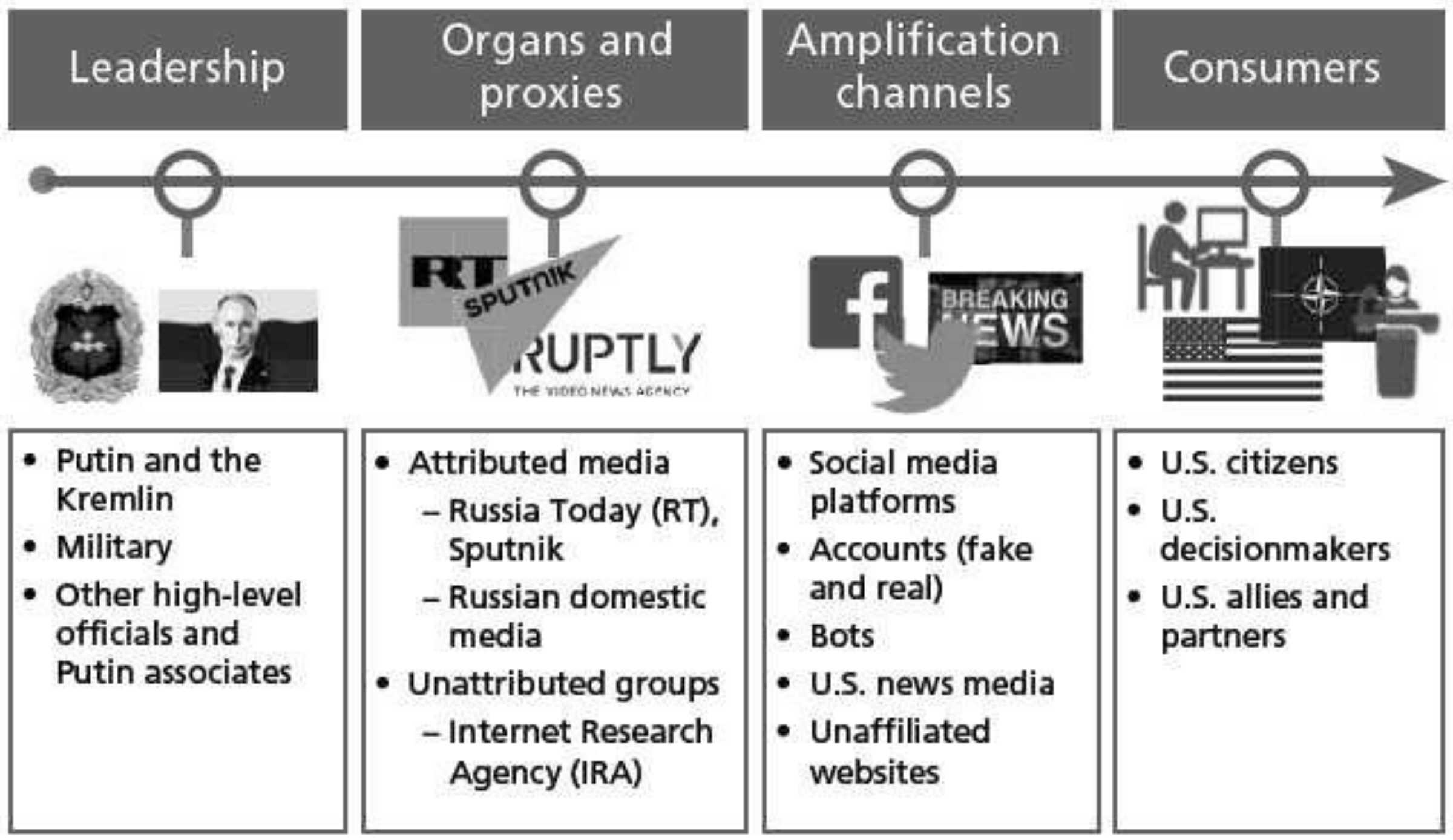

**Fig 50: Layers of the Russian Attack on the 2016 US Presidential Election**

We focus on what the Internet Research Agency (IRA) did as part of the Russian attack, which included many other parts like hacking various Democratic party email accounts and distributing the emails to Wikileaks. The Russians also infiltrated dozens of states voter registries and attempted to disrupt the voting infrastructure. In addition, the Russian effort is ongoing, with massive amounts of propaganda aimed at obfuscating Russian involvement. As such we are quoting from the bipartisan United Senates Senate Report more than would be customary.

First some background on the IRA:

> The IRA is an entity headquartered in St. Petersburg, Russia, which since at least 2013 has undertaken a variety of Russian active measures campaigns at the behest of the Kremlin. The IRA has conducted virtual and physical influence operations in Russia, the United States, and dozens of other countries. The IRA conducted a multi-million dollar, coordinated effort to influence the 2016 U.S. election as part of a broader information campaign to harm the United States and fracture its society.[41]
>
> The IRA is funded and directed by Yevgeniy Prigozhin**,** a Russia oligarch who works to conduct intelligence operations, military activities; and influence operations globally on behalf of the Kremlin.[42]

The Senate Report details plenty of interviews about why the Russians attacked and how they did it.

> As journalist Adrian Chen of The New Yorker reported, the objectives for Russia's troll army are primarily "to overwhelm social media with a flood of fake content, seeding doubt and paranoia, and destroying the possibility of using the Internet as a democratic space." Leonid Volkov, a Russian politician and supporter of opposition leader Alexei Navalny, told Chen, "The point [of Russian disinformation] is to create

[41] Senate Report (2018: 22-23).
[42] 23

> the atmosphere of hate, to make it so stinky that normal people won't want to touch it." He stressed, "Russia's information war might be thought of as the biggest trolling operation in history, and its target is nothing less than the utility of the Internet as a democratic space."[43]

This idea of ruining the democratic capacity of the internet by destroying trust in the legitimacy of any information source and making it difficult to get to the truth by propagating huge numbers of conspiracy theories is one of the most frightening and damaging aspects of the Russian attack.

The employees of the IRA went through rigorous training so that they could influence Americans on social media more effectively:

> According to a former employee interviewed by the news outlet Dozhd, IRA personnel were required to study and monitor tens of thousands of comments in order to better understand the language and trends of internet users in the United States. The ex-troll indicated that they were taught to avoid crude and offensive language that would be off-putting to the typical online reader. 118 According to the former employee, the IRA office dedicated to inflaming sentiments in the United States was prohibited from promoting anything about Russia or President Putin-primarily because, in the IR.A's assessment, Americans do not normally talk about Russia. "Our goal wasn't to tum the Americans toward, Russia. … Our task was to set Americans against their own government: to provoke unrest and discontent, and to lower Obama's support ratings.[44]

And they planned their targets carefully:

> In testifying to the Committee in 2017, Clint Watts outlined three different types of potential real-world targets for Russian influence operators. A class of "useful idiots" refers to unwitting Americans who are exploited to further amplify Russian

[43] 19
[44] 29

> propaganda, unbeknownst to them; "fellow travelers" are individuals ideologically sympathetic to Russia's anti-western viewpoints who take action on their own accord; and "agent provocateurs" are individuals who are actively manipulated to commit illegal or clandestine acts on behalf of the Russian government.[45]

Many websites and Facebook pages were set up to help the Russians with their plans. Most of these looked like a normal website with the usual content. Then after drawing in an audience, the content would change. It would start delivering "payload messages" which were the point of building the audience.

> The tactic of using select payload messages among a large volume of innocuous content to attract and cultivate an online following is reflected in the posts made to the IRA's "Army of Jesus" Facebook page. The page, which had attracted over 216,000 followers by the time it was taken down by Facebook for violating the platform's terms of service, purported to be devoted to Christian themes and Bible passages. The page's content was largely consistent with this fa;ade. The following series of posts from the "Army of Jesus" page illustrates the use of this tactic, with the majority of posts largely consistent with the page's theme, excepting the November 1, 2016 post that represents the IRA's payload content:

- October 26, 2016: "There has never been a day when people did not need to walk with Jesus."
- October 29, 2016: "I've got Jesus in my soul. It's the only way I know .... Watching every move I make, guiding every step I take!"
- October 31, 2016: "Rise and shine-realize His blessing!"
- October 31, 2016: "Jesus will always be by your side. Just reach out to Him and you'll see!"
- November 1, 2016: "HILLARY APPROVES REMOVAL OF GOD FROM THE

[45] 20

PLEDGE OF ALLEGIANCE."

- November 2, 2016: "Never hold on anything [sic] tighter than you holding unto God!"

> This pattern of character development, followed by confidence building and audience cultivation, punctuated by deployment of payload content is discemable throughout the IRA' s content history.

The content critical of Hilary Clinton, Trump's opponent, was not random.

> Hillary Clinton, however, was the only candidate for President whose IRA-posted content references were uniformly negative. Clinton's candidacy was targeted by both the IRA's left and right personas, and both ideological representations were focused on denigrating her.
>
> In contrast to the consistent denigration of Hillary Clinton, Donald Trump's candidacy received mostly positive attention from the IRA' s influence operatives, though it is important to note that this assessment specifically applies to pre-election content. The Committee's analysis indicates that post-election IRA activity shifted to emphasize and provoke anti-Trump sentiment on the left.

This is one of the most telling signs of a professional job done by people with experience. They place getting Trump elected above sowing general discord, but once he is in, they use him as a lightening rod to further their agenda of creating division and chaos.

> The IRA targeted not only Hillary Clinton, but also Republican candidates during the presidential primaries. For example, Senators Ted Cruz and Marco Rubio were targeted and denigrated, as was Jeb Bush. Even after the 2016 election, Mitt Romney-historically critical of Russia and who memorably characterized the country as the United States' "number one geopolitical foe" during a 2012 presidential debate-was targeted by IRA influence operatives while being considered for

> Secretary of State in the Trump administration.[46]

Again, we see evidence of a well-funded and professional organization that switches between propaganda aimed at manipulating supporters of Trump's enemies and propaganda aimed at manipulating supporters of Trump.

There were so many other kinds of social media activities besides attacking certain candidates, and these took place across a range of platforms. One example

> Young Mie Kim, a digital adyertisement research expert from the University of Wisconsin, has closely analyzed the IRA's Facebook ad:vertisements. On the basis of Kim's analysis, three types of voter suppression campaigns on Facebook and Instagram emerge, including: "a) turnout sμppression/election boycott; b) third-candidate promotion; and c) candidate attack, all targeting nonwhites or likely Clinton voters." Kim found no evidence of a comparable voter suppression effort that targeted U.S. voters on the ideological right.

In contrast to the Russian influence on Facebook and Instagram, their Twitter campaign had a military-style organization and distinct clusters of accounts with their own personality traits.

> Clemson researchers, led by Darren Linvill and Patrick Warren, collected all of the tweets from all the IRA-linked accounts between June 19, 2015, and December 31, 2017. … After conducting an analysis of all the content that IRA influence operatives manufactured, the Clemson researchers separated the IRA-affiliated accounts into five categories of social media platform activity.' According to this analysis, "Within each type, accounts were used consistently, but the behavior across types was radically different." Characterizing the IRA Twitter effort as "industrial," the researchers described the campaign as "mass produced from a system of interchangeable parts; where each class of part fulfilled a specialized function." The researchers named the account types: Right Troll, Left Troll, Newsfeed, Hashtag Gamer, and Fearmonger.

[46] 37

- Right Troll. This was the largest and most active group of IRA-affiliated accounts. The 617- Right Troll Twitter accounts tweeted 663, 740 times and cultivated nearly a million total followers. Clemson researchers characterized these accounts as focused on spreading "nativist and right-leaning populist messages." They strongly supported the candidacy of Donald Trump, employed the #MAGA hashtag, and attacked Democrats. Although nominally "conservative," Clemson researchers found that the IRA accounts rarely promoted characteristically conservative positions on issues such as taxes, regulation, and ·abortion, and instead focused on messaging derisive of Republicans deemed "too moderate" (including at the time Senators John McCain and Lindsey Graham). The accounts generally featured very little in the way of identifying information, but frequently used profile pictures of "attractive, young women."
- Left Troll. The second largest classification of IRA-affiliated Twitter accounts, 'consisting of around 230 Twitter profiles that generated 405,549 tweets, was Left Troll. The focus of the Left Troll Twitter accounts was primarily issues relating to cultural identity, including gender, sexual, and religious identity. Left Troll accounts, however, were acutely focused on racial identity and targeting African-Americans with messaging and narratives that mimicked the substance of prominent U.S. activist movements like ~ Black Lives Matter. Left Troll accounts directed derisive content toward moderate Democrat politicians. These accounts targeted Hillary Clinton with content designed to undermine her presidential campaign and erode her support on the U.S. political left.
- News Feed. Designed to appear to be local news aggregators in the United States, News Feed Twitter accounts would post links to legitimate news sources and tweet about issues oflocal interest. Examples of the IRA's news-oriented influence operative accounts on Twitter include @OnlineMemphis and @TodayPittsburgh. About 54 IRA accounts share the characteristics of this classification of Twitter profile, and they were responsible for 567,846 tweets.

- Hashtag Gamer. More than 100 of the IRA's Twitter accounts were focused almost exclusively on playing "hashtag games," a word game popular among Twitter users. At times, these games were overtly political and engineered to incite reactions on divisive social issues from both the left and the right ends of the ideological spectrum.
- Fearmonger. Finally, the IRA's 122 Fearmonger Twitter accounts were specifically dedicated to furthering the spread of a hoax concerning poisoned turkeys during the Thanksgiving holiday of 2014. The Fearmonger Twitter accounts tweeted over 10,000 times.[47]

Again and again, the Russians are trying to manipulate American's emotions in this attack. If you cannot physically harm a population, try to emotionally abuse the entire nation. That is exactly what they have done.

One exception to the Russian's diversity of techniques is that they used very little advertising. The Senate report states:

> Paid advertisements were not key to the IRA's activity, and moreover, are not alone an 'accurate measure of the IRA's operational scope, scale, or objectives, despite this aspect of social-media being a focus of early press reporting and public awareness. According to Facebook, the IRA spent a total of about $100,000 over two years on advertisements-;. a minor amount, given the operational costs of the IRA are estimated to have been around $1.25 million dollars a month.[48]

Instead of advertising, the Russians organized real events in the United States. Some of these were to promote Trump, but many were to simply cause mayhem.

IRA operatives were able to organize and execute a series of coordinated political rallies titled, "Florida Goes Trump," using the Facebook group "Being Patriotic," the Twitter account @March_for_Trurnp, and other fabricated social media personas.

[47] 52-53.
[48] 7

A May 2016, real world event that took place in Texas .illustrates the IRA' s ideological flexibility, command of American politics, and willingness to exploit the country's most divisive fault lines. As publicly detailed by the Committee during a November 1, 2017 hearing, IRA influence operatives used the Facebook page, "Heart of Texas" to promote a protest in opposition to Islam, to occur in front of the Islamic Da'wah Center in Houston, Texas. "Heart of Texas," which eventually attracted over 250,000 followers, used targeted advertisements to implore its supporters to attend a "Stop Islamization of Texas" event, slated for noon, May 21, 2016. Simultaneously, IRA operatives used the IRA's "United Muslims for America" Facebook page and its connection to over 325,000 followers to promote a second event, to be held at the same time, at exactly the same Islamic Da'wah Center in Houston. Again, using purchased advertisements, the IRA influence operatives behind the "United Muslims for America" page beseeched its supporters to demonstrate in front of the Islamic Da'wah Center-this time, in order to "Save Islamic Knowledge." In neither instance was the existence of a counter-protest mentioned in the content of the purchased advertisement. The competing events were covered live by local news agencies, and according to the Texas Tribune, interactions between the two protests escalated into confrontation and verbal attacks. The total cost for the IRA's campaign to advertise and promote the concomitant events was $200~and the entire operation was conducted from the confines of the IRA's headquarters in Saint Petersburg.[49]

[49] 47

The muslim/anti-muslim event cost only $200 out of a $1.25 million per month budget. Imagine this kind of discord multiplied over and again throughout the country in various forms, but all having a single theme – emotion manipulation about divisive topics to sow discord.

One of the most divisive topics in American politics is race, so it is not a surprise to find that the Russians exploited this issue in their attack.

> No single group of Americans was targeted by IRA information operatives more than African-Americans. By far, race and related issues were the preferred target of the information warfare campaign designed to divide the country in 2016. Evidence of the IRA's overwhelming operational emphasis on race is evident in the IRA's Facebook advertisement content (over 66 percent contained a term related to race ) and targeting (locational targeting was principally aimed at "African-Americans in key metropolitan areas with well-established black communities and flashpoints in the Black Lives Matter movement"), as well as its Facebook pages (one of the IRA's top-performing pages, "Blacktivist," generated 11.2 million engagements with Facebook users), its Instagram content (five of the top 10 Instagram accounts were focused on African-American issues and audiences), its Twitter content (heavily focused on hot-button issues with racial undertones such as the NFL kneeling protests), and its YouTube activity (96 percent of the IRA's YouTube content was targeted at racial issues and police brutality).[50]

> The overwhelming preponderance of the video content posted to the IRA's YouTube channels was aimed directly at the African-American population. Most of the videos pertained to police brutality and the activist efforts of the Black Lives Matter organization. Posted to 10 of the IRA's YouTube channels, were 1,063 videos-or roughly 96 percent of the IRA content-dedicated to issues of race and

[50] 38-39

police brutality.[51]

In addition to enflaming racial sentiment, the Russians also attacked the American democracy itself. One worrisome example is the preparations undertaken in the event of a Trump loss. The Intelligence Community Assessment says:

> When it appeared to Moscow that Secretary Clinton was likely to win the presidency the Russian influence campaign focused more on undercutting Secretary Clinton's legitimacy and crippling her presidency from its start, including by impugning the fairness of the election. Before the election, Russian diplomats had publicly denounced the US electoral process and were prepared to publicly call into question the validity of the results. Pro-Kremlin bloggers had prepared a Twitter campaign, #DemocracyRIP, on election night in anticipation of Secretary Clinton's victory, judging from their social media activity.[52]

The Russian attack was never meant to be just about this election, but rather an ongoing process or part of a long term information war.

Not all social media sites reported high levels of manipulation, and some have counter-measures in play to guard against it. For example, the Senate Report states:

> Facebook CEO Sheryl Sandberg testified to the Committee in 2018 that, "Our focus is on inauthenticity, so if something is inauthentic, whether it's trying to influence domestically or trying to influence on a foreign basis-and actually a lot more of the activity is domestic-we take it down." But as the IRA's approach suggests, the current constructs for removing influence operation content from social media are being surpassed by foreign influence operatives, who adapt their tactics to either make their inauthenticity indiscernible, their automated propagation too rapid to control, or their operations compliant with terms of service.[53]

[51] 58
[52] ICA (2017: 12).
[53] 75

We can be sure that with the funding and support the Russian groups have, they will continue to upgrade all the tools at their disposal. Unless we are equally resourceful, we do not stand much of a chance in guarding against this kind of attack. Robert Mueller's report summarizes the power of the Russian social media presence in the US campaign:

> By the end of the 2016 U.S. election, the IRA had the ability to reach millions of U.S. persons through their social media accounts. Multiple IRA-controlled Facebook groups and Instagram accounts had hundreds of thousands of U.S. participants. IRA-controlled Twitter accounts separately had tens of thousands of followers, including multiple U.S. political figures who retweeted IRA-created content. In November 2017, a Facebook representative testified that Facebook had identified 470 IRA-controlled Facebook accounts that collectively made 80,000 posts between January 2015 and August 2017. Facebook estimated the IRA reached as many as 126 million persons through its Facebook accounts.[6] In January 2018, Twitter announced that it had identified 3,814 IRA-controlled Twitter accounts and notified approximately 1.4 million people Twitter believed may have been in contact with an IRA-controlled account.[54]

These are staggering numbers.

One unified message that comes through all these official reports is that the information war is not over, and the Russians are, if anything, intensifying their attack. Here is the Senate report:

> Russian social media manipulation "has not stopped since the election in November and continues fomenting chaos amongst the American populace."293 Committee Members joined witnesses in calling on social media companies to do more to uncover the Russian active measures activities occurring on their platforms. In the wake of the hearing, the Committee publicly and privately pressed social media companies to release more information about the activity of Russian actors on social media in the lead-up to the

[54] Mueller 1: 14-15.

2016 election.[55]

This issue is absolutely urgent and not confined to the United States. Russian active measures campaigns are going on all over the world, manipulating peoples' emotions in an effort to promote Russian interests.

One might wonder whether these measures were effective in furthering the Russian's agenda. Just as with Cambridge Analytica in the previous section, many doubt whether these efforts have any real effects. In response, we present two empirical studies of the effectiveness of the Russian campaign on Twitter. We see that, even when we focus on a single social media platform, the Russian attack had a very high probability of influencing the election. When we consider the combined influence across multiple platforms, this conclusion is even harder to resist.

The first study, by Yuriy Gorodnichenko, Tho Pham and Oleksandr Talavera of the National Bureau of Economic Research, entitled, "Social Media, Sentiment and Public Opinions: Evidence from #Brexit and #USElection," (2018) focuses on the IRA Twitter accounts and their influence on the vote outcomes in the 2016 UK Brexit Election and the 2016 US Presidential Election. They write:

> This study explores the diffusion of information on Twitter during two high-impact political events in the U.K. (2016 E.U. Referendum, "Brexit") and the U.S. (2016 Presidential Election). Specifically, we empirically examine how information flows during these two events and how individuals' actions might be influenced by different types of agents.[56]

And they conclude:

> Since Twitter and other platforms of social media may create a sense of public consensus or support, social media could indeed affect public opinions in new ways. Specifically, social bots could spread and amplify (mis)information thus influencing

[55] 76

[56] Gorodnichenko et al (2018: 21)

> what humans think about a given issue and likely reinforcing humans' beliefs. Not surprisingly, bots were used during the two campaigns we study to energize voters and, according to our simple calculations, bots could marginally contribute to the outcomes of the Brexit and the 2016 U.S. Presidential Election.[57]

One mechanism they uncovered was the political view of the audience for whether the message is received well.

> The degree of influence depends on whether a bot provides information consistent with the priors of a human. For instance, a bot supporting the "leave" campaign has a stronger impact on a "leave" supporter than a "remain" supporter. Similarly, Trump supporters are more likely to react to messages spread by pro-Trump bots. Further examination shows that the sentiment of tweets plays an important role in how information is spread: a message with positive (negative) sentiment generates another message with the same sentiment. These results provide evidence consistent with the "echo chambers" effect in social media; that is, people tend to select themselves into groups of like-minded people so that their beliefs are reinforced while information from outsiders might be ignored. Therefore, social media platforms like Twitter could enhance ideological segmentation and make information more fragmented rather than more uniform across people. Finally, we provide a quantitative assessment of how bots' traffic contributed to the actual vote outcomes. Our results suggest that, given narrow margins of victories in each vote, bots' effect was likely marginal but possibly large enough to affect the outcomes.[58]

Gorodnichenko et al demonstrate that the effect of just the Russian IRA Twitter effort was significant. And even though it was small, it was enough to sway each of the elections.

[57] Gorodnichenko et al (2018: 22)
[58] Gorodnichenko et al (2018: 3-4)

The authors also provide detailed analyses of the amount of change in vote totals associated with IRA disinformation on Twitter, and the numbers are mind-blowing: *1.76% for Leave in Brexit and a whopping 3.23% for Trump in the US election.*

> Our analysis in Section 2.5 indicates that a percentage point increase in the share of pro-"leave" tweets in total tweets is associated with a 0.85 percentage point increase in the share of actual pro-"leave" votes. Hence, the difference between actual and counterfactual traffic could translate into 1.76 percentage points of actual pro-"leave" vote share. Thus, while bots nearly offset each other, the difference could have been sufficiently large to influence the outcome given how close the actual vote was.[59]

> Specifically, our analysis in Section 2.5 suggests that a percentage point increase in the share of pro-Trump tweets in total tweets is associated with a 0.59 percentage point increase in the share of actual pro-Trump votes. Therefore, the observed difference between actual and counterfactual pro-Trump tweet shares suggests that 3.23 percentage points of the actual vote could be rationalized with the influence of bots.[60]

Their analysis, although a single study, gives us good reason to think that these efforts might have actually been successful. We know of no studies at all that show no effect on polls or votes from social media emotion manipulation campaigns like those mounted by Cambridge Analytica and the Russian IRA.

The second analysis is by Damian J. Ruck, Natalie Manaeva Rice, Joshua Borycz, and R. Alexander Bentley – "Internet Research Agency Twitter activity predicted 2016 U.S. election polls." (2019). Ruck et al's analysis is also confined to Twitter, but it looks at the relationship between Russian disinformation on Twitter and changes in opinion polls for the two candidates. It does not look at Brexit, nor does it look at votes. It does establish a particularly strong connection between Russian Twitter disinformation and the increase in Trump's polls. This relationship is termed 'prediction' in the paper, but it is often known in econometrics as *Granger causation*. It has become a common tool across the sciences.[61] Granger causation is stronger than mere correlation, and it shares some nice properties with causation proper. We have already seen Granger causation's cousin, *transfer entropy*, back in Chapter 3. Commenters often make false and embarrassing claims about this topic because they do not understand the word 'prediction' or the underlying concept of Granger causation.[62]

[59] Gorodnichenko et al (2018: 19)
[60] Gorodnichenko et al (2018: 21)

Ruck et al found a striking predictive relationship between Russian Twitter messages being retweeted and increases in Trump's poll numbers. Ruck et al report:

> Here we have (a) examined the timing of the IRA Twitter activity, which suggests a strategic release in parallel with significant political events before the 2016 election and (b) used vector autoregression (VAR) to test if the success of IRA activity on Twitter predicted changes in the 2016 election opinion polls. On a weekly time scale, we find that multiple time series of IRA tweet success robustly predicted increasing opinion polls for one candidate, but not the other. The opinion polls do not predict future success of the IRA tweets. The findings proved robust to many different checks. The result … a one percent poll increase for the Republican candidate for every 25,000 weekly re-tweets of IRA messages.[63]

Notice that Ruck et al found *not* that tweets predicted Trump's poll increases, but rather that *re-tweets* did. That is when someone re-tweeted the disinformation originally tweeted by a Russian account. So Russian Twitter accounts don't predict poll changes unless others fall for them and spread them along. It is also significant that re-tweeted Russian disinformation *did not* predict lower Clinton polls. They clarify their results in important ways:

> Here we have tested prediction, not causality.
>
> Here we have presented evidence that social media disinformation can measurably change public opinion polls.
>
> We use macro-level data to establish a link between exposure to IRA disinformation and changes in U.S. public opinion. However, using aggregated data means we cannot know the extent to which the participants in election polls were exposed to IRA disinformation. This may not matter once social contagion (Centola, 2010) and media ecosystem effects (Benkler, et al., 2018) are taken into consideration.[64]

---

[61] See Stokes & Purdon (2017) for example.
[62] For example, Eric Boehm, in his article "No, Russian Bots Didn't Cause Trump's Poll Numbers To Increase 1 Percent Per 25K Retweets," (2019) castigates journalist Ken Dilanian for confusing correlation and causation in Ruck's paper, but Dilanian actually used the right term, 'predicts' in his Tweet. Boehm is so clueless that he even displays the tweet in his article so that anyone who knows anything about the topic can tell that Boehm has no business discussing these matters in public. See also the equally pathetic analysis by Philip Bump (2019) in the *Washington Post*. Bump complains about the strength of the correlation found by Ruck et al, but there is no mention of Granger causation or the scientific standards used in the paper. Bump does not seem to understand that r-squared values are far more crude than Granger causality.
[63] Ruck et al (2019: 7-8).

Note the locution 'can measurably change' to state the conclusion for which Ruck has provided evidence. Ruck et al has provided solid statistical evidence that at least one of the goals of the Russian attack – electing Trump – was a success. Together with the experiment done by David Goldstein (described in the previous section) and the Gorodnichenko et al study, we have very good reason to believe that the tactics work. That the kinds of things done by corporations like Cambridge Analytica that sell election meddling to the highest bidder and government-sponsored information warfare like that inflicted on the United States by Russia in 2016 are often successful. And it is not a coincidence that these business interests and national interests are coming together – they are using the same tactics, and in many cases, they are training one another and cooperating with one another.

## 7.12 Extensions

[64] Ruck et al (2019: 8).

The scientific study of various aspects of social media is now booming after a considerable lag. This is an area where books feel dated after just a few years, but there are still some earlier work worth reading. See:

McMahon, C. (2019). *The Psychology of Social Media,* Routledge.

Sheldon, P., Rauschnabel, P., and Honeycutt, B. (2019). *The Dark Side of Social media: Psychological Managerial and Social Perspectives,* Elsevier.

Woolley, S. C. & Howard, P. N. (2019). *Computational Propaganda: Political Parties, Politicians, And Political Manipulation On Social Media.* Oxford.

Zuboff, Shoshana. (2019). *The Age of Surveillance Capitalism: The Fight for a Human Future at the New Frontier of Power.* Public Affairs Books.

Many of these exploit the overlap between social networks and network science to offer various quantitative and formal accounts of social networks.

One important area of research that is somewhat controversial and certainly struggling with conflicting results is the impact of social media on our emotions and well being. The following articles offer a glimpse of this rapidly moving area:

Shakya, H. B., & Christakis, N. A. (2017). "Association of Facebook Use With Compromised Well-Being: A Longitudinal Study," *American Journal of Epidemiology*, 185: 203–211, https://doi.org/10.1093/aje/kww189

Hampton, K. N. (2019). "Social Media and Change in Psychological Distress Over Time: The Role of Social Causation," *Journal of Computer-Mediated Communication* 24: 205–222, https://doi.org/10.1093/jcmc/zmz010

Liu, D., Baumeister, R. F., Yang, C., Hu, B. (2019). "Digital Communication Media Use and Psychological Well-Being: A Meta-Analysis," *Journal of Computer-Mediated Communication* 24: 259–273, https://doi.org/10.1093/jcmc/zmz013

Viner, R. M, Gireesh, A., Stiglic, N., Hudson, L. D., Goddings, A., Ward, J. L., Nicholls, D. E. (2019) "Roles of cyberbullying, sleep, and physical activity in mediating the effects of social media use on mental health and wellbeing among young people in England: a secondary analysis of longitudinal data," *The Lancet Child & Adolescent Health* 3: 685-696,

Bekalu, M. A., McCloud, R. F., Viswanath, K. (2019). "Association of Social Media Use With Social Well-Being, Positive Mental Health, and Self-Rated Health: Disentangling Routine Use From Emotional Connection to Use," *Health Education & Behavior*

We have been unable to find any summary of research on mental and physical health effects of social media, but we hope that these sorts of studies are on the horizon.

For a nice discussion of propaganda that is informed by social and political philosophy, see Stanley's *How Propaganda Works* (2015). See Woolley, S. C. & Howard, P. N. (2019). *Computational Propaganda* for coverage of the internet and social media. See Briant (2019b) for a compelling account recent changes in the nature of propaganda. Zuboff (2019) is a must read for anyone interested in how social media works, and it is an instant classic on surveillance capitalism.

The literature on artificial emotion recognition is scattered, but see:

Scherer K. R., Meuleman B. (2013). "Human Emotion Experiences Can Be Predicted on Theoretical Grounds: Evidence from Verbal Labeling. PLoS ONE 8(3):e58166. doi:10.1371/journal.pone.0058166

Konar, A. & Chakraborty, A. (2015). *Emotion Recognition: A Pattern Analysis Approach*. John Wiley and Sons.

Harley, J. M., (2016). "Measuring Emotions," in *Emotions, Technology, Design, and Learning*

Burleson, W. (2017). "Affect Measurement: A Roadmap Through Approaches, Technologies, and Data Analysis," in *Emotions and Affect in Human Factors and Human-Computer Interaction*, Gonzalez-Sanchez, J.,

Cambria, E., Das, D., Bandyopadhyay, S., and Feraco A. (eds.) (2017). *A Practical Guide to Sentiment Analysis*. Springer International.

Pozzi, F. A., Fersini, E., Messina, E., Liu B. (eds.) (2017). *Sentiment Analysis in Social Networks*, Elsevier

The technology on personality trait identification via social media has continued to improve since the original Kosinski et al (2013) paper. For some directions see:

Given the vast array of machine learning techniques being explored right now and the power of personality-based microtargeting, this literature is sure to explode.

For more on Cambridge Analytica, see the Dave Smith's excellent, "Weapons of Micro Destruction," (2018) which provides a great explanation of the machine learning algorithms involved. See also Briant (2017a, 2017b, 2017c, 2019a). Kaiser (2019) and Wylie (2019) are each recent books written by Cambridge Analytica whistleblowers.

For more on the Russian information war on the United States, see Jamieson, K. H. (2018). *CYBERWAR: How Russian Hackers and Trolls: Helped Elect a President, What we don't, can't, and do know*. Oxford. See also Stengel (2019).

There are a ton of interesting cases of emotion manipulation on social media that go beyond these two examples. See the classic:

Kramer A., Guillory J., and Hancock J. (2014). "Experimental evidence of massive-scale emotional contagion through social networks," PNAS 111: 8788-8790, in which Facebook chronicles a bunch of emotionally abusive experiments they did on their users without permission or oversight. Another famous case is from dating site OkCupid:

Rudder, C. "We Experiment on Human Beings!" *OkTrends: Dating Research from OkCupid*, July 28th, 2014. Available: https://www.gwern.net/docs/psychology/okcupid/weexperimentonhumanbeings.html

Some other resources on this topic:

Bond, R., Fariss, C., Jones, J. *et al.* (2012). "A 61-million-person experiment in social influence and political mobilization," *Nature* 489: 295–298. https://doi.org/10.1038/nature11421

Ienca, M. & Vayena, E. (2018). "Cambridge Analytica and Online Manipulation," Scientific American March 30, 2018.

Lee, S. Qiu, L., Whinston, A. (2018) "Sentiment Manipulation in Online Platforms: An Analysis of Movie Tweets," *Production and Operations Management* 27: 393-416.

Benigni M.C., Joseph K., Carley K.M. (2019). "Bot-ivistm: Assessing Information Manipulation in Social Media Using Network Analytics," in Agarwal N., Dokoohaki N., Tokdemir S. (eds) *Emerging Research Challenges and Opportunities in*

The US Senate Intelligence Committee Report we found to be the most helpful in understanding Russian Active Measures, which is why we mostly followed it in our presentation. The Mueller Report (penultimate in the list) has interesting details, but so much of it is redacted that it is hard to get an overall sense of what happened or to follow the various threads of the story. The ICA Report (third from bottom) is "a declassified version of a highly classified assessment that has been provided to the President and to recipients approved by the President. … [T]he conclusions in the report are all reflected in the classified assessment, the declassified report does not and cannot include the full supporting information, including specific intelligence and sources and methods."[65] So it does not have any redactions, and offers the best narrative of any of these three reports. Nevertheless, the ICA lacks the details contained in the Senate Intelligence Report.

[65] ICA (2017: 1).

# *Chapter 8*

# Emotion Network Security

The security of communication networks is a huge topic but nowhere in that gigantic literature is a mention of *emotion communication network security*. The reason is that most work on security in digital technology presupposes a quantitative theory of the topic in question. Until now, there has been no quantitative theory of emotion communication, so it was impossible to apply security techniques to emotion communication systems.

The information theory of emotion communication presented in Chapters 2 – 5 can be applied to the security of emotion communication networks in social media. The theoretical work has already been done, it is just a matter of applying it by using the quantitative theory of emotional information in social media networks.

With information-theoretic metrics for emotion communication, we can design social media that constantly monitors emotion information dynamics to identify and prevent misuse. Understanding the pattern of emotion information being emitted by each human in a social network can be used to identify bots and to block personality targeting, which have been used recently in high-profile cases.[1] Users of the site would be able to check to see that no emotion manipulation is occurring in the emotion network.

[1] Ienca & Vayena (2018).

## 8.1 Emotion Security and Encryption

*Security* is different from *encryption*. Security is about the functioning of communication networks in general, so *emotion security* applies to emotion communication networks. *Encryption* is the study of how to alter messages so that they are decoded only by their intended audience. As such, encryption is one part of security, which is more general. Hence, emotion encryption would be the study of how emotions are expressed in special ways so that they can be detected only by certain people. An example of emotion encryption would be using an emoji when messaging between friends so that one's parents do not understand which emotions are being communicated. We are confident that emotion encryption is a real phenomenon and probably widespread. But it is not our focus. Instead, we focus on emotion security based on the information theory of emotion communication developed in chapters 3-6. Emotion security is about how to protect one's emotional states and emotional capacities from harmful manipulation and mistreatment.

It turns out that encryption in general has been the main topic of computer or information security throughout its history, but recently attention has turned to other features of network security. Notice that encryption would not help protect anyone against the Cambridge Analytica attack or the Russian attack. The problem was not that some kind of encryption failed or was deciphered. Rather, the problem was that seemingly innocuous parts of a social networking website were utilized by a foreign organization to target particular people for emotional manipulation in order to sway an election. The people targeted had no idea that this was a political emotion manipulation campaign. The problem wasn't that a secret message was discovered (encryption problem); rather, the problem was that an entire network was secretly misused in order to emotionally abuse a large number of innocent people for political gain (not an encryption problem, but a security problem nonetheless).

The major part of computer security other than encryption is how to keep computers and computer networks safe from someone outside breaking in and stealing information or changing things.[2] Notice that network security in this sense would not have helped protect against the prominent attacks either. The problem was not that someone broke into a secure computer system. Instead, companies used data they acquired about individual people and data they were given by social media corporations to manipulate thousands of people's emotions in an effort to make them vote a certain way.

The next three sections present material on quantitative network security measures that could be applied to emotion security on social media now that we have a quantitative theory of information in emotion communication. The three topics are: network information flow, control theory, and statistical mechanics. Each of these is related to information theory and network information theory in particular. Each provides the raw materials that could be used to secure social media from the widespread emotion manipulation we see today. In the final section, we apply these tools in examples drawn from the Cambridge Analytica and Russian attacks.

## 8.2 Information Flow

One form of emotion security that information theory provides is quantitative measures of various aspects of emotion communication networks.[3] These measures allow one to identify the range of normal functioning in people across the social media platform. We already have some *structural* measures that help us pick out bots and fake profiles, like how often they post (e.g., during the 2016 US Presidential election, pro-Clinton twitter bots posted around twice as fast as pro-Trump bots, but they all post in very irregular ways compared to a human). As such it makes sense to be careful about the kinds of tools that are available. The first class of tools are structural – they look at the structure of the network over time (e.g., degree distributions, centrality, communities, etc) and try to distinguish which elements of the network are genuine or acceptable and which are not. Call these *independent structural network security tools*. These are already available and have nothing to do with emotion communication in particular or the quantitative theory offered here. They provide quantitative measures of network features that are independent of our ability to quantify the information in emotion communication. One could have identified cases and even networks of emotion communication before the present study, and so someone could have used some independent structural network tools to analyse aspects of emotion communication networks (e.g., their diameter or whether they are small-world).

---

[2] See Stallings (2016), Allsopp (2017), and Kim (2018) for example.
[3] McLean (1990), Gierlichs et al (2008), Newsome et al (2009), Al-Saleh & Crandall (2010), Colbaugh et al (2013), Enck et al (2014), Yang et al (2016), Benigni et al (2019).

A second class of security tools are also purely structural, but they require some kind of underlying quantitative structure, for example, structural network measures that rely on weights. That is, unless we can already assign some numbers to the edges of the network to indicate the weight of the link, these tools cannot be used. Hence, they are dependent in that they require some other theory to attribute quantitative weights to edges before these tools can be utilized. Call them *dependent structural network measures*. These were unavailable for understanding emotion communication prior to our theory; hence, being able to quantify the information in emotion communication systems has opened up all these tools. For example, there are versions of centrality measures like PageRank that utilize weights. Unless one already has some way of figuring out weights for the edges that are tracking some objective and relevant feature of reality, one cannot use the weighted version of PageRank. Because we can now think of the weights as the amount of information passing between the two nodes, we have the option to use these structural measures to understand the features of emotion communication systems better.

One of the most important uses of independent and dependent structural tools in the fight against the ubiquitous emotion manipulation on social media uses contemporary machine learning algorithms to identify anomalies. That is, once one can keep track of the quantity of emotion information passing around a multiplex network, one can use that information to build even better profiles of normal functioning. The machine learning algorithms in question are designed to take as input the characterization of normal functioning, and then they identify anomalies. That is, the characterization of normal emotion information flow can then used by machine learning algorithms to identify fake profiles and people that are being targeted for artificial emotion manipulation. Anomaly detection in machine learning encompasses a wide range of algorithms, some supervised and some unsupervised. It is hard to say what would be the best option in the absence of any meaningful data, but it will probably turn out that different methods work best for certain tasks, and they might need to work together (e.g., one might use a hidden markov model as a procedural anomaly detector for time-series data about a network together with a nearest neighbour clustering anomaly detector for network data at each time slice).[4]

A whole range of additional machine-learning techniques will be helpful in generating meaningful dependent structural measures. For example, algorithms for sentiment analysis are legion at this point, and they can be used to identify emotions expressed in text, audio, or video. These would almost certainly be used by an automatic system that identifies the amount of emotion information passing around a network. Some of these algorithms have been trained to identify the emotion information in the videos we see rather than in the status updates we write. This information too is essential for understanding the overall flow of specifically emotion information into and out of each one of us.

A third class of security tools use explicitly information-theoretic mathematical techniques, not structural tools. Recall that structural tools are mathematical techniques or algorithms or equations that have information about network structure as input, where network structure is defined as having to do with nodes, edges, or weights. Information theoretic tools are different in that they introduce new concepts that depend on the mathematics of probability distributions, kinds of entropy, rates, distortions, coding techniques, capacities, etc. None of these are needed for calculating structural measures (of course, beyond supplying the weights for the dependent structural measures). The third class of tools uses specifically information theoretic tools. For example, we can measure information capacities and rates in emotion communication networks to identify cases where information is being secretly collected via a side channel. These sorts of security risks can be identified now by using explicitly information-theoretic techniques. These tools were already available to help us with any phenomenon that can be described in information-theoretic terms. Now that we have presented the quantitative theory of information in emotion communication systems, all these information-theoretic security tools are newly available to help us with emotion security on emotion communication systems. (A note of caution that the phrase "information flow" is used in many different ways in the computer security literature – often it refers to ways of preventing secure or private information from being released or stolen or altered. These "information flow" techniques – e.g., non-deducibility – have nothing to do with information theory in our sense.)

[4] See Mehrotra et al (2017) on anomaly detection techniques.

## 8.3 Network Control Jamming

Control theory (or control engineering) is a massive area of research that focuses on how to control physical systems. One of the central textbooks in the field, Dorf and Bishop summarize it well:

> Engineers create products that help people. Our quality of life is sustained and enhanced through engineering. To accomplish this, engineers strive to understand, model, and control the materials and forces of nature for the benefit of humankind. A key area of engineering that reaches across many technical areas is the multidisciplinary field of control system engineering. Control engineers are concerned with understanding and controlling segments of their environment, often called systems, which are interconnections of elements and devices for a desired purpose. The system might be something as clear-cut as an automobile cruise control system, or as extensive and complex as a direct brain-to-computer system to control

> a manipulator. Control engineering deals with the design (and implementation) of control systems using linear, time-invariant mathematical models representing actual physical nonlinear, time-varying systems with parameter uncertainties in the presence of external disturbances.[5]

The last sentence is important. Control theory uses certain kinds of mathematical models that are relatively easy to understand in order to model complex systems whose behaviour we do not understand. 'Control' usually means *the ability to steer the system to any desired state*, and this definition is implicit in the quote above as well.

With the rise of network science, one newish area of control theory is how to control networks. There are a number of important results in the area that have appeared recently, and these are directly relevant to securing the emotion information flowing through our social media from the widespread manipulation we see right now. These results are all pro-control, which means that they are useful for those wanting to establish control over networks, especially social media networks.

We want techniques that secure networks from these kinds of control techniques, and we want to apply them to quantitative models of emotion communication in multiplex networks online. Control techniques are part of control theory (duh), but the topic of how to fight against control techniques is relatively unexplored but still a recognized topic, often called "Anti-control theory" or "Chaos control" or "Chaotification theory", where chaos is contrasted with control (another connection to "Six Degrees of Separation"?). Unfortunately, "chaos theory" is already taken. We think "Control Jamming" is a nice name since it invokes control theory without ugly or confusing terminology and appeals to the process/jamming contrast that appears throughout signal processing and we have already seen in communication jamming and emotion recognition jamming.

[5] Dorf and Bishop (2017: 2)

The basics of control jamming are less than two decades old, and we are unable to find much on *network* control jamming theory, except a handful of papers focusing on neural networks. *Network control jamming* is the subject of how to counter network control techniques, that is how to design and regulate the functioning of networks so as to block or minimize network control techniques. We first survey control jamming theory and then present two results from network control theory. We propose ways of countering the example network control techniques.

The first point is that much existing literature on chaotification is not about security at all. It is about ways of making an orderly system more chaotic.[6] That is not what we are trying to do. Instead, we are trying to make our social networks *resistant to control* without thereby making them *more chaotic*. Moreover, chaotification focuses on generating mathematical chaos where this is understood in terms of the vast class of mathematical models studied in chaos theory. Finally, chaotification is actually a part of control theory because many of the techniques used in chaotification are control techniques – they focus on how to control a network by driving it into a mathematically chaotic state. So 'chaos' in 'chaotification' is not really an antonym for 'control'. All the more reason to use 'control jamming' for the topic we want to investigate. We are not at all clear on this, but perhaps all control jamming is part of control theory in the sense that control jamming could use control techniques to drive a network toward uncontrollability. Are there any fundamental theorems about how much control can really be limited by control jamming since it uses some modicum of control?

The fundamental problem of chaotification from Chen and Shi's classic paper is:

> Let us consider a general finite-dimensional discrete-time dynamical system, originally neither chaotic or complex, nor ill-behaved or unstable, in the following form:
>
> $x_{k+1} = f_k(x_k)$, for $x_0$ in $R^n$ is given,

[6] See Wang et al (2000), Huang et al (2011), and Ling et al (2019) for examples of chaotification techniques.

where $f^k$ is only assumed to be continuously differentiable, at least locally in a region of interest. In other words, the given system can be linear or nonlinear, time-invariant or time-varying and stable or unstable. The objective is to design a control input sequence, $\{u_k\}$, such that the output of the controlled system,

$$x_{k+1} = f_k(x_k) + u_k,$$

is chaotic, in the sense of Devaney, Wiggins or Li–Yorke.[7]

and the definition of chaos from Devaney is:

A map $f : S \rightarrow S$ is said to be *chaotic*, if:

(i) the map f has sensitive dependence on initial conditions, in the sense that there exists $\delta>0$, such that for any x in S and any neighbourhood N of x in S, $d(f^m(x), f^m(y)) > \delta$ for some y in N and some $m \geq 0$,

(ii) the map f is topologically transitive, in the sense that for any pair of nonempty open subsets $U, V \subset S$, there exists an integer $m>0$, such that $f^m(U) \cap V \neq \varnothing$:, and

(iii) the periodic points of the map f are dense in S.[8]

where a periodic point is a point $x^*$ in S such that for some m, $x^*=f^m(x^*)$, but $x \neq f^k(x^*)$ for $1 \leq k < m$.

It seems to us that chaotification is just a normal control problem where the desired state of the system in question is a mathematically chaotic one. Indeed, one could *use network control theory* to achieve chaotification in networks. So chaotification is a kind of *control*. We want *control jamming*, which could be used on traditional control techniques or on chaotification. *Chaotification jamming* would then be part of *control jamming*, as it should be. We want something very different, and it does not seem that the chaotification literature holds much promise for our work.

Look instead at a couple of major results in network control theory. Much of the work here is based on a series of results suggesting that one need not control every node in a network to control the network, instead, networks can be controlled in full as long as one has control over certain nodes, but not all of the nodes. These nodes are called, *driver* nodes. Much of network control theory is focused on identifying driver nodes, getting control over them, and proving formal results about how these relate to control of the network. For us, the focus in quantitative information passing through emotion communication networks online. So nodes will often be things like people, but of course not all of them really are people. So Facebook pages and Twitter feeds are good examples. There are really two ways to control a node in this sense: (i) establish a fake node, like a fake Facebook profile or (ii) establish control over a real node, like a real person's Facebook profile.

[7] Chen & Shi (2006: 2440).
[8] Chen & Shi (2006: 2435)

We see each of these in reality. Social media bots are typical examples of (i), but these have become much more sophisticated. In a case from December 2019, a large network of Facebook pages with created pictures using "deep fake" algorithms from artificial intelligence began producing typical "click-bait" content (e.g., celebrity news, personal interest stories, gossip, cat videos, top ten lists, etc.).[9] Only after gaining thousands of followers do these profiles switch to their true content, which is often corporate or nationalist propaganda or divisive content that is xenophobic, misogynist, homophobic, racist, transphobic, or religious. Examples of (ii) can be hacking into a person's account (but this is rarely effective for long), or altering the content that person experiences. Very precisely targeted advertising with no oversight is one way of establishing control of a node. For example, sending voting notices with the wrong date to all the black people in an election precinct as a means of voter suppression.[10] "Catfishing" is another method, which involves setting up fake accounts to pretend to be a person's friends or acquaintances in an effort to manipulate them.[11]

[9] See Nimbo et al (2019).

[10] From US Representative Alexandria Ocasio-Cortez questioning of Facebook CEO Mark Zuckerberg during his testimony to the United States House of Representatives Financial Services Committee on 23 October 2019.

[11] Of all the terminology in this book, this one has probably the most interesting backstory—it comes from a justification for creating fake dating profiles. The story is that tanks of carp used to die in transport, but if one inserts a catfish into the tank, the catfish keeps the carp riled up enough that they survive transportation better. The fake profile is supposed to be the catfish, and the idea is that it is good for ordinary people to be faked like this.

In a classic paper, "Controllability of Complex Networks," Yang-Yu Liu, Jean-Jacques Slotine & Albert-Laszlo Barabasi characterize the problem of network control and prove an important result about the relationship between the minimum number of driver nodes to control a network and the degree distribution for the nodes of that network.[12] They begin characterizing the network control problem and an in-principle solution that is computationally intractable. Their result is that the solution can be approximated well, and this approximation is linked to the degree distribution for the network.

Let us see how it works. Begin with a directed weighted network of N nodes. The nodes can have states at various times, and all the states at a time are captured by a vector **x**(t) and an NxN adjacency matrix A. Let M≤N be the number of controlled network nodes, called *driver* nodes. The driver nodes are controlled as a collective by the input vector **u**(t), which describes the signal sent to each controlled node at time t, and an NxM input matrix. The system as a whole develops over time according to the interaction between A and **x** on the one hand and B and **u** on the other. Liu et al make a simplifying assumption that this relationship is linear, and so offer the following equation:

$$\frac{d\boldsymbol{x}(t)}{dt} = A\boldsymbol{x}(t) + B\boldsymbol{u}(t)$$

where the d**x**(t)/dt notation is familiar for derivatives in calculus but should be interpreted just to mean the change in the network's node's statuses over time. This is a very basic equation for describing the development of a generic dynamical system.

The *Kalman controllability rank condition* is a mathematical theorem that specifies when the network is controllable by the driver nodes: the system can be driven to any desired final state in finite time if and only if the NxNM controllability matrix C = (B, AB, $A^2B$, …, $A^{N-1}B$) has full rank. That is, iff rank (C)=N. The rank function on matrices returns the maximum number of linearly independent columns (or rows) of the matrix. Full rank means that the rank of the matrix is equal to the minimum of its number of columns and rows (i.e., all columns and rows are linearly independent). The fundamental problem in network control theory is to identify the minimum number of driver nodes $N_D$ to control the network as a whole. This problem can also be posed for subnetworks of a network.

[12] Liu et al (2011). See also Gao et al (2014), Ding et al (2014), Gates et al (2016), Menichetti et al (2016), Leitold et al (2017), Cremonini & Casamassima (2017),

One obvious line to pursue for control jamming is to manipulate the features of the network so that rank(C) is not full (i.e., is $<N$). There are so many complex calculations involved in the rank operation on big matrices that it is hard to describe all the ways this could be done. We can call these *rank draining techniques* since they result in the rank no longer being full. Essentially these all involve introducing dependencies among the rows and columns in the controllability matrix for the network in question. One can also consider the space of all controllability matricies for all subnetworks of the network in question, and one could try to insure that the fewest of these have full rank as well, prioritizing the larger subnetworks over the smaller ones.

For large networks, it is computationally intractable to calculate these values and so also to engage in comprehensive rank draining. Liu et al studied networks that have been randomized in various ways to see how this impacts their controllability. It turns out that controllability is largely invariant with respect to degree distribution in a network. That is, if we keep fixed the number of in-links and out-links for each node but randomly distribute the edges otherwise, then $N_D$ is unaffected. Thus, we can insure that our networks have maximum resistance to controllability of the network by manipulating the node degree distribution. This could be accomplished by certain recommendations made to users within the network that encourage healthy (i.e., uncontrollable) degree distributions in neighborhoods of each node and in the network as a whole. These "healthy network" suggestions could be seen by the users of the network, and promoting them would increase the health of their local area of the network. In fact, these healthy network suggestions could be used for local rank draining techniques, which cannot be used on the network as a whole because those calculations are beyond what our computers can do right now. But we can compute neighbourhood or small community ranks and utilize rank draining on them.

Another class of network control jamming techniques utilze the relationship between control and observation. By "observation" control theorists mean inferring internal states by considering how ones controlled inputs to the system affect its outputs. We have already seen definitions of control, but Dorf and Bishop do a good job on the pair:

> A system is *completely controllable* if there exists an unconstrained control u(t) that can transfer any initial state x(0) to any other desired location x(t) in a finite time, 0≤t≤T.
> A system is *completely observable* if and only if there exists a finite time T such that the initial state x(0) can be determined from the observation history y(t) given the control u(t), for 0≤t≤T.

Observable and detectable vs controllable and stabilizable.

Can use observability as an independent variable to manipulate controllability through the Kahnman decomposition. That is, we can investigate how to manage the network in ways that do not manipulate anyone's emotions, but render the network less controllable. There are two ways to do this. One is to think of controllability and observability of a network as a tradeoff. We want some observability otherwise users cannot have a decent experience, but with that observability comes controllability, which we want to minimize. One question is how to make minor changes in the observability of the network that make it less controllable? The other method is to manipulate the detectability / stability of the network. Controllability and observability are limited by this factor, and the less stable/detectable the network states are, the less limited observability and limited controllability can be exploited. So it makes sense to look at both of these options when designing emotionally secure social media networks.

Right now we know that Facebook will allow all sorts of access to its data for the right price. That means essentially unlimited observability. And we know from the Kahlman decomposition, that means essentially unlimited controllability, provided the right set of driver nodes.

In sum, network control jamming can be pursued using these results in at least two ways: (i) by limiting observability which has a complex relationship to controllability, (ii) decreasing detectability and stability in the network so as to render controlled and observed nodes less effective.

Overall, we have looked at control jamming techniques that are based on the fundamental problem of network control: identifying and controlling the driver nodes. We have seen the following kinds of network control jamming:

- Rank draining techniques that utilize Kahnman's result on controllability matrices.
- Degree distribution techniques that utilize Liu et al's result on driver nodes.
- Limiting access to information about the network that allows the identification of driver nodes.
- Limiting the observability of a network so as to decrease the chances of identifying driver nodes.
- Decrease the detectability of a network so as to render observability less effective.
- Decreasing the stabilizability of a network so as to render controllability less effective.

The specific role that emotion information plays in all of this is the following. Quantitative emotion communication information has never been available before, so our theory allows the description new mathematical models that describe the amount and kind of emotion information going into and out of a each node in a social network. This quantitative information can be used by anyone to identify driver nodes in social networks and drive the emotion states of the people in those networks to whatever is desired. That is, the theory of emotion communication presented here provides a huge number of tools for those who might want to control our emotions online even more than they already are. However, in this section, we have shown not only how to block these uses of the tools we have proposed but also how to use some of these tools to make our social networks less susceptible to control. By identifying these mechanisms of control and having the information theory of emotion communication as a framework, we can devise ways of blocking these kinds of emotion manipulations.

## 8.4 Statistical Mechanics: The Ising Model

A third kind of emotion network security tool comes from the connection between information theory and statistical mechanics.[13] We have already seen this connection in Chapter Six when we discussed techniques for deriving information-theoretic features of complex networks. The key to the link between the two fields is the definition of entropy in terms of the probability distribution on the space of possible network configurations. If we can think of emotion communication systems in the manner of statistical mechanics, then it opens tools for emotion security that are based on the security measures available within the statistical mechanics framework.[14]

One of the most familiar models in statistical mechanics is the Ising Model, which is a mathematical structure that can be easily defined but exhibits complex behaviour, especially phase transitions. This might be an unfamiliar use of 'model', which often refers to some mathematical idealization we make to study a complex system. Instead, we need techniques to study the Ising Model because we cannot always expect exact solutions for it. Mean field theory is a good example of such a technique; it uses all sorts of averages in complex ways.

[13] Bianconi (2013), and Cimini et al (2019).
[14] des Mesnards, & Zaman (2018).

We are going to use the Ising Model to identify large-scale manipulation of emotions on social media. However, it was studied in detail by Laurence Ising in 1925 as an aid to understanding ferro-magnetic substances and how they change with temperature. It has since been used in a staggering variety of domains to model all sorts of phenomena. The key is to have a collection of particles interacting with each other in the presence of an external magnetic field. Each particle can be either +1 or -1 (originally for spins of magnetic dipoles), where the positive direction is aligned with the positive direction of the external field. There is a parameter for how coupled the particles are, which means how much the particles affect each other. A bigger coupling constant makes pairs of particles tend to line up, or have the same state. Ising models have been described for one dimension (a line of points), two dimensions (a surface of points in a lattice), three dimensions (a volume of points in a lattice), and more; in addition, Ising models have been generalized in many different ways to accommodate more complex underlying networks beyond simple lattice structures.

We begin with a simple one-dimensional Ising model where the particles are interpreted as people and the states are fear, so +1 is fear and -1 is no fear. The coupling parameter describes how likely it is for neighboring particles to be the same – either both having fear or both having no fear. Because the model is one-dimensional, we imagine a line of people stretching from 1 to m. We want to understand how each person i's emotional states change over time and how they all behave together. We begin by defining the *energy* of the system:

$$H(\{x(i)\}) = -\sum_{i=1}^{m-1} j(i)x(i)x(i+1) - \sum_{i=1}^{m} h(i)x(i)$$

where x(i) is the value of particle (either +1 or -1) person i from 1 to m, <x(i)> is the sequence of values, one for each person from 1 to m, j(i) describes the coupling strength between person i and person i+1, and h(i) is the external field on person i. We have not said yet how to interpret it, but the external field is going to be the coordinated attack on the emotions of the people in this system. And we are going to use the Ising model as a tool for detecting these kinds of coordinated influence campaigns. As such, the emphasis on phase transitions that normally accompanies the Ising model is not a primary focus for us.

A *configuration*, x(i), is a particular sequence of values, one for each person. If there are three people (i.e., m=3), then a partition would be <fear, fear, no fear> or <+1, +1, -1>. There are $2^m$ possible configurations. The probability of observing a particular configuration x(i) is:

$$p\big(x(i)\big) = \sum_{x(i)} \frac{e^{\left(-H\left(\frac{\{x(i)\}}{kT}\right)\right)}}{Z}$$

where e is the natural logarithm base (i.e., 2.718…), k is the Bolzmann constant (again!), T is the *temperature* of the system, and Z is the *partition function*:

$$Z = \sum_{x(i)} e^{(-H(\{x(i)\}))}$$

The temperature introduces an element of randomness – how likely is it that a person will randomly flip from having fear to having no fear, or vice versa. Temperature in this emotion system seems to use like moodiness – how much someone's emotions vary over time on average. Likwise, emotion temperature can be measured for an entire population of people, which is just a measurement of the probability that any given person will randomly switch emotion states due to unknown pressures. We think that this extension of the concept of temperature to emotions and systems of emotion communication in particular is a fruitful consequence of using methods of statistical mechanics to investigate emotions.

Using these equations, we can calculate a number of features of the system including the overall magnetization (for us, this is something like the overall fear activation – how much fear is in the system as a whole). The coupling parameter seems like it might measure emotion contagion, but that is not right. Contagion is where one animal having an emotion causes others who can observe it to feel the same emotion. Coupling, on the other hand, is not directed; it is the tendency of pairs to line up, so it measures mutual influence, not directed influence. And coupling pushes pairs of animals to each have no fear just as much as it pushes pairs to each have fear. We have never seen the absence of an emotion implicated in contagion processes, so that is another difference between the two.

If we can measure some of the features of an Ising model independently, then we can calculate others, like the external field. That is ultimately what we want to use these equations to do – identify nefarious influences on our emotional states in social media. The external field is going to represent these nefarious influences like coordinated inauthentic behaviour.

One main use of the Ising model is to illustrate a critical phase transition in the underlying magnetic matter with respect to temperature. That is, below a certain critical temperature, the system described by the Ising model has a positive significant magnetic field (even in the absence of an external field). But if we raise the temperature past the critical point, $T_C$, then the magnetic field abruptly disappears. In order to see this kind of phase transition, we need to consider two-dimensional Ising models where the particles are arranged in a lattice with x and y axis (depicted in Figure 51).

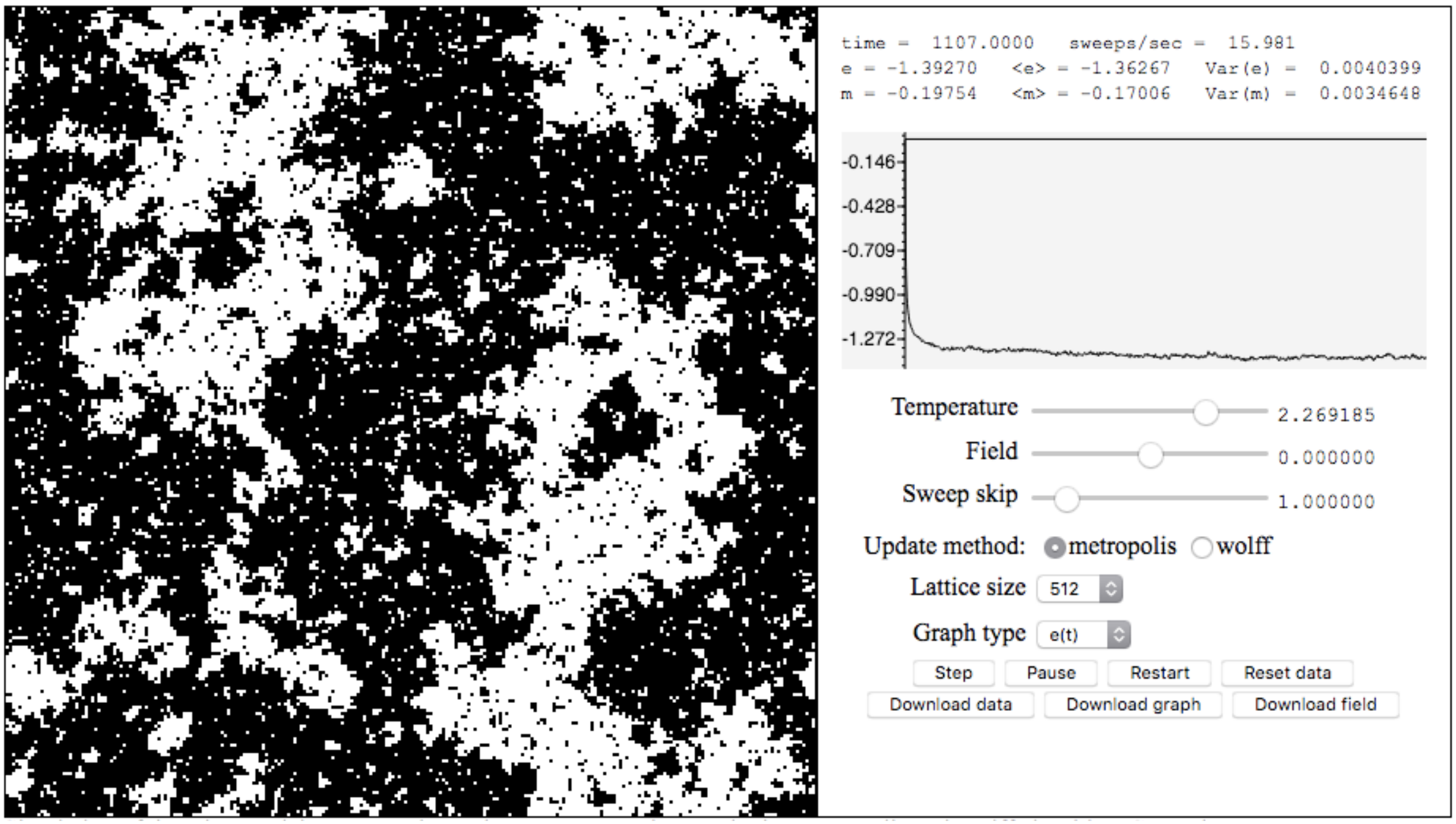

**Fig 51: Ising model near critical temperature**

In this online Ising model simulation by Matt Birmbaum on GitHub, which is free for the public to try out. We encourage readers to go to https://mattbierbaum.github.io/ising.js/ and try it out for themselves. The inputs to the system are temperature and field (each with a slider on the right). Sweep skip just changes how often the simulation updates. We have pictured a 512x512 lattice in two dimensions. The image shows these 262,144 (i.e., 512x512) particles arranged in a lattice so that white pixels are +1 for that particle and black pixels are -1 for that particle. The image shows a variety of black and white pixels in complex shapes with considerable clustering. At lower temperatures, the clusters of positive (white) and negative (black) particles have smoother boundaries and show an obvious lack of spontaneous switching.

In our interpretation, these images show 262,144 people interacting with one another with respect to their fear states. Black pixels are no fear and white pixels are fear. Two people standing next to each other affect each others' emotional states according to the coupling strength, which is a measure of how much pressure there is for any two people to have the same emotional state (both fear or both no fear). At the temperature in figure 51, there is a considerable chance that any pixel will randomly flip from fear to no fear or vice versa.

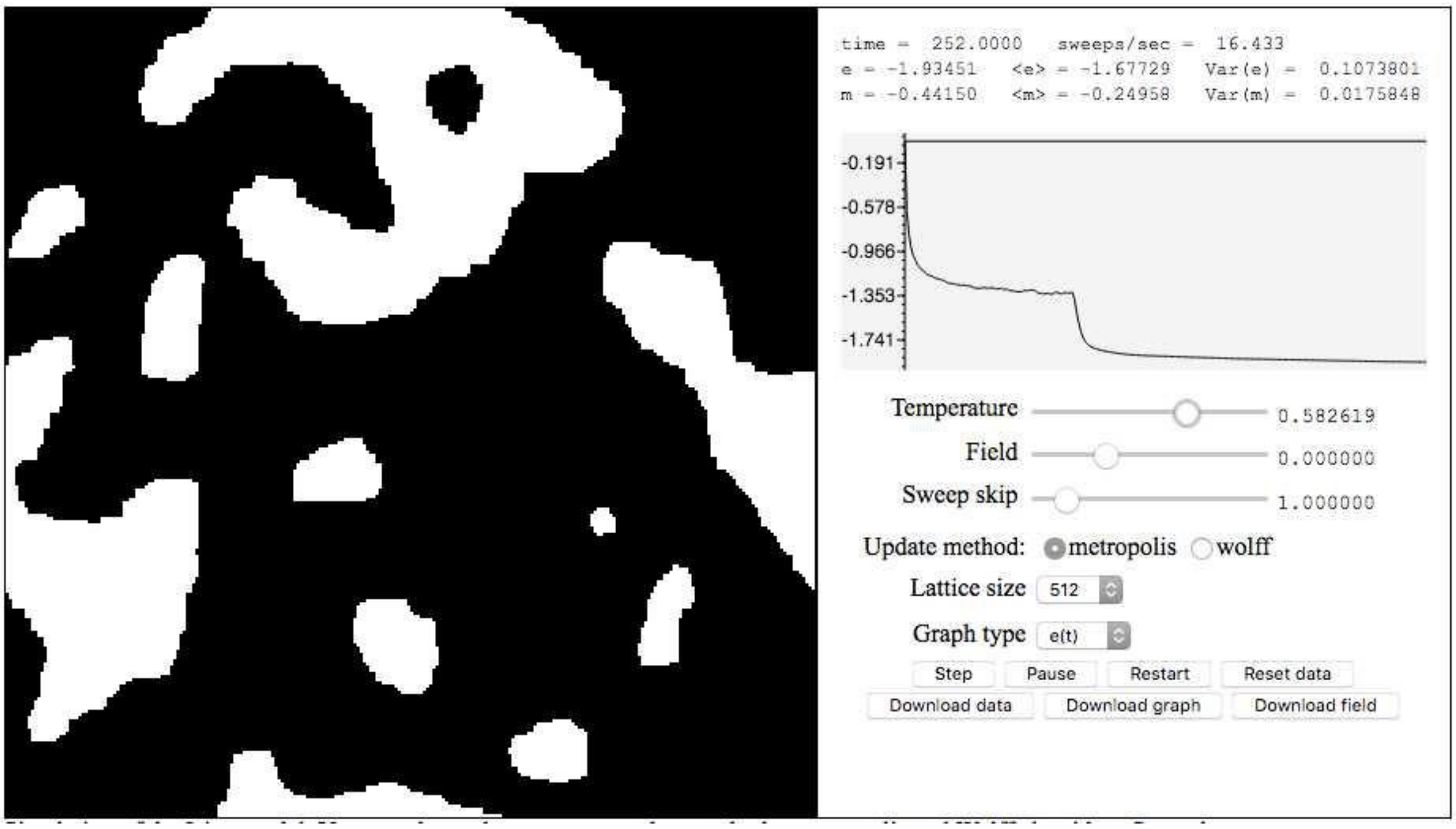

**Fig 52: Ising model at low temperature**

As the temperature goes down (figure 52), the chances of flipping like this are considerably lower.

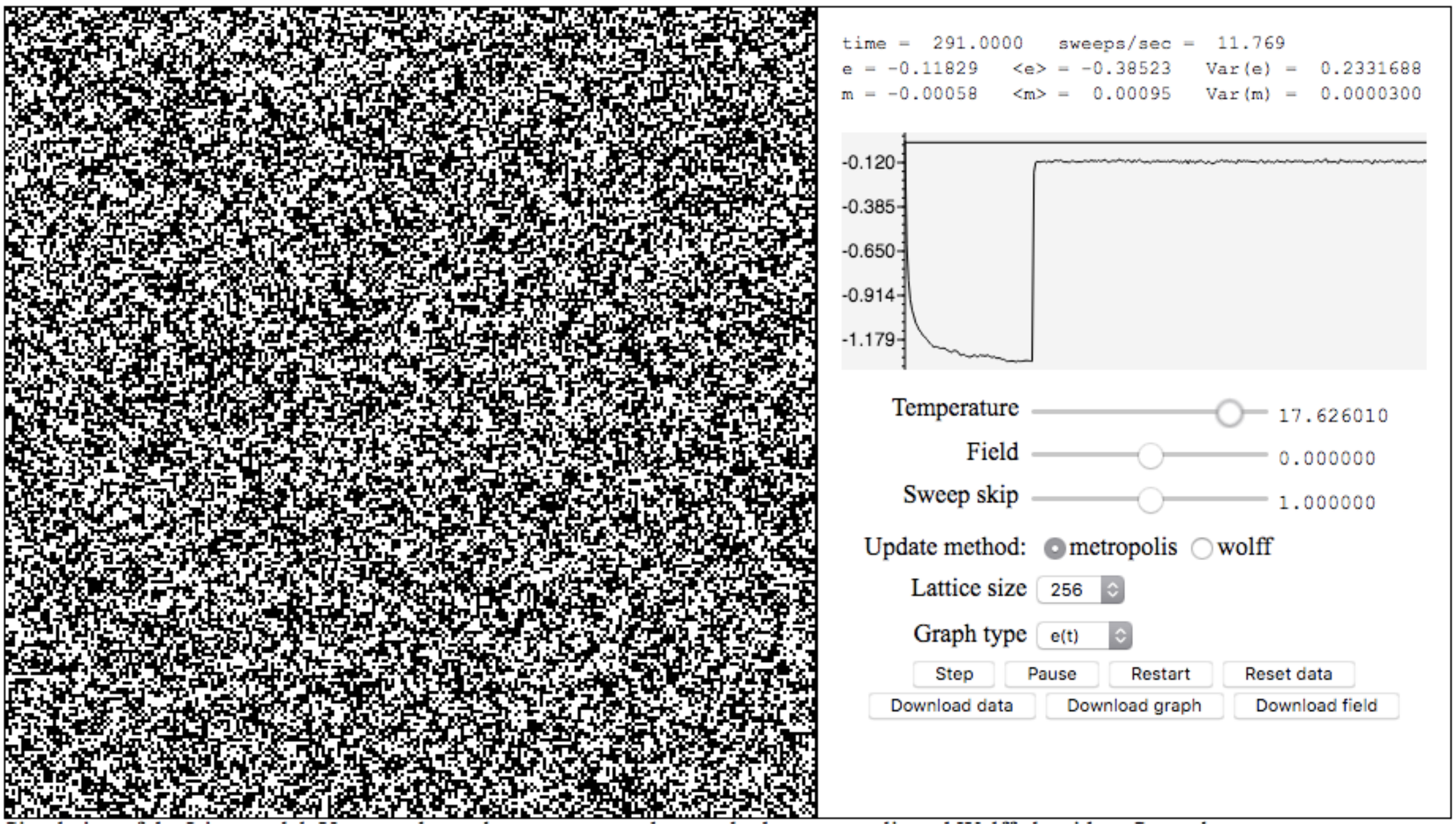

**Fig 53: Ising model at high temperature**

As the temperature goes up, as in figure 53, the chances of flipping are so high that the entire image looks like noise, with no discernable patterns or collections, which means the randomness of the temperature swamps the coupling force. Note that the external field is zero in all three of these figures. Figure 54 shows the kind of phenomena one can see with the presence of the external field.

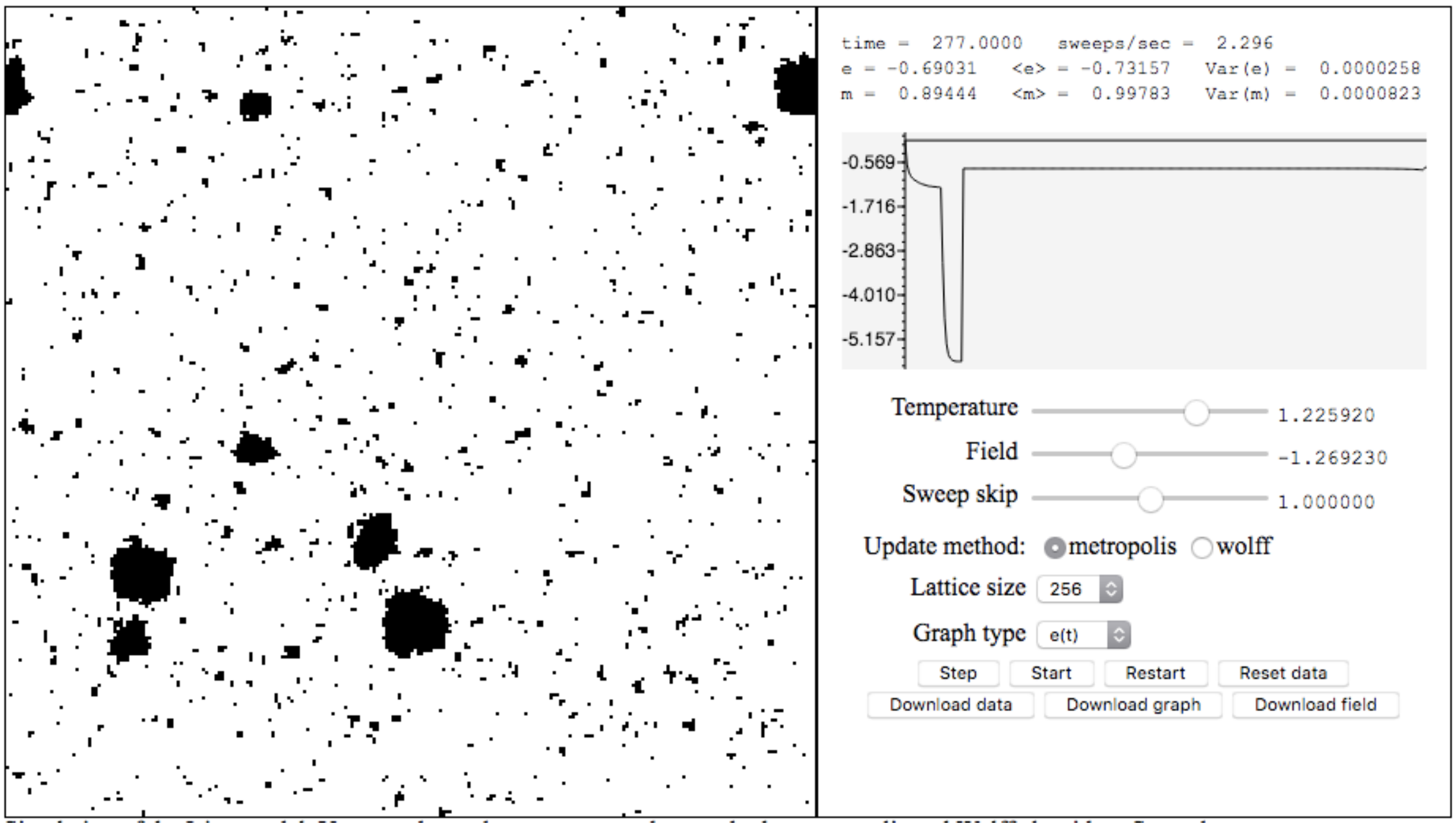

**Fig 54: Ising model with external field**

We present two recent uses of the Ising model to social media for the purposes of understanding its influence on us and how to protect ourselves online.[15] The first is "Detecting Influence Campaigns in Social Networks Using the Ising Model," by Nicolas Guenon des Mesnards and Tauhid Zaman from 2018. They present a version of the Ising Model that is considerably more complex than the one above because it is defined on an underlying complex network rather than a lattice. Their goal is to identify bots on social networks using a structural feature, heterophilia, which is the tendency to associate with those who are unlike you. The insight is that on Twitter, bots retweet humans, but not other bots, and humans retweet other humans, but not bots. So humans display homophillia and bots display heterophillia. Their Ising model incorporates observable features of nodes and interactions between nodes along with unobservable features like whether the node is a human or a bot. des Mesnards and Zaman propose a new algorithm that utilizes specific ways of estimating their Ising model to derive an algorithm that predicts whether each node is a bot or a human based on its observable features

15 See also Loossens et al (2019) for an Ising model applied to phase changes in multiple groups of neurons in a brain that result in the person coming to have a certain emotion.

and its interations with other nodes. The fact that bots are specifically heterophillic and humans are homophillic is built into the mathematics of their model through the coupling function. They show that their algorithm detects bots better than the existing standard.

The other recent study we review is "Ising Model of User Behavior Decision in Network Rumor Propagation," by Chengcheng Li, Fengming Liu , and Pu Li, also from 2018.[16] Li et al formulate a familiar Ising model in order to represent the way in which rumors propagate through social networks. They present a familiar lattice of nodes, where each node is a user in the social network. The +1 value represents the users willingnes to gossip (i.e., spread rumors) and the -1 represents unwillingness to gossip. Each user's proximity to other users have an effect through the coupling parameter. The most interesting thing about Li et al's approach is that they use the external field to represent the influence of blogging on the members of the social network. They factor the external field parameter into two variables to mark how hot the rumor is and to mark how attractive the rumor is. They then use the equations defining the Ising model to define the ways in which rumors spread.

Li et al introduce some sophisticated elements as well, including a utility function for decision making, game theory for equilibria in strategies, von Neuman entropy to study the phase transitions of the network as a whole, and the users self-identity as an additional factor in rumor propagation. Overall, their results suggest that how users identify themselves has a significant impact on the dynamics of rumor spreading through networks.

These two examples illustrate how versitile Ising Models are. In conjunction with the quantititative theory of the information in emotion communication presented here, we can use the Ising model to identify coordinated emotion manipulation attacks on social media. The key is to use the Ising model for this purpose is to let the values of nodes be emotion states, just as above, and we calculate the coupling constant and temperature from empirical data. These are no doubt difficult to estimate, so more investigation into these elements of the model are no doubt in order. However, with these in place, we can calculate the probability of each node being in each state, and this can be measured empirically as well. These are the crucal elements in the model because if they do not match up, then we know that there is an external field affecting the system. By looking at the pattern of differences between calculated and measured probabilities, we can estimate the extent and strength of the field. Of course, the field is going to incorporate all sorts of known influences on the emotions of the people involved, like the effects of blogs, advertising, news feeds, etc. Once these are estimated and factored out, we have a better idea of whether there is some kind of additional influence on the emotions of the people involved.

[16] Li et al (2018).

This sort of emotion security system would be most effective if it were deployed at multiple levels – small groups, small communities, larger communities, etc. After all most of the really alarming emotion manipulation online is coordinated but localized, so it does not affect everyone in the network. The Ising Model for detecting undue emotional influence on social media could be an important tool in detecting a range of emotional abuse online.

## 8.5 Emotion Security on Social Media

We have outlined three kinds of emotion security techniques for social media:

1. The first is based on monitoring the patterns of information flowing into and out of each node (person) using information theory,
2. The second is based on ways of making emotion communication networks harder to control by using the state space representation of networks and their decomposition into controllability and observability, and
3. The third is based on detecting emotion manipulation in social media using the Ising model from statistical mechanics to detect pernicious influences on our emotions online.

Each of these three categories encompasses numerous ways of improving emotion network security. Moreover, they can be combined so as to strengthen one another. For example, the information-theoretic measures detailed in 1 could be used as part of an Ising model to better capture the way emotion information changes across a network over time.

Many of the reports, studies and analyses of Cambridge Analytica and Russian active measures make suggestions about what to do in the future to protect against this kind of activity. To be honest, we found most of these to be nearly useless because they are so broad. Many of them can be summarized as "change the laws so that this can't happen." But that is no strategy at all. So that what cannot happen? Advertising? Social media use? Personality attribution via machine learning algorithms? If one does not know what is wrong, then it is difficult to fix it. We operate under a couple of assumptions. First, all of our data is already available, so blocking its release is not a viable option. Second, it us unlikely that Western democracies are going to be able to enact effective laws that forbid the activity of Cambridge Analytica and its descendants. Instead, our view is that humanity is in desperate need of a new class of security measures to help protect our emotional lives on social media.

We present three kinds of changes to protect our emotional lives from the forces of online emotion manipulation.

- Changes to our governments and economies.
- Changes in the kinds of explicit social media networks available.
- Changes in how we monitor and protect our emotional lives across social media platforms.

We start with the most difficult to implement: changes to government and economic structures.

*Changes to governments and economies.* It should be clear from Zuboff's work that surveillance capitalism is a huge force, not it is highly vulnerable to strengthening individual data rights. If Google were not able to essentially take whatever data they want from you, but instead had to pay you for it, then that would change the game. This could only happen if laws changed governing "terms of service" and similar documents that corporations use to access your data.

Fundamental changes in the legal (and often constitutional) framework surrounding personal data and human rights in dozens of major countries around the world is not likely to put it mildly. At least it is not likely *soon*. But we have elections *now*. The emotional life of humanity is bought and sold and used and abused online all day, every day, right now. Sure we could try to change all these laws, but we think efforts are better spent right now on something more immediate first. Then focus on fixing laws and constitutions.

Another obvious move is to eliminate election "bottlenecks" like the electoral college in the United States presidential election. Any aspect of an election that narrows who decides the election down to a small subset of the electorate is now a national security risk. Because of the network control techniques we outlined in Section 8.3, healthy elections are those that depend on as many people as possible. That would mean treating gerrymandering as a threat to national security as well. Anything that shifts the voting power away from the whole and toward some subset is dangerous and subversive now. We have always known that this feature of electoral systems diminishes the representation had by some of the electorate. That seems like a violation of rights, but this sort of argument has not been persuasive enough to quell the practices. Now, however, any diminishment in the way citizens are represented in government is a national security risk of the highest order, akin to leaving all the ballot boxes out overnight for anyone to tamper with.

It is obvious that elections are complicated, and there are many pressures reflected in way they are run. Moreover, it is not possible to have a perfect voting system.[17] But governments at every level (local, state, national) should insure that their elections are as hard to control as possible. Structures like gerrymandered districts or the electoral college are effectively placing a dangerous network structure at the heart of a voting system. As we have seen, these sorts of network structures can be controlled very easily. In the 2016 US election, it was enough to target a relatively small number of people in a few crucial districts in a handful of swing states to be able to change the entire result. This sort of phenomenon is similar to the controllability of networks, where having control of a few nodes in the network is enough to control the entire network. Indeed, it would not be very difficult to establish some sort of even rough isomorphism between election structures and complex networks so that one could apply the network control results we presented in Section 8.3 to election structures (i.e., to identify which people in which precincts to manipulate so as to drive the overall outcome whichever way you want). We are not going to venture down this technical path, but the intuitive point is obvious to anyone familiar with both topics.

[17] See Felsenthal & Nurmi (2018) for a survey of voting paradoxes.

Indeed, even an institution like the United States Senate, which solidifies huge inequalities in representation, where California's 37,000,000 people get the same number of senators as Wyoming's 568,000 people. At 309,000,000 people (at time of apportionment), the average state's senators represent around 2% of the country; California's represent 12% whereas Wyoming's represent only .18%. Two orders of magnitude in representation is a huge inequality (or bottleneck as we said above); the variation in district population for the United States House of Representatives is nowhere even close to this unequal. It is a colossal weakness in the United States' democratic structure that can be (and certainly is) exploited by all manner of hostile governments and corporations to weaken the country from the inside. Replacing the Senate with some kind of direct democratic system, where citizens vote on measures all together (or even a indirect system with considerable diversity), would make the Representative branch of the US Government far safer, not to mention more effective and representational. However, this is just an example, there are so many other aspects of the US government that are more exploitable than the structure of the US Senate (Citizens United?). Nevertheless, eliminating election bottlenecks is now a matter of national security and this should be a top priority for every government around the world, but we think that if you want to make the most difference in this issue right now, then your efforts would be better spent on the next two suggestions.

*Changes in the kinds of explicit social media networks available.* Healthy social media is going to be a growth industry in the 2020s, and we have offered a number of insights in Section 8.1-8.3 as to how this might be done. The first set of recommendations pertain to the information-theoretic tools for describing how emotion information normally flows through networks and using anomaly detection algorithms (mostly the best ones are in machine learning now) to identify coordinated attacks, microtargeting, influence campaigns, etc. Existing social networks like Facebook or LinkedIn could implement these for themselves and use them to identify and remove bad actors on their networks. Moreover, they could display this information to their users on their websites so that users could see for themselves where emotion information traffic is normal and where it is abnormal in their own neighbourhood of the network.

Another batch of changes focus on techniques from Section 8.2, the control jamming techniques. *Control jamming* is the process of making networks harder to control. We proposed a host of options for control jamming, including limiting the controllability and observability of the network by altering its technical features. For example, the controllability of the network is determined by the rank of its control matrix. By changing the network, one can change the rank of its control matrix. Likewise, the other control jamming techniques we proposed are similar – they specify how to alter a network to make it less controllable.

But how does this help? Figuring out what the network *could be like* does nothing for us! That's right. But it is easy for social networks to monitor these rank draining options and offer them to their users as "network health" suggestions for friends. Indeed, the social network need not label them at all; they could just be part of the suggestions. Establishing new links – the right new links – can actually make the network harder to control. The measurements of controllability, observability, stabilizability, and detectability (explained in Section 8.3) that underlie the rank draining suggestions could be carried out at various scales as well. Some suggestions promote local network health while others improve the health of larger communities. Social networks could also display network heath measures for their users, and by doing this, they would allow users to pressure one another to establish "network health" links (edges) so as to improve network health from a control perspective.

The Ising models introduced in Section 8.4 could be used to detect emotion manipulation on social media networks as well. These mathematical models provide us with a wide array of measures that can be calculated and used to predict how emotions behave dynamically in social networks. How an Ising model can be used depends on what we can measure or predict independently, and which features of the real world system can be incorporated in to the model. The model includes a temperature, a parameter describing how each emoter's state alters its direct neighbors, and an external field, among other things. These are interpreted in the following ways. The *temperature* of a group of emoters, which is the average probability that an individual's emotion state with change without external influence. The *influence parameter* is the probability that being in a certain emotional state causes those interacting with one to come to have the same emotional state. So this is effectively an emotion *contagion* parameter. The external field is interpreted as a systematic influence on the emotions of those in the group that goes beyond temperature and contagion. If one has good ways of identifying other elements of the Ising model, then it can be used to identify coordinated emotion manipulation. Using Ising models to effectively measure emotion manipulation can be used on many different scales from the local to global. These Ising emotion manipulation detectors would be able to pick up influences from advertising, propaganda, bots, microtargeting, influence campaigns, and information warfare. To be useful for detecting any one of these things, measurements of the others would need to be used to factor them out. This entire area of research is new, so there will no doubt be myriad discoveries in coming years.

*Changes in how we monitor and protect our emotional lives across social media platforms.* The emotion information flow methods of Section 8.1 could also be used to develop personal emotion flow monitors that work not on a single social network but across all of a person's networks and across their personal traffic on the internet. The idea would be that a person downloads an app that runs on all their social media devices (phone, laptop, tablet). The app uses the quantitative theory of emotion information to calculate a variety of measures of emotion information flow for that particular person. That is, the person would be able to see how emotion information flows into them and out of them across Facebook, Reddit, Instagram, YouTube, and all the rest. It would also monitor the advertising the person was getting so as to identify probable microtargeting, disinformation, and propaganda. These phenomena all have distinguishing characteristics in how they affect the flow of emotion information into and out of each person online.

Another reasonable way to use network control jamming techniques is to have personal monitors that keep track of social network health across all your social networks. As we mentioned, there are multi-layer network control results about how to control them by taking control over certain nodes *in certain layers*. By using these techniques, companies and governments can seize control by, for example, influencing you via Facebook pages that are working together with Instagram accounts and Twitter accounts to influence you by using the complicated mathematical techniques of multi-layer control theory. By working across layers, it is easier to control you in certain situations. Only a personal, multi-layer network control jammer would be able to identify and fight against this sort of online emotion manipulation (which we already know is happening).

Social media dominates every one of our lives. Whether we are on these webistes or not makes no difference because *second hand social media* affects even those who eschew the internet entirely. Today the internet *is* social media because of surveillance capitalism, which tracks all your activity online in order to sell you goods, services, or opinions more effectively. Our collective emotional health requires that humanity's internet users care enough about emotional safety online to demand emotionally healthy social media.

# *Conclusion*

In the eight main chapters, one finds a new quantitative theory of emotion information as it flows through communication channels and networks. It unifies dozens of research areas and points the way toward even more new research directions. Here we look at a few of the most important and promising of these:

Animal Communication Science

Theory of Mind (ToM) in Cognitive Psychology

Bayesian Theory of Mind (BToM) in Cognitive Science

Computational Models of Emotions in Artificial Intelligence

Machine Learning

Affective Science

Signal Systems

Signal Processing

Formal Semantics and Pragmatics

Information Theory

Evolutionary Biology

Artificial Emotion Recognition Systems

Control Theory

Statistical Mechanics

Dynamical Systems

Complex Networks

Control Jamming

Following up in each of these areas will no doubt help improve the models for emotion communication presented here. For what it is worth, we think that the most interest is in the overlaps. For example, information theory and control theory or statistical mechanics and control jamming.

The original aspects of the quantitative theory of information in emotion communication have been presented in several steps or layers throughout the book. These include:

- The signal theory of emotions and the content of emotion messages in Chapter 2
- The theory of emotion communication channels in Chapter 3
- The theory of coded emotion communication channels in Chapter 4
- The theory of coded emotion communication networks in Chapter 5
- The theory of coded emotion communication multiplex networks in Chapter 6
- The theory of Quantitative Interpersonal of Emotion Regulation (QIER) in Chapter 6

- The Theory of Theory of Mind (ToToM) in Chapter 6
- The theory of social media influence on emotional life in Chapter 6
- The theory of information-based social media security in Chapter 8
- The theory of emotion network control jamming in Chapter 8
- The theory of emotion Ising models to identify emotion manipulations on social media in Chapter 8

Each of these theories has been presented minimally in outline with minimal details. Each of them could be expanded in multiple directions.

Again, the real money is in the overlaps. For example, we can use control theory and its concepts of controllability and observability to think about the relationship between interpersonal emotion regulation and Theory of Mind (ToM), two hot but independent topics in psychology. Recall that controllability is the capacity to make a system reach any given internal state given controls and system inputs, and observability is the capacity to predict a system's internal states given controls and system outputs. Theories of interpersonal emotion regulation describe a kind of controllability – the capacity to make a person reach an intended emotional state. ToM describes a kind of observability – the capacity to predict a person's mental states given their behaviour and interactions with the environment. Controllability and observability are mathematical *duals* in control theory, so theories of controllability and observability are intimately related. So related, in fact, that it makes sense to have a unified theory of each so as to focus on their structural connections (e.g., the analogy between stabilizability and detectability and reciprocal theorems pertaining to them). As such, it makes sense to study interpersonal emotion regulation and emotion Theory of Mind – manipulating others emotions and detecting others emotions – in a single framework as a control process and as a observation process. The Kahlman decomposition uses state space representations to identify the underlying controllability subspace and observability subspace. These are related to one another in complex ways. Likewise, the underlying structures that permit us to regulate each others emotions are related in complex ways to the underlying structures that permit us to attribute emotions to each other. This connection between these two literatures has so far been unnoticed because emotion communication has not been explained in information theoretic terms until now.

Another obvious direction would be to formulate more hypotheses and connections between information theory and emotion communication systems. We proposed the following hypotheses along the way (from Chapters 3 and 4):

- *Shannon hypothesis:* in emotion communication systems, the rate of information is below the channel capacity to allow for the potential for ever-better error correction.
- *Huffman hypothesis*: would be that emotion behavior displays a Huffman-type pattern where simpler actions are associated with more common emotions and less simple actions are linked to less frequent emotions.
- *Prefix hypothesis*: no individual emotion expression forms the initial part of another more complex emotion expression.
- *Higher-order-encoding hypothesis*: emotional communication systems show increases in coding efficiency by encoding several emotional states at once in individual codewords.
- *Hard-decision detection hypothesis*: emotion communication systems sort inputs into discrete categories (e.g., detect fear vs detect no fear) using thresholds (as in the error correction example above, which uses Hamming distance).
- *Soft-decision detection hypothesis*: emotion communication systems sort inputs by using more sophisticated measures that use more of information present and are less likely to end up with ambiguities and arbitrary decoding choices.
- *Maximum Likelihood Decoding (MLD) hypothesis*: emotion communication systems use MLD algorithms, which pick the emotion that corresponds to the emotion behavior with the greatest probability given which emotion behavior was observed.

- *Nearest Codeword Decoding (NCWD) hypothesis*: emotion communication systems use NCWD algorithms, which pick the emotion that corresponds to the emotion behavior most like the observed emotion behavior.
- *Convolutional hypothesis*: emotion communication systems display convolutional coding.
- *Turbo hypothesis*: emotion communication systems display turbo coding.
- *Flexibility hypothesis*: using an emotion communication system requires as little information about probability distributions as possible, since these values can change dramatically across populations as they encounter new environments.
- *Reliability hypothesis*: emotion communication systems utilize reliability assessments in detecting and processing emotion information.
- *Interference hypothesis*: the actions of multiple emotion sources impact the probabilities of detection for each source.
- *Cooperation hypothesis*: sources work together to improve emotion communication.
- *Network-coding hypothesis*: emotion communication systems display *network coding*, which exploits network effects that go beyond what one finds in a single channel.

We want to emphasize that these are just the most obvious handful of hypothesis about basic topics. There are so many more predictions to make and experiments to perform.

Overall, the most important application is to securing our social media from the rampant and destructive manipulation of all our emotions in the interest of making sales or winning elections. The fact that we can now quantify emotion information that is communicated over social media networks gives us powerful new tools to fight online emotion manipulation. These are clearly the most urgent aspects of this work to investigate, develop, and implement. We are convinced that the health of democracies across the world depends on establishing safe and healthy online emotion communication networks.

# *Work Cited*